\documentclass[twocolumn]{aastex631}

\usepackage{amsmath}
\usepackage{multirow}
\usepackage{times}
\usepackage{xcolor}
\usepackage{url}
\usepackage{wrapfig}
\usepackage{bm}
\usepackage{comment}

\received{5$^{\rm th}$ April 2022}
\revised{8$^{\rm th}$ September 2022}
\accepted{17$^{\rm th}$ September 2022}

\submitjournal{ApJS}

\shorttitle{A Homogeneous Survey of 70 Planets with HST WFC3 G141}
\shortauthors{Edwards \& Changeat, et al. 2022}

\graphicspath{{./}{Figures/}}

\begin{document}

\title{Exploring the Ability of HST WFC3 G141 to Uncover Trends in Populations of Exoplanet Atmospheres Through a Homogeneous Transmission Survey of 70 Gaseous Planets}

\correspondingauthor{Billy Edwards}
\email{billy.edwards@cea.fr \\ billy.edwards.16@ucl.ac.uk}

\author[0000-0002-5494-3237]{Billy Edwards}\thanks{These authors contributed equally to this work.}

\affil{Université Paris-Saclay, Université Paris Cité, CEA, CNRS, AIM, 91191, Gif-sur-Yvette, France}

\affil{Department of Physics and Astronomy, University College London, Gower Street, London, WC1E 6BT, United Kingdom}

\author[0000-0001-6516-4493]{Quentin Changeat}\thanks{These authors contributed equally to this work.}
\affil{Department of Physics and Astronomy, University College London, Gower Street, London, WC1E 6BT, United Kingdom}
\affil{European Space Agency (ESA), ESA Baltimore Office, 3700 San Martin Drive, Baltimore, MD 21218, United States of America}
\affil{Space Telescope Science Institute (STScI), 3700 San Martin Drive, Baltimore, MD 21218, United States of America}

\author[0000-0003-3840-1793]{Angelos Tsiaras}
\affil{INAF, Osservatorio Astrofisico di Arcetri, Largo E. Fermi 5, 50125 Firenze, Italy}
\affil{Department of Physics and Astronomy, University College London, Gower Street, London, WC1E 6BT, United Kingdom}

\author[0000-0002-9616-1524]{Kai Hou Yip}
\affil{Department of Physics and Astronomy, University College London, Gower Street, London, WC1E 6BT, United Kingdom}

\author[0000-0003-2241-5330]{Ahmed F. Al-Refaie}
\affil{Department of Physics and Astronomy, University College London, Gower Street, London, WC1E 6BT, United Kingdom}

\author[0000-0002-7771-6432]{Lara Anisman}
\affil{Department of Physics and Astronomy, University College London, Gower Street, London, WC1E 6BT, United Kingdom}

\author[0000-0001-9166-3042]{Michelle F. Bieger}
	\affil{College of Engineering, Mathematics and Physical Sciences,
		University of Exeter,
		North Park Road, Exeter, UK}

\author{Am\'elie Gressier}
\affil{LATMOS, CNRS, Sorbonne Universit\'e UVSQ, 11 boulevard d’Alembert, F-78280 Guyancourt, France}
\affil{Sorbonne Universit\'es, UPMC Universit\'e Paris 6 et CNRS, UMR 7095, Institut d'Astrophysique de Paris, Paris, France}
\affil{LESIA, Observatoire de Paris, Université PSL, CNRS, Sorbonne Universit\'e, Universit\'e de Paris, Meudon, France}

\author[0000-0002-5418-6336]{Sho Shibata}
\affil{Institute for Computational Science, Center for Theoretical Astrophysics \& Cosmology,
		University of Zurich, Winterthurerstr. 190, 8057 Zurich, Switzerland}

\author[0000-0002-9372-5056]{Nour Skaf} 
\affil{LESIA, Observatoire de Paris, Université PSL, CNRS, Sorbonne Universit\'e, Universit\'e de Paris, Meudon, France}
\affil{Subaru Telescope, National Astronomical Observatory of Japan, 650 North A'ohoku Place, Hilo, HI 96720, USA}
\affil{Department of Physics and Astronomy, University College London, Gower Street, London, WC1E 6BT, United Kingdom}
		
\author{Jeroen Bouwman} 
\affil{Max-Planck-Institut für Astronomie, Königstuhl 17, 69117 Heidelberg, Germany}

\author[0000-0002-4525-5651]{James Y-K. Cho}
	\affil{Center for Computational Astrophysics,
		Flatiron Institute,
		162 Fifth Avenue, New York, NY 10010, USA}
		
\author[0000-0002-5658-5971]{Masahiro Ikoma}
\affil{Division of Science, National Astronomical Observatory of Japan, 2-21-1 Osawa, Mitaka, Tokyo 181-8588, Japan}
\affil{Department of Astronomical Science, The Graduate University for Advanced Studies (SOKENDAI), 2-21-1 Osawa, Mitaka, Tokyo 181-8588, Japan}

\author[0000-0003-2854-765X]{Olivia Venot}
\affil{Universit\'{e} de Paris Cit\'{e} and Univ Paris Est Creteil, CNRS, LISA, F-75013 Paris, France}
		
\author[0000-0002-4205-5267]{Ingo Waldmann}
\affil{Department of Physics and Astronomy, University College London, Gower Street, London, WC1E 6BT, United Kingdom}

\author{Pierre-Olivier Lagage}
\affil{Université Paris-Saclay, Université Paris Cité, CEA, CNRS, AIM, 91191, Gif-sur-Yvette, France}

\author[0000-0001-6058-6654]{Giovanna Tinetti}
\affil{Department of Physics and Astronomy, University College London, Gower Street, London, WC1E 6BT, United Kingdom}

\begin{abstract}

We present the analysis of the atmospheres of 70 gaseous extrasolar planets via transit spectroscopy with Hubble's Wide Field Camera 3 (WFC3). For over half of these, we statistically detect spectral modulation which our retrievals attribute to molecular species. Among these, we use Bayesian Hierarchical Modelling to search for chemical trends with bulk parameters. We use the extracted water abundance to infer the atmospheric metallicity and compare it to the planet's mass. We also run chemical equilibrium retrievals, fitting for the atmospheric metallicity directly. However, although previous studies have found evidence of a mass-metallicity trend, we find no such relation within our data. For the hotter planets within our sample, we find evidence for thermal dissociation of dihydrogen and water via the H$^-$ opacity. We suggest that the general lack of trends seen across this population study could be due to i) the insufficient spectral coverage offered by HST WFC3 G141, ii) the lack of a simple trend across the whole population, iii) the essentially random nature of the target selection for this study or iv) a combination of all the above. We set out how we can learn from this vast dataset going forward in an attempt to ensure comparative planetology can be undertaken in the future with facilities such as JWST, Twinkle and Ariel. We conclude that a wider simultaneous spectral coverage is required as well as a more structured approach to target selection.

\end{abstract}




\section{Introduction} \label{sec:intro}

The exoplanet field has rapidly expanded, with thousands of planets currently-known today and thousands more anticipated in the coming decade. The vast number of detected worlds has allowed us to begin to further characterise a diverse selection. While direct imaging has provided high-quality thermal emission spectra for a handful of planets \citep[e.g.][]{samland_di,zhou_di,wang_hr8799}, the bulk of atmospheric characterisation has been undertaken using transit or eclipse spectroscopy. Ground-based high resolution observations have been used to detect atomic metals and their ions \citep[e.g.][]{birkby_hr,ehrenreich_wasp76,kawauchi,kesseli_w76,prinoth_w189,yan_k20} as well as evidence of high-speed winds in the terminator region \citep[e.g.][]{seidel_wind,cauley_w33}. 

While lower resolution space-based data is not capable of distinguishing individual absorption or emission lines, molecular species have been detected via their broadband features, giving insights into the atmospheric diversity of extrasolar planets \citep[e.g.][]{tinetti_water,swain2008methane}. Although most research papers have focused on individual objects, some have begun to conduct population style studies \citep[e.g.][]{cowan_agol,sing,tsiaras_30planets,emission_pop}. For instance, the infrared array camera (IRAC) on-board Spitzer was used extensively before the end of the observatory's life in 2019 and Spitzer eclipses and phase-curves have been used to search for trends in the day-night temperatures of hot-Jupiters \citep[e.g.][]{garhart_spitzer,baxter_spitzer,taylor_spitzer,keating_cowan,may_spitzer_pc}. While the Space Telescope Imaging Spectrograph (STIS) has studied a number of planets, leading to a detection of a variety of spectral features in the visible and UV \citep[e.g.][]{von_essen_w76,evans_wasp121_t2}, Hubble's Wide Field Camera 3 (WFC3) has been the workhorse of infra-red space-based spectroscopy, initially using staring mode observations \citep[e.g.][]{berta_gj1214,mandell_w12_w17_w19,ranjan_spec}. The later development of the spatial scanning technique \citep{mccullough_wfc3_scan} led to far greater efficiencies and thus more precise spectra \citep[e.g.][]{deming_hd209,Kreidberg_GJ1214b_clouds,changeat_edwards_k9}.

Using this technique, groups of planets began to be analysed in transmission and \citet{sing} combined Hubble STIS/WFC3 and Spitzer IRAC data of 10 hot-Jupiters, finding a range of atmospheric feature sizes indicative of different cloud levels within the selected planets. The dataset from \citet{sing} was later used in a number of different retrieval studies \citep{barstow_pop,pinhas}. 19 HST WFC3 G141 transmission spectra were analysed by \citep{iyer_pop} who also concluded that clouds were common in the atmospheres of hot-Jupiters. \citet{tsiaras_30planets} conducted a larger population analysis which included 30 planets. The study used data purely from HST WFC3 G141, finding that around half the datasets showed significant evidence for atmospheric features. \citet{tsiaras_30planets} did not search for trend between chemistry and the bulk characteristics of a planet but their data was later used by \citet{fisher}, who also included data for the TRAPPIST-1 system from \citep{de_Wit_2018} and several other studies \citep{mandell_w12_w17_w19,huitson_w19,Kreidberg_GJ1214b_clouds,knutson_hd97658}. In their study, they found no trends between water abundance and planet mass or temperature. However, the analysis of 19 planets by \citet{welbanks_pop} suggested a mass–metallicity trend where the water abundance increased with decreasing mass. They also noted that the metallicities implied by these water abundances were generally below those of the giants planets in our Solar System \citep{atreya_ss_met}. Additionally, many other studies have performed retrieval analyses of planets by taking spectral data from the literature \citep[e.g.][]{barstow_pop,pinhas,cubillos_pop,kawashima_pop}. For smaller planets within the sub-Neptune or Neptune regime ($\sim$2-6R$_\oplus$), a study of 6 planets showed a strong correlation between the amplitude of the water feature and the equilibrium temperature of the planet or its bulk mass fraction of H/He \citep{crossfield_kreidberg}. 

Population studies have also been undertaken in emission. For instance, data from Spitzer IRAC channels 1 and 2 have been utilised to study tens of planets \citep[e.g.][]{garhart_spitzer,baxter_spitzer,keating_cowan}. \citet{mansfield_metric} presented a simplistic metric which was designed to indicate whether the spectrum showed evidence for a thermal inversion by measuring if the water feature in the HST WFC3 G141 band was in absorption or emission, applying it to 19 planets and comparing the values to a fiducial model. Finally, \citet{emission_pop} presented an analysis of 25 hot and ultra-hot Jupiters in emission with HST and Spitzer and, by using atmospheric retrievals, observationally uncovered an apparent link between the abundance of optical absorbers in their atmospheres and the temperature structure.

In many studies in the literature, data from multiple instruments in combined to expand the wavelength coverage. However, there are many potential issues when trying to infer atmospheric properties based off these merged datasets. Firstly, the wavelength region probed determines the sensitivity of the data to each molecular opacity. Therefore, studying the same planet but with different instruments can often lead to differing constraints on the abundance of a species \citep{pinhas,aresIII}. Applying this to different planets implies that, if the datasets are not homogeneous, then the cause of any trends seen in the retrievals cannot be determined: the underlying abundances could be different or the datasets could have differing sensitivities to these molecules. Combining instruments can also lead to inconsistencies, as the datasets are not necessarily compatible \citep{yip_lc,yip_w96}. Hence, while a longer spectral baseline may give precise atmospheric abundances \citep[e.g.][]{wakeford_w39}, these constraints could be wholly inaccurate. Hence, by combining a menagerie of datasets, biases can be introduced onto the analysis of a single planet as well as a population as a whole. Several works have attempted to overcome the vertical offsets often seen between datasets by adding an offset parameter to the retrieval model \citep[e.g.][]{luque_w74,murgas_offset,wilson_w103,yan_h12,yip_w96,mcgruder_w96}. However, it is unclear how well this works and the issue of varying sensitivity could still apply and temporal changes \citep[e.g.][]{giovanni2020,saba_w17} provide an additional challenge.

Here we conduct a spectroscopic population study of 70 gaseous exoplanetary atmospheres, using a methodology which is standardised and applied uniformly to all targets to try to ensure the extraction of robust trends. In an attempt to avoid the aforementioned biases, we restrict ourselves to using data from HST WFC3 G141 only. To further seek homogeneity, the data were extracted using the same pipeline: Iraclis \citep{tsiaras_hd209,tsiaras_55cnce}. By performing standardised atmospheric retrievals using the TauREx 3 code \citep{al-refaie_taurex3} within the Alfnoor pipeline \citep{changeat2020alfnoor}, we search for trends within these datasets. We attempt to find correlations between the water abundance recovered and the planet's bulk parameters, such as mass and temperature, comparing our findings to those from literature. We also investigate the amplitude of the spectral features seen, searching for trends with the planet's temperature, surface gravity and, for smaller planets, H/He mass fraction. At each stage we attempt to understand the limitations of our approach, including the potential biases that could be introduced. As we stand at the dawn of a new era of increased data quality, consideration of the these will be crucial to avoid misinterpreting these datasets.

\section{Observations}

To ensure homogeneity in our study, we wished all data to be analysed with a single pipeline: Iraclis \citep{tsiaras_hd209}. The pipeline has previously been used in a number of studies and so we acquired a number of spectra from these. Many of these were taken from the population study  by \citet{tsiaras_30planets} and the papers resulting from the Ariel Retrieval of Exoplanets School \citep[ARES,][]{edwards_ares,skaf_ares,aresIII,guilluy_aresIV}. We constrain ourselves to planets which are likely to possess an atmosphere containing significant amounts of hydrogen and helium \citep[R $>$ 2 R$_{\oplus}$, ][]{fulton}. Therefore, we do not include HST WFC3 data of 55\,Cnc\,e \citep{tsiaras_55cnce}, LHS\,1140\,b \citep{edwards_lhs}, GJ\,1132\,b \citep{aresV,libby_gj1132,swain_gj1132} or TRAPPIST-1\,b-h \citep{de_Wit_2018,gressier_t1h,garcia_t1h}. 

\begin{figure}
    \centering
    \includegraphics[width=\columnwidth]{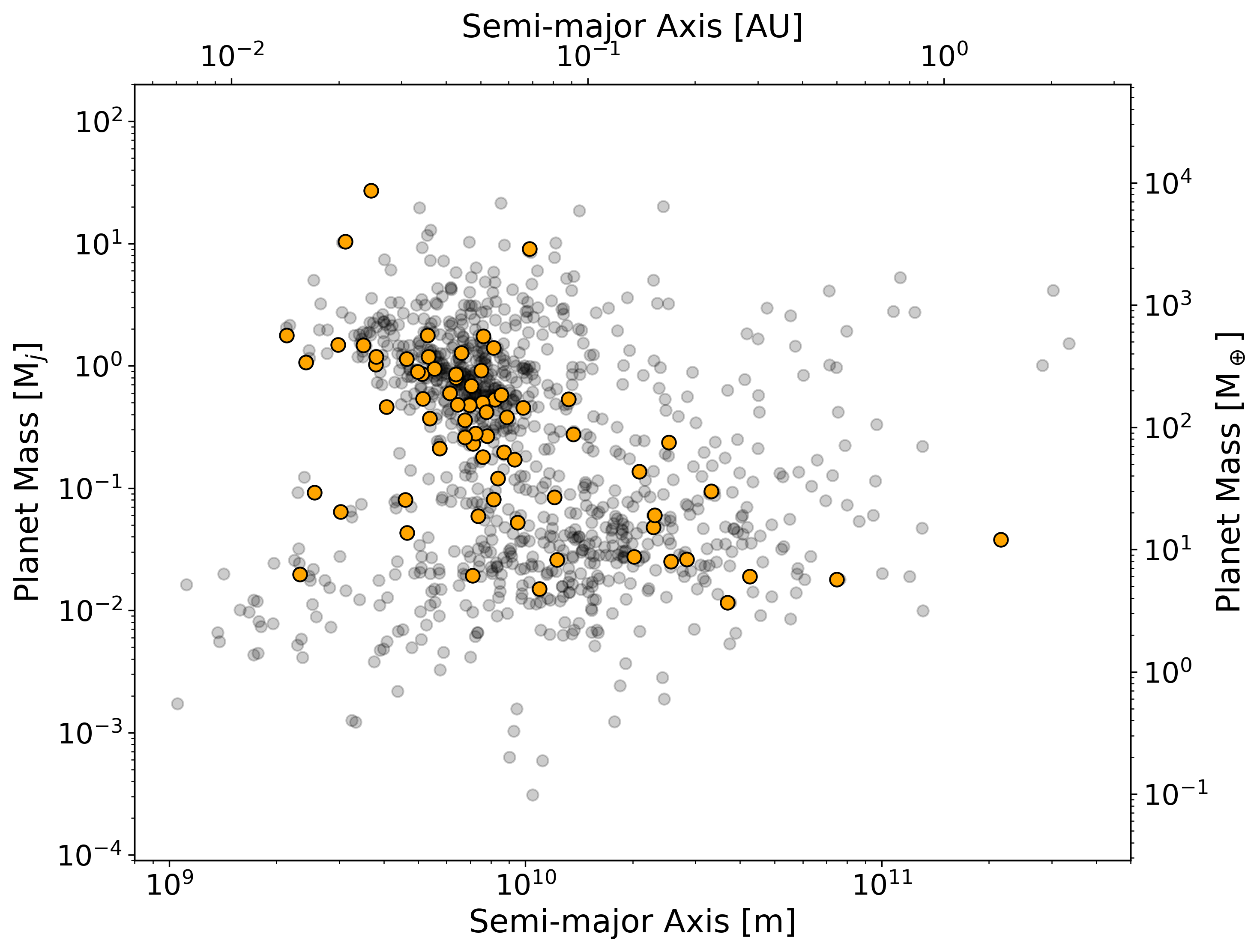}
    \caption{For the planets studied here, the distribution of their semi-major axis and mass (gold). The entire currently-known transiting population with mass measurements is shown in grey.}
    \label{fig:sma_mass}
\end{figure}

A list of all sources of previous datasets analysed with Iraclis are given in Table \ref{tab:previous_data} along with the proposal numbers and principal investigators of the observing proposals. Meanwhile, the observations analysed in this work are given in Table \ref{tab:new_data}. These account for 28 new planets though we note that many of these datasets have been analysed using other pipelines \citep{ranjan_spec,mandell_w12_w17_w19,huitson_w19,Kreidberg_GJ1214b_clouds,knutson_hd97658,ranjan_spec,evans_wasp121_t1,Kreidberg_w107,Kreidberg_w103,spake_w107,carter_w6,libby_k51,chachan_k79,guo_hd97658, alam_hip41378f,brande_toi674,glidic_c1} meaning that only 16 datasets have not previously been published at the time of writing. The distribution of our targets, in terms of the planet's semi-major axis and mass, is shown in Figure \ref{fig:sma_mass}.

\begin{table}[]
    \centering
    \caption{References for data taken from other studies which utilised the Iraclis pipeline.}
    \begin{tabular}{cccc}\hline \hline
    Planet Name & Proposal ID & Proposal PI & Reference\\\hline  \hline
    GJ 436 b & 11622 & Heather Knutson & T18 \\
    GJ 3470 b & 13665 & Bj\"{o}rn Benneke & T18 \\
    HAT-P-1 b & 12473 & David Sing & T18 \\
    HAT-P-3 b & 14260 & Drake Deming & T18 \\
    HAT-P-11 b & 12449 & Drake Deming & T18 \\
    HAT-P-12 b & 14260 & Drake Deming & T18 \\
    HAT-P-17 b & 12956 & Catherine Huitson & T18 \\
    HAT-P-18 b & 14260 & Drake Deming & T18 \\
    HAT-P-26 b & 14260 & Drake Deming & T18 \\
    HAT-P-32 b & 14260 & Drake Deming & T18 \\
    HAT-P-38 b & 14260 & Drake Deming & T18 \\
    HAT-P-41 b & 14767 & David Sing & T18 \\
    HD 3167 c & 15333 & Ian Crossfield & G20 \\
    HD 106315 c & 15333 & Ian Crossfield & G20 \\
    HD 149026 b & 14260 & Drake Deming & T18 \\
    HD 189733 b & 12881 & Peter McCullough & T18 \\
    HD 209458 b & 12181 & Drake Deming  & T18 \\
    KELT-7 b & 14767 & David Sing & P20 \\
    KELT-11 b & 15255 & Knicole Colon & C20 \\
    K2-18 b & 14682 & Bj\"{o}rn Benneke & T19 \\
    WASP-12 b & 13467 & Jacob Bean & T18 \\
    WASP-17 b & 14918 & Hannah Wakeford & S21\\
    WASP-29 b & 14260 & Drake Deming & T18 \\
    WASP-31 b & 12473 & David Sing & T18 \\
    WASP-39 b & 14260 & Drake Deming & T18 \\
    WASP-43 b & 13467 & Jacob Bean & T18 \\
    WASP-52 b & 14260 & Drake Deming & T18 \\
    WASP-62 b & 14767 & David Sing & S20 \\
    WASP-63 b & 14642 & Kevin Stevenson & T18 \\
    WASP-67 b & 14260 & Drake Deming & T18 \\
    WASP-69 b & 14260 & Drake Deming & T18 \\
    WASP-74 b & 14767 & David Sing & T18 \\
    WASP-76 b & 14260 & Drake Deming & E20 \\
    WASP-79 b & 14767 & David Sing & S20 \\
    WASP-80 b & 14260 & Drake Deming & T18 \\
    WASP-96 b & 15469 & Nikolay Nikolov & Y20 \\
    WASP-101 b & 14767 & David Sing & T18 \\
    WASP-117 b & 15301 & Ludmila Carone & A20 \\
    WASP-127 b & 14619 & Jessica Spake &  S20 \\
    XO-1 b & 12181 & Drake Deming & T18 \\ \hline
    \multicolumn{4}{c}{A20: \citet{anisman_w117}; C20: \citet{changeat_k11}}\\
    \multicolumn{4}{c}{E20: \citet{edwards_ares}; G20: \citet{guilluy_aresIV}}\\
    \multicolumn{4}{c}{S20: \citet{skaf_ares}; P20: \citet{aresIII}}\\
    \multicolumn{4}{c}{S21: \citet{saba_w17}; T18: \citet{tsiaras_30planets}}\\
    \multicolumn{4}{c}{T19: \citet{tsiaras_h2o}; Y20: \citet{yip_w96}}\\\hline\hline
    \end{tabular}
    
    \label{tab:previous_data}
\end{table}

\begin{table}[]
    \centering
    \caption{Proposal information for data analysed in this study using the Iraclis pipeline.}
    \begin{tabular}{ccc}\hline \hline
    \multicolumn{3}{c}{Spatial Scanning}\\\hline \hline
    Planet Name & Proposal ID & Proposal PI \\\hline \hline
    GJ 1214 b$^*$ & 13021 & Jacob Bean \\
    HAT-P-2 b & 16194 & Jean-Michel Desert\\
    \multirow{2}{*}{HD 97658 b$^*$} & 13501 & Heather Knutson \\
    & 13665 & Bj\"{o}rn Benneke \\
    HIP 41378 b & 15333 & Ian Crossfield \\
    HIP 41378 f$^*$ & 16267 & Courtney Dressing \\
    HD 219666 b & 15698 & Thomas Beatty \\
    KELT-1 b & 14664 & Thomas Beatty \\
    K2-24 b & 14455 & Erik Petigura \\
    LTT 9779 b & 16457 & Billy Edwards\\
    TOI-270 c & 15814 & Thomas Mikal-Evans\\
    TOI-270 d & 15814 & Thomas Mikal-Evans\\
    TOI-674 b$^*$ & 15333 & Ian Crossfield \\
    V1298 Tau b & 16083 & Kamen Todorov\\
    V1298 Tau c & 16462 & Vatsal Panwar\\
    WASP-6 b$^*$ & 14767 & David Sing \\
    WASP-18 b & 13467 & Jacob Bean \\
    WASP-19 b & 13431 & Catherine Huitson \\
    WASP-103 b$^*$ & 14050 & Laura Kreidberg \\
    WASP-107 b$^*$ & 14915 & Laura Kreidberg \\
     \multirow{2}{*}{WASP-121 b$^*$} & 14468 & Thomas Mikal-Evans \\
     & 15134 & Thomas Mikal-Evans \\ 
    WASP-178 b & 16450 & Joshua Lothringer \\\hline \hline
     & & \\\hline \hline
    \multicolumn{3}{c}{Staring}\\\hline\hline
    Planet Name & Proposal ID & Proposal PI \\\hline \hline
    CoRoT-1 b$^*$ & 12181 & Drake Deming \\
    HAT-P-7 b & 12181 & Drake Deming \\
    Kepler-9 b & 12482 & Jean-Michel Desert \\
    Kepler-9 c & 12482 & Jean-Michel Desert \\
    Kepler-51 d$^*$ & 14218 & Zach Berta-Thompson \\
    TrES-2 b$^*$ & 12181 & Drake Deming \\
    TrES-4 b$^*$ & 12181 & Drake Deming \\ 
    WASP-19 b$^*$ & 12181 & Drake Deming \\ \hline \hline
    \multicolumn{3}{c}{$^*$Data previously published with another pipeline}\\
    \hline \hline
    \vspace{-9mm}
    \end{tabular}
    \label{tab:new_data}
\end{table}

\begin{table}[]
    \centering
    \caption{Proposal information for data for which a good fit could not be achieved with Iraclis and the reference for the spectrum analysed in this study. \vspace{-2mm}}
    \begin{tabular}{cccc}\hline \hline
    Planet Name & Proposal ID & Proposal PI & Reference\\\hline \hline
    Kepler-51 b & 14218 & Zach Berta-Thompson & L20\\
    Kepler-79 d & 15138 & Daniel Jontof-Hutter & C20\\\hline 
    \multicolumn{4}{c}{C20: \citet{chachan_k79}, L20: \citet{libby_k51}}\\\hline\hline
    \end{tabular}
    \label{tab:lit_data}
\end{table}

We detail the methodology of our Iraclis analysis in the Appendix as well as discussing the results of each individual fitting and comparisons to previous works. For two planets where reasonable fits could not be obtained with the Iraclis pipeline, we took values from the literature which are noted in Table \ref{tab:lit_data}. We also attempted to fit the transit observation of K2-33\,b \citep[PN: 14887, PI: Bj\"{o}rn Benneke,][]{benneke_prop14887} but it did not have a post-egress orbit and reliable constrains on the transit depth could not be achieved. Furthermore, a staring mode observation of WASP-18\,b \citep[PN: 12181, PI: Drake Deming, ][]{deming_12181_prop} was analysed but the precision achieved on the transit depth was far lower than that of the scanning mode observation. It was therefore discarded. Similarly, we also analysed the staring mode data of GJ\,1214\,b \citep[PN: 12251, PI: Zach Berta-Thompson, ][]{berta_gj1214} but did not use it in our final analysis due to the better sensitivity offered by the scanning mode data.

In total we analyse the HST WFC3 G141 transmission spectra of 70 planets, 68 of which have been reduced with the Iraclis pipeline. We note that neither of the spectra that were taken from the literature, and therefore not reduced using the Iraclis pipeline, led to detections of atmospheric features.

\section{Data Analysis}

Having created a database of HST WFC3 G141 spectra, we set about analysing them in search of trends within the population. The analysis included Bayesian retrievals as well as studying the strength of the 1.4 $\mu$m water feature.

\subsection{Retrieval Setup}
\label{sec:ret_setup}

Atmospheric retrievals were performed on the transmission spectra using the population analysis tool Alfnoor \citep{changeat2020alfnoor}. Alfnoor extends the capabilities of the publicly available retrieval suite TauREx 3 \citep{al-refaie_taurex3}\footnote{\url{https://github.com/ucl-exoplanets/TauREx3_public}} to populations of exo-atmospheres. The atmospheres of the planets analysed here were simulated to range from 10$^{-4}$ to 10$^6$ Pa (10$^{-9}$ to 10 Bar) and sampled uniformly in log-space by 100 atmospheric layers. For the spectra taken from other studies, the star and planet parameters are given in Tables \ref{tab:star_para_lit} and \ref{tab:planet_para_lit}. For spectra derived here, the star parameters we used the values listed in Table \ref{tab:star_para_new} while the planet parameters are given in Table \ref{tab:planet_para_new}. 

In our retrievals we assumed that all planets possess a primary atmosphere with a solar ratio of helium to hydrogen (He/H$_2$ = 0.17). To this we added trace gases and included the molecular opacities from the ExoMol \citep{Tennyson_exomol}, HITRAN \citep{gordon} and HITEMP \citep{rothman} databases.

The key molecular absorption within the WFC3 range is H$_2$O. However, in a free chemical retrieval, the other molecules chosen can affect the resulting abundance of H$_2$O \citep[e.g.][]{changeat_k11}. Therefore, we attempted several different retrievals to test the robustness of our results. These were:

\begin{enumerate}
    \item \textbf{Standard retrieval}: In this setup we included the opacities of H$_2$O \citep{polyansky_h2o}, CH$_4$ \citep{exomol_ch4}, CO \citep{li_co_2015}, CO$_2$ \citep{rothman_hitremp_2010}, HCN \citep{Barber_2013_HCN} and NH$_3$ \citep{Yurchenko_2011_NH3}. On top of this, we also included Collision Induced Absorption (CIA) from H$_2$-H$_2$ \citep{abel_h2-h2, fletcher_h2-h2} and H$_2$-He \citep{abel_h2-he} as well as Rayleigh scattering for all molecules. We modelled two sets of clouds. Firstly, as a uniform opaque deck, fitting only the cloud-top pressure (i.e. grey clouds). Additionally, we added wavelength dependent Mie scattering using the approximation from \citet{lee_mie}.
    
    \item \textbf{Optical absorbers}: A number of previous WFC3 studies have found evidence for hydrides or oxides \citep[e.g.][]{evans_wasp121_t1,skaf_ares,aresIII}. Hence, in this setup, we included all the opacity sources from our standard retrieval with the addition of  TiO \citep{McKemmish_TiO_new}, VO \citep{mckemmish_vo}, FeH \citep{wende_FeH} and H$^-$ \citep{john_1988_h-,lothringer_h_minus,edwards_ares}.

    
    \item \textbf{Equilibrium Chemistry:} For these retrievals we used the equilibrium chemistry code GGchem \citep{woitke_ggchem} via the recently developed TauREx plugin \citep{taurex3_chem}. As with the free chemistry retrievals, we included Rayleigh scattering and CIA as well as both simple grey clouds and Mie scattering. We ran these retrievals with optical absorbers and without, with the free parameters being the atmospheric metallicity and C/O ratio.
    
    \item \textbf{Fixed C/O Equilibrium Chemistry:} As the HST WFC3 band only really allows for the confident detection of H$_2$O, chemical equilibrium retrievals are essentially fitting two free parameters (metallicity and C/O ratio) to a single observable (the H$_2$O abundance). Therefore, the results are generally highly degenerate, particularly given the lack of sensitivity to carbon-bearing species. Hence, we attempted several retrievals with the C/O ratio fixed to various values in an attempt to see whether the metallicity could be well constrained.
    
    \item \textbf{Flat model:} In this retrieval, no molecular opacities were included. Instead, the only fitted parameters were the planet's temperature and radius as well as the pressure of a grey cloud deck. CIA and Rayleigh scattering were also included. To quantify the significance of our molecular detections, we compare the Bayesian evidence \citep{Kass1995bayes} from each of the retrievals to this flat model. We use this as a baseline from which to calculate the significance of any apparent atmospheric detections.
    
\end{enumerate}

In each free chemistry case, all molecular abundances were allowed to vary from log(VMR) = -1 to log(VMR) = -12. Higher mixing are not expected in the majority of these atmospheres and these would also necessitate accounting for self-broadening of the molecular lines \citep{anisman_lb2,anisman_lb}. For the equilibrium chemistry retrievals, the metallicity was allowed to vary from 0.1 to 100 and the C/O ratio had bounds of 0.1 and 2. For the Mie clouds we followed the methodology of \citet{tsiaras_30planets} and fixed Q$_0$ to 50 who found that uncertainty induced by either varying or fixing Q$_0$ is negligible given the quality of the data at hand. We set a log-uniform prior of $\chi_0$ ranging from 10$^{-40}$ to 10$^{-10}$, particle size from 10$^{-5}$ to 10 $\mu$m and cloud-top pressure from 10$^{-4}$ to 10$^6$ Pa \citep{lee_mie}.

For each planet, the equilibrium temperature was calculated from:
\begin{equation}
    T_{eq} = T_s \left( \frac{(1-A) R_s^2}{4 a^2 e} \right)^\frac{1}{4},
\end{equation}
where $T_s$ is the host star's temperature, $R_s$ is the host star's radius, $a$ is the planet's semi-major axis and the albedo and heat redistribution factor are set to $A$ = 0.2 and $e$ = 0.8, respectively. An isothermal temperature–pressure profile was assumed. While this is an oversimplification and can lead to retrieval biases \citep{rocchetto}, the restrictive wavelength range does not allow for the differentiation of an isothermal from a more complex profile. The temperature bounds of the retrieved were set to $\pm$500 K of the planet's equilibrium temperature while the planet's radius was allowed to vary between $\pm$50 \% of its literature value. The planet's mass was fixed to the value in Tables \ref{tab:planet_para_lit} and \ref{tab:planet_para_new} \citep{Changeat_2020_mass}. 

Finally, we explored the parameter space using the nested sampling algorithm Multinest \citep{Feroz_multinest,buchner_multinest} with 1000 live points and an evidence tolerance of 0.5.

\subsection{Atmospheric Detectability}

\citet{tsiaras_30planets} introduced the Atmospheric Detectability Index (ADI), which compares the Bayesian evidence of an atmospheric retrieval to a flat model. Using this metric, they concluded that 16 of the 30 planets analysed had detectable atmospheres before searching for trends with different bulk parameters, finding a correlation with the planet's radius. Given that we have taken the planets studied in \citet{tsiaras_30planets} and expanded upon their sample, we also explored the detectability of atmospheres and searched for links with planet parameters.

We used the Bayesian evidence to determine the preferred atmospheric model, comparing this to the evidence from the flat model to calculate the significance of any atmospheric detection. Instead of using the ADI, we instead transformed the difference in the Bayesian evidence into a sigma detection and used these values to search for trends with planetary parameters. However, we note that these systems of identifying atmosphere detections are analogous, with an ADI of 3 being equivalent to a 3$\sigma$ detection \citep{tsiaras_30planets}.

\subsection{Search For Atmospheric Trends}

Large gaseous planets are thought to initially form via solid core accretion before undergoing runaway gas accretion and, in the case of the planets studied here, migration is also likely \citep{Mizuno_1980, Bodenheimer_1986, Ikoma_2000}.  In the core-accretion model, lower mass planets are incapable of accreting substantial gaseous envelopes, instead preferentially accreting higher-metallicity solids \citep{mordasini_2012,fortney_2013}. Therefore the metallicity, the ratio of the elements heavier than helium to all the elements, can act as a key test of this theory and studies of methane content of the gaseous planets within our own Solar System are in agreement with the predictions of the core-accretion scenario. Over the last decade, exoplanet observations have expanded the search for a mass-metallicity trend to other planetary systems. Previous observational studies with HST have found some indications of a mass-metallicity trend within exoplanets atmospheres \citep[e.g.][]{wakeford_h26,welbanks_pop}. Furthermore, by comparing the bulk characteristics of exoplanets to structural evolution models, there is evidence that a exoplanet mass-metallicity trend is likely but could differ for that seen in our own solar-system \citep{thorngren_2016}.

The targets studied here cover a wide mass range, from the ultra-low density Kepler-51 b (M = 0.0166 M$_{\rm J}$) to the brown dwarf KELT-1\,b (M = 27.23 M$_{\rm J}$). We utilised both our free chemistry retrievals, and those conducted assuming equilibrium chemistry, to search for an enrichment trend. The planet metallicities extracted by the chemical equilibrium retrievals were compared to the trend found in \citet{thorngren_2016}. They found the strongest correlation was not between the planet's mass and the planet's metallicity but between the mass and the ratio of the planet-to-star metallicity. Hence we, like them, calculated the host star metallicity from:

\begin{equation}
    Z_{S} = 0.014 \times 10^{[Fe/H]},
\end{equation}
using the Fe/H values given in Tables \ref{tab:star_para_lit} and \ref{tab:star_para_new}. We then divided our retrieval metallicities by these values to ascertain the ratio of the metallicities and search for a trend against the planet's mass. 

In the free chemistry case, we attempted to constrain a multitude of molecular species. However, due to the wavelength coverage of HST WFC3 G141, only the abundance of water can be convincing constrained in each case. Taking the water abundance from the preferred atmospheric model (i.e. one with or without optical absorbers) we followed the methodology of \citet{welbanks_pop} to get the ratio of water to hydrogen with respect to solar values. For each planet, the expected water-based metallicity was determined by computing the theoretical abundance at 1e-3 Pa (0.1 Bar) in thermochemical equilibrium assuming C/O = 0.54 and a metallicity equivalent to the that of the host star (Fe/H). The expected water to hydrogen ratio was then compared to the retrieved one to give the relative level of enrichment.

In addition to searching for trends with mass, we also investigated the dependence of the retrieved abundances of molecules on temperature. For these, we computed the expected abundance of H$_2$O, CH$_4$, TiO, VO, FeH and H- using GGchem over a temperature grid of 100-3000\,K at pressures between 1e2 to 1e5 Pa (1e-3 to 1 Bar).

\subsection{Bayesian Hierarchical Modelling}

So far, atmospheric retrievals have been limited to a case-by-case basis, where each observations yield their own atmospheric parameters of interest (such as molecular abundance in the atmosphere). With 70 observations available in our study, we would like to seek trends within our samples. The conventional approach is to fit a trend to a set of error bars, where the mean and sigma values are those computed from the individual posterior distributions. The mean and sigma fall short when attempting to capture the statistics presented by the rich and often non-Gaussian posterior distribution. Bayesian Hierarchical Modelling (BHM) is a principled way to estimate the (hyper-)parameter of the trends that may exist within a population. BHM does this by first treating the posterior distribution from each observation as a sub-model and together these sub-models help to infer the hyper-parameters of the global trend across different datasets. The multi-stage approach accounts for the planet-to-planet variability presented in each observation and properly propagates the uncertainty from each observation to the next layer in the hierarchical model \citep{gelman_bayes}.

While Bayesian retrievals have been common in the exoplanetary field for sometime, and are now the standard methodology, BHM has not been so widely utilised, perhaps partly due to the general lack of sufficient number of datasets. However, a number of studies have employed it \citep[e.g.][]{hogg_ecc_hm, wang_aerosols_hm, wolfgang_mr_hm}, including works focused on seeking trends in exoplanet atmospheres \citep{keating_cowan,lustig_yaeger_hm}.

When searching for temperature-related trends using the abundances from our retrievals we compared two models: a linear trend and a flat trend (i.e. a null hypothesis). We again utilised Multinest for this fitting and used the Bayesian evidence from these fits to determine which gave the best representation of the data. We discuss our implementation of BHM in more depth in Appendix 3.


\section{Results}

\begin{figure*}
    \centering
    \includegraphics[width=0.95\textwidth]{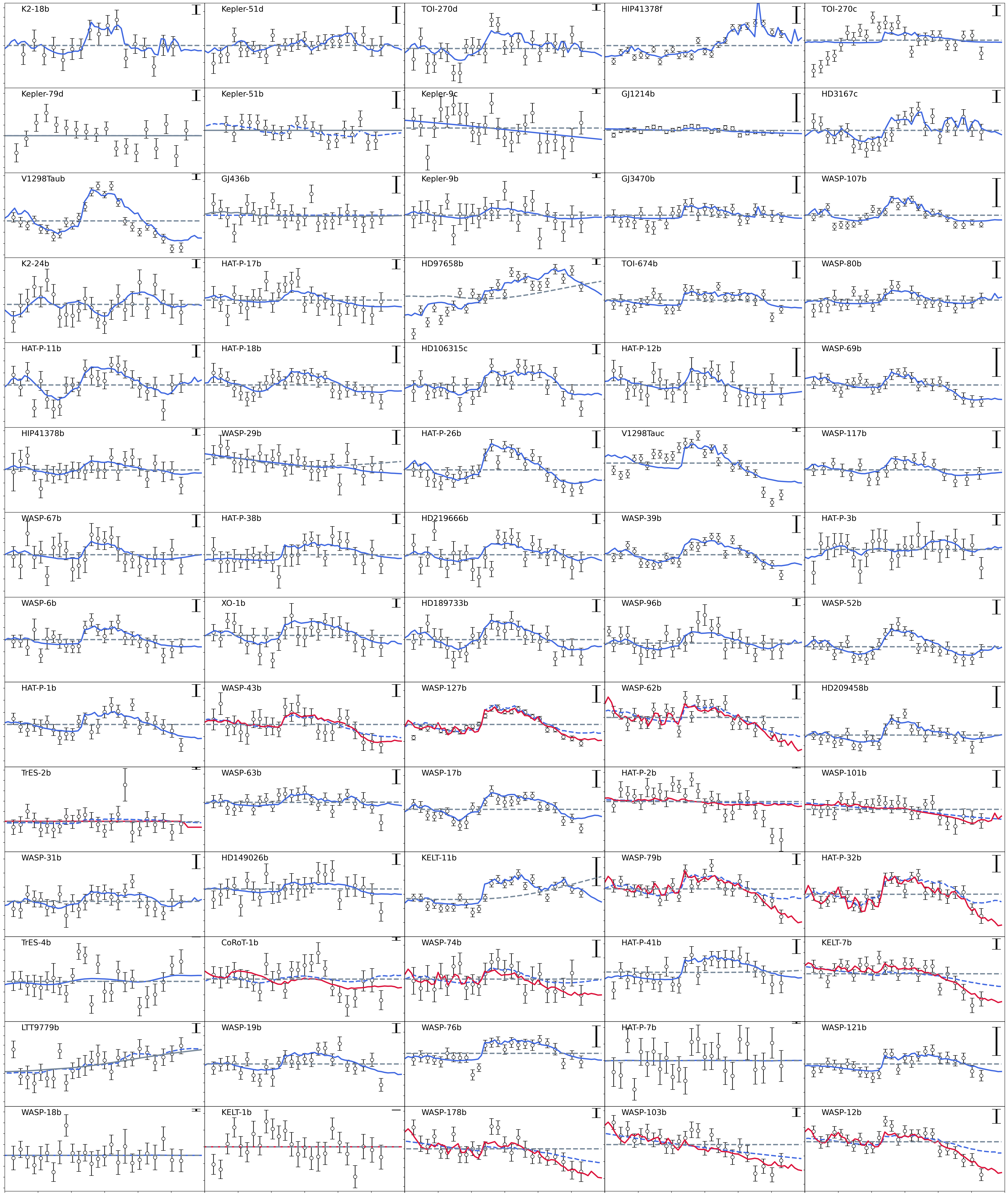}
    \caption{HST/WFC3/G141 datasets utilised in this study. For each planet, the preferred model is denoted by the solid line while the dotted lines show other models which provide a poorer fit. Models including optical absorbers (red) are only shown if they provide a preferable fit to the data than those without (blue). In each case, the grey line denotes the flat model. The error bar in the top right corner of each plot highlights the transit depth variation due to one scale height of atmosphere. Planets have been ordered in terms of their equilibrium temperature, from coolest (top left) to hottest (bottom right).}
    \label{fig:all_spec}
\end{figure*}

For each transmission spectrum analysed, we determined the best fitting models using the Bayesian evidence of our retrievals. For the free chemistry cases, four models were compared as well as a flat model. These spectra, and their best fitting models, are shown in Figure \ref{fig:all_spec}. In each case, two or three models are shown: the flat model, the preferred free chemistry model that does not include optical absorbers, and, for planets above 1500\,K, the preferred free chemistry model that does include optical absorbers. The preferred overall model is shown by the solid line while dashed lines show the other models. In the following sections, we place the results of these retrievals in the context of the findings of previous studies.

\subsection{Atmospheric Detectability}

We find that, of our sample of 70 planets, 37 have strong evidence ($>3\sigma$) of atmospheric modulation. We note that, for several planets (KELT-1\,b, HAT-P-2\,b, WASP-18\,b) the error bars are too large to detect even a completely clear atmosphere. Therefore, we conclude that, of those for which an atmosphere could have been statistically detected, we find evidence for atmospheric detections on 57\% of the population studied here, similar to the 53\% from \citet{tsiaras_30planets}.

To explore the concept further, we computed the expected SNR on a single scale height of atmospheric signal for each planet, comparing this to the achieved detection significance. We find that while there is evidently a correlation between the anticipated SNR and the significance of the detection, for some planets no detection is made even though the precise of the observations is high enough to expect one. GJ\,1214\,b provides the perfect example of this as it has the highest predicted SNR based on a H/He dominated atmosphere yet there is not strong evidence for spectral modulation due to an atmosphere \citep{Kreidberg_GJ1214b_clouds}\footnote{We note that the spectrum recovered here has slightly more spectral modulation than seen in \citep{Kreidberg_GJ1214b_clouds}, as discussed in Appendix 2.}. 

In Figure \ref{fig:detect_sig} we also look to see if planet radius, temperature and surface gravity have an effect on the chances of detecting an atmosphere. We see that large ($>$1 R$_{\rm J}$), hotter ($>$1000 K) planets generally have a better chance of a detection, even if the SNR is low. Meanwhile, for cooler ($<$1000 K), smaller ($<$1 R$_{\rm J}$), a non-detection at a high SNR is more prevalent. When comparing the detection rate for planets with similar temperatures but with different surface gravities, there is some indication that those with a lower gravity more regularly have atmospheric detections. However, it is clearly correlated with the expected SNR which is generally larger for those with a smaller surface gravity within our sample. The effect of the differing precision in the data with respect to the atmosphere's size therefore makes it difficult to distinguish if this is indeed due to this bulk parameter.

\begin{figure}
    \centering
    \includegraphics[width=\columnwidth]{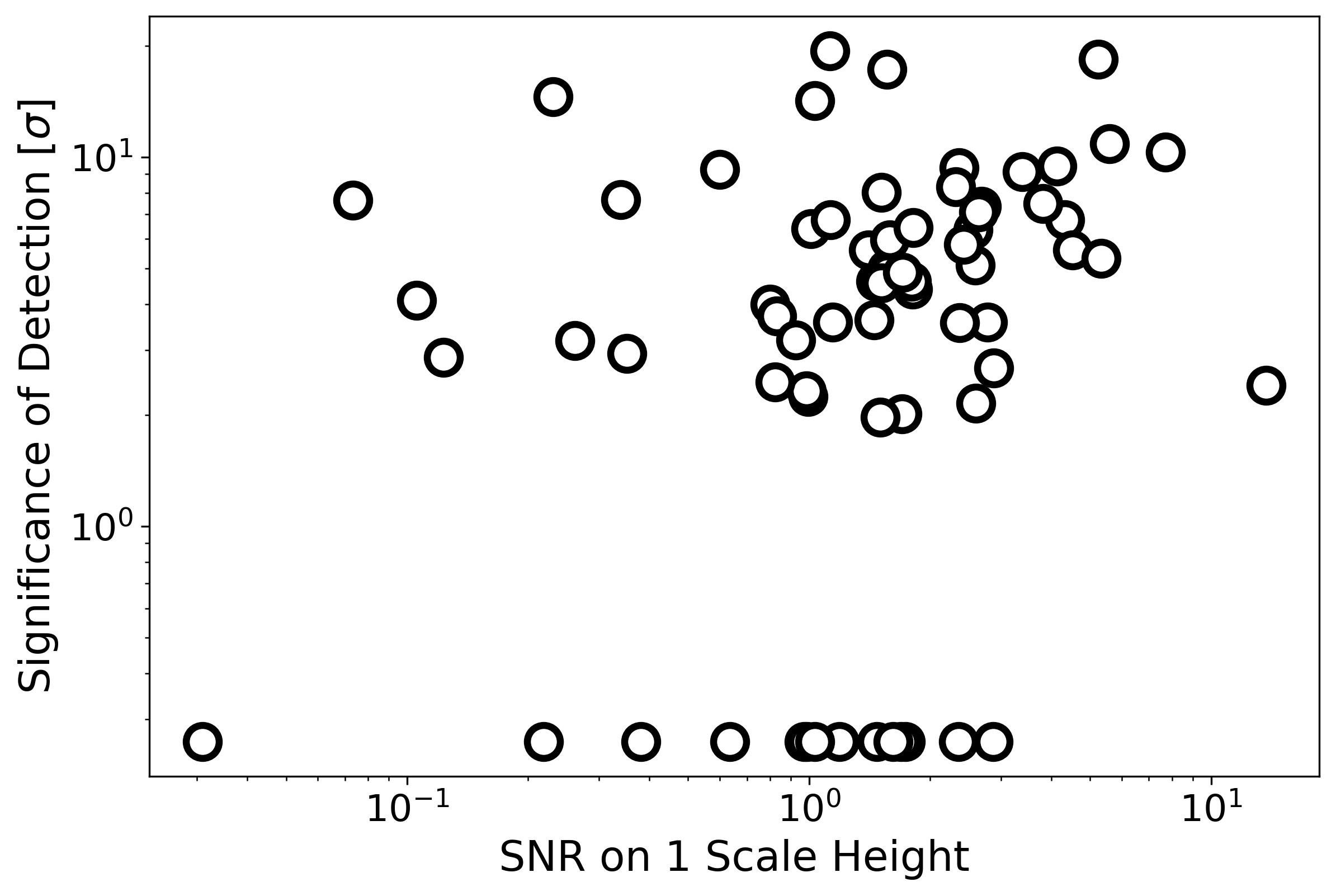}
    \includegraphics[width=\columnwidth]{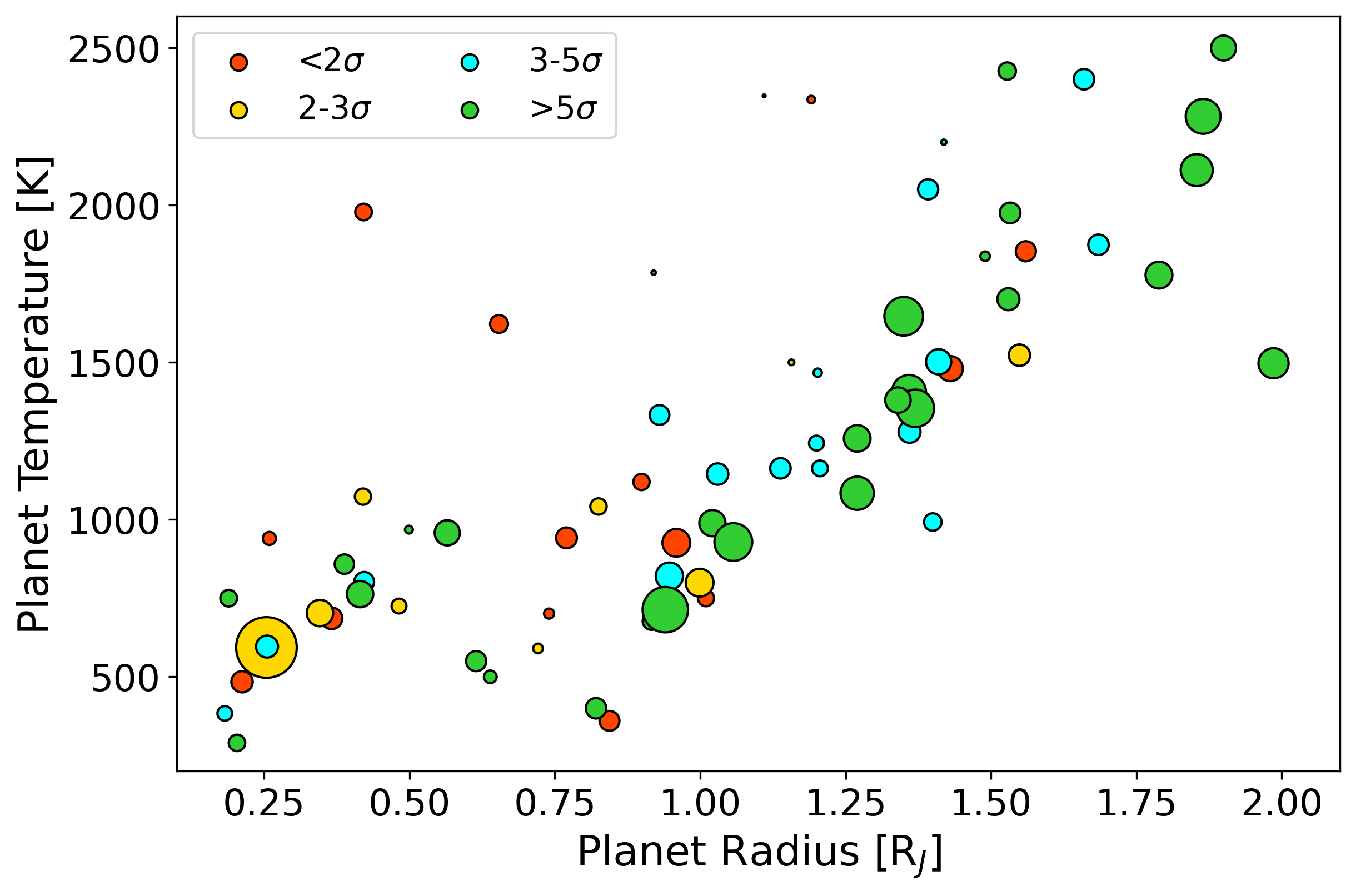}
    \includegraphics[width=\columnwidth]{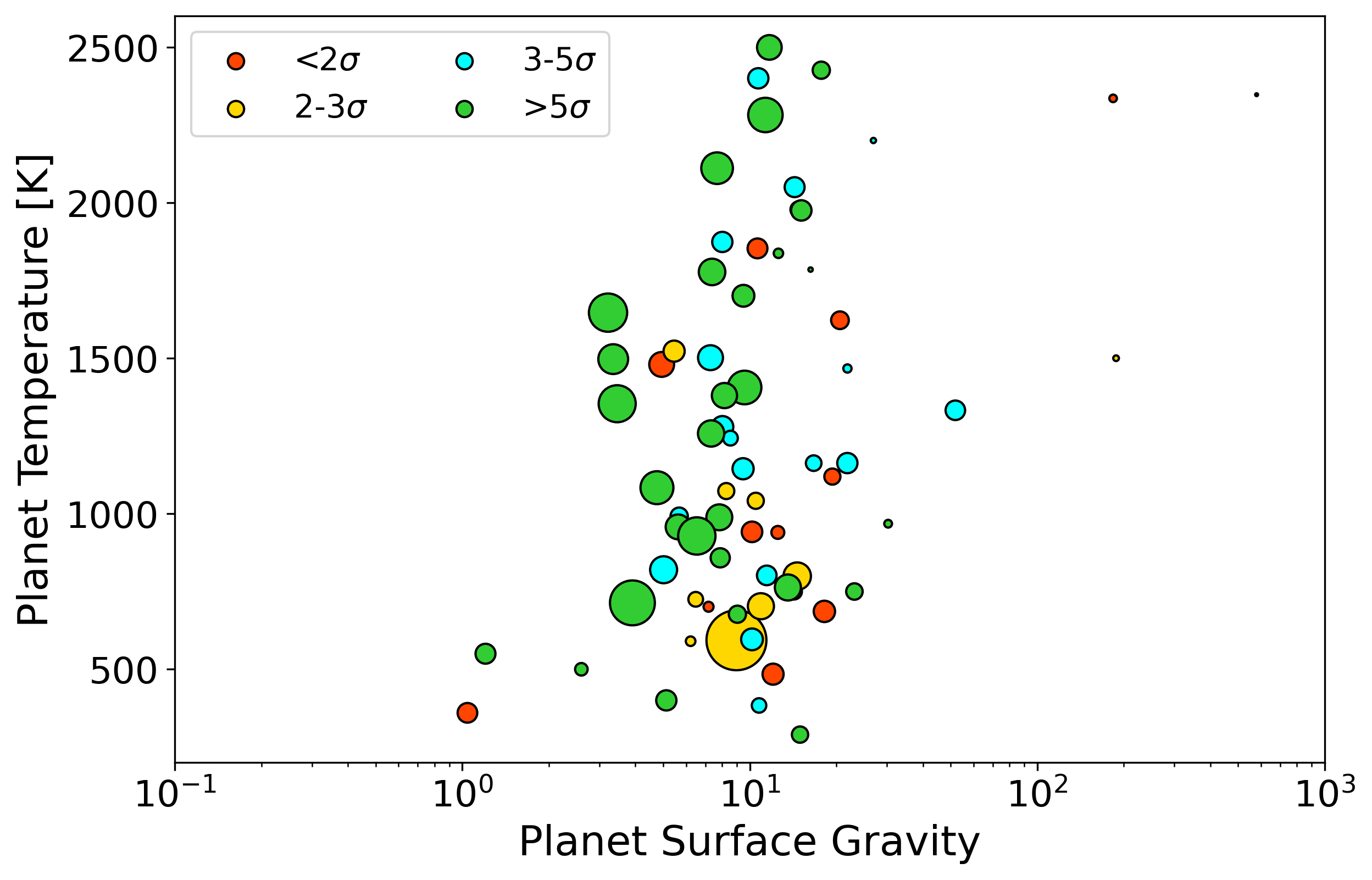}
    \caption{\textbf{Top:} The SNR on the expected spectral modulation due to 1 scale height of a H/He dominated atmosphere against the significance of the atmospheric detection from this work. While there is a positive correlation, in some cases no atmosphere is detected despite very precise observations. \textbf{Middle:} scatter plot of planet radius against planet temperature where the colour of the points indicates the significance of the atmosphere detection and the size shows the SNR on 1 scale height of atmosphere. Non-detections when detections might be expected happen more often for cooler, smaller planets while large, hot planets generally have a better chance of an atmospheric detection, even at low SNRs. \textbf{Bottom:} scatter plot of planet surface gravity against planet temperature where the colour and size again represent the detection significance and SNR respectively. No clear trend can be distinguished given the variance in the precision of the data will affect the significance of the detection.}
    \label{fig:detect_sig}
\end{figure}

\begin{figure*}
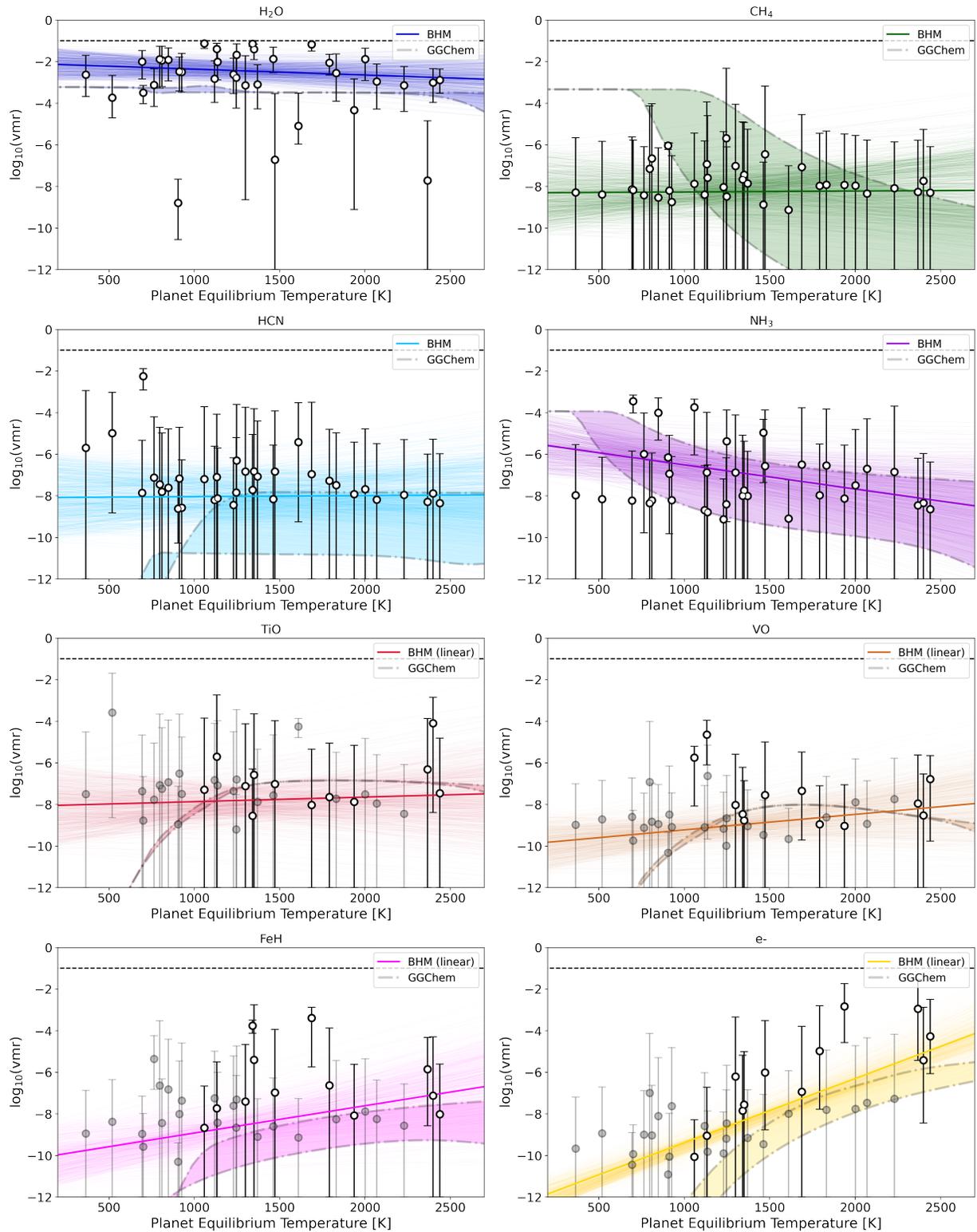

    \centering
    \includegraphics[width=0.95\columnwidth]{H2O_temp.png}
    \includegraphics[width=0.95\columnwidth]{CH4_temp.png}
    \includegraphics[width=0.95\columnwidth]{HCN_temp.png}
    \includegraphics[width=0.95\columnwidth]{NH3_temp.png}
    \includegraphics[width=0.95\columnwidth]{TiO_temp.png}
    \includegraphics[width=0.95\columnwidth]{VO_temp.png}
    \includegraphics[width=0.95\columnwidth]{FeH_temp.png}
    \includegraphics[width=0.95\columnwidth]{e-_temp.png}
    \caption{Retrieved abundances of H$_2$O, CH$_4$, HCN, NH$_3$, TiO, VO, FeH and e- against planet equilibrium temperature. In some cases, only an upper bound on the presence of the molecule could be placed and, for these, the error bar extends to log$_{\rm 10}$(vmr) = -12, the lower bound of our priors. The upper bound is shown by the dashed line. The filled regions bounded by dash-dot grey lines indicate the predicted abundances from GGchem chemical equilibrium models (assuming C/O = 0.54 and solar metallicity) across 1e2 to 1e5 Pa (1e-3 to 1 Bar). For the lower four plots, black data points indicate planets for which the retrieval model with these optical absorbers are preferred while grey point represent those which the fit is preferable without them. In all cases, abundances are only plotted if the associated models provided a $>$3 sigma detection compared to a flat model. The thick coloured line on each plot indicates the linear trend from the BHM while the thinner colours lines represent the traces from the fit that were within the 1$\sigma$ errors of the best-fit model. Only the fit to the abundance of e- gave a Bayesian evidence which was greater than the null hypothesis.}
    \label{fig:chem_abund}
\end{figure*}

\subsection{Search For Trends Between Chemistry and Temperature}

We sought to go beyond the work of \citet{tsiaras_30planets} by also searching for trends in the chemistry, not just the atmospheric detectability. The planets studied in this work vary in equilibrium temperature by over 2000\,K and, as temperature and chemistry are unequivocally intertwined, we searched for evidence of this within our data. Figure \ref{fig:chem_abund} shows the retrieved abundances for a number of absorbers, each plotted against temperature. Comparing the findings of our retrievals to chemical equilibrium models with GGchem we notice a number of things.

Firstly, the retrieved water abundance is almost always above that which is predicted. Indeed, many of the abundances are in fact constrained by our priors, with the upper bound placed at a volume mixing ratio of 10\%. These high water abundances contradict the finds of \citet{emission_pop}, where this molecule was generally found to be sub-solar. Secondly, methane is never constrained, despite being predicted at abundances that would be detectable with HST WFC3 at cooler temperatures. An absence of methane despite it being predicted has also be found by previous works \citep[e.g.][]{benneke_k2-18,anisman_w117,carone_w117,baxter_2021} and clouds, which are found across the majority of the planet studied here, have been suggest as a mechanism for methane depletion \citep{molaverdikhani_ch4}. Additionally, we rarely find evidence for HCN and NH$_3$, with two of the three ``detections" of the latter species being questionable due to the large abundance in the case of K2-24\,b and the high equilibrium temperature in the case of WASP-121\,b. We do detect NH$_3$ in the atmosphere of HD 106315 c (T$_{\rm eq}$ $\sim$ 850 K), a result which has also previously been found by \citet{guilluy_aresIV}, although they note that the model with NH$_3$ is only preferred to $<$2 sigma to one without this molecule.

For the optical absorbers we considered, there are a number of planets for which models with these species are preferred, and a couple where there abundances are constrained in our retrievals. While no planet had a lower 1 sigma abundance limit of TiO greater than 10$^{-9}$, we see evidence for VO in WASP-103\,b. As in other low-resolution studies of the planet \citep{skaf_ares}, a large abundance of FeH was found in the atmosphere of WASP-127\,b. However, compared to \citet{skaf_ares}, we do not find as high abundances of FeH for WASP-62\,b and WASP-79\,b, potentially due to the inclusion of the H- opacity, as noted for WASP-79\,b in \citet{rathcke_w79}. For this absorption we followed the procedure described in \citet{edwards_ares} and retrieved the abundance of e-, with the strongest constraints on this species being in the atmospheres of KELT-7\,b, WASP-12\,b, WASP-79\,b, WASP-103\,b and WASP-178\,b, all of which are planets with equilibrium temperatures higher than 1800 K. The abundances predicted with GGchem for TiO, VO and FeH are relatively low and potentially below the detection limit of HST WFC3 G141. Hence, any trend in retrieved abundances is hard to draw out. However, for e-, the predicted abundance increases strongly after 1500\,K and the retrieved abundances for the hottest planets appear to follow this trend.

\begin{figure}
    \centering
    \includegraphics[width=0.95\columnwidth]{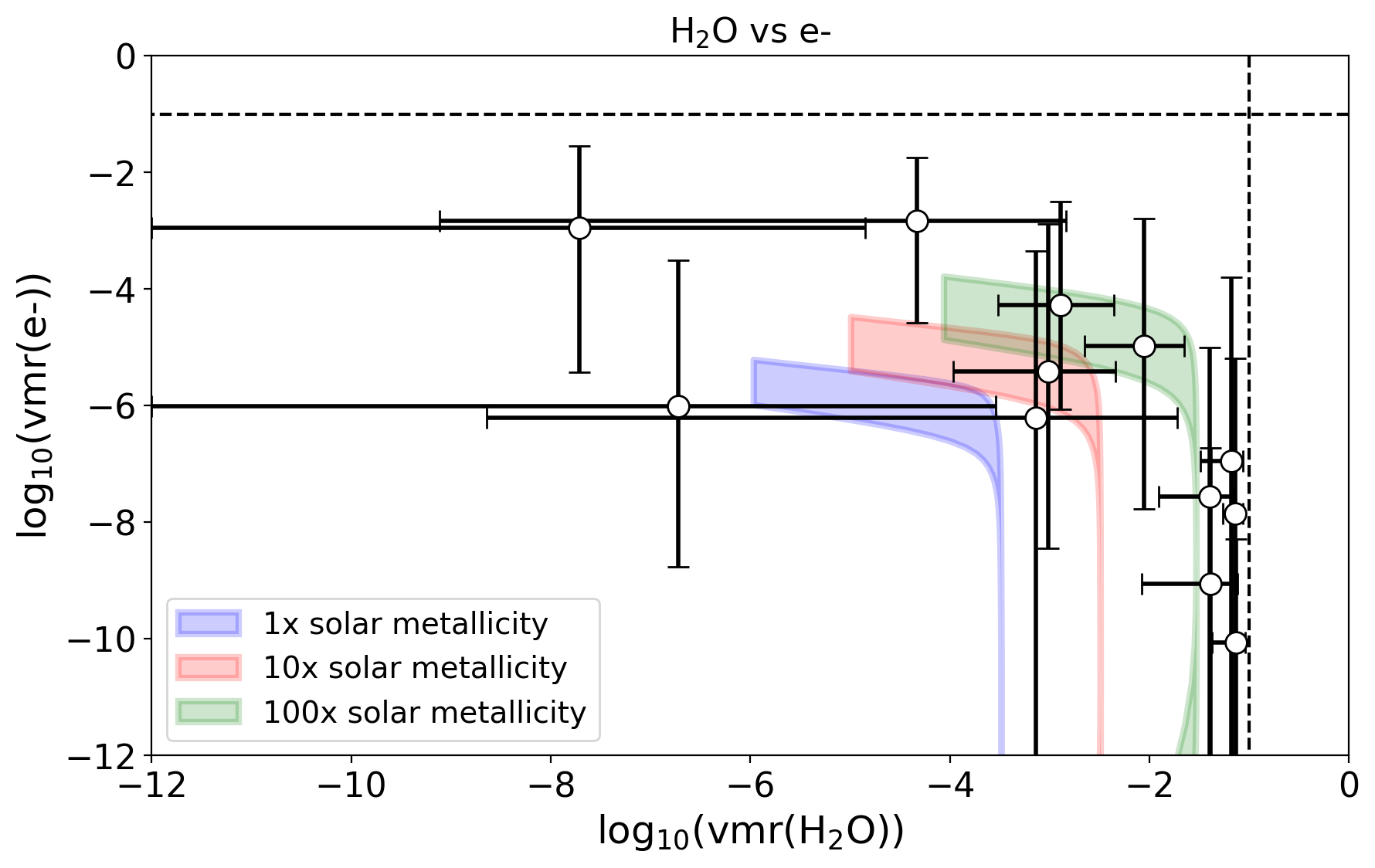}
    \caption{Comparison of the retrieved abundances of H$_2$O and e-. Black data points indicate planets for which the retrieval model with these optical absorbers were preferred to $>$3 sigma compared to the retrieval without optical absorbers. Grey point represent those which are still best fit with optical absorbers present, but at a lower significance compared to the models without them. In all cases, abundances are only plotted if the associated models provided a $>$3 sigma detection compared to a flat model. The blue, red and green filled regions indicate the abundances derived from GGchem at different metallicities at pressure of 1e2 to 1e5 Pa. The dotted lines indicate the upper bound used as a prior for each molecule. The lower bound in each case was log$_{10}$(vmr)=-12. }
    \label{fig:h2o_e-}
\end{figure}

In Figure \ref{fig:chem_abund} we also show the results of our fittings with BHM. In each case, the best-fit linear model is shown as well as the traces that were within 1$\sigma$ of this. We computed the significance of these trends by comparing the Bayesian evidence of this fit to one without a slope (i.e. a model of constant abundance with temperature). Of all the species, only for e- did the fitting with a slope provide a preferable fit to the data and, even in this case, the significance was relatively low (2.28$\sigma$). The emergence of the H- ion comes from the thermal dissociation of H$_2$ at high temperatures. As H$_2$O can also thermally dissociate, albeit at higher temperatures than H$_2$, we compared our retrieved abundances of H$_2$O and e- to explore whether a correlation can be seen in the data. We plot these in Figure \ref{fig:h2o_e-} and a general trend of decreasing water abundance with increasing e- can be seen. The over-plotted models are from GGchem, showing the expected trend in the data for atmospheres of 1x, 10x and 100x solar metallicity. These again show that the water abundance retrieved is generally super-solar.

\subsection{Constraints on Formation via Elemental Ratios}

The ratio of elements within the atmospheres of gaseous planets should provide an indication of the formation and evolutionary processes that have shaped the world into what we observe today. In particular, the metallicity and the ratio of carbon to oxygen have been proposed as key tracers of where and how giant planets collect gas and solids in evolving protoplanetary disks \citep[e.g.][]{oberg_co,mordasini_2016, Brewer_2017, Booth_2017,madhu_formation,Eistrup_2018,turrini_ariel_2018, Cridland_2019,shibata_2020}. Hence, in our chemical equilibrium retrievals, we attempted to constrain both the metallicity and C/O ratio of the planets.

\begin{figure}
    \centering
    \includegraphics[width=\columnwidth]{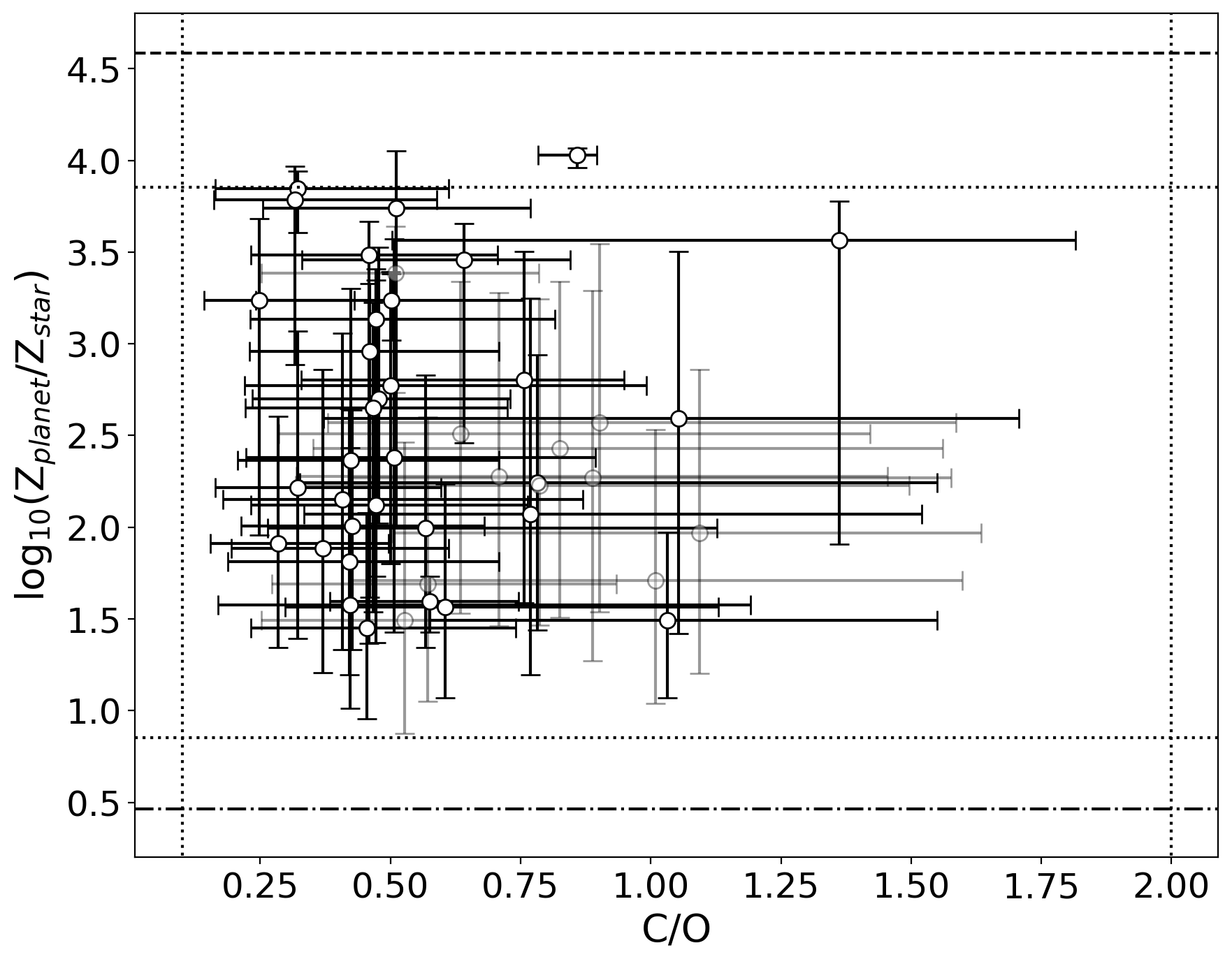}
    \includegraphics[width=\columnwidth]{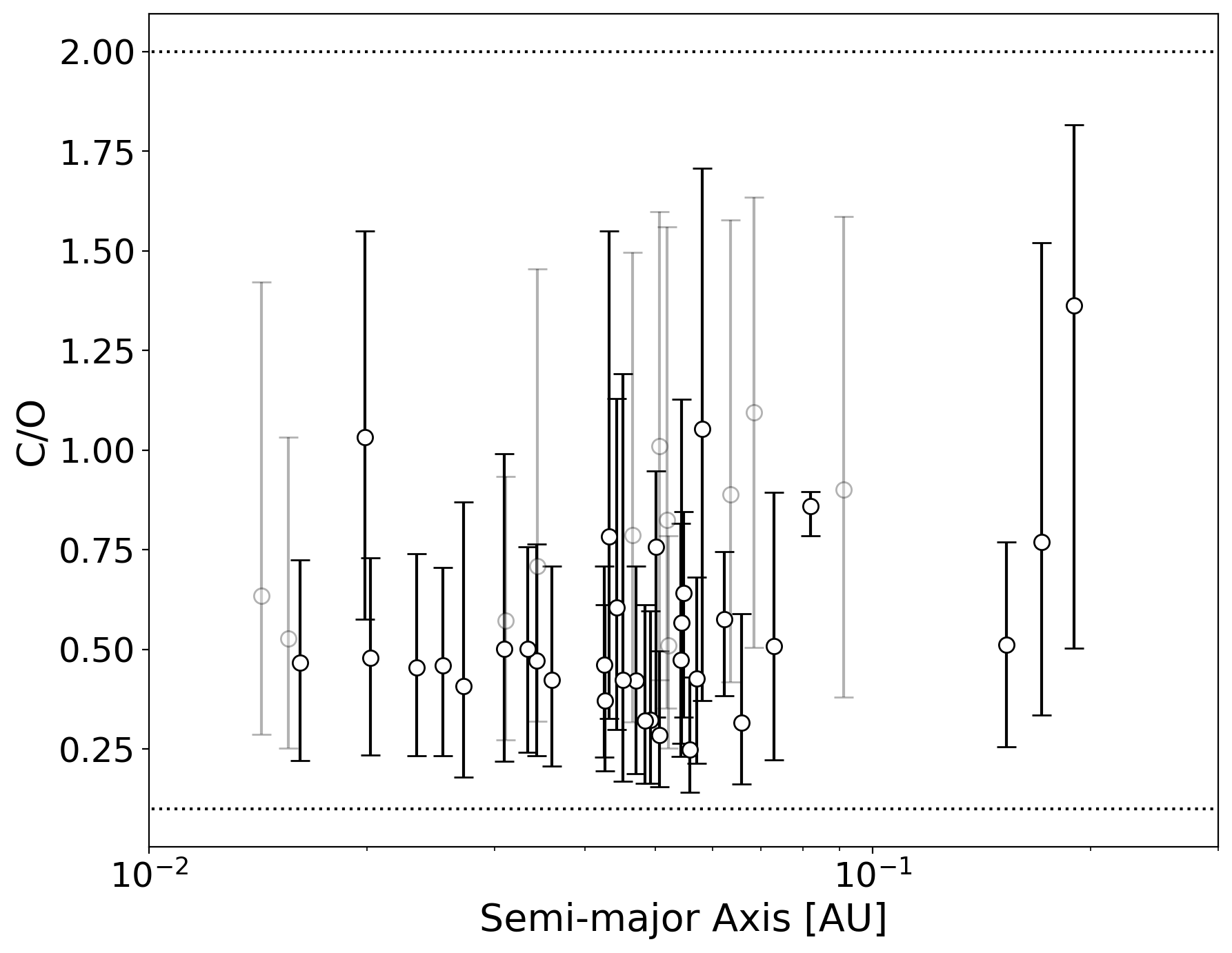}
    \caption{The retrieved carbon to oxygen ratio for planets within our sample against the retrieved metallicity (top) and planet's semi-major axis (bottom). We note that the C/O ratio is poorly constrained in most cases due to the narrow wavelength coverage offered by HST WFC3 G141. In both plots, the dotted line highlights the prior range assuming a solar metallicity. The dashed line indicates the upper bound on the metallicity used for the planet around the star with the highest metallicity. The dash-dot line is the lower bound for the planet around the lowest metallicity star.}
    \label{fig:ret_co}
\end{figure}

In Figure \ref{fig:ret_co}, we show the retrieved C/O ratio against the retrieved metallicity as well as the reduced semi-major axis of the planet. We note that the C/O ratio is generally poorly constrained and thus a conclusive trend cannot be extracted. Such a results is expected given that the wavelength coverage of WFC3 G141 only offers very limited sensitivity to carbon bearing species. While the datasets analysed here offer the possibility of detecting and constraining CH$_4$ and HCN, which has been proposed as an indicator of high C/O atmospheres \citep{venot_co_rich}, determining the presence and abundance of CO and CO$_2$ is much harder as their features are weak in this band. The study of KELT-11\,b by \citet{changeat_k11} provides a good example of the complexity of constraining carbon-bearing species using only data from HST WFC3 G141.

Despite the large uncertainties on the retrieved C/O ratio, we noted that the majority of planets were not consistent with C/O ratios larger than 1. However, the constraints are not precise enough to distinguish between different formation scenarios which generally predict values between 0.5 and 0.9 \citep[e.g.][]{turrini_ariel_2018}. Hence we can only conclude that HST WFC3 G141 data alone is not enough to accurately determine elemental ratios and thus confidently distinguish between formation scenarios. Furthermore, we note that work by \citet{turrini_formation} and \citet{pacetti_formation} suggests other ratios may be more important. These definitely cannot be constrained with HST WFC3 G141 data alone but may be achievable with data where a wider spectral coverage has been achieved with a single instrument \citep[e.g.][]{gardner_jwst,edwards_exo,tinetti_ariel}.

\subsection{Search For A Mass-Metallicity Trend}

\begin{figure*}
    \centering
    \includegraphics[width=0.95\textwidth]{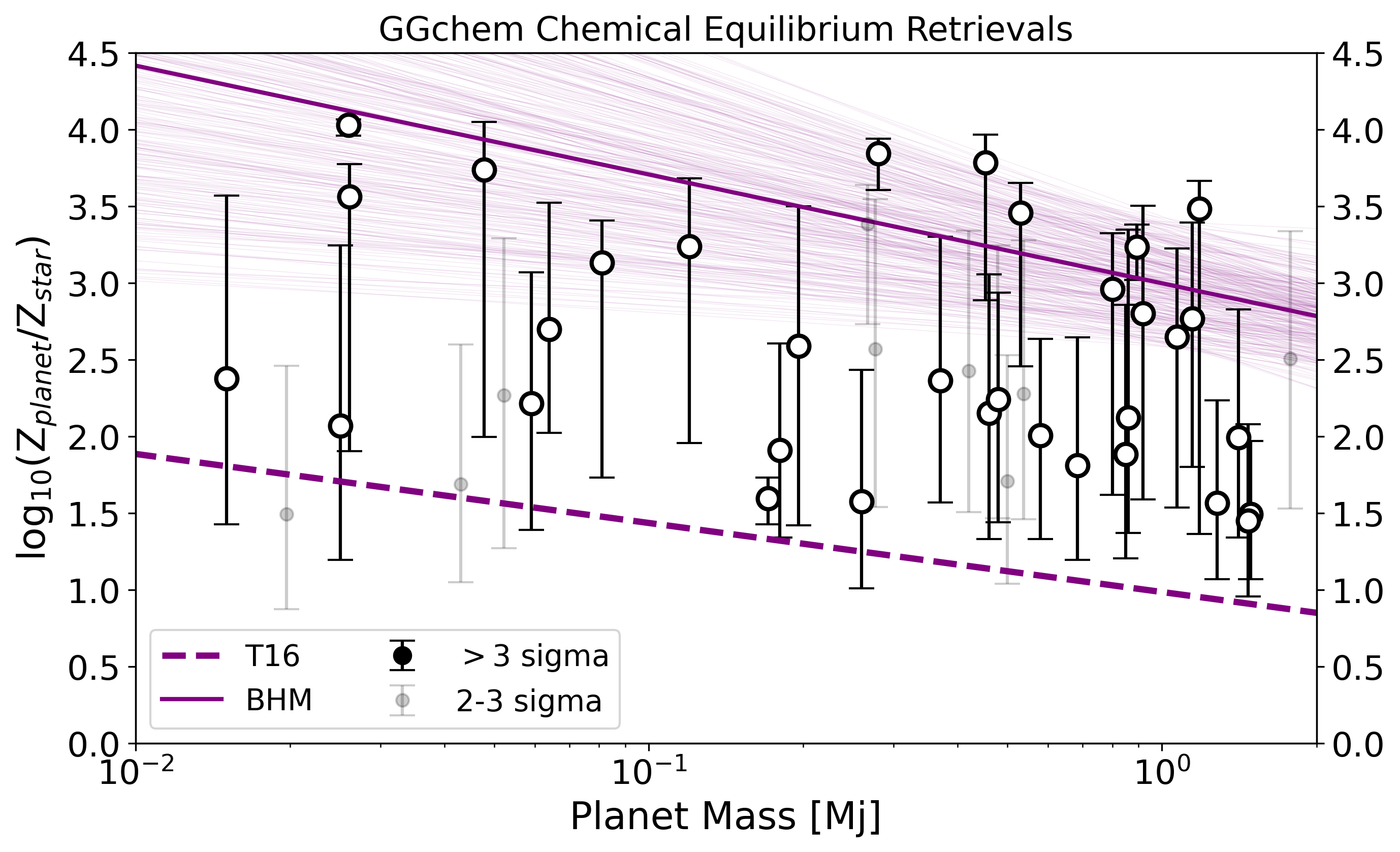}
    \includegraphics[width=0.95\textwidth]{EqChemMet_Thorngren_All.png}
    \caption{Retrieved metallicity from our GGChem runs with the model from \citet{thorngren_2016} also shown. Top: BHM results when only using retrievals which give a $>3\sigma$ atmospheric detection. Bottom: BHM results when using retrievals which give a $>2\sigma$ atmospheric detection. In both cases, the BHM found a best-fit trend line which had a negative slope. However, in both cases, the null hypothesis (metallicity is not dependent upon mass) was preferred.}
    \label{fig:met_bhm}
\end{figure*}

\begin{figure*}
    \centering
    \includegraphics[width=0.95\textwidth]{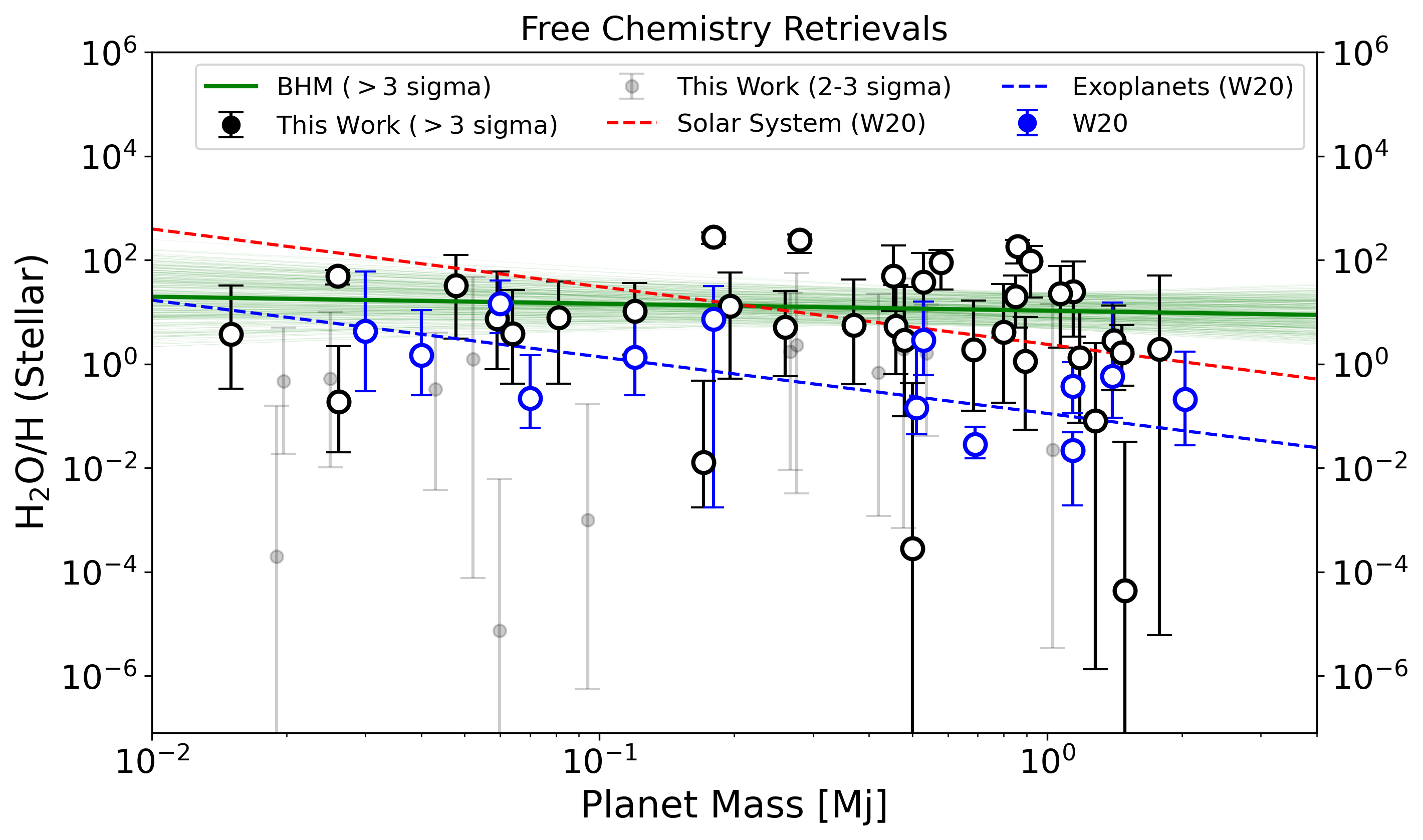}
    \includegraphics[width=0.95\textwidth]{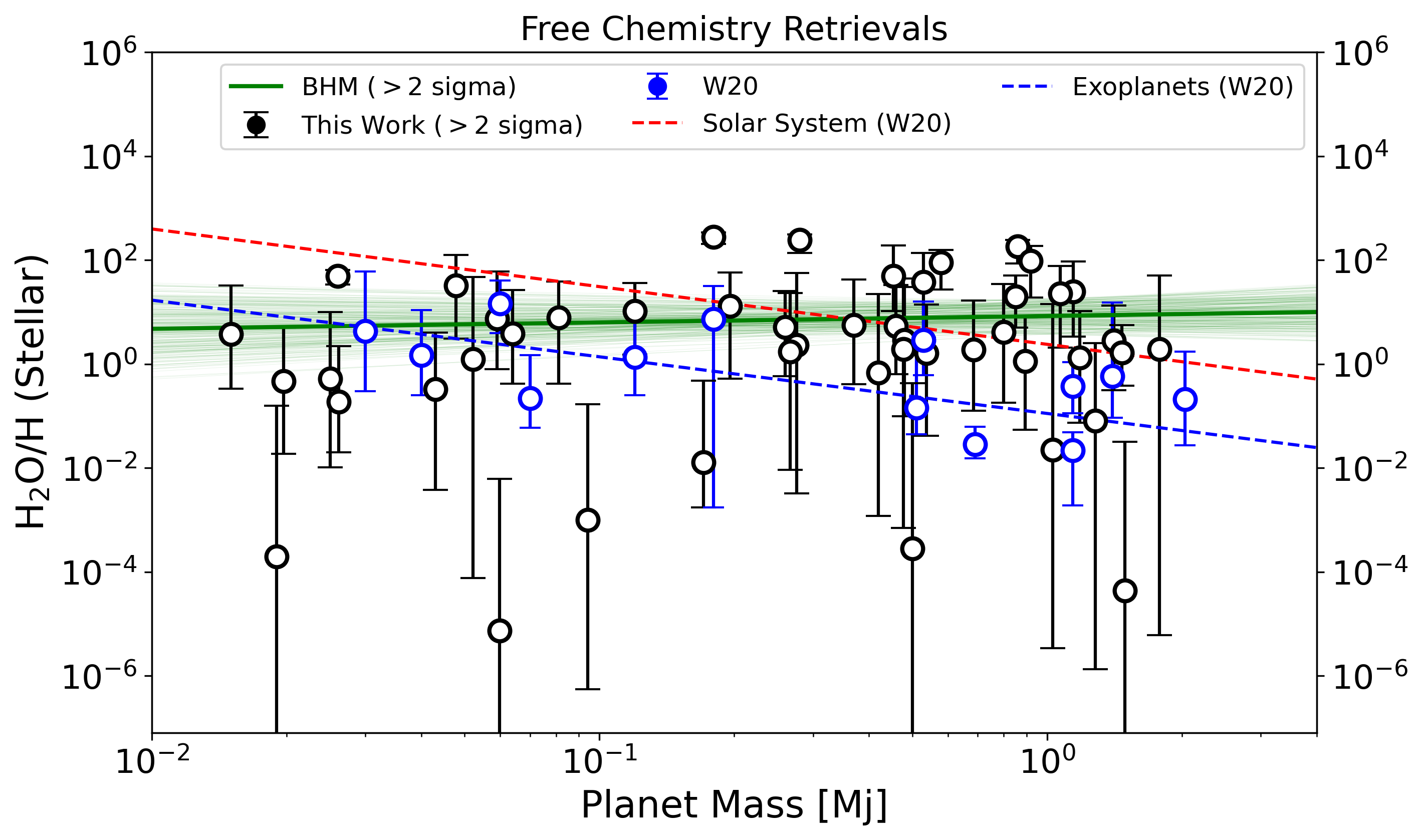}
    \caption{Ratio of H$_2$O to H with respect to stellar metallicity values for the planets from our study. The recovered values from \citet{welbanks_pop} are also shown, as are their best-fit trends. Top: BHM results when only using retrievals which give a $>3\sigma$ atmospheric detection. Bottom: BHM results when using retrievals which give a $>2\sigma$ atmospheric detection. In the former case, the BHM found a best-fit trend line which had a negative slope while the latter had a positive slope. However, in both cases, the null hypothesis (metallicity is not dependant upon mass) was preferred. Additionally, either sign of slope, or indeed no slope, is also within the 1$\sigma$ errors of each fit.}
    \label{fig:h2o_r_bhm}
\end{figure*}

Several previous studies of exoplanetary atmospheres have sought to find trends between the mass of the planet and the metallicity of its atmosphere \cite[e.g.][]{wakeford_h26,welbanks_pop}. Here we explore this across our population in two ways, firstly using the GGchem chemical equilibrium retrievals where the metallicity is a fitted parameter and, secondly, using the methods employed in \citet{welbanks_pop}. In Figure \ref{fig:met_bhm}, we compare our results to those of \citet{thorngren_2016}, where the plotted parameter is the ratio of the planet's metallicity to the star's. We find that our retrieved planet metallicities lead to ratios which are generally far above those found by \citet{thorngren_2016}. While the best-fit model when fitting a linear trend led to a negative slope, the BHM analysis does not find strong evidence for a mass-metallicity trend as the null-hypothesis (i.e. a model of constant metallicity with mass) is preferred when fitting a trend to the retrieved metallicities. We give the preferred models, and associated Bayesian evidence, in Table \ref{tab:met_bhm}.

\begin{table}[]
    \centering
    \begin{tabular}{cccc} \hline \hline
    m & c & s & ln(E)\\ \hline 
    -0.71 $^{+ 0.59 }_{- 0.6 }$ & 3.0 $^{+ 0.44 }_{- 0.41 }$ & -0.14 $^{+ 0.13 }_{- 0.12 }$ & -154.51 \\
- & 2.58 $^{+ 0.34 }_{- 0.34 }$ & -0.12 $^{+ 0.13 }_{- 0.11 }$ & -152.86 \\ \hline
-0.09 $^{+ 0.18 }_{- 0.18 }$ & 2.57 $^{+ 0.32 }_{- 0.34 }$ & -0.14 $^{+ 0.11 }_{- 0.1 }$ & -204.46 \\
- & 2.48 $^{+ 0.3 }_{- 0.29 }$ & -0.17 $^{+ 0.13 }_{- 0.07 }$ & -200.9 \\ \hline \hline
    \end{tabular}
    \caption{Results from our BHM fittings in search of a mass-metallicity trend from our GGChem retrievals. Top: using retrievals where the GGChem model provided a $>3\sigma$ detection of an atmosphere. Bottom:  using retrievals where the GGChem model provided a $>2\sigma$ detection of an atmosphere. Neither linear model, fitted in log-log space, is preferred to the null hypothesis (i.e. no trend between mass and metallicity).}
    \label{tab:met_bhm}
\end{table}

\begin{table}[]
    \centering
    \begin{tabular}{cccc} \hline \hline
    m & c & s & ln(E)\\ \hline 
    -0.13 $^{+ 0.48 }_{- 0.43 }$ & 1.03 $^{+ 0.35 }_{- 0.34 }$ & -0.31 $^{+ 0.3 }_{- 0.23 }$ & -165.1 \\
- & 1.11 $^{+ 0.28 }_{- 0.26 }$ & -0.22 $^{+ 0.2 }_{- 0.13 }$ & -161.56 \\ \hline
0.12 $^{+ 0.41 }_{- 0.39 }$ & 0.93 $^{+ 0.33 }_{- 0.36 }$ & -0.14 $^{+ 0.15 }$ $_{- 0.12 }$ & -232.99 \\
- & 0.9 $^{+ 0.28 }_{- 0.29 }$ & -0.19 $^{+ 0.19 }_{- 0.15 }$ & -229.08 \\ \hline \hline
    \end{tabular}
    \caption{Same as Table \ref{tab:met_bhm} except it is for the mass-metallicity fit based upon the water-to-hydrogen ratio.}
    \label{tab:h2o_r_bhm}
\end{table}


Meanwhile, in Figure \ref{fig:h2o_r_bhm} we show the metallicity derived from the H$_2$O abundance, over-plotting the data from \citet{welbanks_pop} as well as their best-fit to the data for comparison. To maintain consistency with their study, we also plot the results of retrievals where the detection significance is 2-3 sigma. However, we notice their trend is evidently not reflected in our data. All planets studied in \citet{welbanks_pop} are also included in our population and one can also notice differences in masses, depending upon which reference studies were used, as well as in the retrieved water to hydrogen ratios. The latter of these could be due to a number of factors but the major cause is likely to be a difference in the datasets used. \citet{welbanks_pop} used data from a variety of instruments, while we used only HST WFC3 G141. Each instrument gives access to differing absorption features and thus provides the observer with different sensitivities. Observing the same planet but with different instruments could provide contrasting findings on its composition. Therefore, the same is true when observing a selection of planets with disparate datasets and so the trend they see may be caused by these differing sensitives to given molecules.  

On the other hand, the lack of an obvious trend derived here could well imply biases in our retrievals, with the high water abundances derived, and the large planet-to-star metallicity ratios, potentially providing evidence for this. Alternatively, we could conclude that HST WFC3 alone is simply not sensitive enough to be able to accurately constrain atmospheric metallicities, even through the form of the H$_2$O abundance, as the uncertainty on this parameter is generally very high. Our BHM analysis found that, as with the GGChem metallicites, the null hypothesis is preferred over a mass-metallicity trend, with the results given in Table \ref{tab:h2o_r_bhm}. Hence, we find no evidence for a mass-metallicity trend within our datasets.

\subsection{Comparison of Free Chemistry and Chemical Equilibrium Retrievals}

\begin{figure}
    \centering
    \includegraphics[width=\columnwidth]{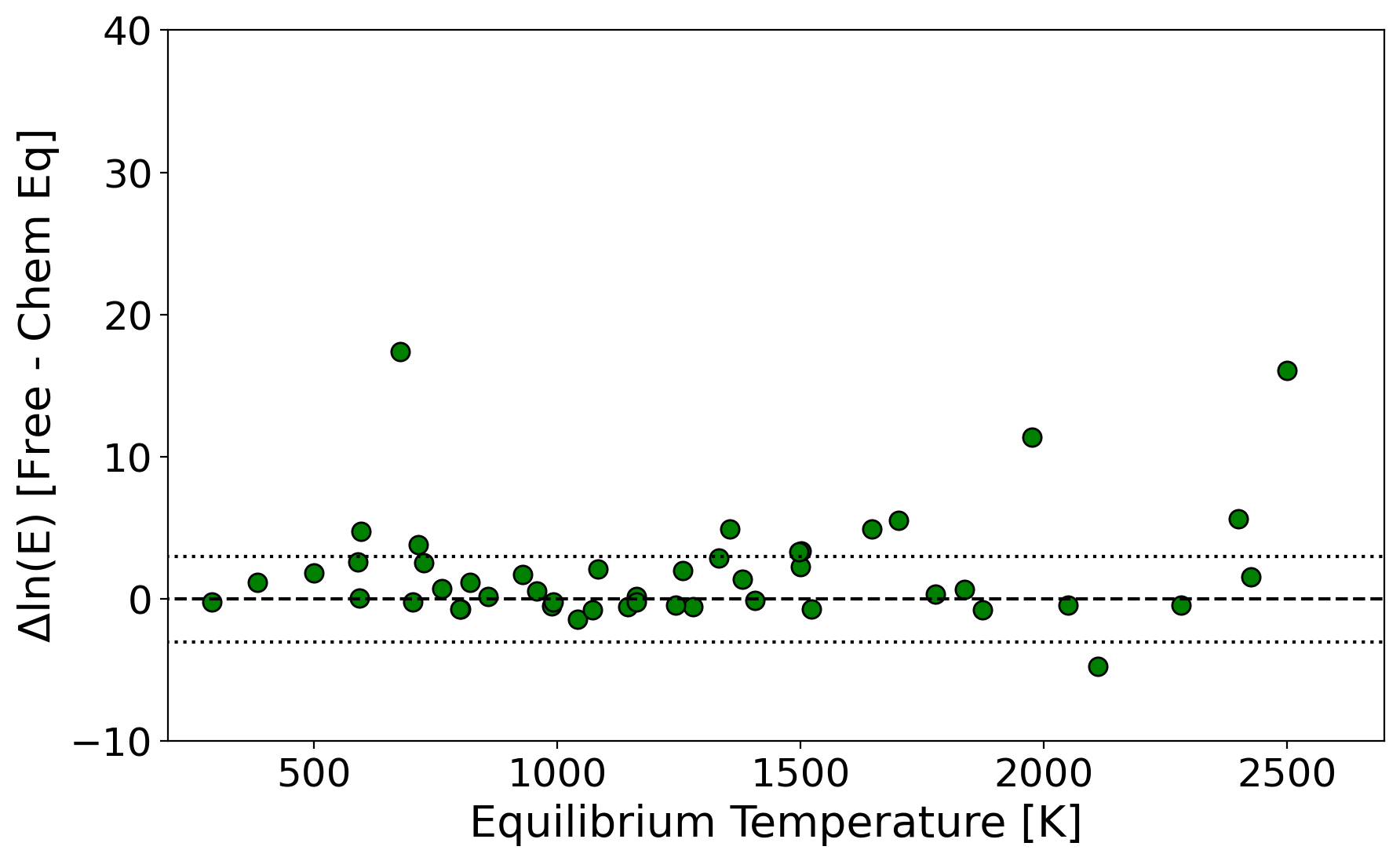}
    \caption{Comparison of the Bayesian evidence for retrievals with free chemistry and those using the GGChem chemical equilibrium scheme. }
    \label{fig:free_ce_comp}
\end{figure}

To assess which models may be best for fitting the population as a whole, we compared the Bayesian evidence of our free chemistry and chemical equilibrium retrievals. The latter has fewer free parameters and thus is penalised less while the former is more agile as the relative abundances of molecules is completely inhibited. By comparing the evidence for both, we find that both provide fits of a similar quality for lower temperatures but, for planets above 1500 K, the free chemistry model often provides a statistically preferable fit to the data as shown in Figure \ref{fig:free_ce_comp}. Such a finding could be suggestive of disequilibrium chemistry, a claim which has also been made in previous studies \citep[e.g.][]{baxter_2021,keating_cowan,roudier_diseq}. However, this finding is in opposition to the noted dearth of methane detected in our free chemical retrievals. Given that equilibrium models predict significant amounts of methane at these lower temperatures, it is strange to find they fit the data as well as models which don't infer the presence of this molecule. It also contrasts the results from \citet{baxter_2021} whose analysis of Spitzer data suggested that cooler planets were not in chemical equilibrium.

\subsection{Amplitude of Absorption Features}
 
The key spectral feature within the HST WFC3 G141 band is the 1.4 $\mu$m water feature and, instead of performing retrievals in an attempt to recover the abundance of this molecule, several studies have instead measured the size of the feature in relation to other bands within WFC3's spectral range \citep[e.g.][]{stevenson_wfc3, Fu_2017,wakeford_clouds,dymont_fs}. We fitted the feature using the process described in Appendix 3. 

The 1.4 $\mu$m feature size was utilised by \citet{crossfield_kreidberg} to imply a number of trends within the atmospheres of sub-Neptunes based off spectra of six planets. Using the most massive planet from their sample, HAT-P-11\,b (25.7M$_\oplus$), as the upper boundary in terms of mass, and the largest of their sample, HAT-P-26\,b (6.3\,R$_\oplus$), as the radii limit, we have extended this to sixteen gas dwarf planets. The two key correlations found by \citet{crossfield_kreidberg} were with the planet's H/He mass fraction and it's equilibrium temperature. One interpretation of the trend seen with the former would be that planets with a low H/He mass fraction would have a high mean molecular weight, leading to a smaller than predicted scale height which was calculated assuming $\mu$ = 2.3. Meanwhile, a reduction in the size of the H$_2$O amplitude with decreasing temperature was postulated to be due to hazes for cooler ($<$850 K) planets.

\begin{figure}
    \centering
    \includegraphics[width=\columnwidth]{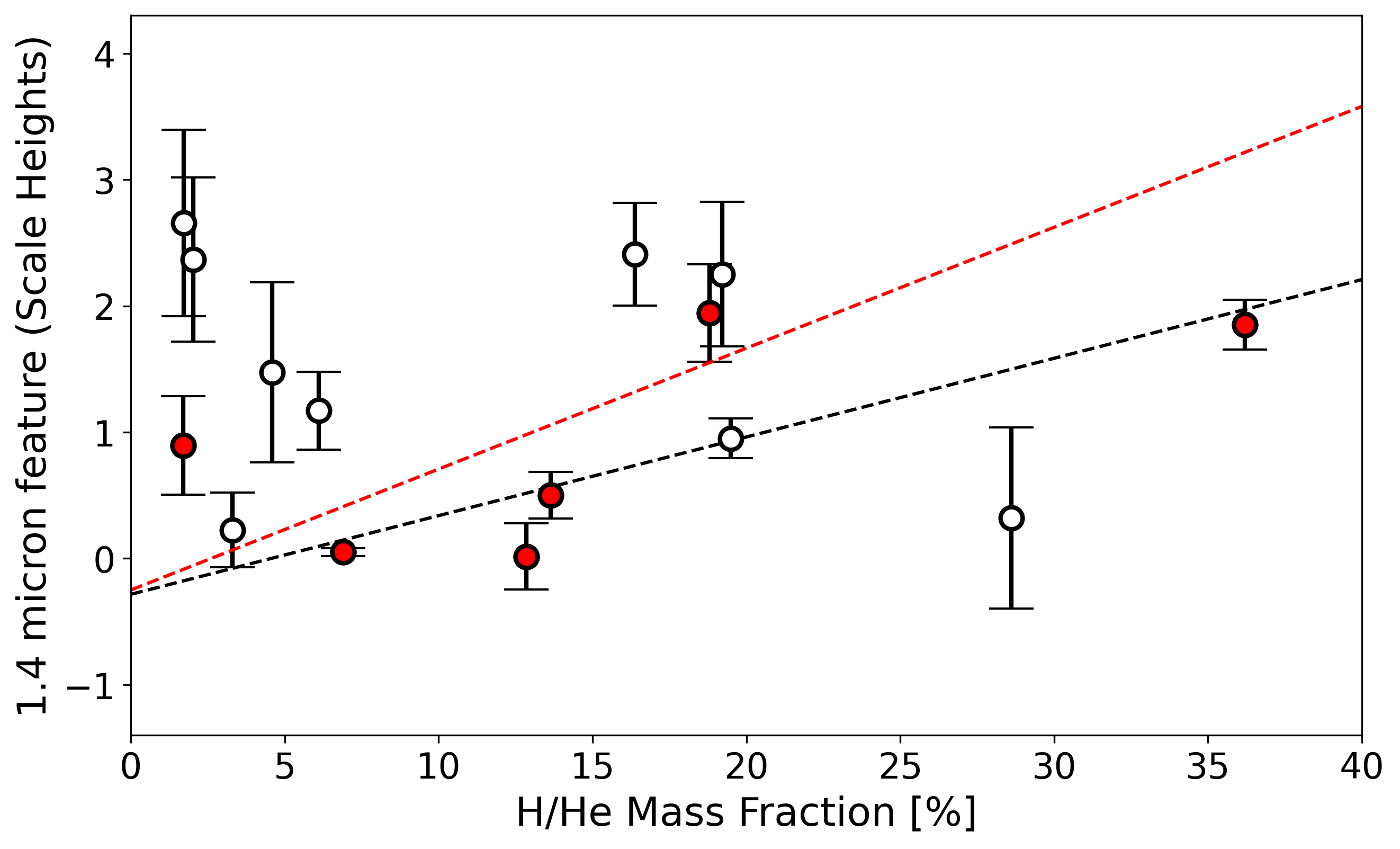}
    \includegraphics[width=\columnwidth]{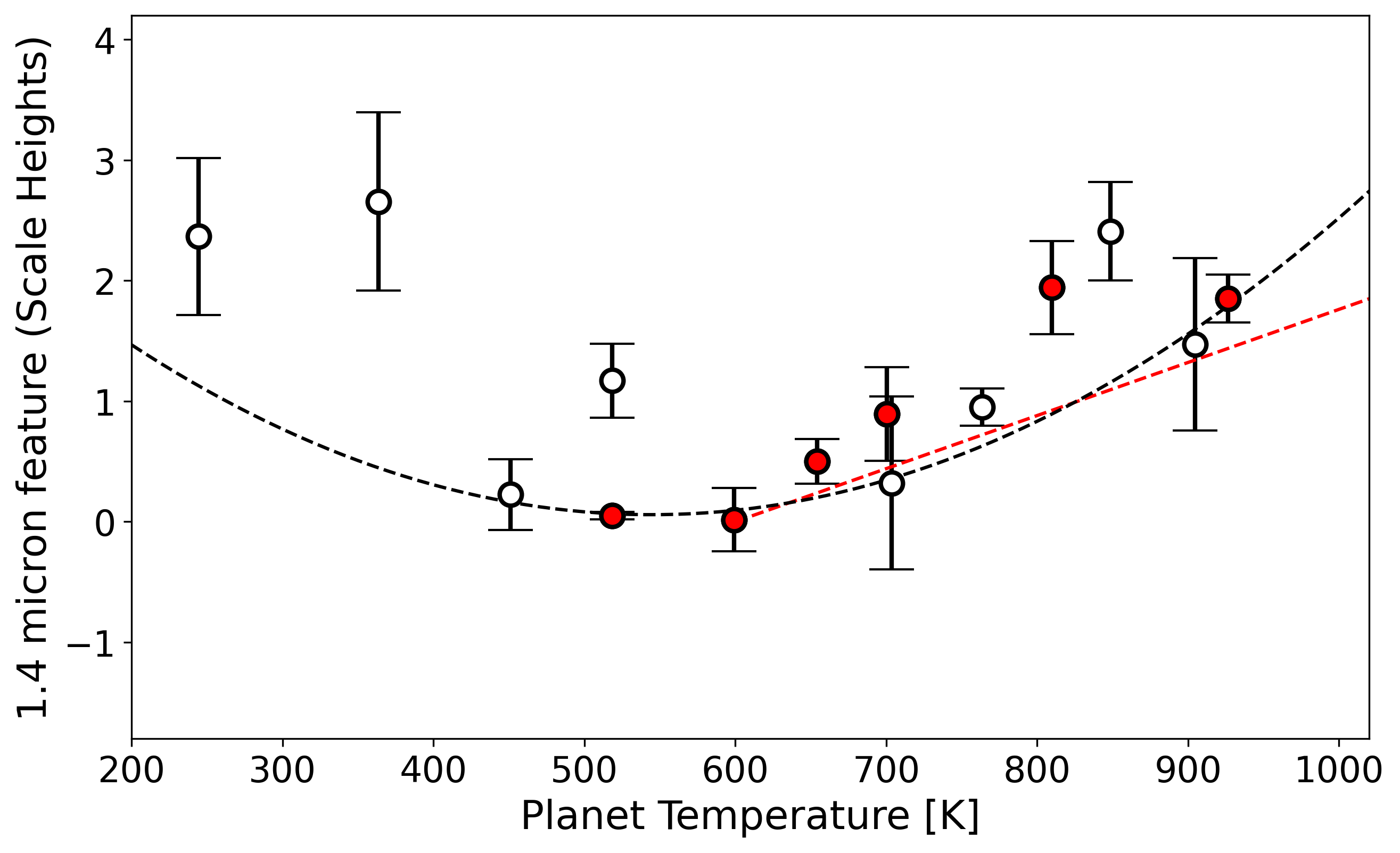}
    \includegraphics[width=\columnwidth]{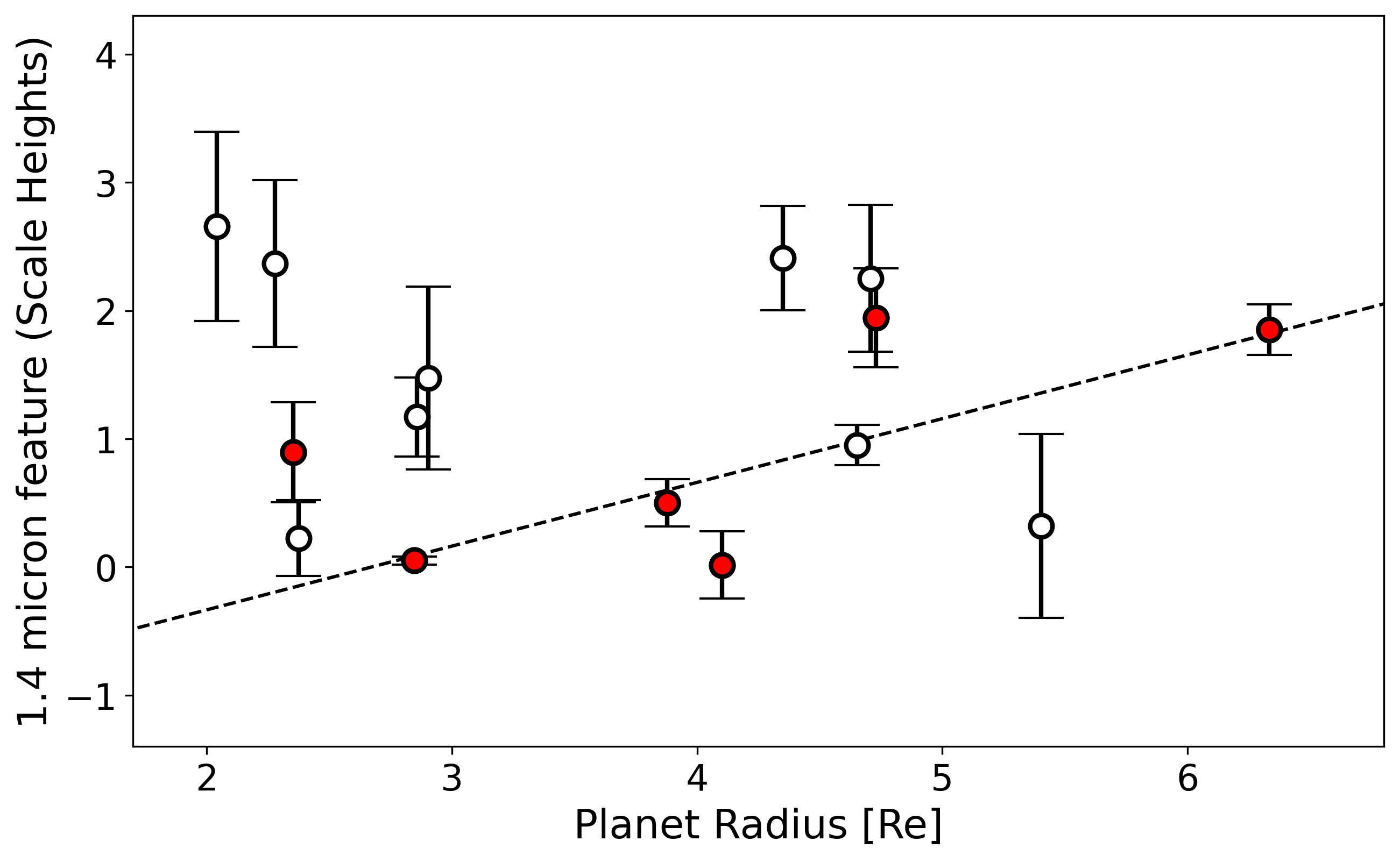}
    \includegraphics[width=\columnwidth]{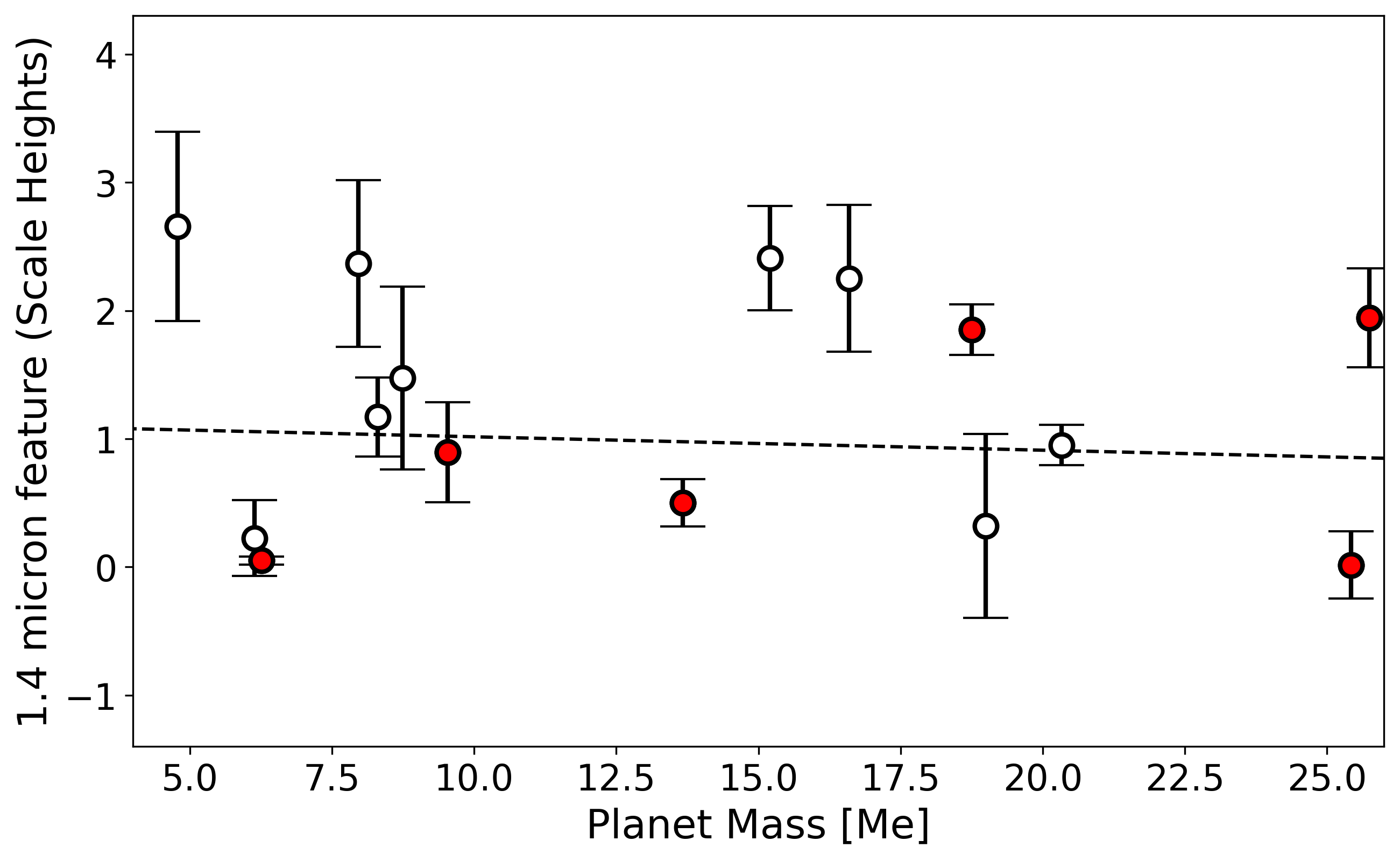}
    \caption{Comparison of the 1.4 $\mu$m feature height to different bulk parameters for the Sub-Neptunes within our study. Red data points indicated those studied by \citet{crossfield_kreidberg} with their trends also shown in red. Our trends are shown in black and are linear fits, except for with temperature where a second-order polynomial fit was used.}
    \label{fig:sub_nept_feature}
\end{figure}

We updated these plots but find these correlations no longer hold, as shown in Figure \ref{fig:sub_nept_feature}. We now find a decreasing feature size with increasing H/He mass fraction, a result which is counter intuitive. Furthermore, we find that, at temperatures cooler than 550\,K, the feature size increase with decreasing temperature, a result that concurs with work done by \citet{yui_clouds}. We highlight the planets that were in the \citet{crossfield_kreidberg} sample and suggest the trends were seen due to a selection bias. For these trends we find that the are driven significantly by GJ\,1214\,b as the feature size is extremely well constrained but also very close to zero.

We also attempted to draw out other trends looking across two parameters simultaneously, but found no significant correlations. From this we imply that either a) no simplistic correlations exist or b) the sample is not large enough, or has not been selected carefully enough, to allow for any trends to be teased out. Such a finding highlights the importance of a structured, hypothesis-based selection of planets when attempting population studies and the need for dedicated exoplanet atmospheric survey missions (e.g. Twinkle, Ariel).

\begin{figure*}
    \centering
    \includegraphics[width=0.9\textwidth]{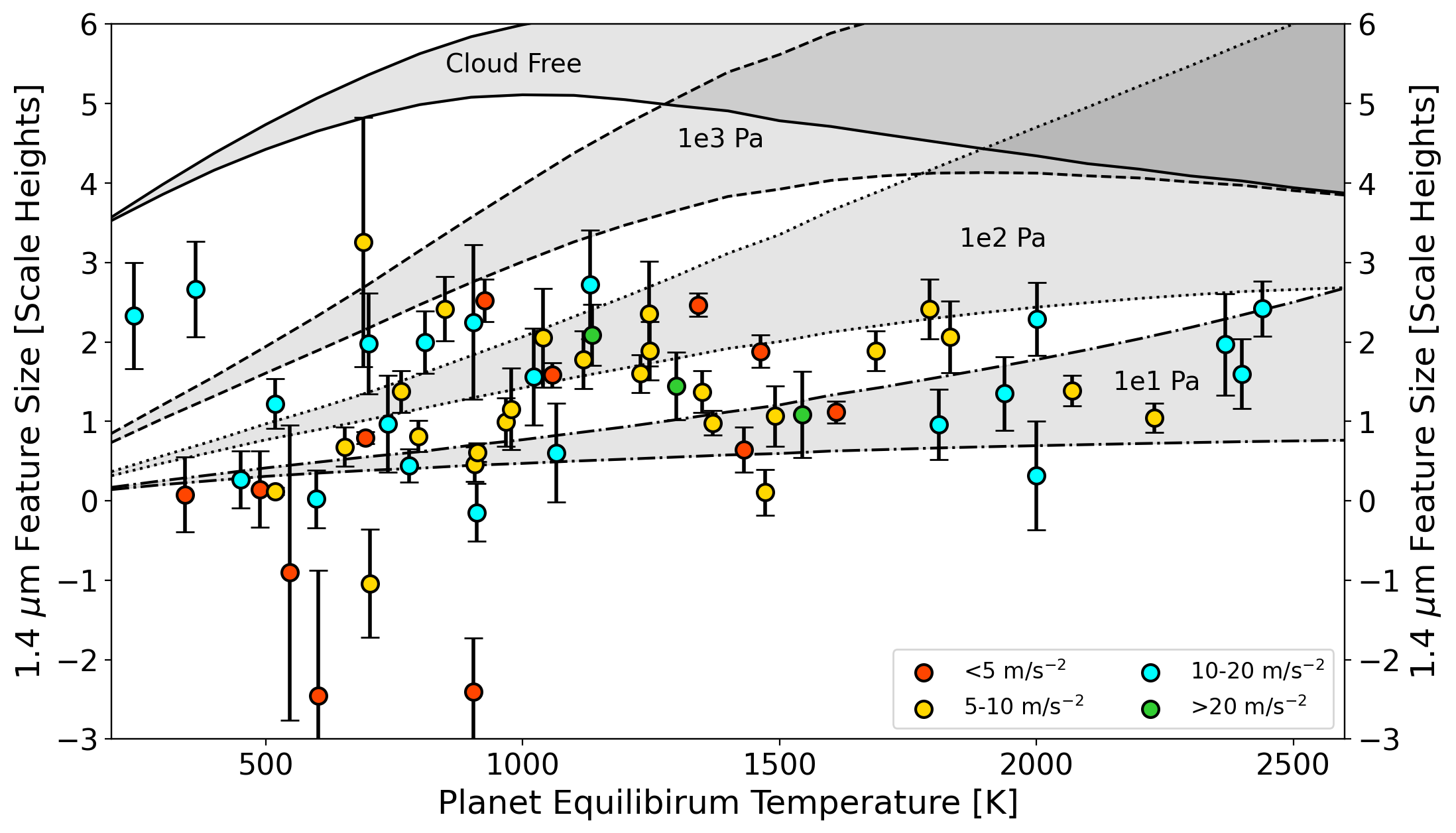}
    \includegraphics[width=0.9\textwidth]{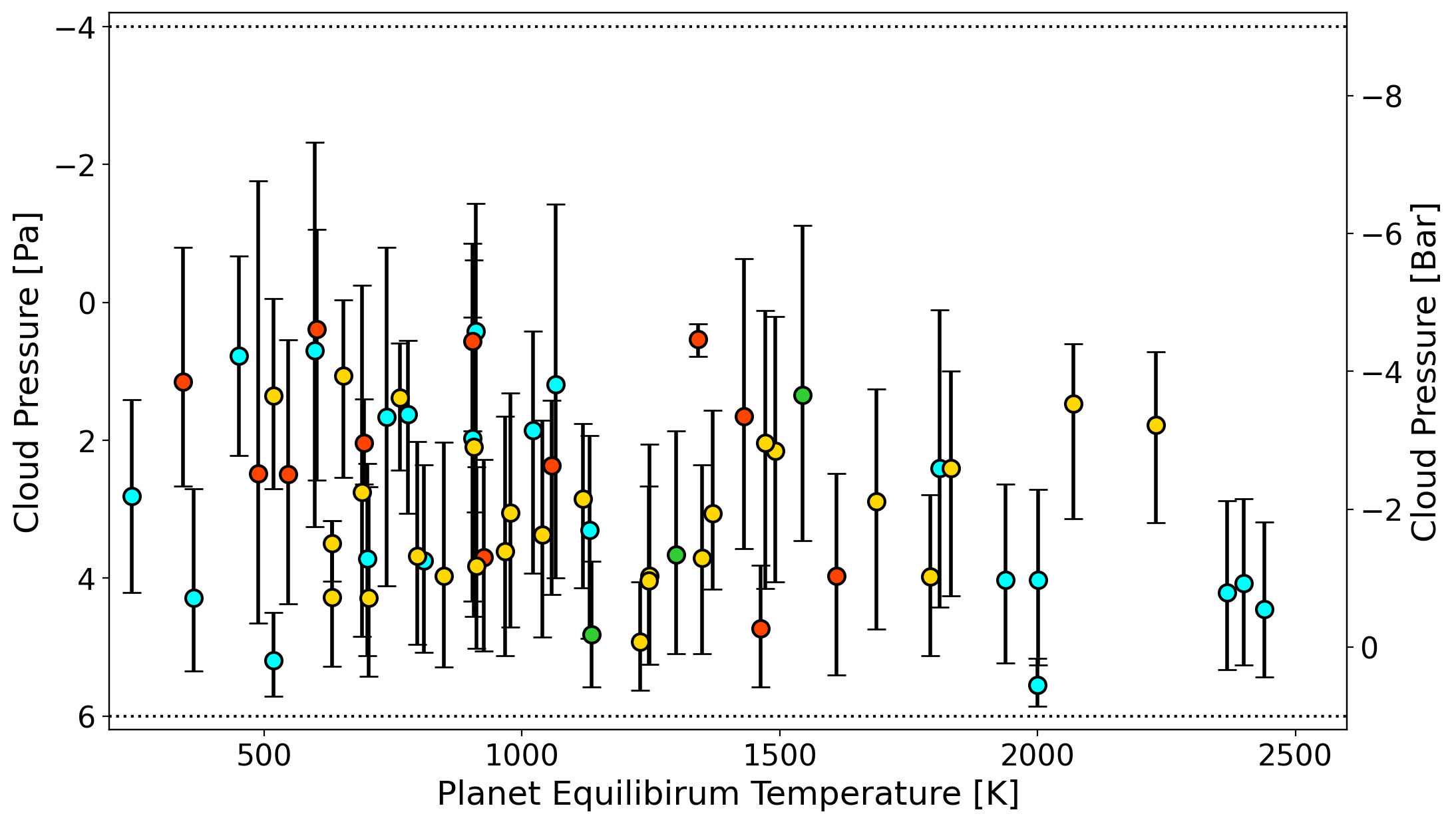}
    \caption{Size of 1.4$\mu$m feature against planet temperature and sub-divided by the planet's surface gravity. At lower temperatures, the surface gravity may have an impact on the cloudiness of the planet, and thus the feature size, but, at higher temperatures, there appears to be no obvious correlation. Grey regions highlight the expected feature size for solar metallicity atmospheres for a given cloud pressure and a surface gravity between 10-20m/s$^2$. Across the entire population, the mean 1.4 $\mu$m feature size is 0.81 scale heights in the HST/WFC3/G141 while the median was 1.37. The feature size can be directly correlated to the retrieved cloud pressure from our retrievals (bottom) and we note that the 1.4 $\mu$m feature size is not a good predictor of the feature size across wider wavelength ranges due to the stronger absorption seen at longer wavelengths. The priors for the cloud pressure in our retrievals is shown by the dotted lines.}
    \label{fig:feature_size}
\end{figure*}

We extended the analysis of the feature size to all planets in our study and Figure \ref{fig:feature_size} displays the recovered 1.4 $\mu$m amplitude against temperature. Previous studies have suggested that the surface gravity of a planet, along with its temperature, affect the cloud coverage which in term would affect the WFC3 feature size \citep[e.g.][]{stevenson_wfc3,bruno_w67_h38}. Hence, we divide the population up by its surface gravity to search for such a trend. While some differences in feature size are seen in planets with similar temperatures, the surface gravity doesn't appear to be driving this in all cases: in some, the feature size is identical while, where there are differences, these are not universal. In general, the distribution of feature size is similar to those of previous studies \citep[e.g.][]{Fu_2017}, with an increase at around 1200\,K. In Figure \ref{fig:feature_size}, we also plot the retrieved cloud pressure. Here we see that, in the same temperature range that the feature size increases, the cloud pressure retrieved is deeper in the atmosphere. Therefore, the knowledge gained by performing the spectral feature analysis is also readily available from retrieval analyses.

The spectral feature size has often been used to infer the presence of clouds and proposed as a way of guiding observers as to the spectral modulation that could be expected when planning future observations. Across the population, we recover an average (weighted mean) feature size of 0.81 scale heights but we noted that the median value is much higher (1.37 scale heights). We noted that such a low value for the weighted mean is driven by the feature size of GJ\,1214\,b (0.12$\pm$0.04) and if one removes this from the weighted mean calculation, an average value of 1.21 scale heights is obtained. All these values are comparable value to previous studies: 1.4 H \citep{Fu_2017} and 0.89 H \citep{wakeford_clouds}. As demonstrated in Figure \ref{fig:feature_size}, the amplitude of this feature is far below what would be expect from a clear, solar metallicity atmosphere. There would also appear to be some correlation with temperature, with hotter planets (T$>$1500\,K) generally having larger feature sizes than cooler ones (T$<$700\,K). Additionally, we see a trend at around 1200 K where the feature size increases before decreasing again. A similar feature was found by \citet{Fu_2017} and our retrievals find these planets have a lower cloud top pressure (see Figure \ref{fig:feature_size}).

Extending the models used to derive the 1.4 $\mu$m feature size in the HST WFC3 G141 data, we estimate the amplitude of features seen in observations with future instruments by studying the minimum and maximum transit depth across their spectral coverage. For JWST NIRISS GR700XD (0.6-2.8 $\mu$m, \citet{doyon_niriss}) and JWST NIRSpec G395H (2.8-5.1 $\mu$m, \citet{birkmann_nirspec}) we predict average feature sizes of 3.68 and 2.59 scale heights respectively. Combining both instruments, observing with the JWST NIRSpec PRISM (0.6-5.3 $\mu$m), or with Twinkle (0.5-4.5 $\mu$m, \citet{edwards_exo}), gives an amplitude of 3.74 scale heights. Finally, the expected amplitude across the spectral coverage of Ariel (0.5-7.8 $\mu$m, \citet{tinetti_ariel}) is 4.60 scale heights.

Therefore, while clouds and hazes obviously need to be accounted for during the planning of observations with future facilities, current data shows the expected amplitude should, on average, be greater than a single scale height. However, we note also that the methodology used to measure the amplitude of absorption features is somewhat flawed. While the parameters in Equation \ref{eq:feature_size} allow the spectrum to be modulated to fit the data, and account for the features seen, the final fit does not provide a robust analysis of the nature of the atmosphere: this can only be achieved by running a Bayesian retrieval which models the passage of starlight through the atmospheric layers to explain the spectrum via base atmospheric proprieties such as temperature, composition and clouds.

Hence, for comparison, we also took our preferred spectral retrieval models and computed the amplitude of features seen across these wavelength bands. For the spectral ranges 0.6-2.8\,$\mu$m, 0.5-5.3\,$\mu$m and 0.5-7.8\,$\mu$m, we found a median feature size of 3.32, 3.66 and 4.38 scale heights respectively. These are roughly 1 scale height larger than the feature amplitude fit suggests. However, we note that these predictions may also be biased as some species, such as CO$_2$, are not constrained by our Hubble WFC3 observations. Therefore, the best-fit value of these is essentially an average of the prior range. When considering the spectrum across these longer wavelengths, this may induce features which are not present if the molecules actually exist at values lower than the value ``retrieved". 

In any case, both models point towards features several scale heights in amplitude being observable with future facilities, largely due to the strong absorption that occurs at longer wavelengths. As data is collected with these facilities, our understanding of the cloudiness of extrasolar planets will be enhanced and therefore allow us to more confidently predict the data quality. Nevertheless, we will never truly know until we look.


\section{Discussion and Conclusions}

The Hubble Space Telescope has been at the forefront of exoplanet atmospheric characterisation over the last two decades. While many different instruments on this facility have been used, WFC3 has perhaps been the mostly widely utilised due to its sensitivity to water. In this work we have presented a population study of atmospheres, each studied with the WFC3 G141 grism as the planet transits its host star. 

Of the 70 planets studied, we found strong evidence ($>$3 sigma) for atmospheric features on 37 of them, with some evidence (2-3 sigma) for spectral modulation on an additional 14 planets. We note that for several planets (e.g. WASP-18\,b), the derived spectrum has error bars that are several scale heights in size, meaning no atmospheric constraints could be expected. As noted by other studies, clouds are ever present and are muting the size of the features seen. While clouds should certainly not be discounted from observational planning, future instruments will probe longer wavelengths, where the absorption of molecules such as H$_2$O and CO$_2$ are stronger, and so may not be as affected, particularly given that the signal-to-noise ratio should be higher for these datasets.

\begin{figure}
    \centering
    \includegraphics[width=\columnwidth]{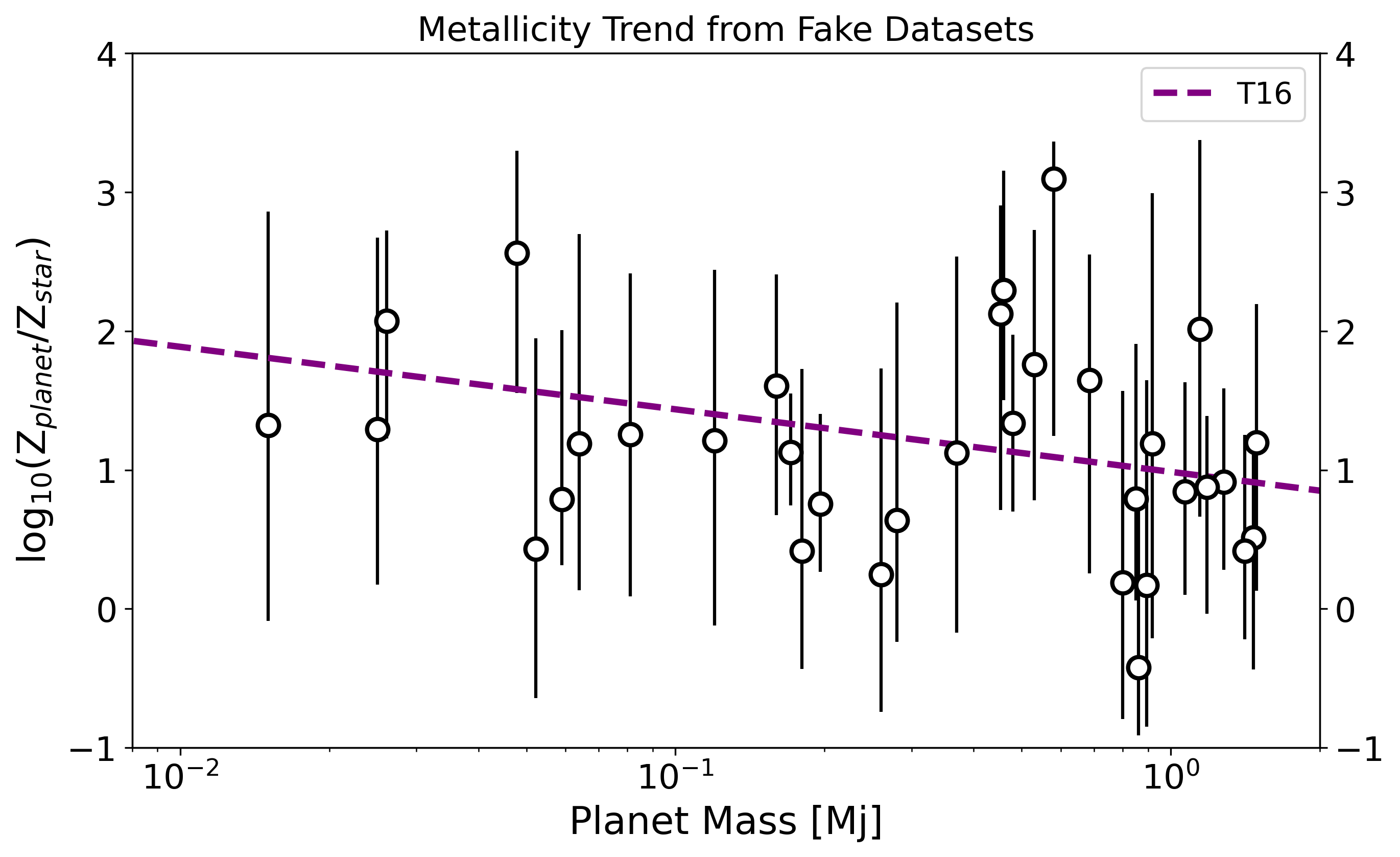}
    \caption{Retrieved metallicity when simulated data is created using the model from \citet{thorngren_2016}. From our fake datasets, which utilised the error bars from the real observations as well as the retrieved 10 bar radius and cloud pressure, we can recover the input trend, albeit it a marginal detection due to the size of the uncertainties on the planet's metallicity. The finding contrasts the real data, where no obvious trend was uncovered.}
    \label{fig:fake_met}
\end{figure}

In this work we have largely struggled to draw out trends in the population, despite employing BHM which exploits the full richness of the posterior distributions from our Bayesian retrievals. Nevertheless, these null results are not without value as they can inform us of how to approach atmospheric studies in the future. Upon viewing our results, it may be concluded that HST WFC3 G141 data alone is insufficient to draw out detailed trends from a population of objects. To explore this, we created simulated datasets of the planets for which our retrievals suggested there was evidence for spectral modulation due to an atmosphere. For each planet, we utilised the error bars, the retrieved cloud pressure, and retrieved 10 bar radius from the real observations to generated fake datasets, inputting the mass-metallicity trend suggested by \citet{thorngren_2016} and performing retrievals to see if our simulated data would recover the trend. We added Gaussian scatter to the datasets and the retrieved planet-to-star metallicity ratios are shown in Figure \ref{fig:fake_met}. We find that our simulated data is capable of recovering the input trend which suggests that, in our analysis of the real data, we do not recover a trend because i) one does not exist across the whole population, ii) there are biases in our retrieval analysis which are obscuring the true trend or iii) both.

In our analysis, on both the real data and the simulated, the constraints placed on the atmospheric chemistry are poor. Such a result is not overly surprising; all those who have worked with HST WFC3 G141 understand that its limited wavelength coverage means the conclusions that can be drawn from the analysis of it are similarly restricted. The inability to place tight constraints on the planet's metallicity or C/O ratio are a by-product of a dataset which is only truly sensitive to a single feature of a single molecule: the 1.4 $\mu$m water feature. As such, studies in the literature have often combined datasets from multiple instruments to extend the wavelength coverage and unlock additional spectral features. However, such an approach has serious implications for the analysis, particularly when attempting to draw trends from a population.

Firstly, the instruments experience significant systematics. While the correction of these with different pipelines usually leads to uniform spectral features, offsets in the transit depth are common \citep[e.g.][]{guo_hd97658,luque_w74,murgas_offset,yip_w96,mcgruder_w96}. When analysing data from a single instrument, this has little effect as the retrieved planetary radius is generally just slightly smaller or larger. However, which combining instruments, these offsets can cause wild differences in the retrieved atmospheric parameters \citep{yip_lc}. While an offset can be fitted for in the retrieval, without spectral overlap between instruments one cannot be sure of the compatibility of datasets. Currently, only HST WFC3 G102 offers spectral overlap with HST WFC3 G141 but this filter has rarely been used. Combining both the WFC3 IR grisms, with data from HST STIS, can be done with some confidence due to this spectral overlap \citep[e.g.][]{wakeford_h26}, although the data is not taken in the same epoch so temporal changes may be an issue \citep[e.g.][]{saba_w17}.

Yet, even if one can confidently combine data from multiple instruments, another factor for consideration emerges. If one compares two planets having utilised different instruments in each case, one cannot be sure if any differences seen are because the planets are distinct or if the instrument's simply offer discrete views of these atmospheres. An example of this is demonstrated in \citet{pinhas} for HD\,209458\,b. If HST WFC3 G141 data is considered alone, they retrieved water abundance of log$_{\rm 10}$(H$_{\rm2}$O) = -3.26$^{+0.75}_{-0.75}$ which is comparable to the value obtained here of -3.10$^{+0.95}_{-1.17}$. With the addition of STIS data, the value retrieved was -4.66$^{+0.39}_{-0.30}$ which is not compatible to within 1$\sigma$ and takes the water abundance of HD \,209458\,b from being solar to distinctly sub-solar. The debate here is not about the true value; the STIS data may bring you closer to it by avoid some degeneracies, but about the change the new data brings about in the result. Consider another planets, for example HAT-P-38\,b or TOI-674\,b, for which we retrieved a similar water abundances to HD\,209458\,b (log$_{\rm 10}$(H$_{\rm2}$O) = -3.07$^{+1.12}_{-2.27}$ and -3.12$^{+0.78}_{-1.04}$, respectively). Comparing the retrieved abundances from the WFC3 G141 data alone would lead us to conclude the atmospheres were similarly enriched. Comparing the HD\,209458\,b abundance retrieved using the STIS data as well would lead us to believe they had differing water abundances. As neither HAT-P-38\,b nor TOI-674\,b have STIS data at the time of writing, it is impossible to know if the retrieved water abundances would also change if such data was also added to our retrievals. While analysing HST WFC3 G141 data alone has its limitations, such as potentially introducing biases into the retrievals, but these limitations are at least uniform across the population. Therefore, when searching for trends the recovered correlations may be biased but at least the relative trend between planets would be consistent. However, when utilising differing instruments between planets one cannot tell if the changes seen are due to the instruments used or due to the atmospheres actually differing in composition. If one looks at which planets have been studied by both STIS and WFC3, this further enforces this point: higher mass planets have more often have both datasets whereas lower mass planets have generally only been studied with WFC3 G141. Therefore, if one finds evidence for a mass-metallicity trend, which is based on the H$_{\rm 2}$O abundance in the atmosphere, with inhomogeneous datasets, one cannot be sure the trend is not caused by the addition of these datasets for the higher mass planets. As such, if one attempts a population study where the instruments each member is studied with varies, one must account for this bias when inferring the presence, or lack, of trends. For example, conducting retrievals on both datasets (e.g. with/without STIS or Spitzer) and comparing the results \citep[e.g.][]{pinhas,aresIII,yip_w96}.

The ability to seek out these correlations is further impaired by the choice of targets, with those studied in this work essentially being a random collection of worlds as they have been observed via a variety of proposals, each with different aims. As such, drawing comparison becomes difficult because, in all likelihood, more than one parameter will be affecting the chemistry. Therefore, one may be tempted to split the targets into sub-groups in an attempt to uncover trends but doing so after the observation is required is risky; with the dataset available, if one tries hard enough ``trends'' can uncovered due to the large uncertainties in the derived parameters and the sparsity of the data once more than a single parameter is used to divide the population. Such an effect can be seen in the relation suggested between temperature and atmospheric feature sized previously proposed for sub-Neptunes. In this case, and in all other attempts to draw conclusions, the fact that the precision of the data with respect to the expected atmospheric signal (i.e. the SNR) also differs across the population, makes it yet harder to definitively draw comparisons and to imply the atmospheric conditions or conclude on why an atmospheric detection has not occurred. 

In addition to free chemistry retrievals, we fitted chemical equilibrium models to our datasets in an attempt to draw out trends in metallicity and the C/O ratio. However, as noted in the methodology, these retrievals are essentially fitting two free parameters (metallicity and C/O ratio) to a single observable (the H$_2$O abundance). As such, the retrieved values are not necessarily reliable. Taking again the example of HD\,209458\,b and HAT-P-38\,b, for which we recovered an almost identical water abundances in the free chemistry experience, we can see the issue clearly. While we recovered a metallicity of log$_{10}$(Z$_{\rm P}$) = -0.04$^{+0.83}_{-0.62}$ for HD\,209458\,b, the value for HAT-P-38\,b, 1.59$^{+0.25}_{-0.65}$, is different to greater than 1$\sigma$ (despite the large uncertainties) as the model preferred a higher C/O ratio for this planet. Many previous studies have fitted for these without additional data, or with optical data which doesn't add an additional molecular observable, but our results suggest these findings should be taken cautiously. Spitzer IRAC data has often been added to HST WFC3 G141 to help further constrain the C/O ratio by given sensitivity to carbon-bearing species (e.g. CH$_4$, CO, CO$_2$) but these studies are then exposed to the potential offset risks that we have discussed previously.

Hence, our current ability to extract population-level trends in exoplanet atmospheres are limited by a number of factors. In short, to truly perform population studies of exoplanetary atmospheres, one must achieve a wide spectral coverage with a single instrument and ensure that the planets are selected in a robust manner. JWST \citep{gardner_jwst} will provide better opportunities for this than Hubble, particularly on the spectral coverage front, although for brighter targets multiple instruments will still need to be combined to get the wavelength coverage necessary to accurately constrain both refractory elements and carbon-bearing species. However, while certain JWST proposals are designed as miniature population studies, an organised, well-structured survey of a hundred or more worlds is unlikely to occur due to the time required for such a survey and the proposal-based nature of time allocation. Therefore, it is likely that the population of planets studied will be somewhat random, with the added complexity of different instruments being used and so some of the hurdles discussed here will still be relevant. Additionally, the SNR achieved will vary and so comparative studies will have to be careful when drawing conclusions, even if the same instruments are used.

To overcome these hurdles, one requires missions with dedicated exoplanet surveys which will allow researchers to pose and, hopefully, answer specific questions on the nature of exo-atmospheres. By allowing a large population of targets to be selected with these questions in mind, Twinkle and Ariel will be better placed to provide demographical insights into the atmospheres of exoplanets, revolutionising our understanding of them in the process \citep[e.g.][]{edwards_ariel,changeat2020alfnoor,tinetti_ariel2,edwards_ariel2,TwinkleSPIE}. However, these missions will of course have their own limitations in terms of data quality. For instance, Ariel will only provide photometric data at visible wavelengths and so may struggle to disentangle the spectral features of optical absorbers. Furthermore, these surveys will recoup a higher science yield if they are constructed upon robust prior knowledge instead of undertaking a blind search for trends.

Therefore, we must strive to utilise each facility in ways which complement their respective capabilities. It is undeniable that JWST, and the continued use of Hubble and other facilities, will provide critical insights into the nature of a diverse set of worlds, with JWST in particular facilitating extraordinary sensitivity and thus offering the chance to probe for extremely small signals (e.g. secondary atmospheres). The knowledge gained must then be leveraged to inform us of how to use these dedicated surveys to best understand the population at large. By striving for these meticulous chemical surveys we will then truly begin to understand the demographics of exoplanet atmospheres.


\section*{Acknowledgements}

BE is a Laureate of the Paris Region fellowship programme which is supported by the Ile-de-France Region and has received funding under the Horizon 2020 innovation framework programme and the Marie Sklodowska-Curie grant agreement no. 945298. QC is funded by the European Space Agency under the 2022 ESA Research Fellowship Program. NS acknowledges support from the PSL Iris-OCAV project, and from NASA (Grant \#80NSSC19K0336). OV acknowledges funding from the ANR project `EXACT' (ANR-21-CE49-0008-01), from the Centre National d'\'{E}tudes Spatiales (CNES), and from the CNRS/INSU Programme National de Plan\'etologie (PNP). This project has also received funding from the European Research Council (ERC) under the European Union's Horizon 2020 research and innovation programme (grant agreement No 758892, ExoAI) and from the Science and Technology Funding Council (STFC) grant ST/S002634/1 and ST/T001836/1 and from the UK Space Agency grant ST/W00254X/1. 

We thank the referee of this manuscript for taking the time to read our work and provide feedback. Their constructive comments guided the direction of the paper, thereby improving the quality of the final results.\\

\textbf{Computing:} We acknowledge the availability and support from the High Performance Computing platforms (HPC) from the Simons Foundation (Flatiron), DIRAC and OzSTAR, which provided the computing resources necessary to perform this work. This work utilised the Cambridge Service for Data Driven Discovery (CSD3), part of which is operated by the University of Cambridge Research Computing on behalf of the STFC DiRAC HPC Facility (www.dirac.ac.uk). The DiRAC component of CSD3 was funded by BEIS capital funding via STFC capital grants ST/P002307/1 and ST/R002452/1 and STFC operations grant ST/R00689X/1. DiRAC is part of the National e-Infrastructure. Additionally, this work utilised the OzSTAR national facility at Swinburne University of Technology. The OzSTAR program receives funding in part from the Astronomy National Collaborative Research Infrastructure Strategy (NCRIS) allocation provided by the Australian Government. \\

\clearpage

\textbf{Software:} Iraclis \citep{tsiaras_hd209}, TauREx3 \citep{al-refaie_taurex3}, ExoTETHyS \citep{morello_exotethys}, PyLightcurve \citep{tsiaras_plc}, Astropy \citep{astropy}, h5py \citep{hdf5_collette}, emcee \citep{emcee}, Matplotlib \citep{Hunter_matplotlib}, Multinest \citep{Feroz_multinest,buchner_multinest}, Pandas \citep{mckinney_pandas}, Numpy \citep{oliphant_numpy}, SciPy \citep{scipy}, corner \citep{corner}.\\

\textbf{Data:} This work is based upon publicly-available observations taken with the NASA/ESA Hubble Space Telescope obtained from the Space Telescope Science Institute, which is operated by the Association of Universities for Research in Astronomy, Inc., under NASA contract NAS 5–26555. These were obtained from the Hubble Archive which is part of the Mikulski Archive for Space Telescopes. We are thankful to those who operate the Hubble Space Telescope and the corresponding archive, the public nature of which increases scientific productivity and accessibility \citep{peek2019}.

For each observation, the associated proposal number and principal investigator are given in Tables \ref{tab:previous_data}, \ref{tab:new_data} and \ref{tab:lit_data}. Where a NASA ADS entry could be found for the proposal, they are given here: 11622 \citep{knutson_prop11622}, 12181 \citep{deming_12181_prop}, 12449 \citep{deming_prop12449}, 12473 \citep{sing_prop12473}, 12482 \citep{desert_prop12482}, 12881 \citep{mccullough_prop12881}, 12956 \citep{huitson_prop12956}, 13021 \citep{bean_prop13021}, 13431 \citep{huitson_prop13431}, 13467 \citep{bean_prop13467}, 13501 \citep{knutson_prop13501}, 13665 \citep{benneke_prop13665}, 14050 \citep{kreidberg_w103_wfc3_w18_spit_prop}, 14218 \citep{berta_prop14218}, 14767 \citep{sing_14767_prop},  14455 \citep{petigura_prop14455}, 14468 \citep{evans_14468_prop}, 14619 \citep{spake_prop14619}, 14642 \citep{stevenson_prop14642}, 14664 \citep{beatty_14664_prop}, 14682 \citep{benneke_prop14682}, 14915 \citep{kreidberg_prop14915}, 14918 \citep{wakeford_prop14918}, 15134 \citep{evans_15134_prop}, 15138 \citep{jontof_prop15138}, 15255 \citep{colon_prop15255}, 15301 \citep{carone_proposal},  15333 \citep{crossfield_prop15333}, 15698 \citep{beatty_15698_prop}, 16083 \citep{todorov_prop}, 16194 \citep{desert_prop16194}, 16267 \citep{dressing_prop}, 16450 \citep{lothringer_prop}, 16457 \citep{edwards_prop}, 16462 \citep{panwar_prop}. \\

\textbf{Planet Discovery Papers:} The characterisation of exoplanetary atmospheres cannot occur without first knowing of the planet's existence. We are therefore grateful to all those who have contributed to planet discovery efforts. Thus, the works announcing the discovery of all planets studied here are now given:

CoRoT-1\,b \citep{barge_corot1},
GJ\,436\,b \citep{butler_gj436,gillon_gj436},
GJ\,1214\,b \citep{Charbonneau_GJ1214},
GJ\,3470\,b \citep{bonfils_gj3470},
HAT-P-1\,b \citep{bakos_h1},
HAT-P-2\,b \citep{bakos_h2},
HAT-P-3\,b \citep{torres_h3},
HAT-P-7\,b \citep{pal_h7},
HAT-P-11\,b \citep{bakos_h11},
HAT-P-12\,b \citep{hartman_h12},
HAT-P-17\,b \citep{howard_h17},
HAT-P-18\,b \citep{hartman_h18},
HAT-P-26\,b \citep{hartman_h26},
HAT-P-32\,b \citep{hartman_h32},
HAT-P-38\,b \citep{sato_h38},
HAT-P-41\,b \citep{hartman_h41},
HD\,3167\,c \citep{vanderburg_hd3167},
HD\,106315\,c \citep{rodriguez_hd106,crossfield_hd106},
HD\,149026\,b \citep{sato_hd149},
HD\,189733\,b \citep{bouchy_hd189},
HD\,209458\,b \citep{charbonneau_hd209,henry_hd209},
HD\,219666\,b \citep{esposito_hd219666},
HD\,97658\,b \citep{howard_hd97658},
HIP\,41378\,b \citep{vanderburg_hip41378},
HIP\,41378\,f \citep{vanderburg_hip41378},
K2-18\,b \citep{montet_k218,crossfield_k218},
K2-24\,b \citep{petigura_k2_24,sinukoff_k2_24},
KELT-1\,b \citep{siverd_k1},
KELT-7\,b \citep{Bieryla_2015},
KELT-11\,b \citep{pepper_k11},
Kepler-9\,b \citep{holman_k9},
Kepler-9\,c \citep{holman_k9},
Kepler-51\,b \citep{steffen_k51,masuda_k51},
Kepler-51\,d \citep{masuda_k51},
Kepler-79\,d \citep{jontof_k79},
LTT\,9779\,b \citep{jenkins_ltt9779},
TOI-270 c \citep{gunther_toi270},
TOI-270 d \citep{gunther_toi270},
TOI-674 b \citep{murgas_toi674},
TrES-2\,b \citep{odonovan_t2},
TrES-4\,b \citep{mandushev_t4},
V1298\,Tau\,b, \citep{david_v1298_b}
V1298\,Tau\,c, \citep{david_v1298_all}
WASP-6\,b \citep{gillon_w6},
WASP-12\,b \citep{hebb_w12},
WASP-17\,b \citep{anderson_w17},
WASP-18\,b \citep{hellier_w18},
WASP-19\,b \citep{hellier_w19},
WASP-29\,b \citep{hellier_2010},
WASP-31\,b \citep{anderson_w31},
WASP-39\,b \citep{faedi_w39},
WASP-43\,b \citep{hellier_2011},
WASP-52\,b \citep{hebrard_w52},
WASP-62\,b \citep{hellier_2012},
WASP-63\,b \citep{hellier_2012},
WASP-67\,b \citep{hellier_2012},
WASP-69\,b \citep{anderson_w69},
WASP-74\,b \citep{hellier_2015},
WASP-76\,b \citep{west_w76-82-90},
WASP-79\,b \citep{smalley_w79},
WASP-80\,b \citep{triaud_w80_dis},
WASP-96\,b \citep{hellier_2014},
WASP-101\,b \citep{hellier_2014},
WASP-103\,b \citep{gillon_wasp103},
WASP-107\,b \citep{anderson_w107},
WASP-117\,b \citep{lendl_wasp117},
WASP-121\,b \citep{delrez_w121},
WASP-127\,b \citep{lam_w127},
WASP-178\,b \citep{hellier_w178,rodriguez_w178},
XO-1\,b \citep{mccullough_xo1}.

\clearpage

\section*{Appendix 1: Light curve fitting with Iraclis}

We carried out the analysis of the transit data using Iraclis, our highly-specialised software for processing WFC3 spatially scanned spectroscopic images \citep{tsiaras_hd209,tsiaras_55cnce,tsiaras_30planets} which has been used in a number of studies \citep[e.g.][]{libby_gj1132,brande_toi674,garcia_t1h}. The reduction process included the following steps: zero-read subtraction, reference-pixels correction, non-linearity correction, dark current subtraction, gain conversion, sky background subtraction, calibration, flat-field correction and bad-pixels/cosmic-rays correction. Then we extracted the white (1.088-1.68 $\mu$m) and the spectral light curves from the reduced images, taking into account the geometric distortions caused by the tilted detector of the WFC3 infrared channel.

We fitted the light curves using our transit model package PyLightcurve \citep{tsiaras_plc} with the transit parameters from Table \ref{tab:planet_para_new}. The limb-darkening coefficients were calculated using ExoTETHyS \citep{morello_exotethys} and based on the PHOENIX 2018 models from \citet{phoenix}. The stellar parameters are also given in Table \ref{tab:star_para_new}.

During our fitting of the white light curve, the planet-to-star radius ratio and the mid-transit time were the only free parameters, along with a model for the systematics \citep{Kreidberg_GJ1214b_clouds, tsiaras_hd209}. It is common for WFC3 exoplanet observations to be affected by two kinds of time-dependent systematics: the long-term and short-term `ramps'. The first affects each HST visit and has a linear behaviour, while the second affects each HST orbit and has an exponential behaviour. The formula we used for the white light curve systematics (Rw) was the following:

\begin{table*}
    \centering
     \caption{Stellar parameters for observations acquired from the literature. For consistency, these match those used in the original studies.}
    \begin{tabular}{ccccccc}  \hline  \hline
    Star & Fe/H & Temperature [K] & log(g) & Radius [R$_\odot$] & Mass [M$_\odot$] & Reference \\  \hline
    
    GJ 436 & 0.02 & 3416 & 4.84 & 0.455 & 0.47 & \citet{lanotte_gj436}\\
    
    GJ 3470 & 0.17 & 3652 & 4.78 & 0.48 & 0.51 & \citet{biddle_gj3470}\\
    
    HAT-P-1 & 0.13 & 5975 & 4.45 & 1.15 & 1.12 & \citet{bakos_h1}\\
    
    HAT-P-3 & 0.27 & 5185 & 4.56 & 0.833 & 0.936 & \citet{torres_2008}\\
    
    HAT-P-11 & 0.31 & 4780 & 4.59 & 0.75 & 0.81 & \citet{bakos_h11}\\
    
    HAT-P-12 & -0.29 & 4650 & 4.61 & 0.701 & 0.733 & \citet{hartman_h12}\\
    
    HAT-P-17 & 0.00 & 5246 & 4.52 & 0.838 & 0.857 & \citet{howard_h17}\\
    
    HAT-P-18 & 0.1 & 4870 & 4.57 & 0.749 & 0.77 & \citet{esposito_h18}\\
    
    HAT-P-26 & -0.04 & 5079 & 4.56 & 0.788 & 0.816 & \citet{hartman_h26}\\
    
    HAT-P-32 & -0.04 & 6207 & 4.33 & 1.219 & 1.16 & \citet{hartman_h32}\\
    
    HAT-P-38 & 0.06 & 5330 & 4.45 & 0.923 & 0.886 & \citet{sato_h38}\\
    
    HAT-P-41 & 0.21 & 6390 & 4.14 & 1.683 & 1.683 & \citet{hartman_h41}\\
    
    HD 3167 & 0.03 & 5286 & 4.53 & 0.835 & 0.877 & \citet{gandolfi_hd3167}\\
    
    HD 106315 & -0.276 & 6256 & 4.24 & 1.31 & 1.079 & \citet{guilluy_aresIV}\\
    
    HD 149026 & 0.36 & 6160 & 4.28 & 1.368 & 1.294 & \citet{torres_2008}\\
    
    HD 189733 & -0.03 & 5040 & 4.59 & 0.756 & 0.806 & \citet{torres_2008}\\
    
    HD 209458 & 0.00 & 6065 & 4.36 & 1.155 & 0.806 & \citet{torres_2008}\\
    
    KELT-7 & 0.139 & 6789 & 4.15 & 1.732 & 1.535 & \citet{Bieryla_2015}\\
    
    KELT-11 & 0.17 & 5375 & 3.70  & 2.69 & 1.44 & \citet{beatty_k11} \\
    
    K2-18 & 0.123 & 3457 & 4.77 & 0.411 & 0.359 & \citet{benneke_k218_dis} \\
    
    
    
    WASP-12 & 0.33 & 6360 & 4.16 & 1.657 & 1.434 & \citet{collins_w12})\\
    
    WASP-17 & -0.25 & 6550 & 4.149 & 1.583 & 1.286 & \citet{southworth_w17}\\
    
    WASP-29 & 0.11 & 4800 & 4.54 & 0.808 & 0.825 & \citet{hellier_2010}\\
    
    WASP-31 & -0.2 & 6302 & 4.31 & 1.252 & 1.163 & \citet{anderson_w31}\\
    
    WASP-39 & -0.12 & 5400 & 4.50 & 0.895 & 0.93 & \citet{faedi_w39}\\
    
    WASP-43 & -0.05 & 4400 & 4.65 & 0.67 & 0.58 & \citet{hellier_2011}\\
    
    WASP-52 & 0.03 & 5000 & 4.58 & 0.79 & 0.87 & \citet{hebrard_w52}\\
    
    WASP-62 & 0.04 & 6230 & 4.45 & 1.29 & 1.28 & \citet{brown_w79_w62} \\
    
    WASP-63 & 0.08 & 5570 & 4.01 & 1.88 & 1.32 & \citet{hellier_2012}\\
    
    WASP-67 & -0.07 & 5200 & 4.5 & 0.87 & 0.87 &\citet{hellier_2012}\\
    
    WASP-69 & 0.144 & 4715 & 4.54 & 0.813 & 0.826 & \citet{anderson_w69}\\
    
    WASP-74 & 0.39 & 5970 & 4.18 & 1.64 & 1.48 & \citet{hellier_2015}\\
    
    WASP-76 & 0.366 & 6329 & 4.20 & 1.756 & 1.458 & \citet{ehrenreich_wasp76} \\
    
    WASP-79 & 0.03 & 6600 & 4.06 & 1.51 & 1.39 & \citet{brown_w79_w62} \\
    
    WASP-80 & -0.13 & 4143 & 4.66 & 0.586 & 0.577 & \citet{triaud_w80} \\
    
    WASP-96 & 0.14 & 5540 & 4.42 & 1.05 & 1.06 & \citet{hellier_2014} \\
    
    WASP-101  & 0.2 & 6380 & 4.35 & 1.29 & 1.34 & \citet{hellier_2014}\\
    
    WASP-117 & -0.11 & 6038 & 4.28 & 1.126 & 1.17 & \citet{lendl_wasp117}\\
    
    WASP-127 & -0.18 & 5750 & 3.90 & 1.39 & 1.08 & \citet{lam_w127} \\
    
    XO-1 & 0.02 & 5750 & 4.51 & 0.934 & 1.027 & \citet{torres_2008}\\ \hline \hline
   
    \end{tabular}
    
    \label{tab:star_para_lit}
\end{table*}{}

\begin{table*}
    \centering
    \caption{Planet parameters for observations acquired from the literature. For consistency, these match those used in the original studies.}
    \begin{tabular}{ccccccccc}  \hline  \hline
    Planet & Mass & Radius & Period & i & a/Rs & e & $\omega$ & Ref\\
    Name & [M$_J$] &[$R_J$]&[days] & [$^\circ$] & & & [$^\circ$] &\\\hline
    GJ 436 b & 0.08 & 0.366 & 2.64389803 & 86.858 & 14.54 & 0.1616 & 327.2 & \citet{lanotte_gj436} \\
    
    GJ 3470 b & 0.043 & 0.346 & 3.3366487 & 88.88 & 13.94 & - & - & \citet{biddle_gj3470} \\
    
    HAT-P-1 b & 0.53 & 1.36 & 4.46529 & 85.9 & 10.247 & -  & - & \citet{bakos_h1} \\
    
    HAT-P-3 b & 0.596 & 0.899 & 2.899703 & 87.24 & 10.59 & - & - & \citet{torres_2008}\\
    
    HAT-P-11 b  & 0.081 & 0.422 & 4.8878162 & 88.5 & 15.58 & 0.198 & 355.2 & \citet{bakos_h11} \\
    
    HAT-P-12 b & 0.211 & 0.959 & 3.2130598 & 89 & 11.77 & - & - & \citet{hartman_h12}\\
    
    HAT-P-17 b & 0.534 & 1.01 & 10.338523 & 89.2 & 22.63 & 0.342 & 201 & \citet{howard_h17}\\
    
    HAT-P-18 b & 0.196 & 0.947 & 5.507978 & 88.79 & 16.67 & - & - & \citet{hartman_h18}\\
    
    HAT-P-26 b & 0.057 & 0.549 & 4.234515 & 88.6 & 13.44 & - & - & \citet{hartman_h26}\\
    
    HAT-P-32 b & 0.86 & 1.789 & 2.150008 & 88.9 & 6.05 & - & - & \citet{hartman_h32}\\
    
    HAT-P-38 b & 0.267 & 0.825 & 4.640382 & 88.3 & 12.17 & - & - & \citet{sato_h38}\\
    
    HAT-P-41 b & 0.8 & 1.685 & 2.694047 & 87.7 & 5.44 & - & - & \citet{hartman_h41}\\
    
    HD 3167 c & 0.0262 & 0.244 & 29.84622 & 89.6 & 46.5 & 0.05 & 178 & \citet{gandolfi_hd3167} \\
    
    HD 106315 c & 0.0459 & 0.444 & 21.05731 & 88.17 & 25.10 & 0.052 & 157 &   \citet{guilluy_aresIV}\\
    
    HD 149026 b & 0.359 & 0.654 & 2.87598 & 90 & 7.11 & - & - & \citet{torres_2008}\\
    
    HD 189733 b & 1.144 & 1.138 & 2.218573 & 85.58 & 8.81 & - & - & \citet{torres_2008}\\
    
    HD 209458 b & 0.685 & 1.359 & 3.524746 & 86.71 & 8.76 & - & - & \citet{torres_2008}\\
    
    KELT-7 b & 1.28 & 1.533 & 2.7347749 & 83.76 & 5.49 & - & - & \citet{Bieryla_2015}\\
    
    KELT-11 b & 0.171 & 1.35 & 4.73613 & 85.3 & 4.98 & 0.0007 & 0 & \citet{beatty_k11}\\
    
    K2-18 b & 0.025 & 0.2033 & 32.94007 & 81.3 & 89.56 & - & - & \citet{cloutier_k218} \\
    
    WASP-12 b & 1.47 & 1.9 & 1.0914203 & 83.37 & 3.039 & - & - & \citet{collins_w12}\\
    
    WASP-17 b & 0.78 & 1.87 & 3.735430 & 86.63 & 8.97 & - & - & \citet{southworth_w17}\\
    
    WASP-29 b & 0.244 & 0.792 & 3.922727 & 88.8 & 12.415 & - & - & \citet{hellier_w18}\\
    
    WASP-31 b & 0.478 & 1.549 & 3.4059096 & 84.41 & 7.99 & - & - & \citet{anderson_w31}\\
    
    WASP-39 b & 0.28 & 1.27 & 4.055259 & 87.83 & 11.647 & - & - & \citet{faedi_w39}\\
    
    WASP-43 b & 1.78 & 0.93 & 0.813475 & 82.6 & 5.124 & - & - & \citet{hellier_2011}\\
    
    WASP-52 b & 0.46 & 1.27 & 1.7497798 & 85.35 & 7.401 & - & - & \citet{hebrard_w52}\\
    
    WASP-62 b & 0.58 &  1.34 & 4.411953 & 88.5 & 9.5253 & - & - & \citet{brown_w79_w62}\\
    
    WASP-63 b & 0.38 & 1.43 & 4.37809 & 87.8 & 6.773 & - & - & \citet{hellier_2012} \\
    
    WASP-67 b & 0.42 & 1.4 & 4.61442 & 85.8 & 12.835 & - & - & \citet{hellier_2012} \\
    
    WASP-69 b & 0.26 & 1.057 & 3.8681382 &  86.71 & 11.953 & - & - & \citet{anderson_w69} \\
    
    WASP-74 b & 0.95 & 1.56 & 2.13775 & 79.81 & 4.861 & - & - & \citet{hellier_2015} \\
    
    WASP-76 b & 0.894 & 1.854 & 1.80988198 & 89.623 & 4.08 & - & - & \citet{ehrenreich_wasp76} \\
    
    WASP-79 b & 0.85 & 1.53 & 3.662387 & 86.1 & 6.069 & - & - & \citet{brown_w79_w62} \\
    
    WASP-80 b & 0.538 & 0.999 & 3.06785234 & 89.02 & 12.63 & - & - & \citet{triaud_w80} \\
    
    WASP-96 b & 0.48 & 1.20 & 3.4252602 & 85.14 & 8.84 & - & - & \citet{hellier_2014}\\
    
    WASP-101 b & 0.5 & 1.41 & 3.585722 & 85.0 & 8.445 & - & - & \citet{hellier_2014}\\
    
    WASP-117 b & 0.276 & 1.021 & 10.02 & 89.14 & 17.39 & 0.302 & 242 & \citet{lendl_wasp117} \\
    
    WASP-127 b & 0.18 & 1.37 & 4.17807015 & 88.2 & 7.846 & - & - & \citet{palle_w127} \\
    
    XO-1 b & 0.918 & 1.206 & 3.941534 & 88.81 & 11.55 & - & - & \citet{torres_2008}\\ \hline\hline
   
    \end{tabular}

    \label{tab:planet_para_lit}
\end{table*}{}

\begin{table*}
    \centering
    \caption{Stellar parameters utilised in this work on observations fitted here with Iraclis.}
    \begin{tabular}{ccccccc}  \hline  \hline
    Star & Fe/H & Temperature [K] & log(g) & Radius [R$_\odot$] & Mass [M$_\odot$] & Reference \\  \hline
    
    CoRoT-1 & -0.3 & 5950 & 4.25 & 1.11 & 0.95 & \citet{barge_corot1}\\
    
    GJ 1214 & 0.29 & 3250 & 5.026 & 0.215 & 0.178 & \citet{cloutier_gj1214} \\
    
    HAT-P-2 & 0.14 & 6290 & 4.16 & 1.64 & 1.36 & \citep{pal_h2} \\
    
    HAT-P-7 & 0.26 & 6350 & 4.07 & 1.84 & 1.47 & \citet{esteves_h7_2015}\\
    
    HD 97658 & -0.23 & 5212 & 4.64 & 0.728 & 0.85 & \citet{ellis_hd97658}\\
    
    HD 219666 & 0.04 & 	5527 & 4.40 &  1.03 & 0.92 & \citet{esposito_hd219666}\\
    
    HIP 41378 & -0.10 & 6320 & 4.294 & 1.273 & 1.16 & \citet{santerne_hip41378} \\
    
    KELT-1 & 0.008 & 6518 & 4.337 & 1.462 & 1.324 & \citet{siverd_k1}\\
    
    Kepler-9 & 0.05 & 5774 & 4.49 & 0.958 & 1.022 & \citet{borsato_k9}\\
    
    Kepler-51 & 0.05 & 5670 & 4.7 & 0.881 & 0.985 & \citet{libby_k51}\\
    
    K2-24 & 0.34 & 5625 & 4.29 & 1.16 & 1.07 & \citet{petigura_k2_24}\\
    
    LTT 9779 & 0.27 & 5443 & 4.35 & 0.949 & 0.770 & \citet{jenkins_ltt9779} \\
    
    TOI-270 & -0.20 & 3506 & 4.872 & 0.378 & 0.386 & \citet{van_eylen_toi270}\\
    
    TOI-674 & 0.17 & 3514 & 5.28 & 0.42 & 0.42 & \citet{murgas_toi674}\\
    
    TrES-2 & -0.15 & 5850 & 4.4 & 0.952 & 0.99 & \citet{kipping_t2}\\
    
    TrES-4 b & 0.14 & 6200 & 4.064 & 1.816 & 1.394 & \citet{torres_2008}\\
    
    V1298 Tau & 0.14 & 4970 & 4.246 & 1.314 & 1.099 & \citet{david_v1298_b}\\
    
    WASP-6 & -0.20 & 5450 & 4.60 & 0.73 & 0.55 & \citet{Stassun_planetparam}\\
    
    WASP-18 & 0.11 & 6431 & 4.47 & 1.26 & 1.46 & \citet{shporer_w18}\\
    
    WASP-19 & 0.15 & 5568 & 4.45 & 1.004 & 0.904 & \citet{wong_w19} \\
    
    WASP-103 & 0.06 & 6100 & 4.22 & 1.436 & 1.22 & \citet{gillon_wasp103} \\
    
    WASP-107 & -0.02 & 4430 & 4.5 & 0.66 & 0.69 & \citet{anderson_w107}\\
    
    WASP-121 & 0.13 & 6460 & 4.24 & 1.458 & 1.353 & \citet{delrez_w121}\\
    
    WASP-178 & -0.06 & 8640 & 4.211 & 1.801 & 1.93 & \citet{rodriguez_w178} \\\hline\hline
   
    \end{tabular}
    
    \label{tab:star_para_new}
\end{table*}{}

\begin{table*}
    \centering
    \caption{Planet parameters utilised in this work on observations fitted here with Iraclis.}
    \begin{tabular}{ccccccccc}  \hline  \hline
    Planet & Mass & Radius & Period & i & a/Rs & e & $\omega$ & Ref\\
    Name & [M$_J$] &[$R_J$]&[days] & [$^\circ$] & & & [$^\circ$] &\\\hline
    
    CoRoT-1 b & 1.03 & 1.49 & 1.5089557 & 85.1 & 4.92 & - & - & \citet{barge_corot1}\\
    
    GJ 1214 b & 0.0257 & 0.2446 & 1.58040433 & 88.7 & 14.85 & - &  - & \citep{cloutier_gj1214} \\
    
    HAT-P-2 b & 9.09 & 1.157 & 5.6334729 & 86.72  & 8.99 & 0.5171 & 185.22 & \citep{pal_h2}\\
    
    HAT-P-7 b & 1.781 & 1.419 & 2.2047 & 83.143 & 4.1545 & - & - & \citet{esteves_h7_2015} \\
    
    HD 97658 b & 0.026 & 0.189 & 9.4897116 & 89.05 & 24.2 & 0.05 & 0 & \citet{ellis_hd97658} \\
    
    HD 219666 b & 0.0522 & 0.420 & 6.03607 & 86.38 & 13.27 & - & - & \citet{esposito_hd219666}\\
    
    HIP 41378 b & 0.0217 & 0.2315 & 15.57208 & 88.75 &21.6721  & - & - & \citet{santerne_hip41378} \\
    
    HIP 41378 f & 0.0378 & 0.8208 & 542.07975 & 89.97 & 231.41717 &  - & - & \citet{santerne_hip41378} \\
    
    KELT-1 b & 27.23 & 1.11 & 1.217514 & 87.8 & 3.62 & 0.0099 & 61 & \citet{siverd_k1}\\
    
    Kepler-9 b & 0.137 & 0.74 & 19.23891 & 88.982 & 31.3 & 0.0609 & 357 & \citet{borsato_k9}\\
    
    Kepler-9 c & 0.0941 & 0.721 & 38.9853 & 89.188 & 49.8 & 0.06691 & 167.5 & \citet{borsato_k9}\\
    
    Kepler-51 b & 0.0116 & 0.615 & 45.1542 & 90 & 60.22 &  0.01 & 53.3 & \citet{libby_k51}\\
    
    Kepler-51 d & 0.0179 & 0.844 & 130.1845 & 90 & 121.98 & 0.01 & 347.4 & \citet{libby_k51}\\
    
    K2-24 b & 0.0598 & 0.48 & 20.88977 & 90 & 28.57 & 0.06 & 0 & \citet{petigura_k2_24}\\
    
    LTT 9779 b & 0.09225 & 0.421 & 0.792052 & 76.39 & 3.877 & - & - & \citet{jenkins_ltt9779} \\
    
    TOI-270 c & 0.01932 & 0.20805 & 5.660183 & 89.35 & 25.741 & 0.031 & 34.4 & \citet{van_eylen_toi270}\\
    
    TOI-270 d & 0.01504 & 0.17888 & 11.38027 & 89.69 & 41.007 & 0.034 & 11.5 & \citet{van_eylen_toi270}\\
    
    TOI-674 b & 0.0743 & 0.468 & 1.97714 & 87.21 & 12.80 & - & - & \citet{murgas_toi674}\\
    
    TrES-2 b & 1.202 & 1.187 & 2.47061892 & 84.07 & 8.06 & 0.018 & 268 & \citet{kipping_t2} \\
    
    TrES-4 b &  0.92 & 1.751 & 3.553945 & 82.81 & 6.03 & - & - & \citet{torres_2008} \\
    
    V1298 Tau b & 0.64$^*$ & 0.911 & 24.13861 & 89.35 & 27.6 & 0.112 & 91 & \citet{david_v1298_b} \\
    
    V1298 Tau c & - & 0.499 & 8.24958 & 88.49 & 13.19 & - & - & \citet{david_v1298_all} \\
    
    WASP-6 b & 0.37 & 1.03 & 3.3610100 & 88.47 & 10.62 & 0.05 & 0 & \citet{Stassun_planetparam}\\
    
    WASP-18 b & 10.2 & 1.24 & 0.94145223 & 	83.5 & 3.48 & 0.0051 & 275 & \citet{cortes_w18}\\
    
    WASP-19 b & 1.069 & 1.392 & 0.788838989 & 78.78 & 3.46 & 0.002 & 259 & \citet{wong_w19}\\
    
    WASP-103 b & 1.49 & 1.528 & 0.925542& 86.3 & 2.978  & - & - & \citet{gillon_wasp103} \\
    
    WASP-107 b & 0.12 & 0.94 & 5.721490 & 89.7 & 18.2  & - & - & \citet{anderson_w107}\\
    
    WASP-121 b & 1.183 & 1.865 & 1.2749255 & 87.6 & 3.754 & - & - & \citet{delrez_w121}\\
    
    WASP-178 b & 1.41 & 1.66 & 3.3448412 & 84.45 & 6.49 & - & - & \citet{rodriguez_w178} \\ \hline
    
    \multicolumn{9}{c}{$^*$Mass taken from \citet{suarez_v1298}.} \\
    
    \hline\hline
   
    \end{tabular}
    \label{tab:planet_para_new}
\end{table*}{}

\begin{equation}
    R_w(t) = n^{scan}_w(1-r_a(t - T_0))(1-r_{b1}e^{-r_{b2}(t-t_)}),
\end{equation}

\noindent where $t$ is time, $n^{scan}_w$ is a normalisation factor, $T_0$ is the mid-transit time, $t_o$ is the time when each HST orbit starts, $r_a$ is the slope of a linear systematic trend along each HST visit and ($r_{b1},r_{b2}$) are the coefficients of an exponential systematic trend along each HST orbit. The normalisation factor we used ($n^{scan}_w$) was changed to $n^{for}_w$ for upward scanning directions (forward scanning) and to $n^{rev}_w$ for downward scanning directions (reverse scanning). The reason for using separate normalisation factors is the slightly different effective exposure time due to the known upstream/downstream effect \citep{mccullough_wfc3_scan}. 

\begin{figure}
\includegraphics[width=0.95\columnwidth]{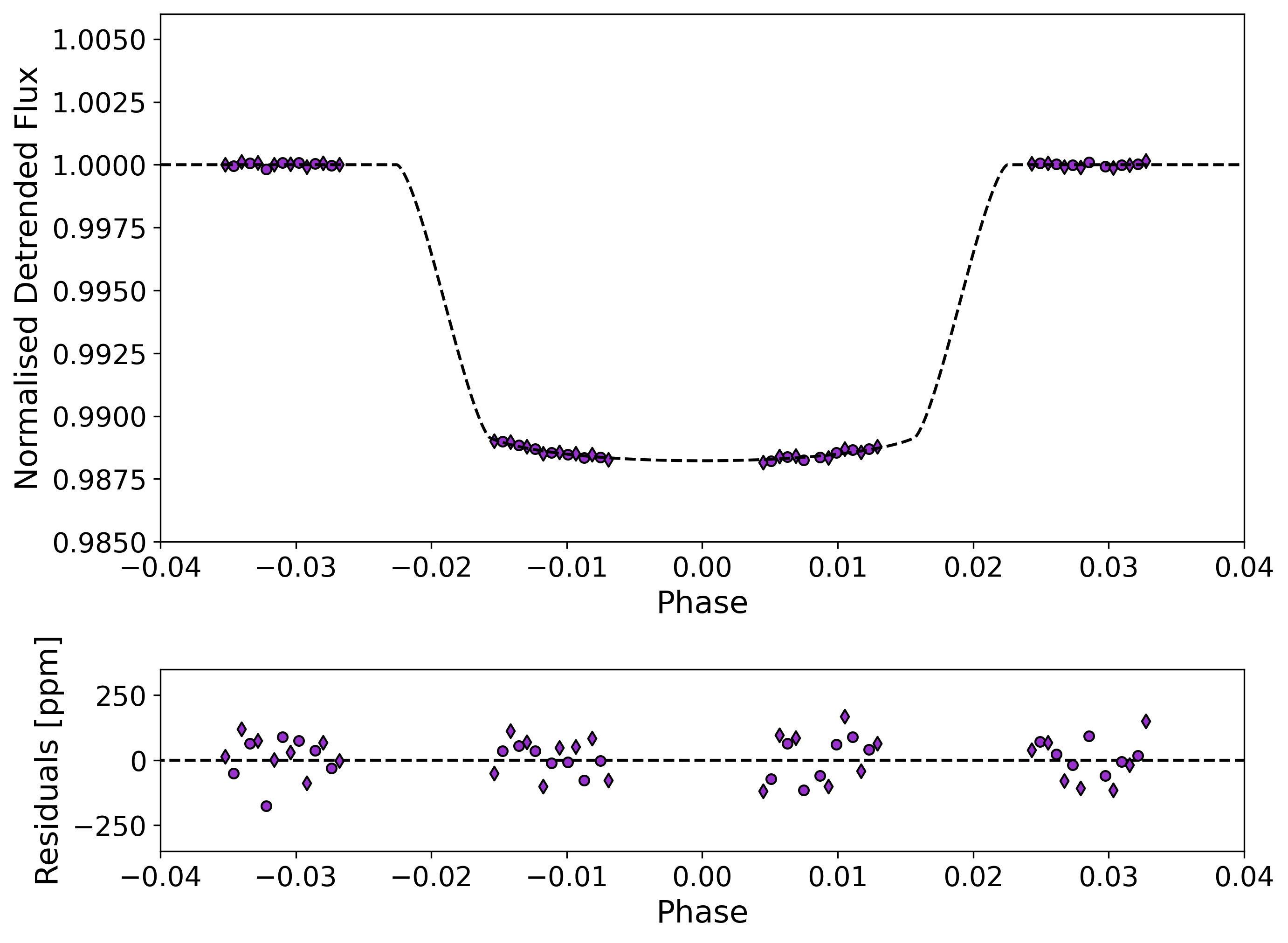}
\caption{White light-curve fit of WASP-178 b with \emph{Iraclis} (Top) and residuals (Bottom). This Figure is a sample from the Figure Set A1 that contains the same information for all the planets.}\label{figset:WLC}
\end{figure}

We fitted the white light curves using the formulae above and the uncertainties per pixel, as propagated through the data reduction process. However, it is common in HST/WFC3 data to have additional scatter that cannot be explained by the ramp model. For this reason, we scaled up the uncertainties in the individual data points, for their median to match the standard deviation of the residuals, and repeated the fitting \citep{tsiaras_30planets}. The only free parameters in our white fitting, other than the HST systematics, were the mid-transit time and the planet-to-star radius ratio. The full set of white light curve fits are given in Figure set A1, with an example given in Figure \ref{figset:WLC}.

\graphicspath{{./}{Figures/}{FiguresAppendix/}}

Next, we fitted the spectral light curves with a transit model (with the planet-to-star radius ratio being the only free parameter) along with a model for the systematics ($R_\lambda$) that included the white light curve \citep[divide-white method,][]{Kreidberg_GJ1214b_clouds} and a wavelength-dependent, visit-long slope \citep{tsiaras_hd209}. 

\begin{equation}
    R_\lambda(t) = n^{scan}_\lambda(1-\chi_\lambda(t-T_0))\frac{LC_w}{M_w},
\end{equation}{}

\noindent where $\chi_\lambda$ is the slope of a wavelength-dependent linear systematic trend along each HST visit, $LC_w$ is the white light curve and $M_w$ is the best-fit model for the white light curve. Again, the normalisation factor we used ($n^{scan}_\lambda$) was changed to $n^{for}_\lambda$ or $n^{for}_\lambda$ for upward or downward scanning directions respectively. Also, in the same way as for the white light curves, we performed an initial fit using the pipeline uncertainties and then refitted while scaling these uncertainties, for their median to match the standard deviation of the residuals.


\begin{table*}
    \centering
    \caption{White light curve depths, and transit mid times, for all observations fitted here. These will be added to the ExoClock database \citep{kokori_3}.}
    \begin{tabular}{cccccc}  \hline  \hline
    Planet Name & White Depth [$\%$] & Mid Time [BJD$_{\rm TDB}$] & Planet Name & White Depth [$\%$] & Mid Time [BJD$_{\rm TDB}$] \\ \hline
GJ1214b & 1.3456 $\pm$ 0.0102 & 2455478.57655 $\pm$ 5.3e-05 & Kepler-9b & 0.5898 $\pm$ 0.0117 & 2456093.7579 $\pm$ 0.000918\\ 
 GJ1214b & 1.3484 $\pm$ 0.0052 & 2456524.80419 $\pm$ 2.8e-05 & Kepler-9b & 0.596 $\pm$ 0.0096 & 2456151.4162 $\pm$ 0.000655 \\ 
 GJ1214b & 1.3491 $\pm$ 0.0064 & 2456516.90183 $\pm$ 3.1e-05 & Kepler-9c & 0.5776 $\pm$ 0.0084 & 2456098.4048 $\pm$ 0.000822 \\ 
 GJ1214b & 1.3652 $\pm$ 0.0058 & 2456480.552823 $\pm$ 1.9e-05 & Kepler-9c & 0.5685 $\pm$ 0.0097 & 2456293.768 $\pm$ 0.00091 \\ 
 GJ1214b & 1.3514 $\pm$ 0.0036 & 2456414.17563 $\pm$ 1.8e-05 & Kepler-51d & 0.9082 $\pm$ 0.0179 & 2457388.20161 $\pm$ 0.001853  \\ 
 GJ1214b & 1.3778 $\pm$ 0.0088 & 2456387.30978 $\pm$ 3.1e-05 & Kepler-51d & 0.9254 $\pm$ 0.0177 & 2457778.751421 $\pm$ 0.001762  \\ 
 GJ1214b & 1.3645 $\pm$ 0.0041 & 2456379.406717 $\pm$ 2e-05 & LTT9779b & 0.207 $\pm$ 0.0102 & 2459377.56157 $\pm$ 0.00018 \\ 
 GJ1214b & 1.361 $\pm$ 0.0045 & 2456366.763689 $\pm$ 1.9e-05 & TOI-270c & 0.4238 $\pm$ 0.0048 & 2458961.21466 $\pm$ 7.2e-05 \\ 
 GJ1214b & 1.3628 $\pm$ 0.0039 & 2456365.183036 $\pm$ 1.7e-05 & TOI-270c & 0.3341 $\pm$ 0.0086 & 2459125.371 $\pm$ 0.007551 \\ 
 GJ1214b & 1.3547 $\pm$ 0.0031 & 2456322.512144 $\pm$ 1.7e-05 & TOI-270c & 0.3881 $\pm$ 0.0077 & 2458904.60695 $\pm$ 0.000162 \\ 
 GJ1214b & 1.3577 $\pm$ 0.0048 & 2456219.786079 $\pm$ 1.9e-05 & TOI-270d & 0.2788 $\pm$ 0.0219 & 2459129.343 $\pm$ 0.009614 \\ 
 GJ1214b & 1.3519 $\pm$ 0.0051 & 2456211.883798 $\pm$ 2.6e-05 & TOI-674b & 1.3174 $\pm$ 0.0079 & 2459056.6099 $\pm$ 0.000775 \\ 
 GJ1214b & 1.3419 $\pm$ 0.0062 & 2456508.99984 $\pm$ 3.6e-05 & TOI-674b & 1.3207 $\pm$ 0.0052 & 2459042.7681 $\pm$ 0.000538 \\ 
 GJ1214b & 1.35 $\pm$ 0.0052 & 2456197.660221 $\pm$ 2.4e-05 & TOI-674b & 1.3064 $\pm$ 0.0067 & 2459040.79261 $\pm$ 7.8e-05 \\ 
 GJ1214b & 1.3414 $\pm$ 0.0039 & 2456203.981787 $\pm$ 2e-05 & TrES-2b & 1.6538 $\pm$ 0.012 & 2455479.53351 $\pm$ 0.000275 \\ 
 GJ1214b & 1.3853 $\pm$ 0.007 & 2455766.20993 $\pm$ 5.4e-05 & TrES-4b & 0.912 $\pm$ 0.0179 & 2455524.5351 $\pm$ 0.000372 \\ 
 GJ1214b & 1.3596 $\pm$ 0.0064 & 2455649.26003 $\pm$ 4.5e-05 & V1298Taub & 0.4706 $\pm$ 0.0134 & 2459119.0243 $\pm$ 0.000387 \\ 
 HAT-P-2b & 0.4729 $\pm$ 0.0025 & 2459204.1089 $\pm$ 0.000612 & V1298Tauc & 0.1102 $\pm$ 0.0087 & 2459505.9037 $\pm$ 0.007106 \\ 
 HAT-P-7b & 0.5388 $\pm$ 0.0138 & 2455761.29153 $\pm$ 0.000304 & WASP-6b & 2.0592 $\pm$ 0.0119 & 2457880.132159 $\pm$ 9.6e-05  \\ 
 HD97658b & 0.0906 $\pm$ 0.0058 & 2457491.0335 $\pm$ 0.001066 & WASP-18b & 0.9409 $\pm$ 0.0063 & 2456896.14839 $\pm$ 0.000278 \\ 
 HD97658b & 0.0846 $\pm$ 0.001 & 2457785.2052 $\pm$ 0.001584 & WASP-19b & 2.0506 $\pm$ 0.0136 & 2456820.79792 $\pm$ 8.1e-05 \\ 
 HD97658b & 0.0933 $\pm$ 0.001 & 2456665.462 $\pm$ 0.001538 & WASP-19b & 2.0249 $\pm$ 0.01 & 2455744.03242 $\pm$ 0.000164\\ 
 HD97658b & 0.0906 $\pm$ 0.0009 & 2456646.485232 $\pm$ 7.1e-05 & WASP-103b & 1.1449 $\pm$ 0.0113 & 2457237.05754 $\pm$ 0.000155 \\ 
 HIP41378b & 0.0388 $\pm$ 0.0023 & 2458133.32 $\pm$ 0.00763 & WASP-103b & 1.1818 $\pm$ 0.0038 & 2457080.64093 $\pm$ 8.4e-05 \\ 
 HIP41378b & 0.0361 $\pm$ 0.0026 & 2458989.7623 $\pm$ 0.000532 & WASP-107b & 2.0561 $\pm$ 0.0042 & 2457910.45444 $\pm$ 5.1e-05\\ 
 HIP41378b & 0.0357 $\pm$ 0.0077 & 2458242.325 $\pm$ 0.016892 & WASP-121b & 1.4518 $\pm$ 0.0027 & 2458191.11649 $\pm$ 4.1e-05  \\ 
 HIP41378f & 0.4709 $\pm$ 0.0015 & 2459355.10091 $\pm$ 0.000241 & WASP-121b & 1.4656 $\pm$ 0.0069 & 2457424.88673 $\pm$ 0.000115 \\ 
 K2-24b & 0.1971 $\pm$ 0.0017 & 2457574.4045 $\pm$ 0.000816 & WASP-121b & 1.4736 $\pm$ 0.0032 & 2458518.77202 $\pm$ 5.8e-05  \\ 
 KELT-1b & 0.5773 $\pm$ 0.0048 & 2457713.61848 $\pm$ 8.2e-05 & WASP-178b & 1.1462 $\pm$ 0.0052 & 2459265.1123 $\pm$ 0.000987 \\  \hline \hline
    \end{tabular}
    \label{tab:white_lc}
\end{table*}{}

\begin{figure*}
    \figurenum{A3}
    \centering
    \includegraphics[width=0.45\textwidth]{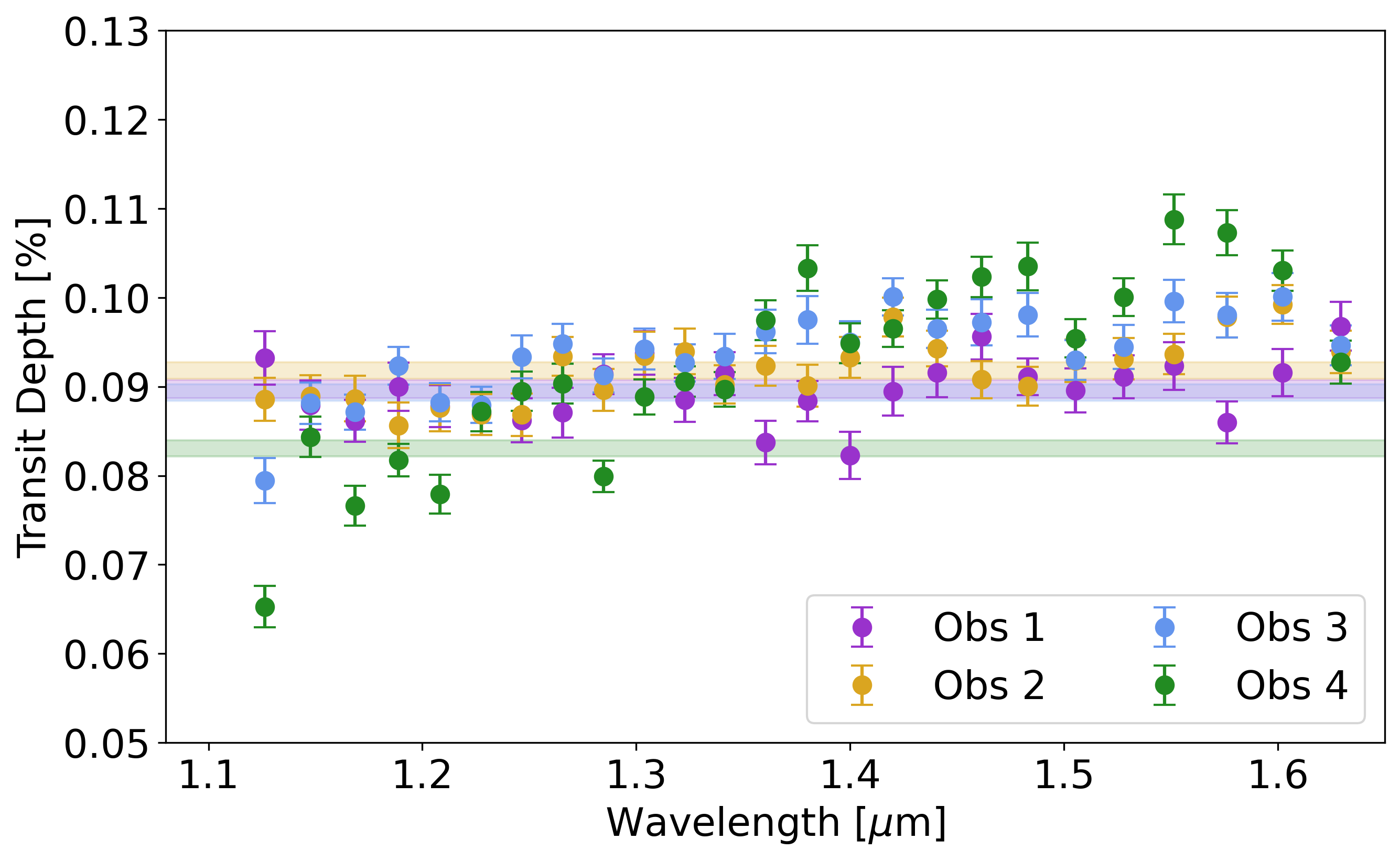}
    \includegraphics[width=0.45\textwidth]{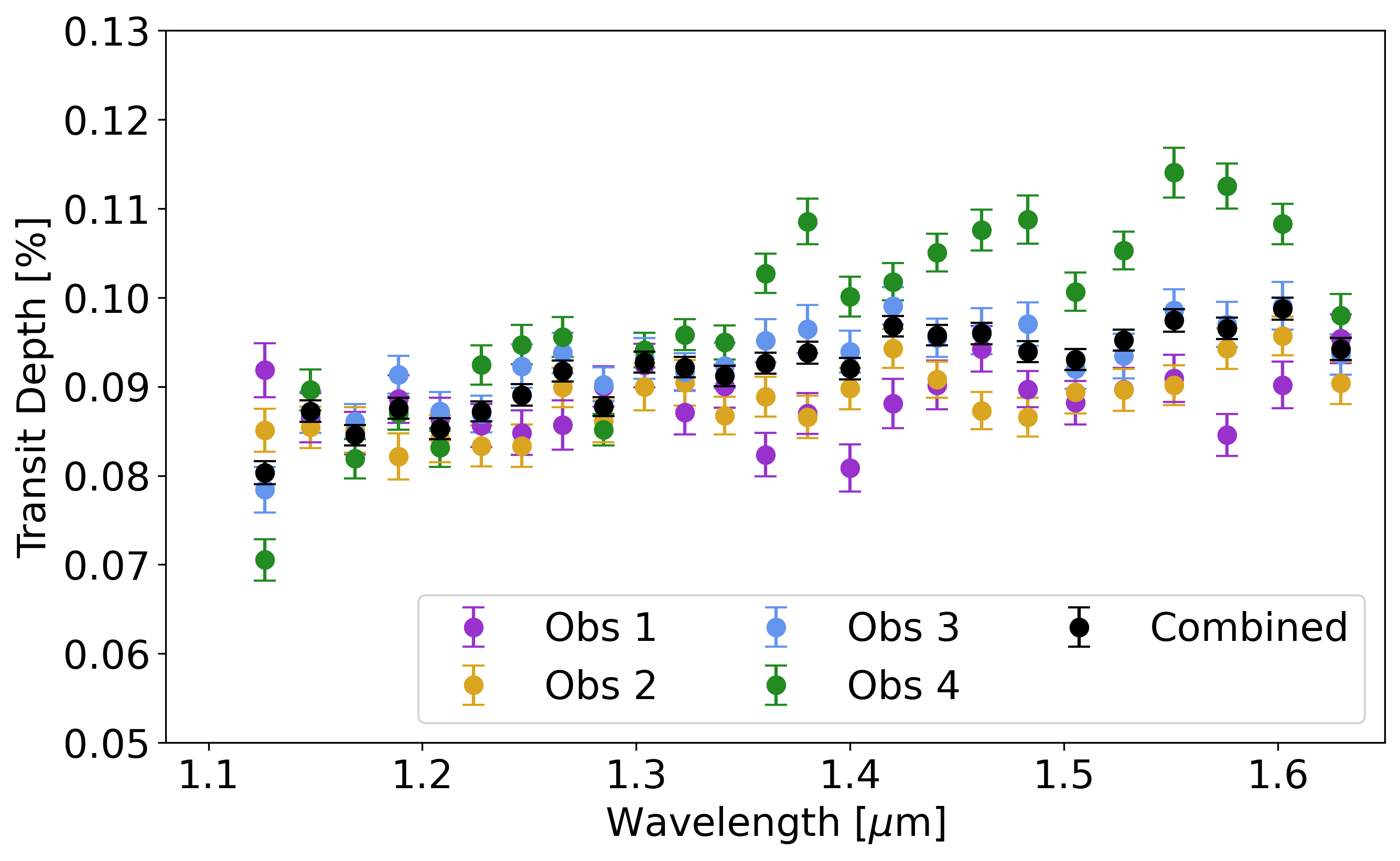}
    \caption{\textbf{Left:} Uncorrected transit spectra for each visit of HD\,97658\,b. The shaded regions indicate the white light curve depth, and 1 sigma uncertainty, for each visit. \textbf{Right:} Normalised transit spectra for each visit of HD\,97658\,b. The averaged transit spectrum, which was used in the analysis here, is given in black. This Figure is a sample from the Figure Set A1 that contains the same information for all the planets.}
    \label{figset:spec_cor}
\end{figure*}

\clearpage

\section*{Appendix 2: Summary of Results for Individual Planets}

Here we present further information for the planets where the spectra were fit with Iraclis for this study. We briefly summarise the data that was taken and compare our results to any we could find in the literature for the same dataset.

\subsection*{CoRoT-1\,b}

The first transit of CoRoT-1\,b was observed in 2008 by \citep{barge_corot1}, unveiling an inflated, 1.5 R$_J$ planet. A ground-based transmission spectrum of CoRoT-1\,b, obtained with the InfraRed Telescope Facility (IRTF), achieved a spectral precision which was comparable to the modulation expected from a single scale height of atmosphere but no spectral features could be discerned \citep{schlawin_c1}. 

HST WFC3 observed the transit of CoRoT-1\,b in staring mode as part of proposal 12181 (PI: Drake Deming, \citet{deming_12181_prop}). The GRISM128 subarray was utilised alongside the SPARS10 sequence and 16 up-the-ramp reads leading to an exposure time of 100.651947 s. This data was previously studied but no evidence for molecular absorption was found due to the high uncertainties on the data \citep{ranjan_spec}. Similar results were also recently found by \citet{glidic_c1}.

We fitted the data with Iraclis and it is worth noting that there is no post-egress orbit, likely due to poor knowledge of the ephemeris at the time of observing, and this leads to larger uncertainties on the transit depth. For the transit spectrum, the decreasing absorption with wavelength is best fit with a cloudless atmosphere and the continuous absorption from H-. The result is consistent with the recovered temperature of around 1800 K, which should lead to the dissociation of molecular species.

\begin{figure}[!h]
    \centering
    \includegraphics[width=\columnwidth]{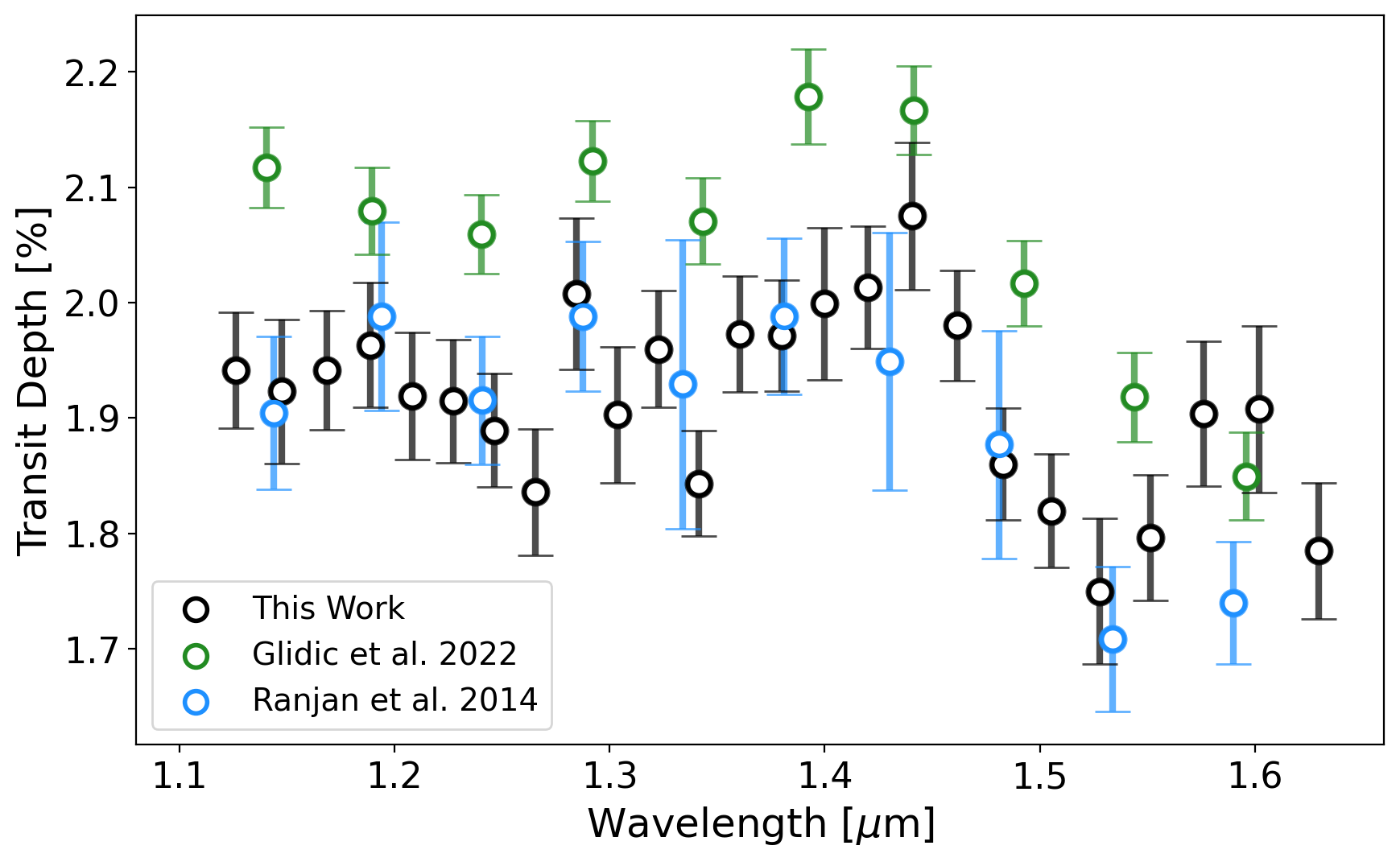}
    \caption{Comparison of the transit spectrum of CoRoT-1\,b obtained here, to those in the literature. While the features within the spectra are similar, there are offsets between different studies.}
    \label{fig:c1_comp}
\end{figure}

The transit spectrum is compared to literature results in Figure \ref{fig:c1_comp}, showing a good consistency with these results in terms of shape but with an offset between the spectrum recovered here and that from \citet{glidic_c1}.

\subsection*{GJ\,1214\,b}

In 2011, three staring mode transits of GJ\,1214\,b were taken for proposal GO-12251 (PI: Zachory Berta). The analysis of this data led to a flat spectrum which was interpreted as suggesting GJ\,1214\,b had an atmosphere with a mean molecular weight $>$ 4. Subsequently, 15 scanning mode transits of GJ\,1214\,b were taken between September 2012 and August 2013 as part of proposal GO-13021 (PI: Jacob Bean). Combining these lead to a spectrum which had uncertainties of around 50 ppm, equivalent to around 0.15 scale heights assuming a hydrogen dominated atmosphere. Yet, despite the high precision of the observations, not indication of an atmosphere could be extracted \citep{Kreidberg_GJ1214b_clouds}. 

\begin{figure}
    \centering
    \includegraphics[width=\columnwidth]{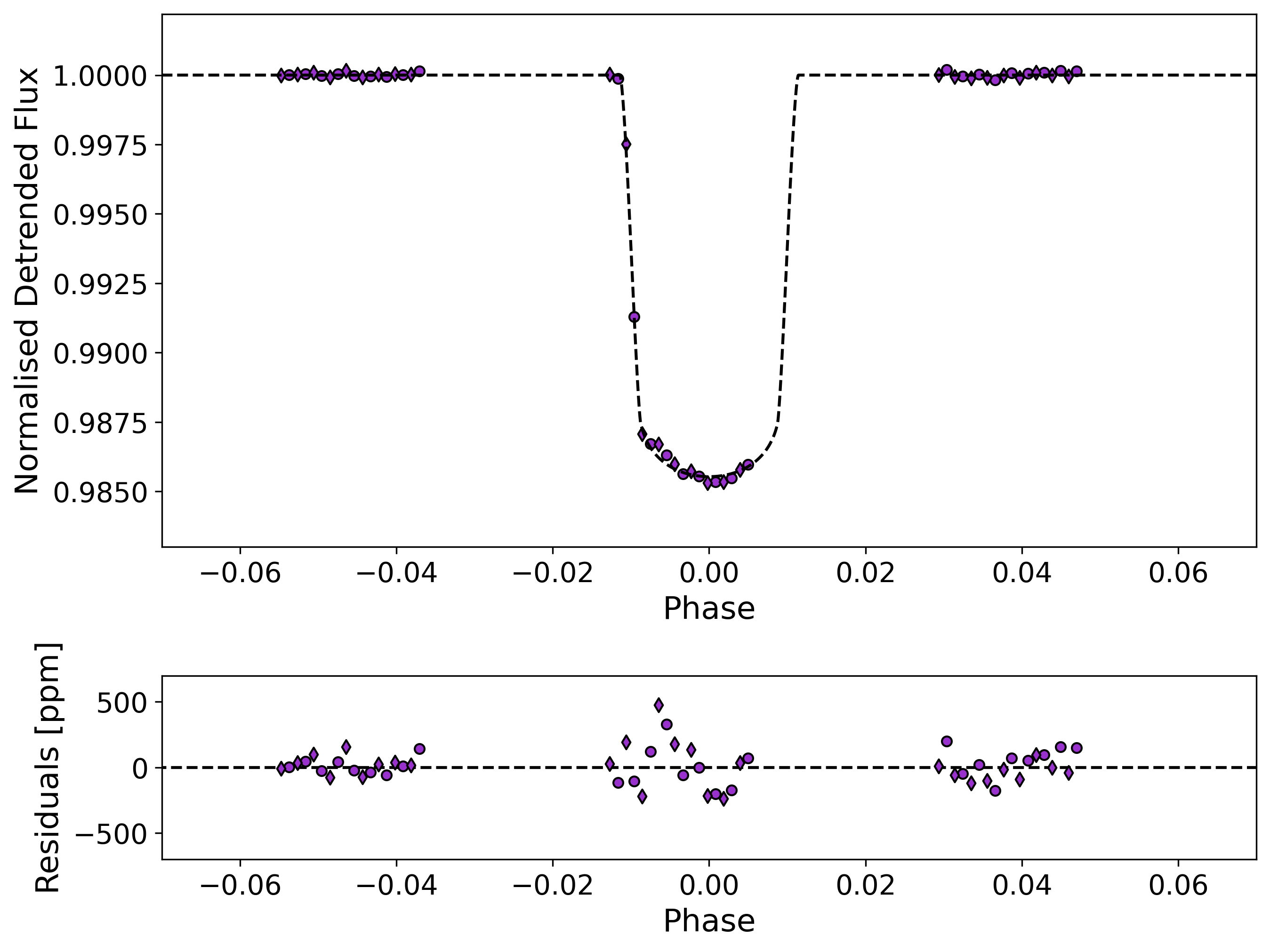}
    \includegraphics[width=\columnwidth]{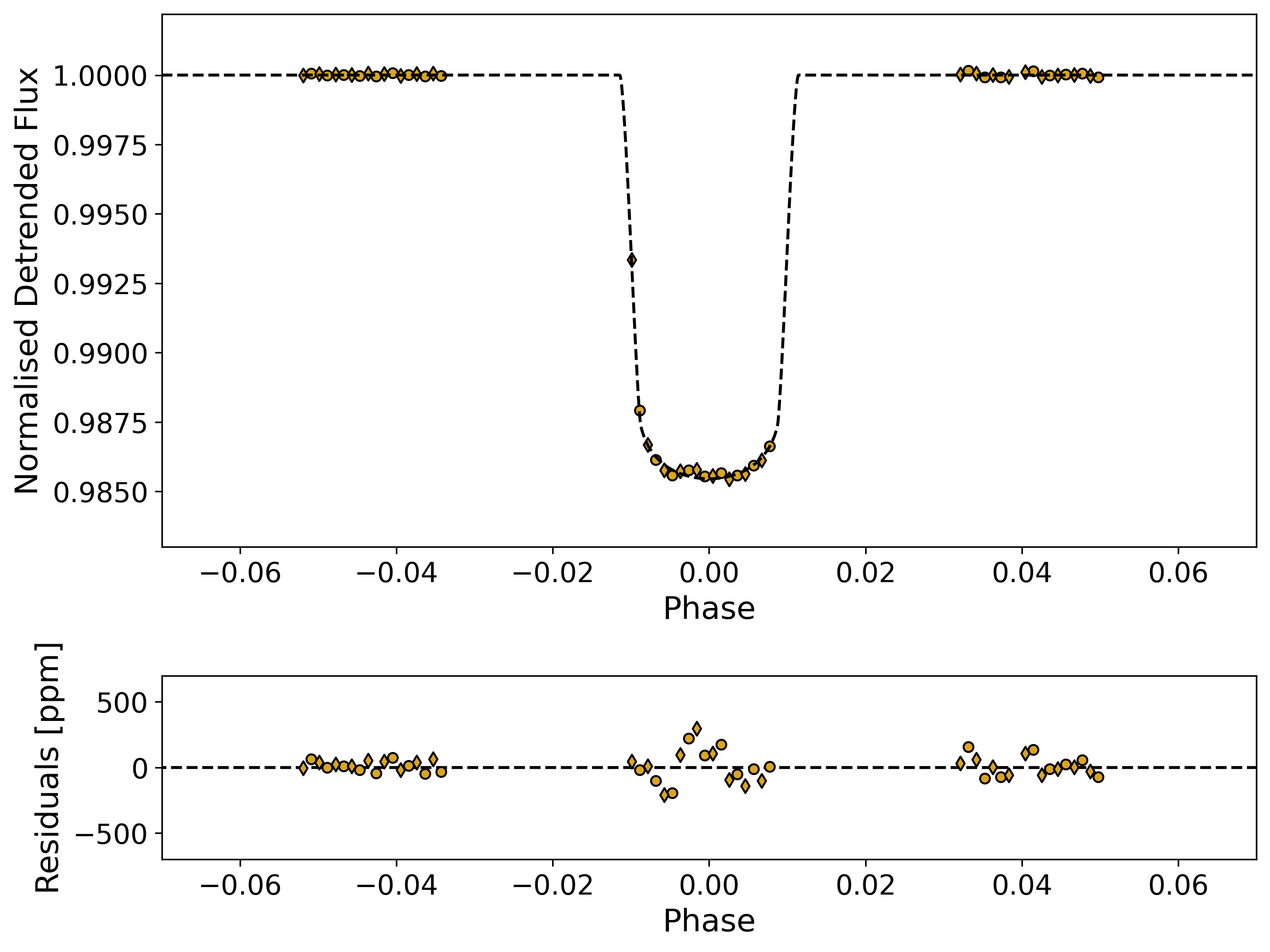}
    \caption{White light curves for scanning mode data of GJ\,1214\,b on 4$^{\rm th}$ (top) and 12$^{\rm th}$ (bottom) August 2013. The residuals of the in-transit orbit are non-gaussian and this could be caused by starspot crossing. }
    \label{fig:gj1214_ss_wlc}
\end{figure}

\begin{figure}
    \centering
    \includegraphics[width=\columnwidth]{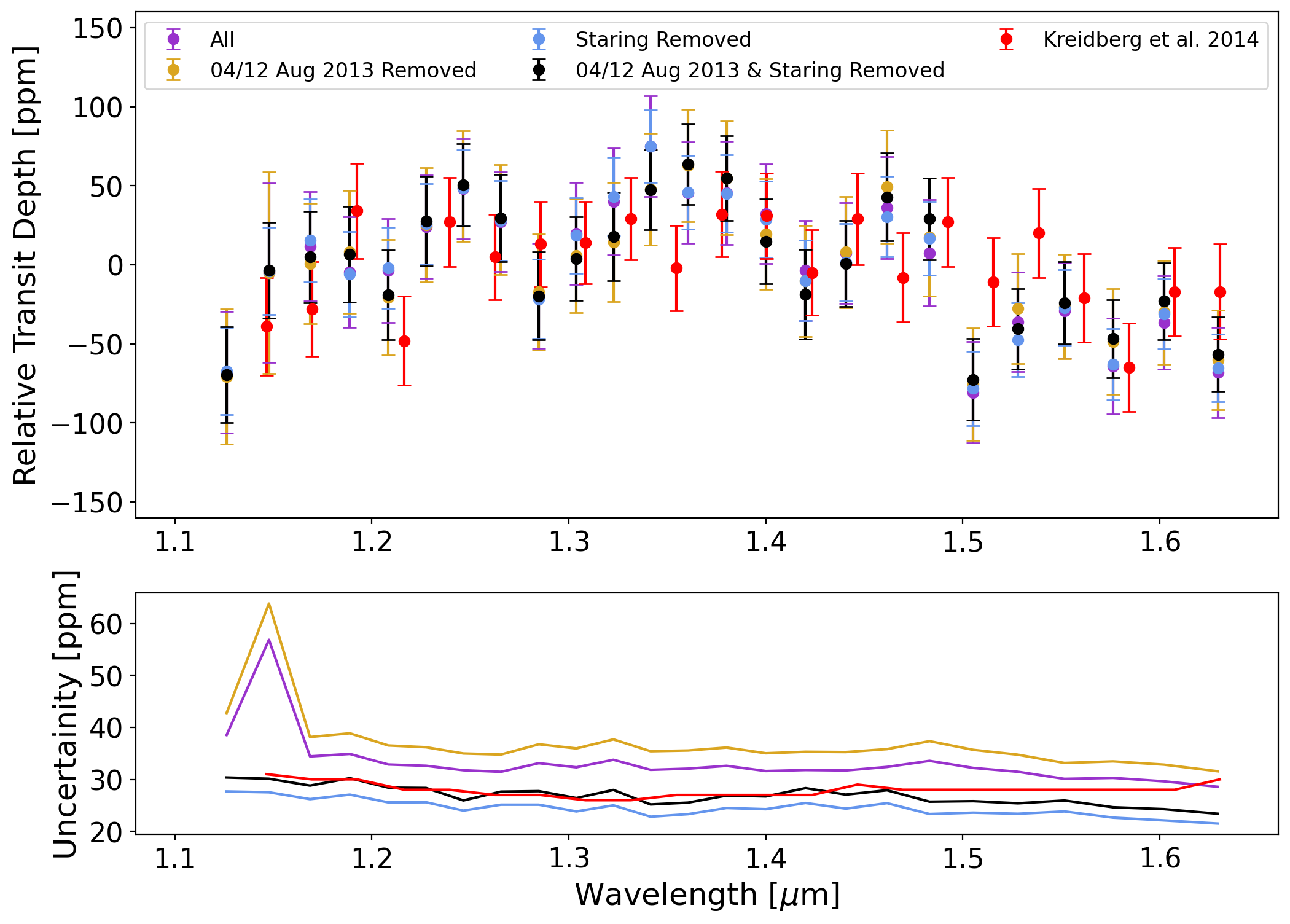}
    \caption{Transit spectra of GJ\,1214\,b using different amalgamations of datasets. All are roughly consistent with one another and the spectrum used for atmospheric analyses in this study is shown in black.}
    \label{fig:gj1214_diff_spec}
\end{figure}

We analysed all available data for GJ\,1214\,b. However, the scanning mode observations provided a higher precision than the staring mode data. As discussed in \citet{Kreidberg_GJ1214b_clouds}, the observation taken on 12$^{\rm th}$ April 2013 was affected by poor pointing and so was excluded from the analysis. \citet{Kreidberg_GJ1214b_clouds} found evidence of a starspot crossing in two observations (4$^{\rm th}$ and 12$^{\rm th}$ August 2013) and so they excluded these from their analysis. We show the white light curves of these transits in Figure \ref{fig:gj1214_ss_wlc}, showing slight bumps in the residuals which could indeed be due to a starspot crossing. We remove these data, along with the staring observations, to compute the final spectrum analysed here. In Figure \ref{fig:gj1214_diff_spec}, we show the final spectrum obtained in this study using different datasets as well as a comparison to the spectrum from \citet{Kreidberg_GJ1214b_clouds}.  
We find that the spectrum has slightly more modulation than found by \citet{Kreidberg_GJ1214b_clouds}. Additionally, our retrievals prefer a model with spectral modulation to the 2.41$\sigma$ level for the free chemistry model and to 2.38$\sigma$ for the chemical equilibrium run. Given the relatively low significance of these atmospheric detections, and the large number of observations that were combined to create the final dataset, we are cautious about inferring the presence of atmospheric features for this planet. Given the size of the error bars derived, the terminator of GJ\,1214\,b is evidently not cloud free. The JWST MIRI phase-curve taking during Cycle 1 should shed more light on the nature of this world.

\subsection*{HAT-P-2\,b}

HAT-P-2\,b was discovered in 2007 by \citet{bakos_h2}. It is a massive hot Jupiter (9.1 M$_{jup}$) that orbits its host star in an highly eccentric orbit (e = 0.52) in about 5.6 days. Due to its large density, the planet is believed to require the presence of a large core. This large mass, combined with the highly eccentric orbit, raises many questions regarding the physics of this planet and its formation. For instance, along the entire orbit, the planet's equilibrium temperature varies from 1240K to 2150K \citet{bakos_h2}. Studying the Rossiter-McLaughlin effect, it was found that stellar spin axis and orbital axis of the planet should be aligned, thus implying that the planet did not evolved through scattering or Kozai migration \cite{Winn_2007_hp2, Loeillet_2008_hp2}.

While being a very interesting planet, the atmosphere of HAT-P-2\,b was not studied with many instruments. A phase-curve observation with Spitzer at 3.6\,$\mu$m, 4.5\,$\mu$m, 5.6\,$\mu$m and 8\,$\mu$m was presented in \cite{Lewis_2013_hp2_spz}, highlighting a very complex atmosphere due to the particular orbital configuration of this planet. The study also suggested the planet might experience a temporary day side thermal inversion near periapse. In a follow-up work, \cite{Lewis_2014_hp2} performed a complementary analysis with GCM models to evaluate the impact of the eccentricity on the chemistry and the thermal structure of this planet, highlighting that dis-equilibrium processes on this planet might be important. 

Recently, a partial phase-curve was acquired with HST using the G141 grism \cite[PN: 16194, PI:][]{desert_prop16194}. The eclipse spectrum from this proposal was presented in \citet{emission_pop} and here we extract the transit spectrum. However, due to the high mass of HAT-P-2\,b, we did not recover a spectrum that was sensitive enough to allow for atmospheric constraints to be made.

\subsection*{HAT-P-7\,b}

HAT-P-7\,b is an inflated hot-Jupiter of 1.4 R$_J$ \citep{pal_h7}, which was studied during the commissioning program of Kepler when the satellite detected the eclipse as part of an optical phase curve \citep{borucki_hatp7}. These measurements indicated that HAT-P-7\,b could have a day-side temperature of around 2650 K which confirmed predictions from \citet{pal_h7,fortney}. This optical eclipse measurement was combined with Spitzer photometry over 3.5 - 8 $\mu m$ to infer the presence of a thermal inversion \citep{Christiansen_hatp7}, suggested by the high flux ratio in the 4.5 $\mu m$ channel of Spitzer compared to the 3.6 $\mu m$ channel. In their paper, chemical equilibrium models associated these emission features with CO, H$_2$O and CH$_4$. A thermal inversion was also reported to provide the best fit to this data by the atmospheric models of \cite{spiegel_hatp7,madhu_hatp7} but all three studies noted that models without a thermal inversion could also well explain the data though only with an extremely high abundance of CH$_4$. Further Kepler phase curves identified an offset in the day-side hot-spot \citep{esteves_h7_2013,esteves_h7_2015} as well as changes in its location \citep{armstrong_hatp7}, highlighting the complex dynamics of hot-Jupiter atmospheres. However, while Spitzer phase curves at 3.5 \& 4.5 $\mu m$ were also best fitted with a thermal inversion on the day-side and relatively inefficient day–night re-circulation, \cite{wong_hatp7} did not find evidence of a hot-spot offset.

We fitted the HST WFC3 data with Iraclis but it is worth noting that there is no post-egress orbit. Combined with the fact these observations were taken in staring mode, this led to a transit spectrum which wherein the uncertainties on the depth were equivalent to nearly 10 scale heights, meaning no atmospheric signal could be discerned.

\subsection*{HD\,97658\,b}

The sub-Neptune HD\,97658\,b was discovered as part of the NASA-UC Eta-Earth Program \citep{howard_hd97658} and four HST WFC3 transit observations have been obtained across two proposals (GO-13501 \citep{knutson_prop13501}, GO-13665 \citep{benneke_prop13665}). These have previously been analysed in \citep{knutson_hd97658,guo_hd97658}. 

The data from GO-13501 was collected using the GRISM256 subarray, the SPARS10 sequence and 4 up-the-ramp reads leading to an exposure time of 14.970785 s. The scan rate was 1.4 ''/s. Meanwhile, the data from GO-13665 used a different observational setup, with an exposure time of 12.795406 s via 16 up-the-ramp reads using the RAPID sampling sequence. The GRISM512 subarray was utilised and a scan rate of 1.4 ''/s.

For the observations from the latter proposal, the spatial scan was not correctly positioned within the subarray window. Therefore, the longer wavelengths were not recovered. Hence, for consistency, we extracted the white light curves of all data sets from 1.088-1.650\,$\mu$m. The change in the white light curve did not affect our spectral bins, which only extend to 1.643\,$\mu$m. We found that the white curve depths were consistent across the first three visits, with the fourth visit being slightly shallower. The fourth visit also displayed a greater slope across the WFC3 bandpass. 

\begin{figure}
    \centering
    \includegraphics[width=0.95\columnwidth]{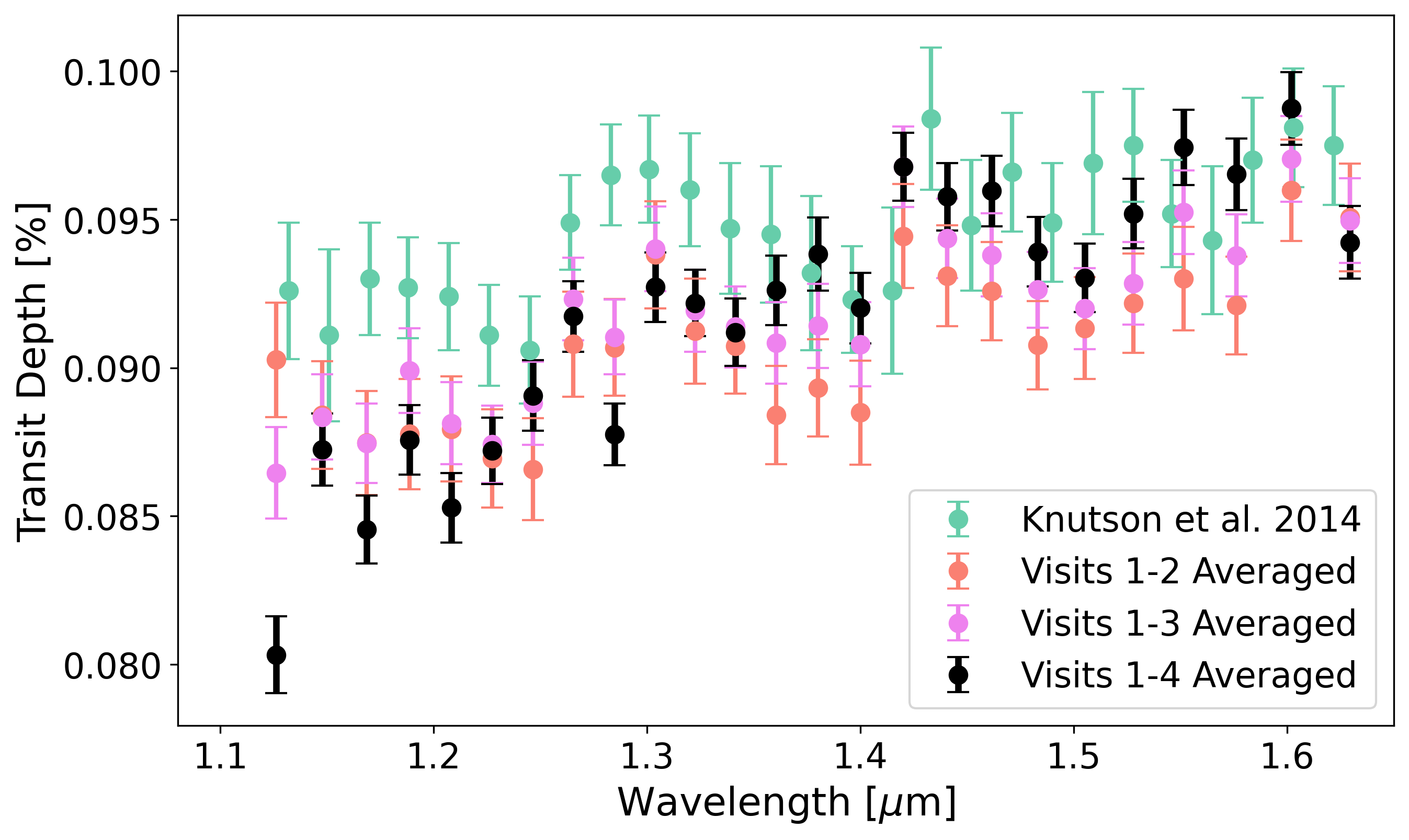}
    \includegraphics[width=0.95\columnwidth]{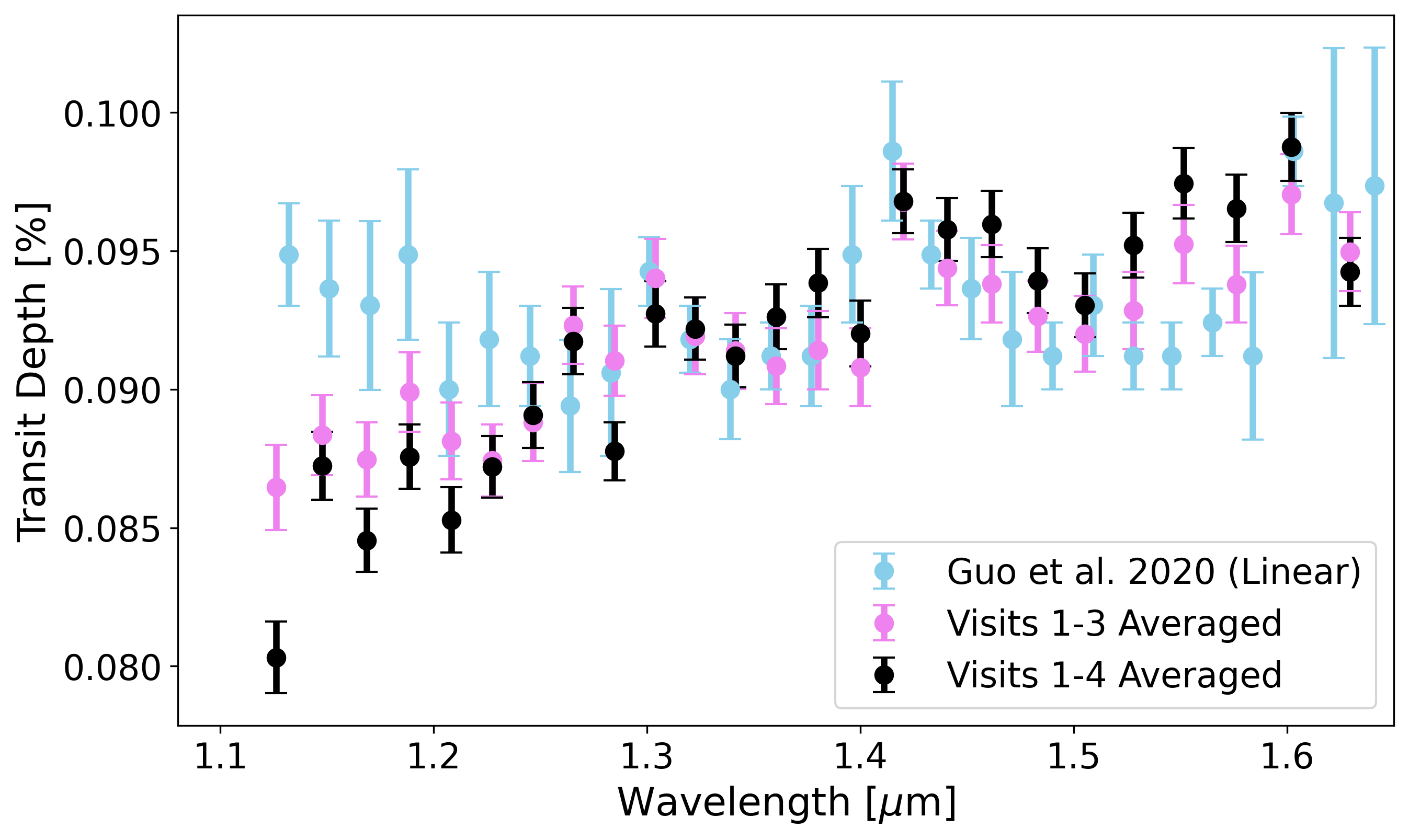}
    \caption{\textbf{Left:} Comparison of the spectrum derived in this work of HD\,97658\,b, to that from \citet{knutson_hd97658} who only analysed visits 1 and 2. Other than a slight offset, the spectra agree well after two visits but the averaged spectrum of all four transits has a stronger slope. \textbf{Right:} Comparison of the spectrum derived in this work of HD\,97658\,b, to that from \citet{guo_hd97658}. While the spectra agree over 1.3-1.5 $\mu$m, outside of this range there is a significant disagreement in the features.}
    \label{fig:hd97658_comp}
\end{figure}

We compared our spectrum to the literature results in Figure \ref{fig:hd97658_comp}. We find that, with the exception of a small offset between the two spectra, our fitting of the first two visits compares well to the results of \citet{knutson_hd97658}. However, we find that our analysis of all four observations differs from the spectrum of \citet{guo_hd97658} at shorter wavelengths, despite excellent agreement across the 1.3-1.5 $\mu$m range. We note that \citet{guo_hd97658} found that the choice of long-term detrending technique affected the white light-curve depth for the HST WFC3 G141 observations of HD\,97658\,b and suggested that a linear trend may not fully explain the systematics seen. We only fitted a linear trend here such that the analysis was consistent across the population. The analysis of this spectrum was complicated by the aforementioned incorrect positioning of the spatial scan and this may have further contributed to this discrepancy. Our retrievals showed a clear preference for the presence of an atmosphere and the free chemistry retrieval provided the best fit to the data.

\subsection*{HD\,219666\,b}

HD\,219666\,b, a hot-Neptune (T$_{\rm eq}$ = 1070 K, R = 4.71 R$_\oplus$), was discovered using data from the first sector of TESS \citep{esposito_hd219666} and was one of the first discoveries by this mission to be announced. With an orbital period of around 6 days, it lies in an under-populated region of the period-radius diagram in an area often coined the ``hot-Neptune desert". It is one of two planets studied in this work that lies within this region, the other being LTT\,9779\,b. \citet{esposito_hd219666} showed that the target was an excellent candidate for atmospheric studies by simulating JWST observations. 

The planet was subsequently observed as part of proposal GO-15698 (PI: Thomas Beatty) which planned to take two transit observations of HD\,219666\,b. The first set of observations were taken in June 2019 but pointing was lost and so the target was revisited in August and November of the same year. However, for the visit in August, the first orbit was taken during the transit of the planet, rather than the third orbit as planned. Evidently the planning of the observations was based on poor orbital ephemeris and this highlights the critical importance of programmes which seek to refine the periods of exoplanets which are good targets for atmospheric characterisation \citep[e.g.][]{edwards_orbyts,kokori,kokori_2}. The extracted white light curve of this visit is shown in Figure  alongside that of the November visit, the only one that could be utilised. 

Despite these issues, the spectrum derived from this single visit led to a confident atmospheric detection. The TauREx free chemistry retrieval recovered a very high volume mixing ratio of water: log$_{\rm 10}$(H$_{\rm 2}$O) = -1.54$^{+0.37}_{-1.08}$. The retrieval was preferred to the flat model to 3.57$\sigma$, suggesting a strong atmospheric detection.

\subsection*{HIP\,41378\,b}

Discovered using data from the K2 mission, HIP\,41378\,b is a sub-Neptune (R = 2.6 R$_\oplus$) with an orbital period of 15.5712 days and an equilibrium temperature of 960 K \citep{vanderburg_hip41378,santerne_hip41378}. Its atmosphere has not previously been studied and the HST transit data was taken as part of GO-15333 (PI: Ian Crossfield). Three successful transit observations of HIP\,41378\,b were taken, each consisting of 7 orbits. Two other visits were attempted with one resulting in a complete loss of data and the other having issues due to a guide star acquisition failure which led to large shifts in the position of the spectrum on the detector. A further two observations were originally planned but were scrapped in favour of observing three transits of TOI-674\,b (see Section \ref{sec:toi674}). For these observations, the GRISM512 subarray was used, with 8 up-the-ramp reads using the SPARS25 sampling sequence, and an exposure time of 138.380508 s, combined with a scan rate of 0.25 ''/s.

From our free chemistry retrievals, we find the spectrum to be compatible with a flat line. However, our chemical equilibrium retrieval provided a preferable fit that the flat model and was preferred by 2.65$\sigma$. The 1.4$\mu$m feature size derived was 2.25 $\pm$ 0.97.

\subsection*{HIP\,41378\,f}

The outermost cureently-know planet in the HIP\,41378 system, HIP\,41378\,f orbits its host star every 542 days \citep{vanderburg_hip41378,santerne_hip41378}. The host star is bright and, as the planet's density is very low (M = 12 $\pm$ 3 M$_{\oplus}$, R = 9.2 $\pm$ 0.1 R$_{\oplus}$, $\rho$=0.09 $\pm$ 0.02 \citet{santerne_hip41378}), it is an excellent target for atmospheric studies. Given its temperature (T$\sim$300 K), HIP\,41378\,f is also far cooler than any other large gaseous planets studied here. The transit spectrum of HIP\,41378\,f required 18 consecutive orbits of HST and was taken by proposal GO-16267 (PI: Courtney Dressing). The data was obtained with a scan rate of 0.419 "/s. The SQ256 subarray was used as well as the SPARS10 reading sequence. The first forward and reverse scans of each orbit were acquired using 7 nondestructive reads while the rest had 9. Direct images were acquired with the F126N at various points in the visit as a guidance check. 

The transmission spectrum of HIP\,41378\,f was analysed in \citet{alam_hip41378f}. However, their derived spectrum differs dramatically from ours. We performed a number of additional fittings to check the credibility of our result. Firstly, we fitted the data using the wavelength bins and limb-darkening coefficients from \citet{alam_hip41378f} but this did not yield a similar spectrum to their work. We tried extracting the data by splitting the nondestructive reads in case of a contamination by a background star but this did not change the derived spectrum. We also sought an independent fit of the data, with CASCADe\footnote{\url{https://gitlab.com/jbouwman/CASCADe}} also being used to analyse the data. The CASCADe pipeline is instrument independent, meaning the manner in which systematic effects are removed from the data is completely different to Iraclis, which was built purposefully for HST WFC3. The initial fit was also done without prior knowledge of the Iraclis result but we ensured the orbital parameters utilised were the same. The resulting CASCADe spectra has highly similar features to the Iraclis fit. We note there is a slight offset between the datasets, but this is common when analysing spectra with different pipelines \citep[e.g.][]{aresV}, as well as a slightly steeper slope in the CASCADe data which gives lower transit depths at shorter wavelengths.

\begin{figure}
    \centering
    \includegraphics[width=0.95\columnwidth]{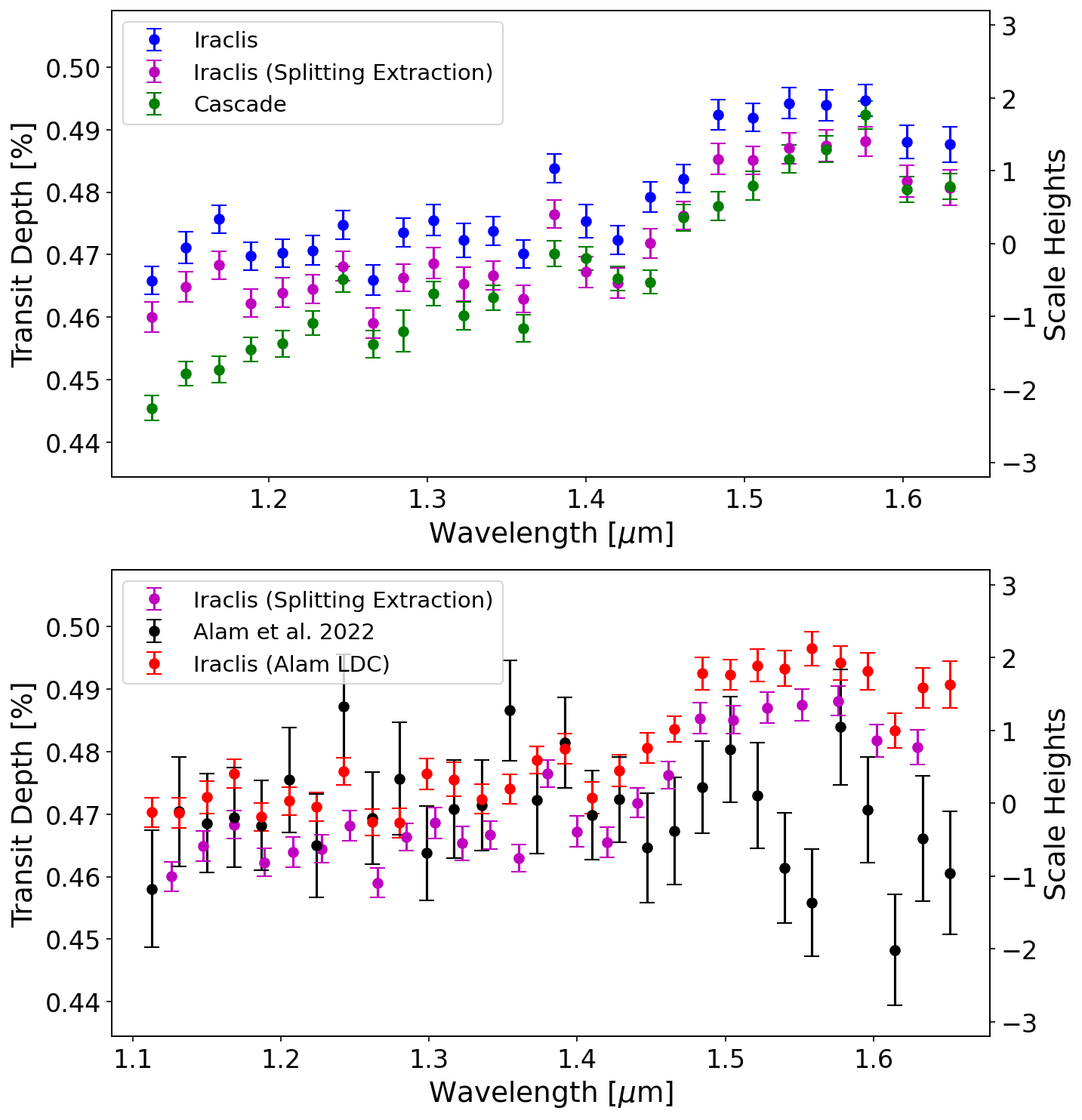}
    \caption{Comparisons of different spectra for HIP\,41378\,f. All spectra derived in this work exhibit features while the result from \citet{alam_hip41378f} is flat. Two pipelines were used here to check the result. They give slightly different mean depths and the Cascade spectrum has a decreasing slope at shorter wavelengths. Utilising the wavelength bins and limb-darkening coefficients from \citet{alam_hip41378f} did not remedy the situation.}
    \label{fig:my_label}
\end{figure}

From comparisons between our white light curve fit and that from \citet{alam_hip41378f}, we noted that data had been removed from their analysis (see Figure 1 from their work), including an entire orbit just after ingress. In \citet{alam_hip41378f} they state that the observations were affected by the SAA which is undoubtedly true: many orbits did not acquire the full number of requested frames. However, they do not discuss why this orbit, and data from others, were not present in their fitting despite the data being taken and having no obvious signs of degradation. We also attempted fits without this orbit but still found a spectrum with the same features as before. We also noted that the uncertainties on the spectrum derived by \citet{alam_hip41378f} were much higher than those from Iraclis and CASCADe. While some of this could be due to the removal of data by \citet{alam_hip41378f}, the residuals on the white light-curve fit are also significantly higher in their work. Here, the residuals of the white light-curve fit had a standard deviation of around 125 ppm while the standard deviation of the residuals from the fit of \citet{alam_hip41378f} was around 500 ppm. While we cannot know for sure the source of this increased scatter on their white light-curve as their pipeline is not public, we speculate that some part of their calibration, reduction, or extraction process must have been sub-optimal. 

Having find no way to not achieve a spectrum without significant features, we attempted to fit the data using TauREx. Our free chemistry retrievals were flexible enough to fit the features seen, but the credibility of the fit is questionable. Nevertheless, we show the best fit model to the data in Figure Y. In the case of the GGChem retrieval, while it is preferred to the flat line fit, it is obviously unable to replicate the features in the data.

There have been suggestions that HIP\,41378\,f could have rings which would inflate the measured radius of the planet, thereby explaining the very low density \citep{akinsanmi_rings,belkovski_hip41378f}. Based on data from K2, their modelling suggested the true radius of HIP 41378\,f could be 3.7 R$\oplus$. The effect of these rings on the transit depth is likely to be chromatic as, at wavelengths were the rings are optically opaque, more stellar light would be blocked. While this could be the source of the features seen, we do not have the modelling capability to pursue this further although frameworks have been proposed \citep{ohno_rings}. If the planet's true radius is 3.7 R$\oplus$, the uncertainties on the transit spectrum would now each be equivalent to around 1.5 scale heights. Therefore, any attempt to determine the effect of rings, if present, may also need to account for the atmospheric contribution. Over such a short spectral range, the solutions to this are likely the be degenerate but we encourage further work into this option as it may be useful for future data taken of this planet, and others like it, with observatories such as JWST.

We note the the slope seen in the spectrum of HIP\,41378\,f is similar to that seen for HD\,97658\,b and LTT\,9779\,b, which might suggest that all these observations are affected by a systematic or astrophysical effect that cannot currently be identified. The narrow wavelength range of HST WFC3 G141 makes it difficult to determine whether strange spectra are a result of poor reduction or are a true representation of the signal. Hopefully, future data of this planet will allow us to uncover the truth about its nature. However, due to the strange nature of the HST WFC3 G141 spectrum, we chose not to include in when fitting for trends within the data using BHM.

\subsection*{K2-24\,b}

Transits of two planets orbiting K2-24, a G3 dwarf, were detected during Campaign 2 of the K2 mission \citep{petigura_k2_24,sinukoff_k2_24}. K2-24\,b, the inner and smaller (R = 5.68 $\pm$ 0.56 R$_\oplus$)of the two planets, orbits the host star in 20.8851 days. With a mass of 19 ± 2 M$_\oplus$ \citep{petigura_k2_24_ttv}, K2-24\,b has a relatively light density which, combined with the brightness of the host star and relatively large transit depth, makes it a good target for atmospheric characterisation \citep{petigura_k2_24}. 

A single transit observation was obtained with HST WFC3 G141 by proposal GO-14455 (PI: Erik Petigura). The data was collected using the GRISM256 subarray, the SPARS10 sequence and 16 up-the-ramp reads leading to an exposure time of 103.128586 s. The scan rate was 0.16 ''/s, leading to a peak signal of around 20,000 counts.

Our chemical equilibrium retrieval was not capable of fitting the spectral features seen in the K2-24\,b data. However, the best-fit free chemistry model was preferred to a flat line with a significance of 2.46$\sigma$. It suggested the presence of NH$_3$, in a high abundance, but did not find evidence for H$_2$O. Due to the shape of the spectrum, the fitting of the 1.4 $\mu$m feature yielded a negative value (-1.04 $\pm$ 0.68). 

\subsection*{KELT-1\,b}

The first low-mass object discovered by the KELT-North survey, KELT-1\,b is a $~27M_{J}$, $1.12R_{J}$ planet with a very short period circular orbit of 29 hours. \citep{siverd_k1} presented spectroscopy, photometry and radial velocity data in order to obtain an equilibrium temperature of $T_{eq} \approx 2400$ K, assuming zero albedo, due to a significant amount of stellar irradiation. 

Its extreme temperature and significant inflation make KELT-1\,b a valuable case-study for short-period atmospheric characterisation. In several early studies it was successfully characterised in eclipse \citep{Beatty_2014,Beatty_2017}, suggesting a monotonically decreasing temperature-pressure profile. However, more recent work has suggested the atmosphere presents indications of a localised thermal inversion associated with VO, FeH and H$^-$ \citep{emission_pop}.

We find that, due to it's high mass which leads to it being classified as a brown dwarf, there is no evidence for atmospheric species in transit spectrum. We note that expected transit depth modulation due to one scale height of atmosphere is far below the size of the error bars, explaining why a flat spectrum is recovered.

\subsection*{Kepler-9\,b and Kepler-9\,c}

The Kepler-9 system was discovered in 2010 and planets b and c were the first planets to be confirmed via transit timing variations \cite{holman_k9}. These TTVs allowed the masses of the planets to be measured \citep[e.g.][]{dreizler_k9} and these masses were later verified by radial velocity observations \cite{borsato_k9}.

The two transit observations of Kepler-9\,b, as well as the two transit observations of Kepler-9\,c, were taken in staring mode using the GRISM256 aperture. The SPARS10 sampling sequence was utilised, with 12 up-the-ramp reads, leading to an exposure time of 73.742661 s. These were taken as part of GO-12482 (PI: Jean-Michel Desert). 

For Kepler-9 b, the free retrievals did not provide more preferable fit to the data than the flat model. However, the GGChem retrieval did provide a preferable fit, albeit to only 1.89$\sigma$. In the case of Kepler-9\,c, the opposite was true: the flat model was preferred to the GGChem retrieval but not to the free chemistry retrievals. However, the preferred free chemistry model did not detect features, rather a slope in the spectrum. Neither atmospheric ``detection" is convincing, but this is unsurprising given the size of the error bars with respect to the expected atmospheric modulation due to a single scale height of atmosphere.

\subsection*{Kepler-51\,b \& Kepler-51\,d}

Each of the three planets which are known to be orbiting Kepler-51 have low densities \citep{masuda_k51}. Kepler-51\,c has only a grazing transit, making constraints on its size difficult. However, Kepler-51\,b and Kepler-51\,d were originally determined to have radii of 7.1$\pm$0.3 R$_\oplus$ and 9.7$\pm$0.5 R$_\oplus$, respectively \citep{masuda_k51}. Their masses, derived from transit timing variations, gave them both densities of less than 0.05 gcm$^{-3}$ \citep{masuda_k51}.

Two HST WFC3 G141 observations were taken of each planet (PN: 14218, PI: Zach Berta-Thompson) and these were analysed by \citet{libby_k51}. In this study, the radii and masses of the planets were updated and they found that the densities were slightly higher than previously thought. However, they were still low at 0.064 gcm$^{-3}$ and 0.038 gcm$^{-3}$, respectively. Despite the low density, and thus large atmospheric scale height, the transmission spectra uncovered by \citet{libby_k51} did not yield any spectral features.

The observations of HST WFC3 G141 were taken with the following settings. The GRISM256 aperture and the SPARS10 readout sequence were utilised, with 15 up-the-ramp reads resulting in exposure time of 103 seconds. Due to the faintness of the host star (J = 13.56), the staring mode was used as it offered a higher efficiency and precision than the now common scanning mode \citep{libby_k51}.

We analysed all four observations but had issues with the extraction of data for one of the Kepler-51\,b visits. We show a comparison between our spectra and those from \citet{libby_k51}. As we could only analyse a single visit of Kepler-51\,b with Iraclis, we use the spectrum from \citet{libby_k51} for this planet. However, we note the good comparison between the single visit and there spectrum and also clarify that we utilise our fitting of the Kepler-51\,d data.

\subsection*{LTT\,9779\,b}

LTT\,9779\,b is an ultra-hot Neptune discovered using data from TESS \cite{jenkins_ltt9779}. With a period of less than a day, LTT\,9779\,b lies within the Neptune desert: there is a dearth of planets between 2 and 10 R$_\oplus$ in these very short orbits. Photometric eclipse observations of LTT\,9779\,b with Spitzer revealed a spectrum which is best-fitted by a non-inverted atmosphere and evidence was found for the presence of CO \citep{dragomir_ltt}. Phase curve observations with TESS and Spitzer have also been made, which found a large ($~$1100 K) day-night brightness contrast and suggestions of a super-solar atmospheric metallicity \citep{crossfield_ltt}. The transit observations from these phase curves were not precise enough to constrain the atmosphere in transmission.

HST WFC3 G141 observations of LTT\,9779\,b's transit were taken as part of proposal GO-16457 \citep[PI: Billy Edwards, ][]{edwards_prop}, with an additional dataset using the G102 grism also taken. The data were taken using the SPARS10 sequence and the GRISM256 aperture. The total exposure time was 103.13 seconds, with 16 samples being taken per exposure. To avoid contamination by a background star, we used this up-the-ramp reads, using Iraclis's splitting mode to extract these individually.

Assuming a H/He dominated atmosphere (mmw = 2.3), the derived spectrum has error bars that are just below the expected atmospheric modulation due to one scale height of atmosphere. The spectrum has an increasing transit depth with wavelength, which is best fitted by the model with no absorbers (flat model) where the slope is caused by CIA. The bluest spectral point is significantly higher than the subsequent data points and the retrieval including optical absorbers suggests this could be due to TiO. However, as the detection is based off of a single data point it is unreliable and the G102 data will be required to ascertain the presence of this molecule. However, we don't analyse it here to maintain homogeneity across the planets studied.

\subsection{TrES-2 b}

TrES-2\,b was among the first transiting exoplanets to be discovered \citep{odonovan_t2}. The 1.2 R$_{\rm J}$ planet orbits its host star in roughly 2.47 days, giving it an equilibrium temperature of around 1700 K. It was among the planets considered for the JWST Early Release Science (ERS) Transiting Exoplanet programme \citep{stevenson_jwst_ers} but ultimately not selected \citep{bean_jwst_ers}. \citet{turner_t2} obtained a ground-based UV transit of TrES-2\,b to search for asymmetries in the light curve but no such phenomena was observed. Given the large transit depth, the planet has also been regularly followed-up from the ground with numerous studies updating the ephemerides and searching for non-linear periods \citep[e.g.]{rabus_t2,raetz_t2,edwards_orbytsII,ozurk_t2}. Additionally, a K-band eclipse was detected by \citet{croll_t2} which, when combined with Spitzer eclipses of the planet \citep{odonovan_t2_spit}, suggested the dayside could be represented by a blackbody and confirmed the planet's orbit was circular.

\begin{figure}
    \centering
    \includegraphics[width=\columnwidth]{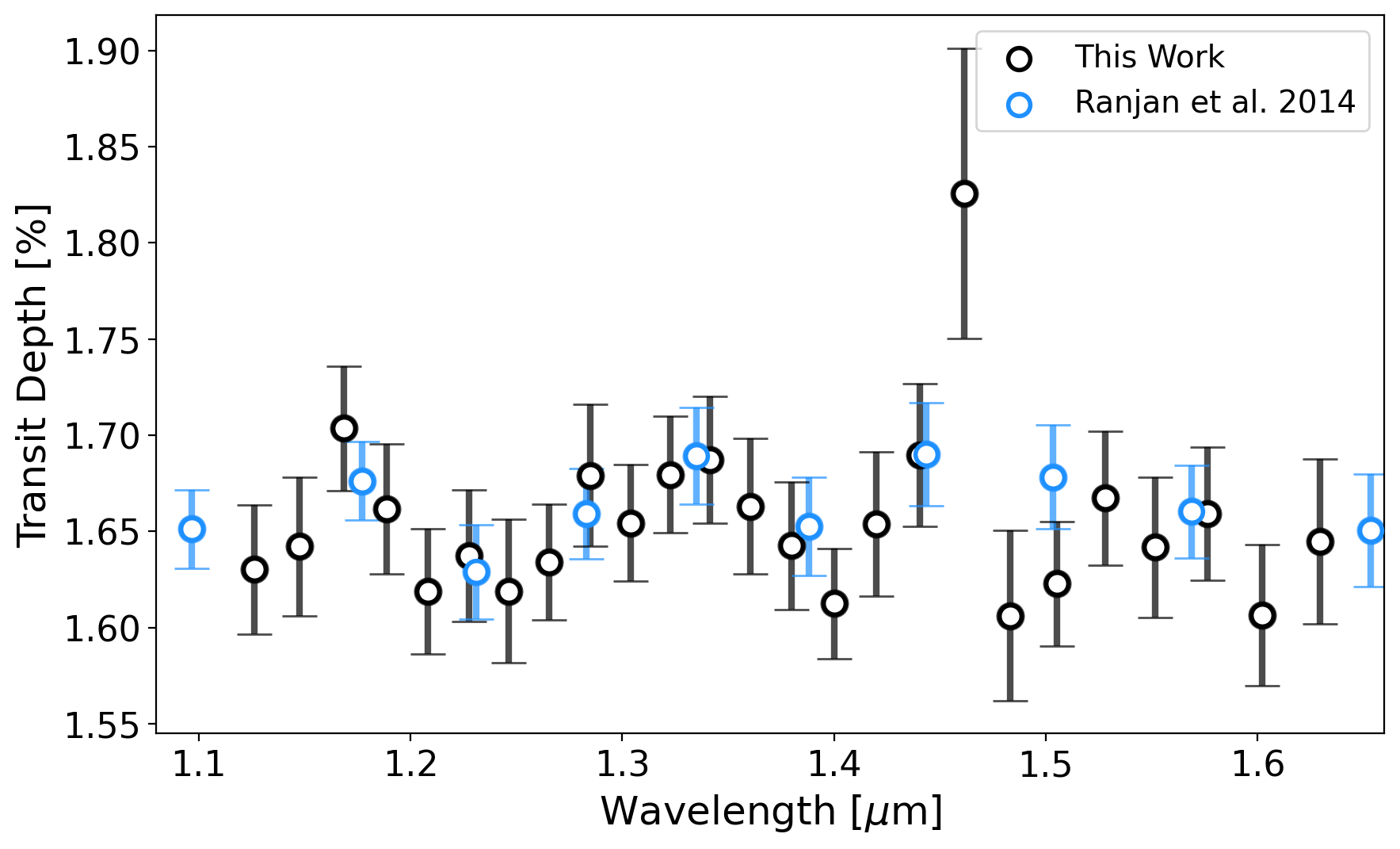}
    \caption{Comparison of the transit spectrum of TrES-2\,b obtained here, to the spectrum from \citet{ranjan_spec}.}
    \label{fig:t2_comp}
\end{figure}

The HST WFC3 G141 transmission spectrum was obtained using the staring mode. The RAPID readout mode was used for the GRISM512 aperture, with 16 sample being taken per exposure to give an exposure time of 12.8 seconds. The data was previously analysed by \citet{ranjan_spec} who found that the spectrum was not precise enough to constrain the atmosphere. We reanalyse this data here, using Iraclis, and yield the same result: the uncertainites on each data point are several scale heights in size. The comparison between our spectrum and the one from \citet{ranjan_spec} is given in Figure \ref{fig:t2_comp}. 

\subsection{TrES-4 b}

TrES-4\,b is a short-period (3.5539268 days), hot Jupiter (R = 1.706 R$_{\rm J}$) \citep{mandushev_t4}. The planet has been observed in eclipse, with data from the Spitzer space telescope suggested a thermal inversion \citep{knutson_t4} while attempts have also been made to measure the thermal emission from the ground \citep{martioli_t4}.

\begin{figure}
    \centering
    \includegraphics[width=\columnwidth]{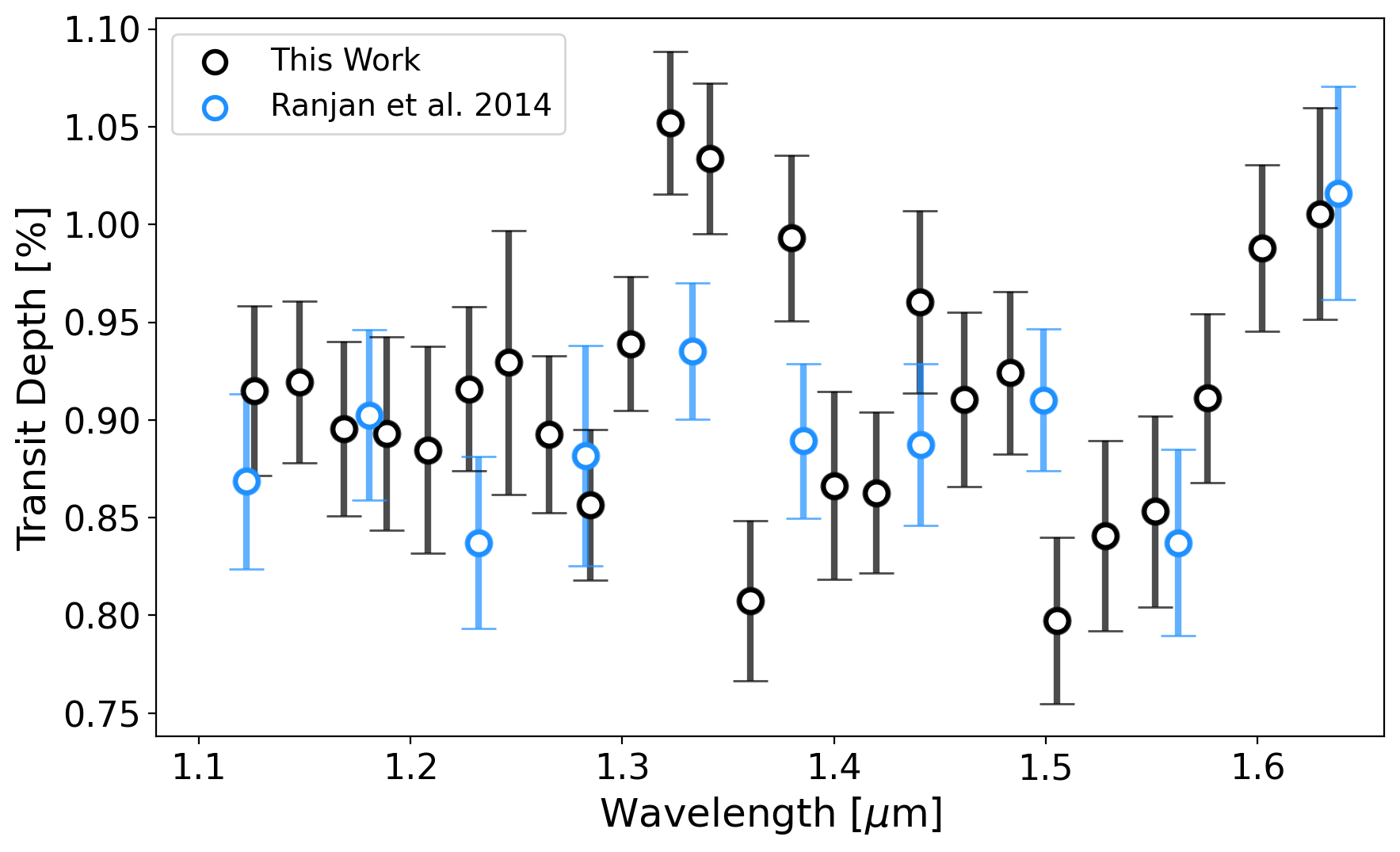}
    \caption{Comparison of the transit spectrum of TrES-4\,b obtained here, to the spectrum from \citet{ranjan_spec}.}
    \label{fig:t4_comp}
\end{figure}

As with TrES-2\,b, the HST WFC3 G141 transmission spectrum was obtained using the staring mode. Again the RAPID readout mode was used for the GRISM512 aperture, with 16 sample being taken per exposure to give an exposure time of 12.8 seconds. The data also was previously analysed by \citet{ranjan_spec} who found that the spectrum was not precise enough to constrain the atmosphere. We reanalyse this data here and find that the recovered error bars are tens of scale heights in size, therefore offering no information on the atmosphere. A comparison of the spectra is shown in Figure \ref{fig:t4_comp}.

\subsection*{TOI-270 c and TOI-270 d}
\label{sec:toi270}

TOI-270\,c is part of a three planet system, consisting of two sub-Neptunes and a Super-Earth \citep{gunther_toi270}, found by TESS. The planets orbit a bright (K=8.25) M-dwarf (T$_{\rm Eff}$ $\sim$3500 K) and the masses of the planets were subsequently measured using radial velocity measurements \citep{van_eylen_toi270}.

Three transit observations were taken of TOI-270\,c, each constituting three orbits which meant the first orbit was not, in this case, discarded. Despite having larger ramps than subsequent orbits, fitting the light curve using these orbits still led to good fits to the data. For these observations, the GRISM512 subarray was used, with 8 up-the-ramp reads using the SPARS25 sampling sequence, and an exposure time of 138.380508\,s, combined with a scan rate of 0.111\,''/s. Additionally, a single transit of TOI-270\,d was acquired using the same setup. These observations were taken as part of GO-15814 (PI: Thomas Mikal-Evans).

\begin{figure*}
    \centering
    \includegraphics[width=0.45\textwidth]{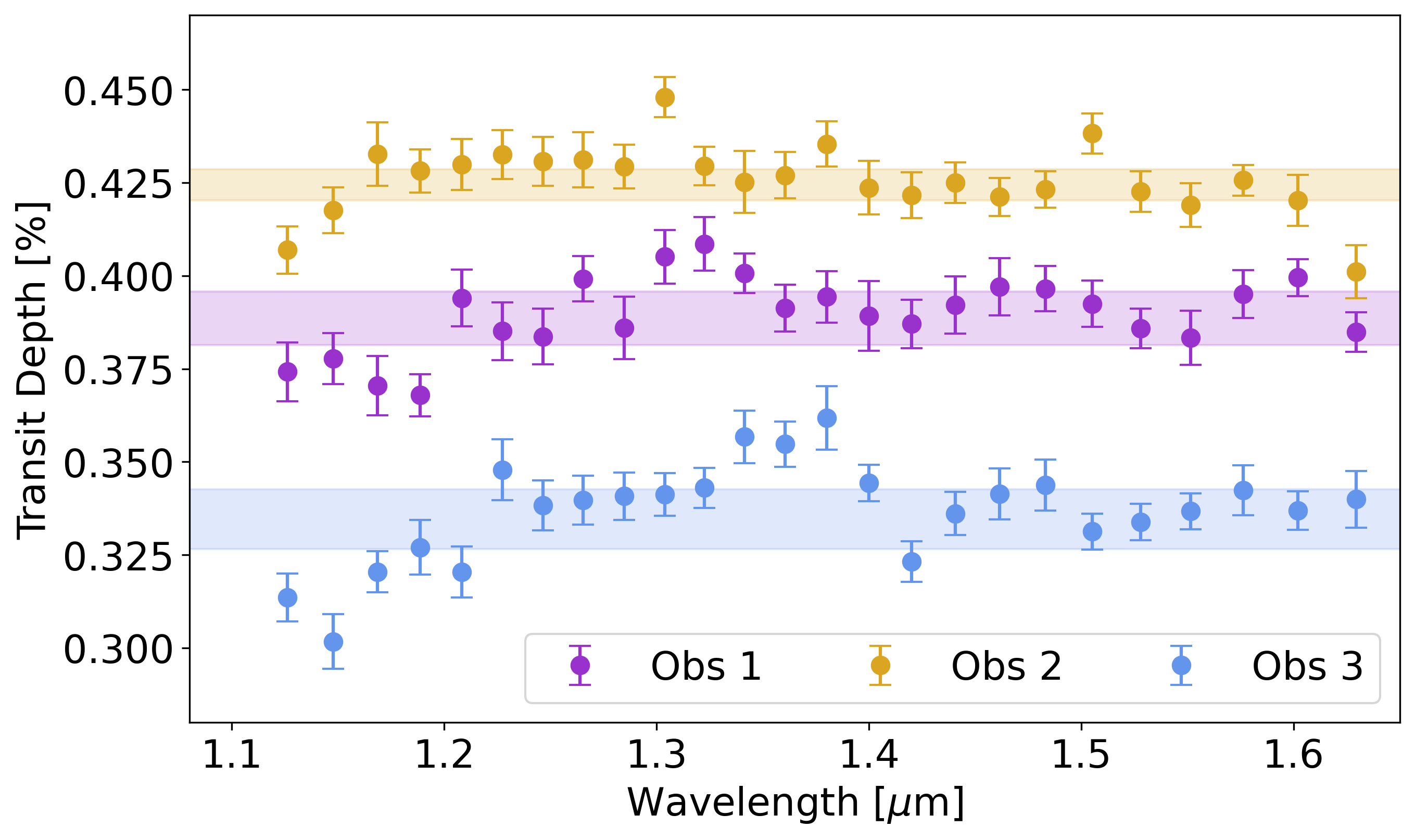}
    \includegraphics[width=0.45\textwidth]{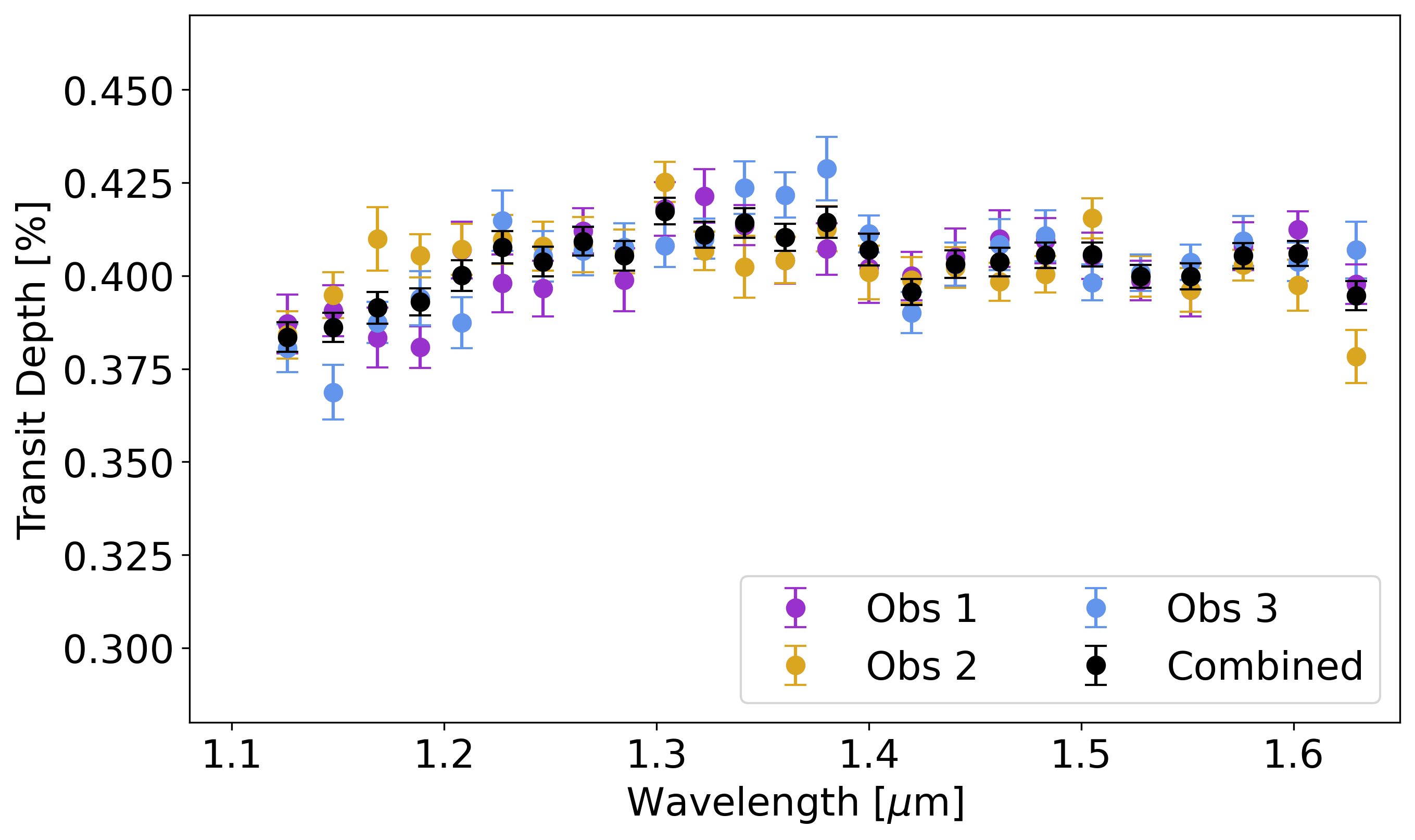}
    \caption{\textbf{Left:} Uncorrected transit spectra for each visit of TOI-270\,c. The shaded regions indicate the white light curve depth, and 1 sigma uncertainty, for each visit. \textbf{Right:} Normalised transit spectra for each visit of TOI-270\,c. The averaged transit spectrum, which was used in the analysis here, is given in black.}
    \label{fig:toi270c_final_spec}
\end{figure*}

We find that the white light curve depth for TOI-270\,c varies drastically, with a $~$1000 ppm difference between visits 2 and 3. On the other hand, the spectral shape of these visits is consistent, as shown in Figure \ref{fig:toi270c_final_spec}. However, the spectral shape is  itself strange: there is a significant drop at bluer wavelengths. Atmospheric scenarios seem unlikely because of the size of the feature ($~3.5$ scale heights) leaving two hypotheses. Firstly, it could be due to the fitting of the data but this seems unlikely as the effect is recovered in all three visits. Furthermore, no significant residuals within any of the fittings and there are no indications that the fittings are, in any way, poor. 

\begin{figure*}
    \centering
    \includegraphics[width=0.45\textwidth]{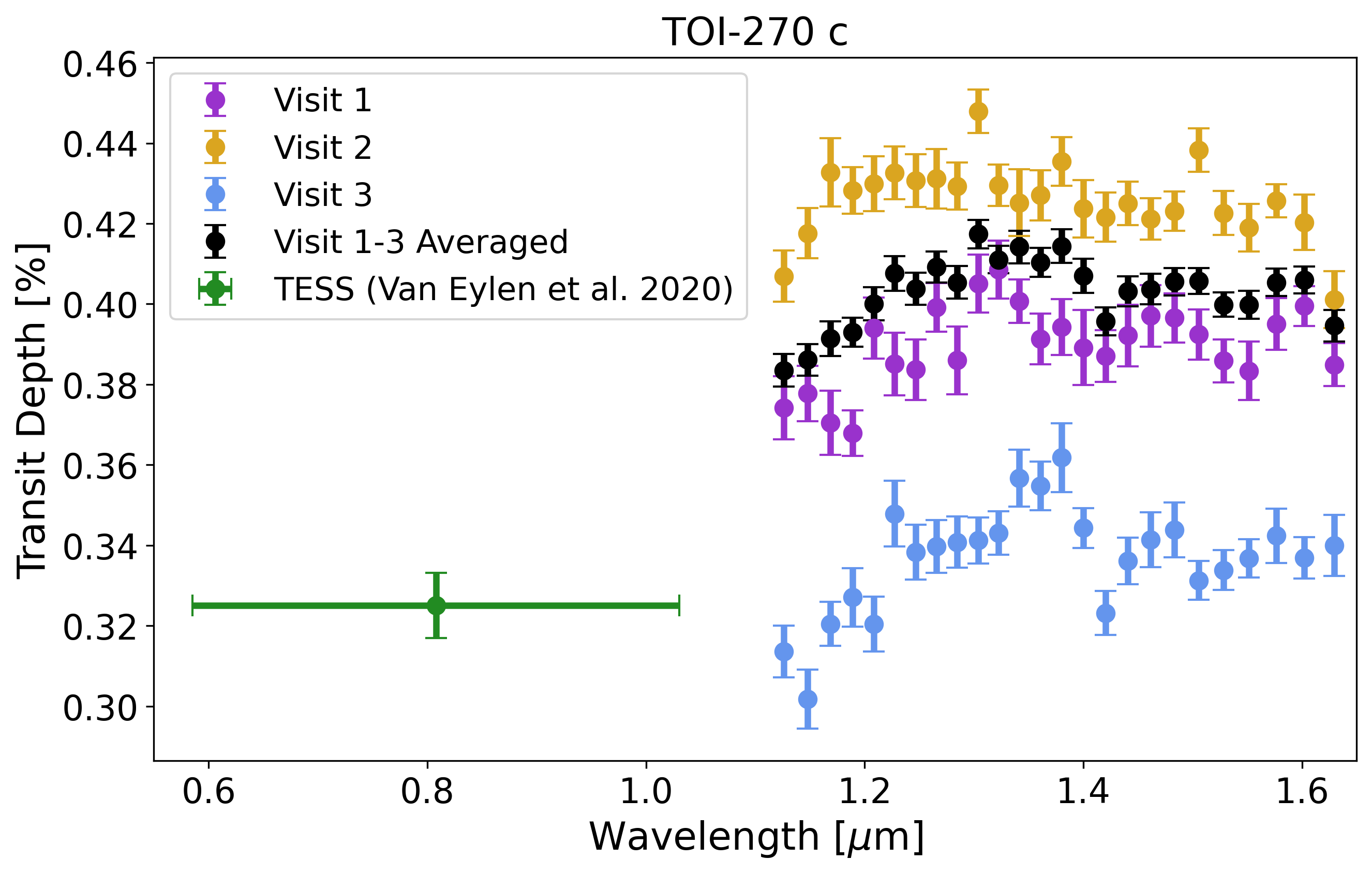}
    \includegraphics[width=0.45\textwidth]{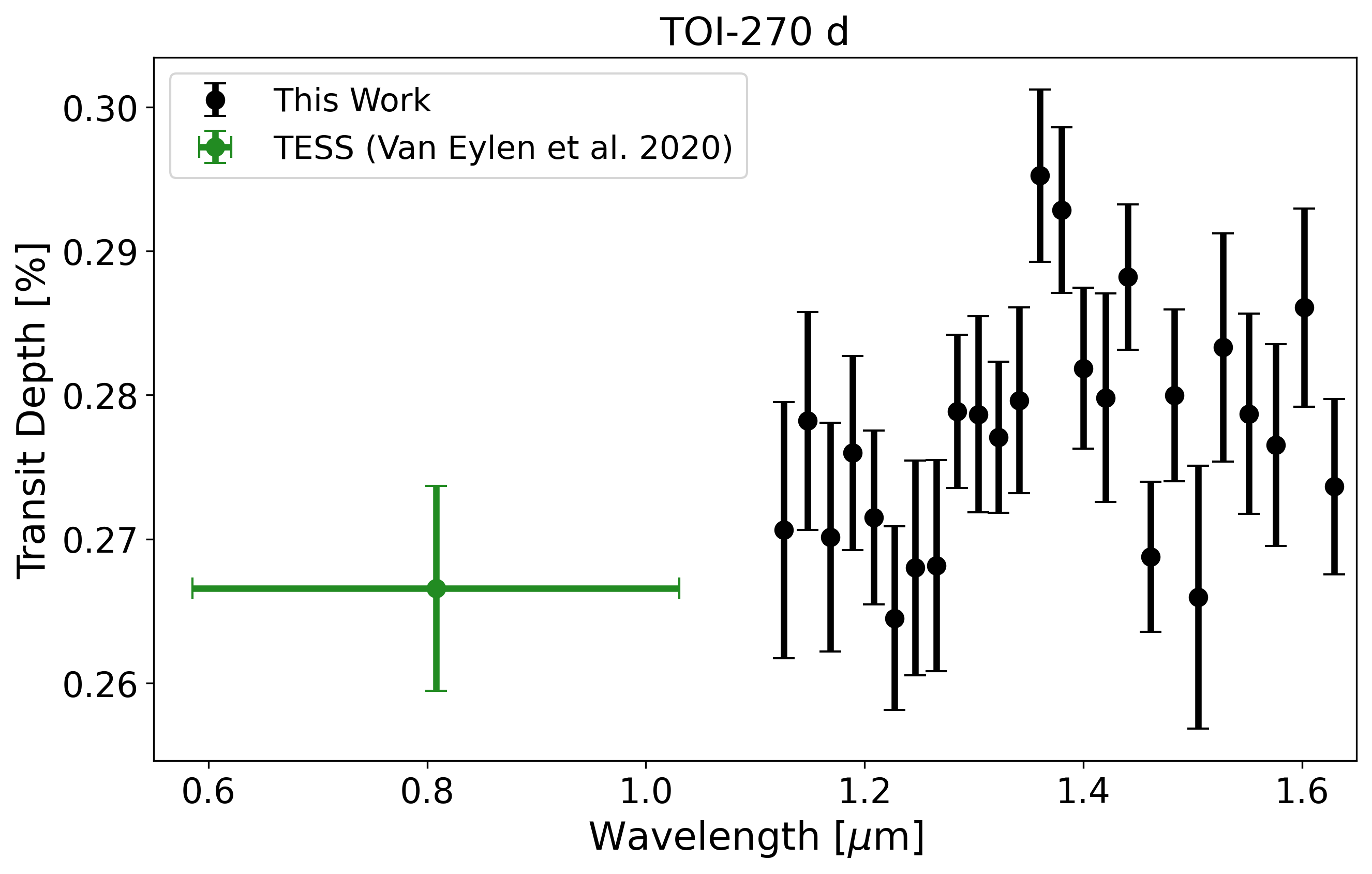}
    \caption{Comparison of the transit depths obtained here to those from \citet{van_eylen_toi270} for TOI-270\,c (left) and TOI-270\,d (right). As previously noted, we find a large variance in the transit depth obtained for each visit of TOI-270\,c and only one of these is of a similar depth to the TESS data. }
    \label{fig:toi270cd_comp}
\end{figure*}

The second possible explanation is that the feature could be due the transit light source effect as stellar spots or facuale can cause significant spectral modulation. While no spot crossing events were found in the light curves, unocculted spots could be to blame and these could also explain the large difference between the averaged HST data and the TESS transit depth from \citet{van_eylen_toi270}. Surprisingly, the HST data for TOI-270\,d has no such feature. The observation of TOI-270\,d was taken between 4 days after the third visit of TOI-270\,c, with the observations of TOI-270\,c data being taken over a period of roughly 7 months. There are 56.61 days between visits 1 and 2, while visits 2 and 3 were separated by 164.16 days. \citet{van_eylen_toi270} estimated that the rotation period of the star was around 58 days, which would imply that approximately the same face of the star was visible for each observation of TOI-270\,c. Nevertheless, the visible portion of the star would have only changed by around 7\% between the third visit of TOI-270\,c and the transit of TOI-270\,d. The transit chords of these planets is different as they have slightly different inclinations, but this would not affect the transit light source effect from unocculted spots.

We attempted atmospheric retrievals on both planets, but could find no model which would adequately fit the data for TOI-270\,c. Meanwhile, for TOI-270\,d, our retrievals uncovered evidence for H$_2$O.

\subsection*{TOI-674\,b}
\label{sec:toi674}

The discovery of the Neptune-sized planet TOI-674\,b (R = 5.25 R$_\oplus$, M = 23.6 M$_\oplus$,) was recently announced \citep{murgas_toi674}. The planet has a period of only 1.977143 days, but orbits an M-dwarf (Teff=3514 K) and thus has a equilibrium temperature of around 650 K. Three transits of TOI-674\,b were taken as part of GO-15333 (PI: Ian Crossfield). These have previously been presented in \citet{brande_toi674} where the discovery of a water feature was announced. Their study also utilised data from TESS and Spitzer, but the retrieved water abundance in their study (log(vmr) = -3.00$\pm$1.00) is consistent with ours (log(vmr) = -3.12$^{+0.78}_{-1.04}$). We note that a small vertical offset is present between our HST WFC3 G141 spectrum and that of \citet{brande_toi674}. As such offsets are a common occurrence, this is not concerning except when combining instruments without wavelength overlap. Finding such an offset further motivates our choice to only consider data from HST WFC3 G141 instead of combining all available datasets without being able to verify their compatibility.



\subsection*{V1298\,Tau\,b \& c}

V1298\,Tau\,b, a warm Jupiter-sized planet with an orbital period of 24.148 days, was detected around a young solar analogue by \citet{david_v1298_b}. Further analysis of the K2 data unveiled three additional planets in the system \citep{david_v1298_all}, with planets b and c having orbital periods of 8.25 and 12.4 days, respectively. However, the period of V1298\,Tau\,e could not be constrained as it only transited once in the K2 data. A second transit was later observed by TESS, which helped constrain potential orbital periods with the highest probability orbit placing V1298\,Tau\,e close to a 2:1 resonance with V1298\,Tau\,b \citep{feinstein_v1298}. 

Measuring the masses of these planets has proved difficult due to the intense activity of the host star. \citet{beichman_v1298} placed a 3$\sigma$ upper-limit on the mass of V1298\,Tau\,b of 2.2 M$_{\rm J}$ while dynamical arguments have placed constraints on the total masses of the planet pairs \citep{david_v1298_all}. Further radial velocity measurements were taken and lead to the first mass measurements in the system \citet{suarez_v1298}. Their work concluded that V1298\,Tau\,b had a mass of 0.64 M$_J$ while V1298\,Tau\,e had a mass of 1.16 M$_J$, making it much denser than typical giant exoplanets. However, the period of V1298\,Tau\,e derived by \citet{suarez_v1298} disagrees to 4$\sigma$ with the value from \citep{feinstein_v1298}. 

\begin{figure}
    \centering
    \includegraphics[width=\columnwidth]{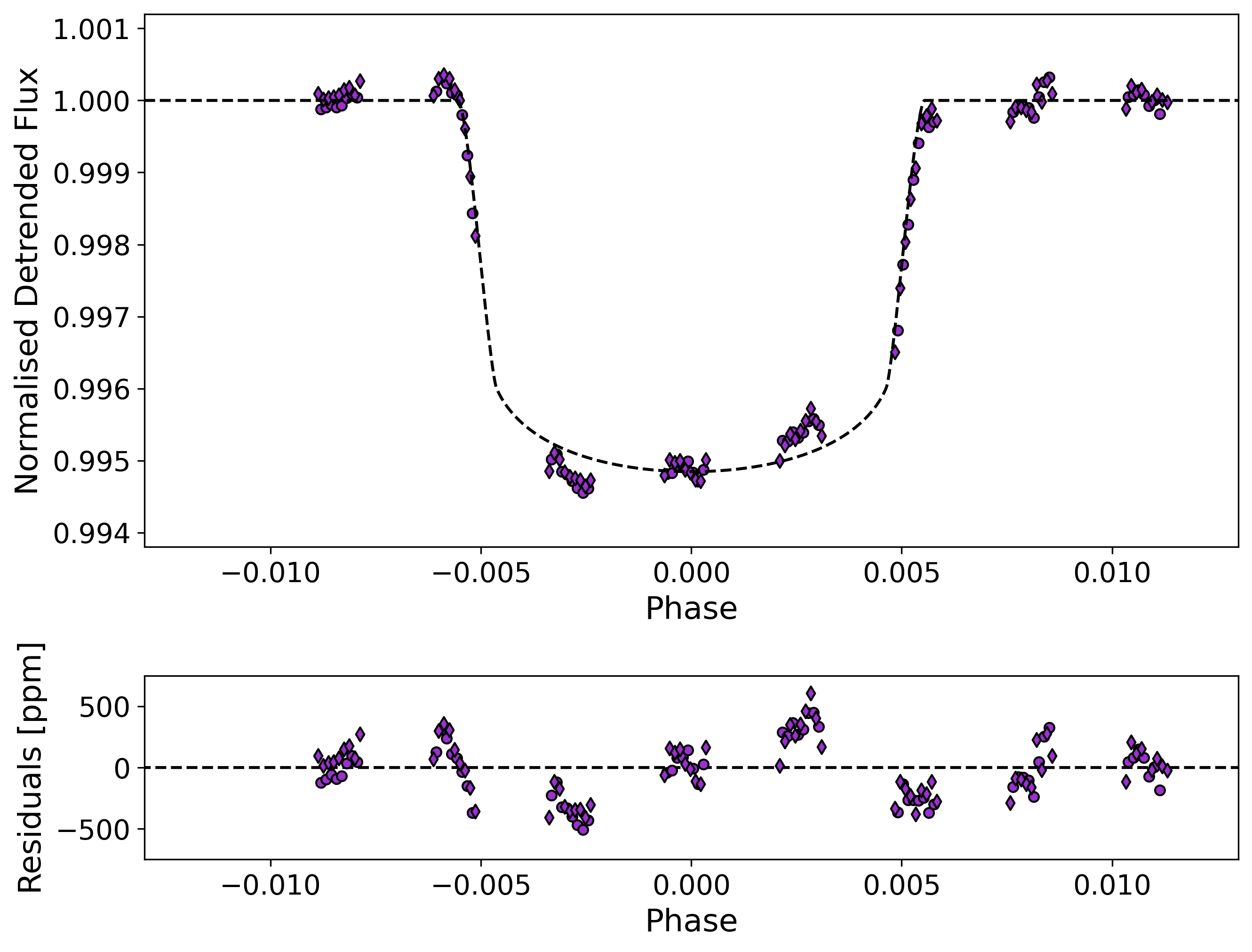}
    \includegraphics[width=\columnwidth]{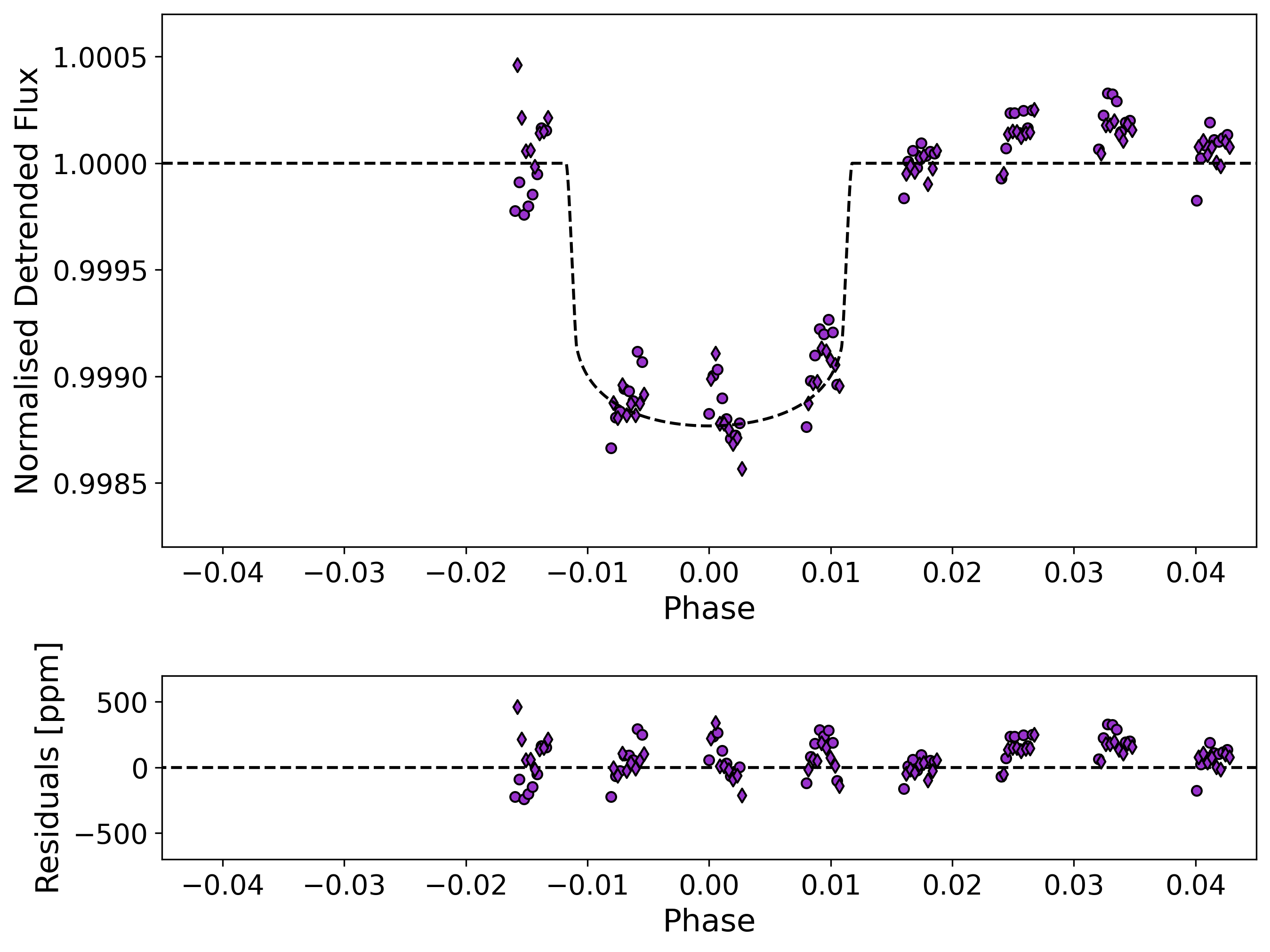}
    \caption{White light curve fits for the observations of V1298\,Tau\,b (top) and V1298\,Tau\,c (bottom). We note the non-Gaussian residuals in each case, potentially due to stellar activity.}
    \label{fig:v1298_wlc}
\end{figure}

Two datasets have been taken with Hubble WFC3 G141 to probe the atmospheres of worlds in the V1298 Tau system. The transit of V1298\,Tau\,b utilised ten Hubble orbits and was taken as part of GO-16083 (PI: Kamen Todorov). For these observations, the GRISM256 subarray was used, with 5 up-the-ramp reads using the SPARS25 sampling sequence, leading to an exposure time of 89.661957\,s,. S scan rate of 0.230\,''/s was used, which gave a scan length of around 170 pixels. The observation of V1298\,Tau\,c (GO-16462, Vatsal Panwar) required only eight orbits but used the same detector setup as the transit of V1298\,Tau\,b.

The white light curves fits for the two planets are shown in Figure \ref{fig:v1298_wlc}. In both cases, non-Gaussian residuals can be seen and these are likely due to the high variability of the host star. Nevertheless, thanks to the divide-by-white method, the residuals on the spectral light curves were Gaussian. The spectra recovered showed clear evidence of spectral modulation and we conducted retrievals to attempt to constrain the chemistry. As note above, there are disagreements about the orbital period of V1298\,Tau\,e. As the total radial velocity signal is a combination of components from all planets in the system, an error in the derivation of the period of one planet can lead to an inaccurate mass measurement of both it and another planet in the system. Additionally, V1298\,Tau\,c doesn't have a measured mass. Therefore, for both planets we attempted retrievals in which we fitted for the mass \citep{Changeat_2020_mass}. For V1298\,Tau\,b, we also ran retrievals with the mass fixed to the value from \citet{suarez_v1298}.

\begin{figure}
    \centering
    \includegraphics[width=\columnwidth]{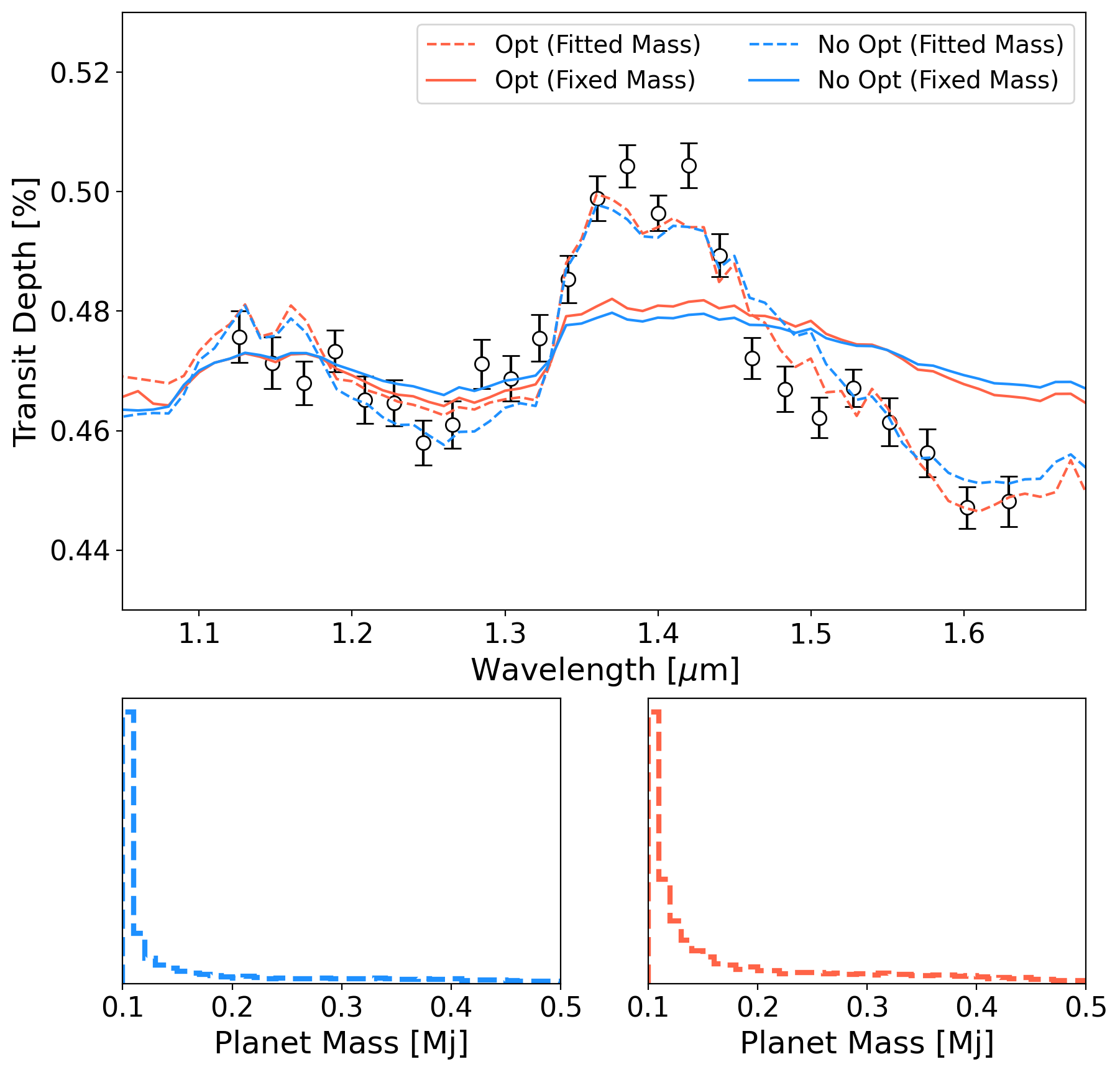}
    \caption{Top: spectrum of V1298 Tau b recovered with Iraclis and the preferred models from our retrievals. Bottom: the probability distributions for the planetary mass when it was fitted for. We note that, while it gives a preferable fit to the data, the fitted mass is extremely low \citep[a value of 0.64 M$_{\rm J}$ was recovered by][]{suarez_v1298}. The optical absorber model is unlikely for this planet given the equilibrium temperature ($\sim$1000 K). }
    \label{fig:v1298_b_spec}
\end{figure}

\begin{figure}
    \centering
    \includegraphics[width=\columnwidth]{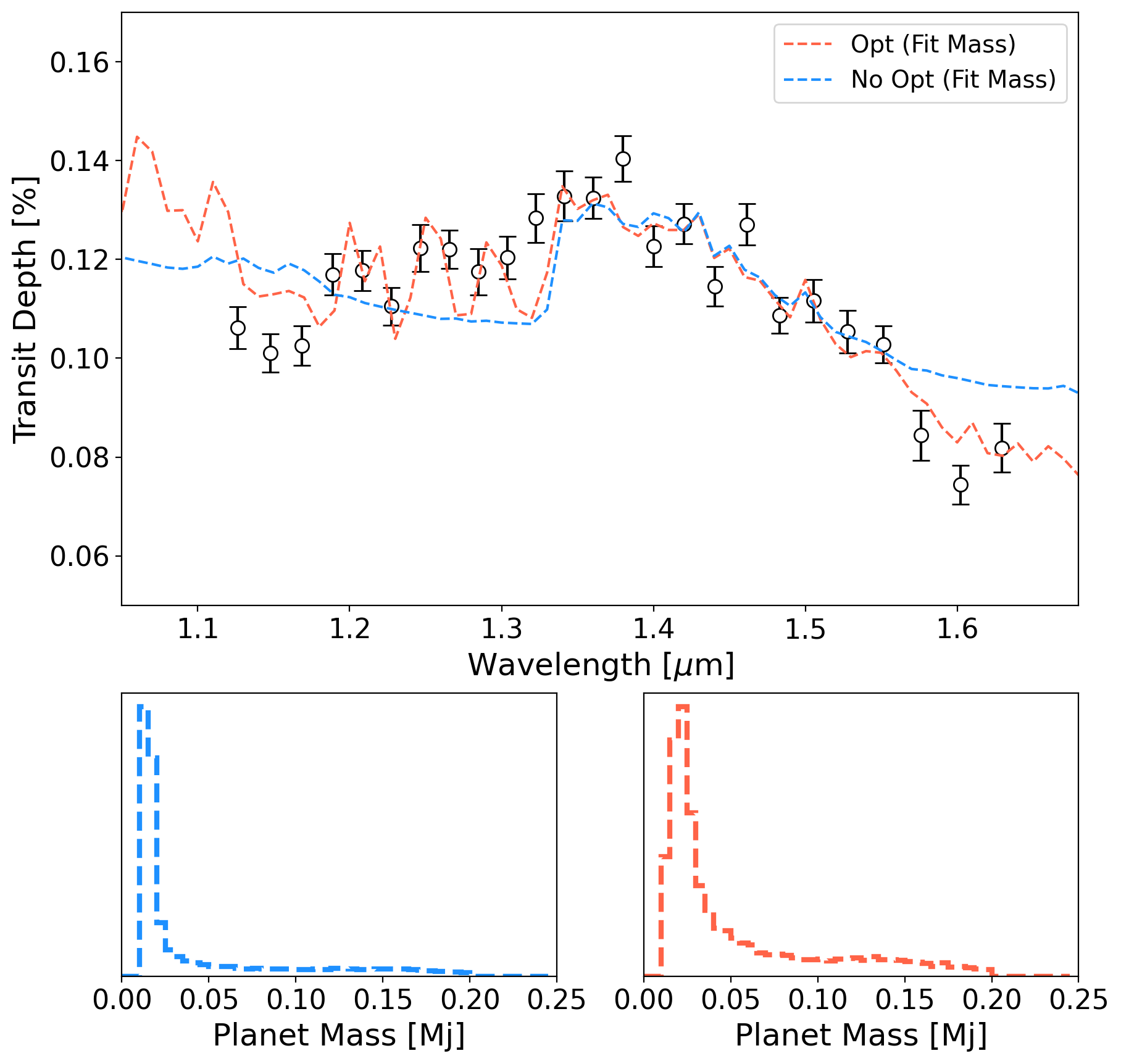}
    \caption{Top: spectrum of V1298 Tau b recovered with Iraclis and the preferred models from our retrievals. Bottom: the probability distributions for the planetary mass when it was fitted for. The optical absorber model is preferred, but it is unlikely that FeH and e- are present in the atmosphere of this planet given the equilibrium temperature ($\sim$700 K).}
    \label{fig:v1298_c_spec}
\end{figure}

In the case of V1298\,Tau\,b, the fixed-mass retrievals did not lead to models which well-fitted the data. As can be seen in Figure \ref{fig:v1298_b_spec}, the retrievals were unable to recreate the strong water seen at 1.4 $\mu$m. When we fitted for the mass, the retrieval was subsequently able to fit this feature but the retrieved mass was extremely low. We had placed a lower bound on the mass at 0.1 M$_{\rm J}$, which is far below the 0.64 M$_{\rm J}$ from \citet{suarez_v1298}. 

For V1298\,Tau\,c, the both retrievals fitted for the mass as there are currently no constraints on it from radial velocities or transit timing variations. As seen in Figure \ref{fig:v1298_c_spec}, the retrieval without optical absorbers struggles to fit the spectrum. However, when they are included a combination of FeH and e- help to create the features seen yet the cool temperature of the planet makes the existence of these species unlikely. 

Given the issues with fitting of these spectra it is possible that the stellar activity has affected the spectrum recovered with Iraclis. Iraclis can fit the long-term trend using a linear or quadratic model, but in these cases a different model might be preferable (e.g. a sinusoid). However, we leave such an exploration for future work which should also aim to resolve the discrepancy in the period of V1298\,Tau\,e to increase the accuracy and confidence in the mass measurements. Some transit timing variations have already been seen in the system and future transit measurements may also led to mass constraints. Due to the poor fitting of the spectra of V1298\,Tau\,b and c, we do not include them in the primary analyses of this paper (e.g. any of the trend fitting) but include them for completeness and to highlight the issue with the data.

\subsection*{WASP-6\,b}

Discovered by \citet{gillon_w6}, WASP-6\,b has a mass around half that of Jupiter but is inflated and has an equilibrium temperature of 1200 K. The atmosphere of the planet has been widely studied, with \cite{nikolv_w6} presenting the HST STIS and Spitzer IRAC transmission spectrum which showed signs of scattering in the optical. The same data was later analysed by \citet{sing} before \citet{carter_w6} present the HST WFC3 G141 spectrum. \citet{carter_w6} also took data with the VLT and other ground-based spectra of WASP-6\,b have been also been obtained, all of which have concluded the atmosphere is hazy due to a lack of spectral features due to Na or K \citep{jordan_w6,carter_w6}. 

\begin{figure}
    \centering
    \includegraphics[width=\columnwidth]{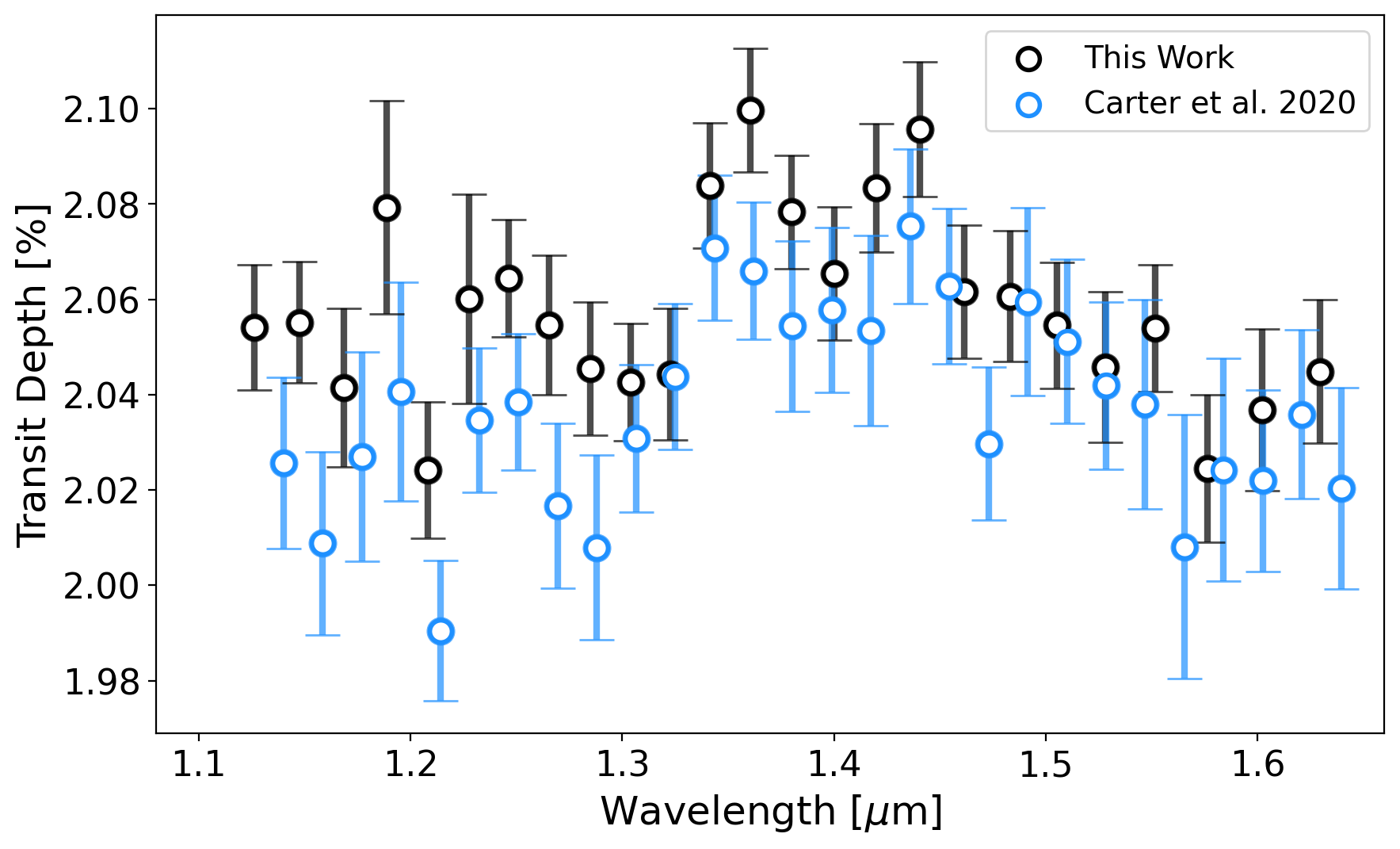}
    \caption{Comparison of the transit spectrum of WASP-6\,b obtained here to the spectrum from \citet{carter_w6}.}
    \label{fig:w6_comp}
\end{figure}

The HST WFC3 G141 data for WASP-6\,b was taken for proposal GO-14767 \citep[PI: David Sing, ][]{sing_14767_prop}. For these observations, the GRISM512 subarray was used, with 8 up-the-ramp reads using the SPARS25 sampling sequence, and an exposure time of 138.380508\,s, combined with a scan rate of 0.06\,''/s. The data was previously presented in \citet{carter_w6} where it was combined with data from the VLT. The analysis by \citet{carter_w6} again found the atmosphere to be hazy and they also noted that corrections for the stellar heterogeneity could have significant effect on the Na and K abundances. We find a highly similar HST WFC G141 spectrum to this work, as shown in Figure \ref{fig:w6_comp}.

\subsection*{WASP-18\,b}

WASP-18\,b \citep{hellier_w18} has been thoroughly studied since its discovery in 2008. Spitzer, Hubble WFC3 and ground-based eclipses have been taken \citep{nymeyer_w18_spit,w18_ground_ec, sheppard_w18,arcangeli_wasp18_em, manjavacas_wasp103, iro_w18, gandhi_w18} as well as phase curves with Hubble WFC3 and TESS \citep{arcangeli_w18_phase,shporer_w18}. These have revealed a low albedo, poor redistribution of energy to the night-side and evidence for an inverted day-side temperature-pressure profile. In particular, the analysis from \cite{sheppard_w18} considered a similar data-set to us and detected a strong thermal inversion, associated with the presence of H$_2$O and CO.

The HST transmission spectrum was taken, along with two eclipses, as part of a phase curve (GO-13467, PI: Bean). These observations had an exposure time of 73.74 seconds having used the GRISM256 aperture and the SPARS10 sampling sequence with 12 up-the-ramp reads. Due to the high mass of WASP-18\,b, 1$\sigma$ uncertainties on the recover transit spectrum were equivalent to around 5 scale heights, denying any chance of recovering spectral features.

\subsection*{WASP-19\,b}

WASP-19\,b orbits its bright host star on a very short orbit (0.94 days) and its high temperature, and large size, make it an excellent target for atmospheric studies \citep{hellier_w19}. WASP-19\,b has been the subject of a number of investigations, from both the ground and from space. Work by \cite{anderson_wasp19_spitzer} analysed four Spitzer eclipses, taken across 3.6-8\,$\mu m$, and constructed a spectral energy distribution of the planet’s day-side atmosphere. They found no stratosphere, supporting the hypothesis that hot Jupiters orbiting active stars have suppressed thermal inversions \citep{knutson_therm_inv}. Analysis of the TESS optical phase curve showed moderately efficient day-night heat transport, with a day-side temperature of 2240 K and a day to night contrast of around 1000 K \citep{wong_wasp19_tess}. This study also utilised a host of ground-based observations by \cite{anderson_wasp19_ground,burton_wasp19,abe_wasp19,bean_wasp19}. 

WASP-19\,b has also been studied via transmission spectroscopy. The retrievals of the STIS-G430L, G750L, WFC-G141 and Spitzer-IRAC observations suggest the presence of water at log(H$_2$O) $\approx$ 4 but show no evidence for optical absorbers \citep{sing,barstow_pop,pinhas}. Those results do not match the ground-based transits that were acquired with the European Southern Observatory’s Very Large Telescope (VLT), using the low-resolution FORS2 spectrograph which covers the entire visible-wavelength domain (0.43–1.04\,$\mu m$). When analysing this data, \cite{sedahati_tio_wasp19} detected the presence of TiO to a confidence level of 7.7$\sigma$. However, data from Magellan/IMACS did not find any evidence for TiO or Na as a featureless transmission spectrum was recovered \citep{espinoza_w19}.

Two transit observations of WASP-19\,b have been acquired. The first, in staring mode, as part of Proposal GO-12181 \citep[PI: Drake Deming, ][]{deming_12181_prop} has previously been presented in another studies \citep[e.g.][]{sing}. The scanning mode data were taken with the GRISM512 aperture and consisted of 4 up-the-ramp reads with the SPARS25 sequence. This gave an exposure time of 46.695518\,s scan rate was 0.026\,''/s. These observations were part of a phase curve and, while guide star acquisition issues were incurred, the failure in pointing happened long after the transit had occurred, meaning the data analysed here was unaffected.

\subsection*{WASP-103 b}

WASP-103\,b is an ultra-short period planet ($P$ = 22.2 hr) whose orbital distance is less than 20\% larger than its Roche radius, resulting in the possibility of tidal distortions and mass-loss via Roche-lobe overflow \citep{gillon_wasp103}. Given its size, temperature, and the brightness of its host star it is a great target for atmospheric studies and has been observed with numerous instruments.

WASP-103 b's HST WFC3 emission spectrum was found to be featureless down to a sensitivity of 175 ppm, showing a shallow slope toward the red \citep{cartier_wasp103}. Work by \cite{manjavacas_wasp103}, which performed a reanalysis of the same data-set, found that the emission spectrum of WASP-103\,b was comparable to that of an M-3 dwarf. \cite{delrez_wasp103} obtained several ground-based high-precision photometric eclipse observations which, when added to the HST data, could be fit with an isothermal black-body or with a low water abundance atmosphere with a thermal inversion. However, their Ks band observation showed an excess of emission compared to both these models. More recently, a phase-curve analysis of the planet was taken and reported in \cite{Kreidberg_w103}. The study also utilized the previous HST emission spectra and confirmed a seemingly featureless day-side. A later study on the same data employed a unified phase-curve retrieval technique to obtain a more complex picture of the planet \citep{changeat_2021_w103}. It confirmed the presence of thermal inversion and dissociation processes on the day-side of the planet and found signature of FeH emission. The study also constrained water vapour across the entire atmosphere. Furthermore, ground-based transmission observations found strong evidence for sodium and potassium \citep{lendl_wasp103}. Later observations by \citet{wilson_w103} yielded a featureless spectrum while a comprehensive analysis of 11 transmission spectra by \citet{kirk_w103} found evidence for unocculted regions or the star as well as weak evidence for TiO. \\

Two HST G141 phase curves of WASP-103\,b were obtained \cite[PN: 14050, PI:][]{kreidberg_w103_wfc3_w18_spit_prop} which each contained a single transit and eclipse. We analysed these two transits and observed that the slope in the final spectrum is well fit by VO. Other optical absorbers might be present (TiO, H$^-$) but the data does not allow verification of this. The solution found possesses a wide range of metallicities; sub-solar in nature. We note that \citet{kirk_w103} found that VO could only account for their spectrum if present in extremely high quantities.

\subsection*{WASP-107 b}

A sub-Saturn around a solar-metallicity K6 star, WASP-107\,b was immediately noted as an excellent target for atmospheric studies \citep{anderson_w107}. Soon after its discovery, the transmission spectrum of WASP-107\,b was taken with HST WFC3 G141 as part of proposal GO-14915 (PI: Laura Kreidberg). The data was previously presented in \citet{Kreidberg_w107}, showing strong evidence for the presence of water but a possible methane depletion. The study also noted that the features seen were smaller than would be expected for a cloud-free atmosphere, inferring the presence of high-altitude aerosols. In line with the results from \citet{Kreidberg_w107}, we found strong evidence for water but muted features compared to a clear atmosphere.

\begin{figure}
    \centering
    \includegraphics[width=\columnwidth]{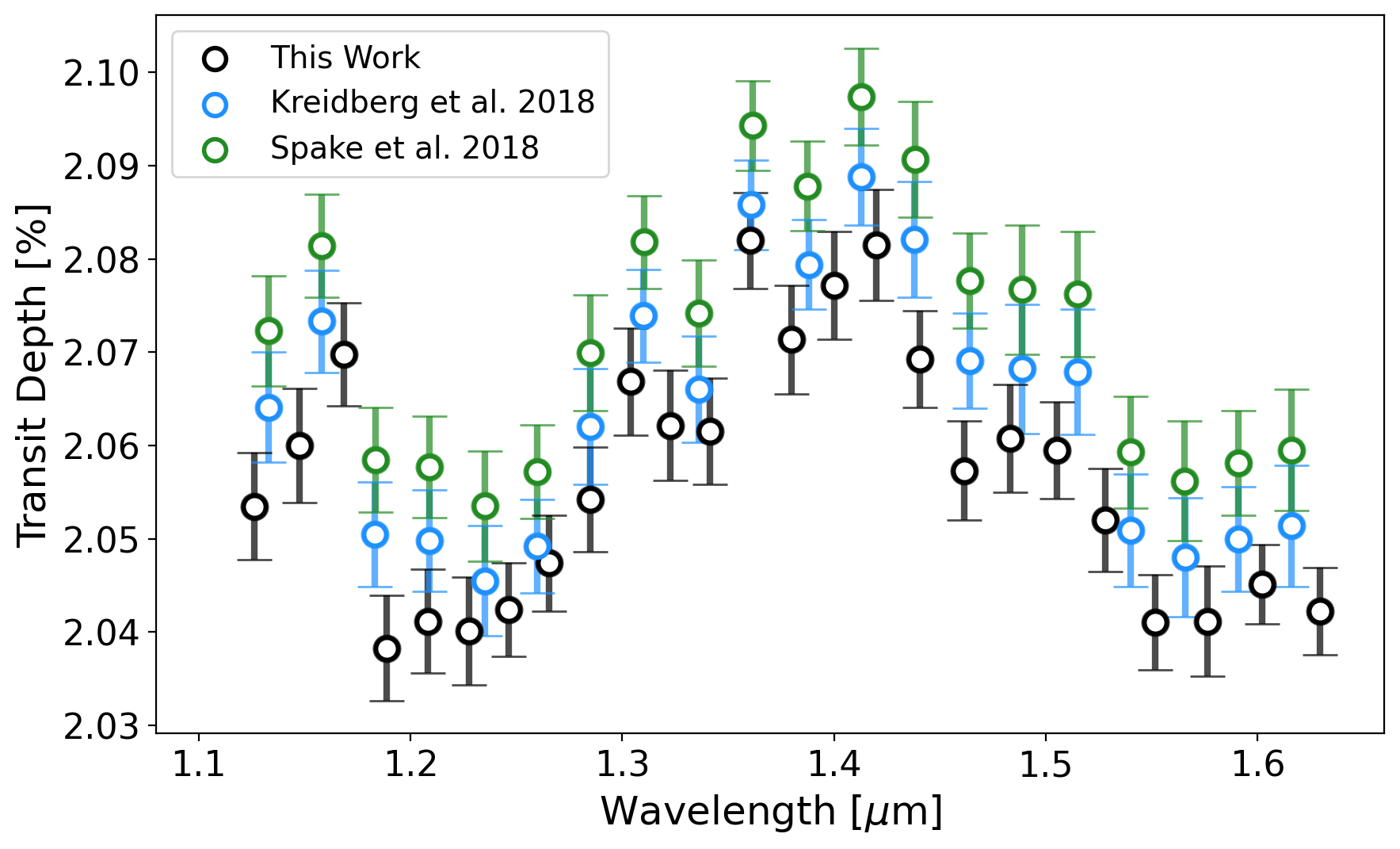}
    \caption{Comparison of the transit spectrum of WASP-107\,b obtained here to the spectrum from \citet{Kreidberg_w107} and \citet{spake_w107}.}
    \label{fig:w107_comp}
\end{figure}

Additionally, an observation was taken with the G102 grism of HST WFC3 (GO-14916, PI: Jessica Spake) which was used to demonstrate that the atmosphere of WASP-107\,b was eroding as the G102 data gave access to the 1.083 $\mu$m Helium line \citep{spake_w107}. High-resolution observations have since confirmed this detection \citep{allart_w107,kirk_w107}. We do not fit the G102 data as part of this study to ensure homogeneity across our datasets. We show a comparison between the spectrum from previous studies and ours in Figure \ref{fig:w107_comp}.

\subsection*{WASP-121 b}

Significant observational time has been spent on WASP-121\,b. In transmission, the analyses of ground-based observations, Hubble STIS and Hubble WFC3 have shown the presence of H$_2$O and optical absorption attributed to VO and/or FeH \citep{evans_wasp121_t1, evans_wasp121_t2}. The authors of these studies note that chemical equilibrium models with solar abundances cannot reproduce the spectrum seen, while free chemical retrievals can only do so by converging to high abundances of VO and FeH. In parallel, high resolution ground-based observations of the transit have put upper limits on the abundances of TiO and VO at the terminator with log(VMR) $<$ -7.3 and 7.9 respectively, suggesting these cannot be causing the inversion seen \citep{merritt_w121}. However, the study highlighted that these limits are largely degenerate with other atmospheric properties such as the scattering properties or the altitude of clouds on WASP-121\,b. Another study found a host of atomic metals, including V, which are predicted to exist if a planet is in equilibrium and has a significant quantity of VO \citep{hoeij_w121}. They too noted the absence of TiO which could support the hypothesis that Ti is depleted via a cold-trap. Furthermore, various high resolution studies have found evidence for absorption due to metallic lines \citep[e.g.][]{borsa_w121,cabot_w121,gibson_w121}.

Here, we fit three transit observations. Two of these were obtained as part of a phase curve \citep[PN: 15134, PI: Thomas Mikal-Evans,][]{evans_15134_prop} which was recently published, showing evidence for a diurnal water cycle \citep{evans_w121_pc}. The other transit \cite[PN: 14468, PI:][]{evans_14468_prop} was previously analysed, both with Iraclis \citep{tsiaras_30planets} and by \citet{evans_wasp121_t1}. We choose to refit this observation to ensure the methodology, parameters and limb-darkening coefficients were the same across all three transit fits.

\subsection*{WASP-178 b}

WASP-178\,b, also known as KELT-26\,b, is an ultra-hot Jupiter orbiting an A1V host star \citep{hellier_w178}. The planet appears to be in a highly misaligned orbit and has a mass of 1.93$^{+0.14}_{-0.16}$ M$_J$ \citep{rodriguez_w178}, making it one of the heaviest planets in our sample. 

A single transit was taken as part of proposal GO-16450 (PI: Joshua Lothringer). For these observations, the GRISM256 subarray was used, with 8 up-the-ramp reads using the SPARS25 sampling sequence, and an exposure time of 138.354034\,s, combined with a scan rate of 0.07597\,''/s. We note that WFC3 data with the UVIS and G102 grisms has also been taken for this planet but was not included in this study due to the need to ensure homogeneity across the planet sample. 

We fitted the data with Iraclis and the preferred atmospheric model for this finds strong evidence for water in the atmosphere. The increasing absorption at shorter wavelengths is best fit with a large abundance of TiO as well as absorption by H-. The result is consistent with the expected chemistry given the equilibrium temperature of around 2400 K, although we note that the retrieved temperature is cooler than expected ($\sim$1250 K). We note that analysis of the HST WFC3 UVIS data by \citet{lothringer_w178_uv} found evidence for SiO. To understand this planet fully, an analysis of all three HST datasets is required.

\clearpage

\section*{Appendix 2: Fixed CO retrievals}

One set of retrievals conducted in this study fitted equilibrium chemistry models to the data, with the C/O ratio and metallicity as free parameters controlling the chemistry. Despite the large uncertainties on the retrieved C/O ratio, we noted that the majority of planets were not consistent with C/O ratios larger than 1. To explore this further we attempted chemical equilibrium retrievals with fixed C/O ratios of 0.5 or 1. 

\begin{figure}[b]
    \centering
    \includegraphics[width=0.95\columnwidth]{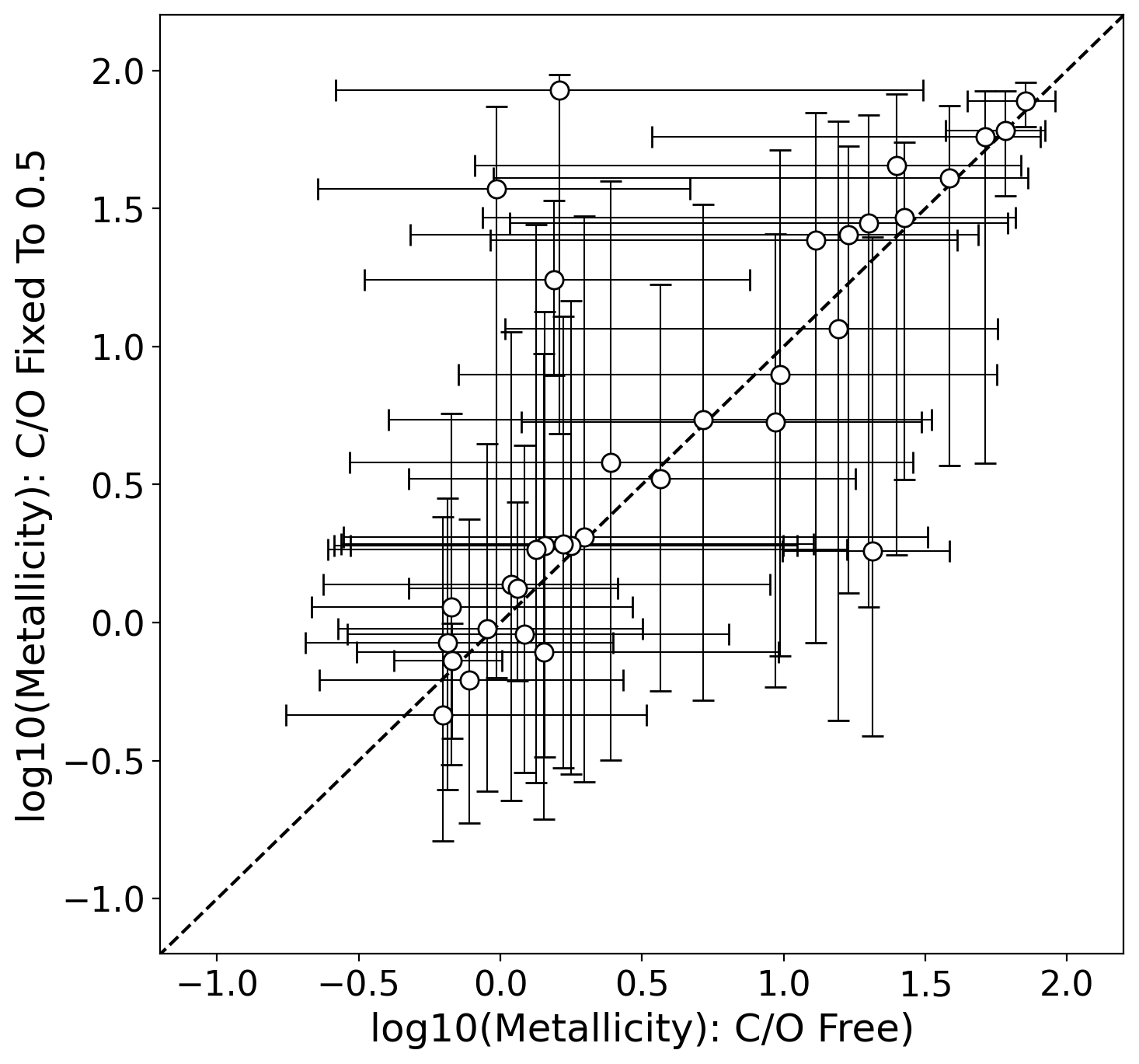}
    \includegraphics[width=0.95\columnwidth]{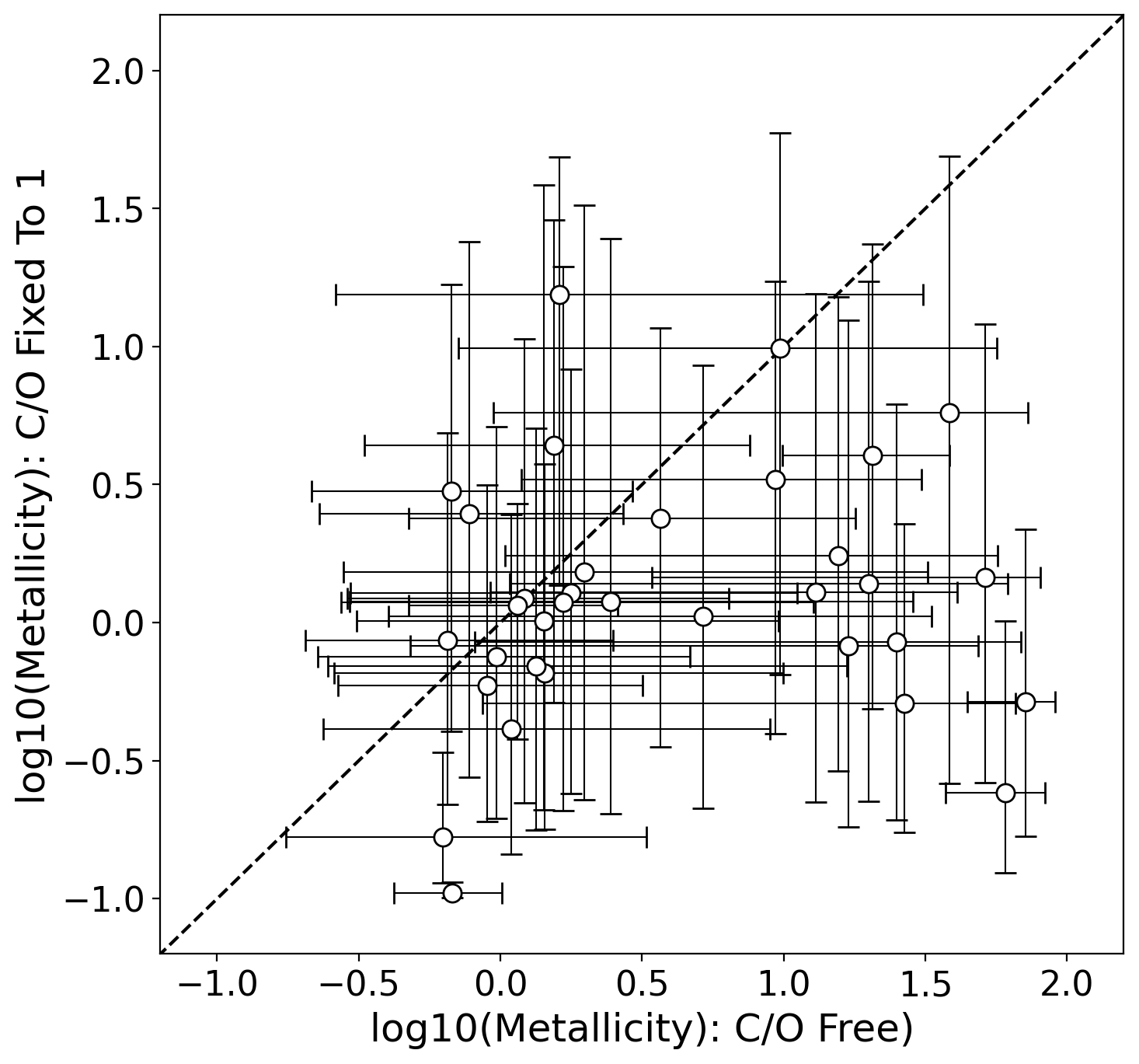}
    \caption{The comparison of the retrieved metallicity when the C/O ratio is a free parameter to when it is fixed to 0.5 (top) and 1 (bottom). We only show planets with our sample which have a strong ($>$3$\sigma$) atmosphere detection.}
    \label{fig:co_fixed_res}
\end{figure}

We then compared the metallicities retrieved in each case, as well as the goodness of the fit. We find that, forcing the C/O to be equal to 1 led to the metallicities of the planets generally being retrieved as solar but that these models provided a poorer fit to the data than when the C/O ratio was free to vary. On the other hand, we find fixing the C/O to 0.5 generally leads to only minor differences in the metallicity and goodness of fit in comparison to the free C/O case. The results of these retrievals are displayed in Figure \ref{fig:co_fixed_res} and provide further suggestions that high C/O ratio atmospheres are not compatible with the spectra derived in this work. However, we caution that the narrow wavelength coverage of HST WFC3 G141 does not provide strong constrains on carbon-bearing species so the robustness of the finding is questionable.

\begin{figure}[b]
    \centering
    \includegraphics[width=0.95\columnwidth]{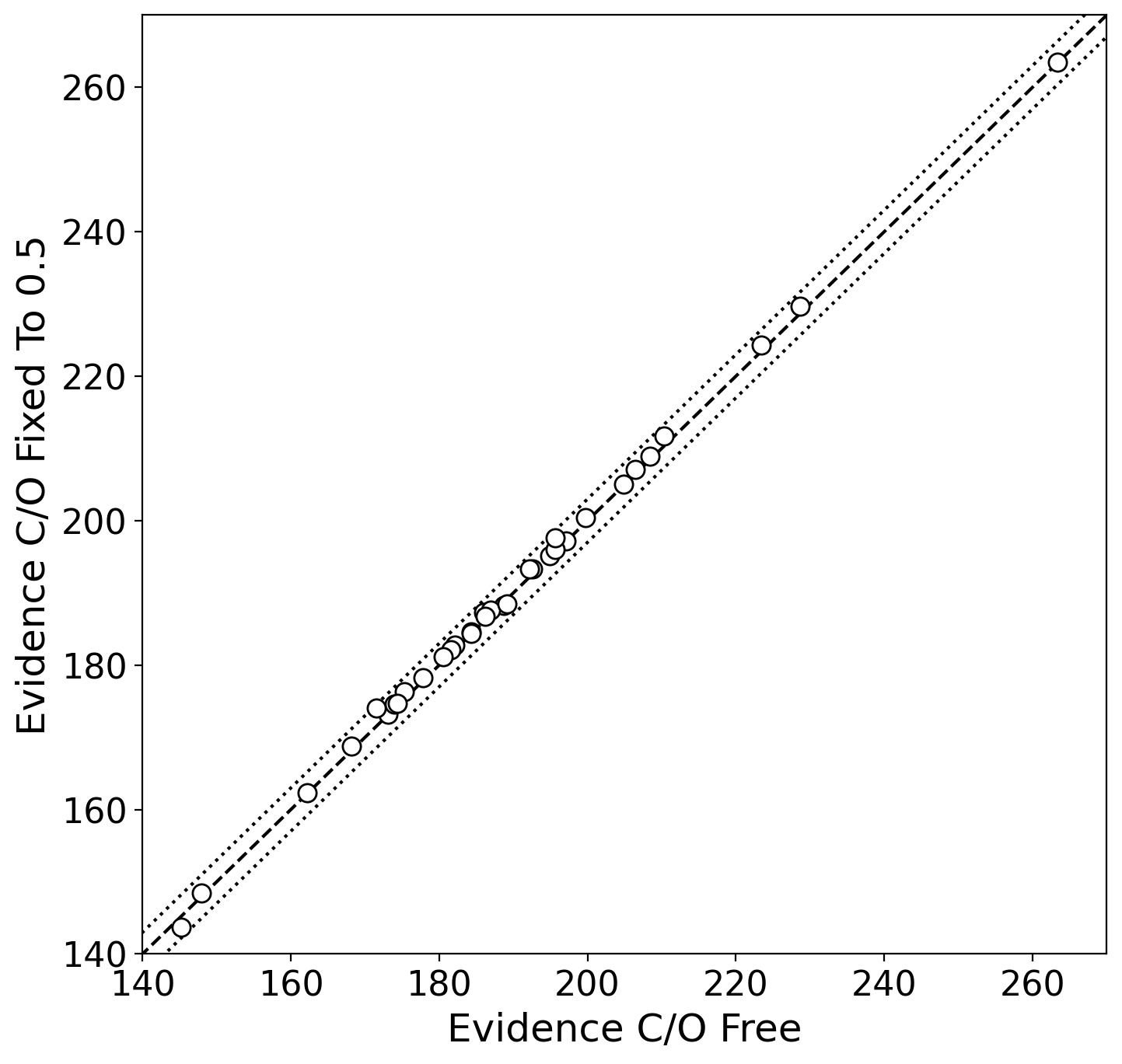}
    \includegraphics[width=0.95\columnwidth]{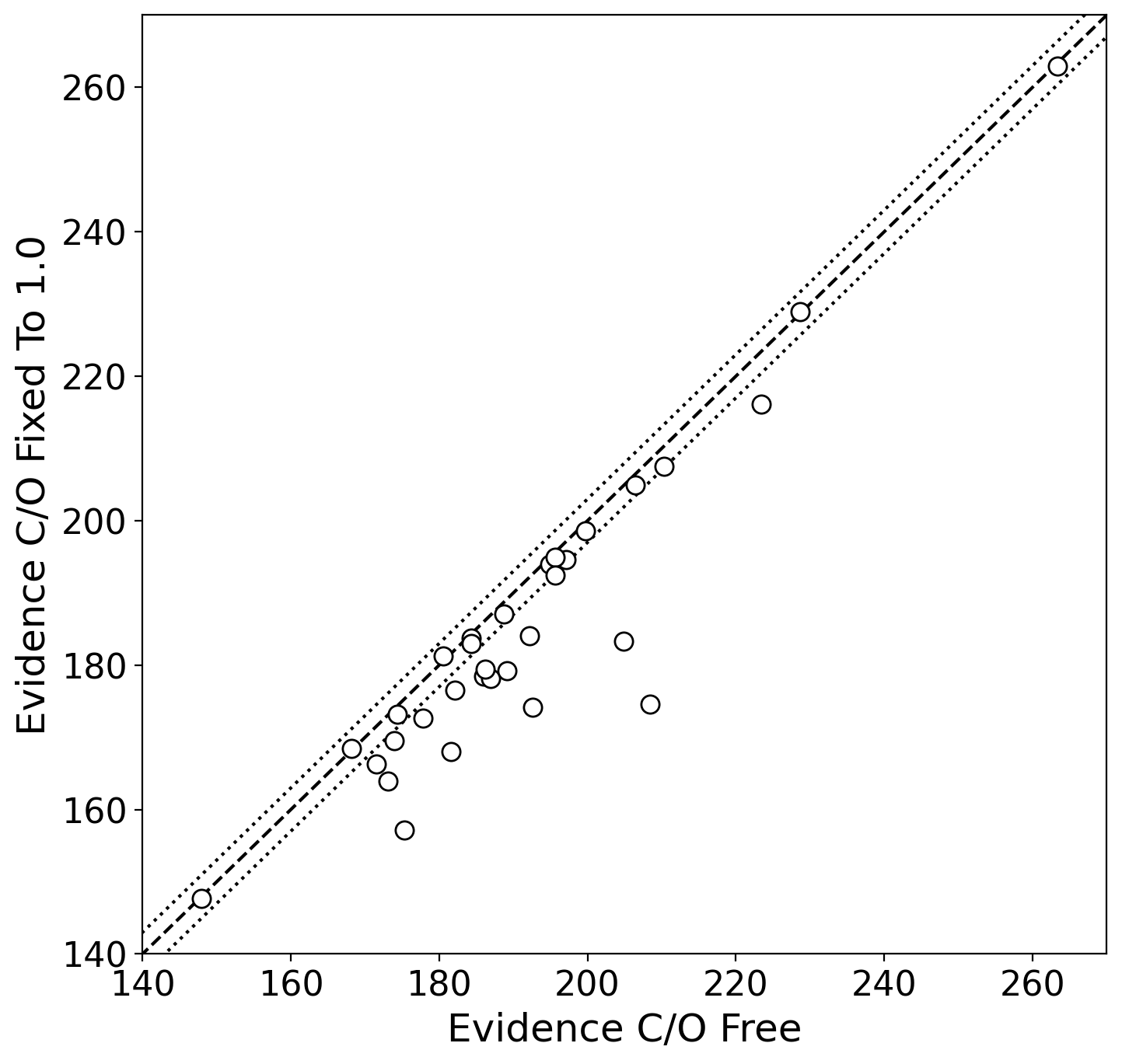}
    \caption{Comparison of the Bayesian evidence for the fixed C/O ratio retrievals when it is fixed to 0.5 (top) and 1 (bottom). The dotted line shows the 3$\sigma$ region.}
    \label{fig:co_fixed_ev}
\end{figure}

\clearpage

\section*{Appendix 3: Bayesian Hierarchical Modelling}
\label{app:bhm}
The following derivation is based primarily the work of \cite{lustig_yaeger_hm} so, while we will provide a brief discussion on the method, interested readers can refer to Section 2.2 in \cite{lustig_yaeger_hm} for more detailed discussion. Consider that we have obtained $N$ exoplanets, each with $M_n$ observations $\bm{D}_n$. For our case $M_n = 1$ but it can be more than 1. For each of these planets we can infer the joint posterior distribution of a pre-defined set of parameters $\bm{\theta}_n$ given $\bm{D}_n$, $P(\bm{\theta}_n|\bm{D}_n)$ via standard Bayesian formulation:
\begin{equation}
    P(\bm{\theta}_n|\bm{D}_n) = \frac{P(\bm{D}_n|\bm{\theta}_n)P(\bm{\theta}_n)}{P(\bm{D}_n)}
    \label{eqn:bayes}
\end{equation}
where $P(\bm{D}_n|\bm{\theta}_n$) is the likelihood of observing $D$ given the specific set of parameters $\theta$. $P(\bm{D})$ is the marginal likelihood over all possible sets of $\bm{\theta}_n$, also known as the evidence. The prior function, $P(\bm{\theta}_n)$ is taken to be flat or uninformative. 

Suppose we would like to derive the population level trend of planetary temperature against some molecular abundance. If we assume the trend can be parameterised by a set of hyper-parameters $\bm{\alpha}$, we can compute the (population level) likelihood of the entire data ensemble $\mathcal{L}_{\alpha}$ as
\begin{equation}
    \mathcal{L}_{\alpha} = P(\{\bm{D}\}_{n=1}^N|\bm{\alpha})
\label{eqn:llh_alpha}
\end{equation}

If we further assume that there is no likelihood covariance between parameters of different exoplanets $n$, we can express Equation \ref{eqn:llh_alpha} as the product of $N$ marginalised integral over parameters $\bm{\theta}_n$.
\begin{align*}
    \mathcal{L}_{\alpha} &= \prod^N_{n=1}\int P (\bm{D}_n,\bm{\theta}_n|\bm{\alpha}_n)d\bm{\theta}_n \\
    &= \prod^N_{n=1}\int P (\bm{D}_n|\bm{\theta}_n,\bm{\alpha})P(\bm{\theta}_n|\bm{\alpha})d\bm{\theta}_n
    \label{eqn:llh_alpha}
\end{align*}
assuming $\bm{D}_n$ is related to $\alpha_n$ through $\bm{\theta}_n$. The first term is simply the likelihood function for each individual observations (nth planet) and the second term act as a re-weighting function. 

We can further manipulate Equation \ref{eqn:llh_alpha} by recognising that the second term is a ratio between the new prior $P_{\alpha}(f_{n,T})$, that depends solely on $\alpha$ and the original prior $P_{0}(f_{n,T})$ , multiples the prior function of the parameters $P(\bm{\theta}_n)$:
\begin{equation}
\mathcal{L}_{\alpha} = \prod^N_{n=1}\int P (\bm{\theta}_n|\bm{D}_n)\frac{P_{\alpha}(f_{T,n})}{P_{0}(f_{T,n})})d\bm{\theta}_n
\label{eqn:llh_expand}
\end{equation}
Note that we have combined $P (\bm{D}_n|\bm{\theta}_n)$ and $P(\bm{\theta}_n)$ to get the posterior $P (\bm{\theta}_n|\bm{D}_n)$. We have also omitted the evidence term here since it is merely a constant in this context.
In our implementation, we have assumed a linear trend for the temperature, hence:
\begin{equation}
    f_T(m, c, X_{mol}) = mX_{mol}+c
\end{equation}
where $m$ and $c$ represent the slope and intercept of the straight line, and $X_{mol}$ represent the molecular abundance of the chemical species. 

We can simplify the integral in Equation \ref{eqn:llh_expand} by summing over all the samples in the posterior traces:
\begin{equation}
    \mathcal{L}_{\alpha} \approx \prod_{n=1}^N \frac{1}{K}\sum_{k=1}^K \frac{P_{\alpha}(f_{T,n,k})}{P_{0}(f_{T,n,k})}
\end{equation}
The new prior function $P_{\alpha}(f_{T,n,k})$, is assumed to be a Gassuian distribution, and therefore we can compute the probability analytically by comparing $f_{T,n,k}$ with $f_T(m, c, X_{mol})$:
\begin{equation}
    P_{\alpha}(f_{T,n,k}) = \mathcal{N}(f_{T,n,k} - f_T(m, c, X_{mol}), \sigma)
    \label{eqn:new_prior}
\end{equation}
Similar to \cite{lustig_yaeger_hm}, we have added an additional term $\sigma$ to account for the variability in the traces. The hyperparameter $\alpha$ has in total 3 free parameters, i.e. $\alpha \equiv [m, c, \sigma]$  that we can infer from the posterior traces

Once the likelihood function $\mathcal{L}_{\alpha}$ is computed, the posterior on the hyperparameter $\bm{\alpha}$ can be inferred simply by referring to Equation \ref{eqn:bayes}:
\begin{equation}
    P(\bm{\alpha}|\{\bm{D}_n\}_{n=1}^N) \propto \mathcal{L}_{\alpha}P(\bm{\alpha})
\end{equation}
where $P(\bm{\alpha})$ is the (hyper-)prior function for the hyperparameters $\bm{\alpha}$. In this paper we have fixed all the hyper-priors as uniform distribution. 

To infer the hyperparameters, we have used MultiNest algorithm \citep{Feroz_multinest,buchner_multinest} to compute the preferred BHM importance sampling model. Using the log-evidence provided by MultiNest, we are able to compare the different models and assess the evidence for a particular trend when compared to a null hypothesis.

\begin{figure*}
    \centering
    \includegraphics[width=0.9\columnwidth]{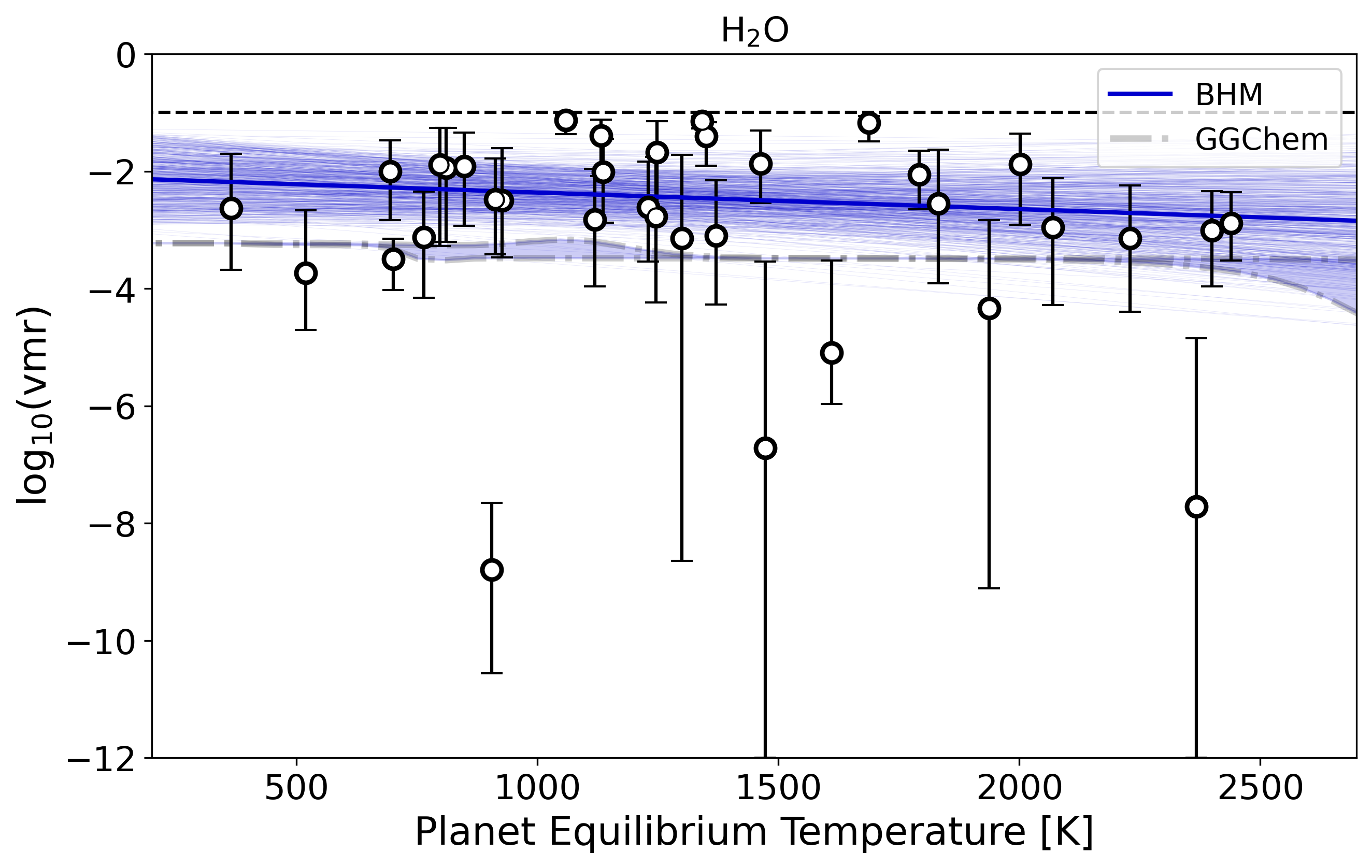}
    \includegraphics[width=0.9\columnwidth]{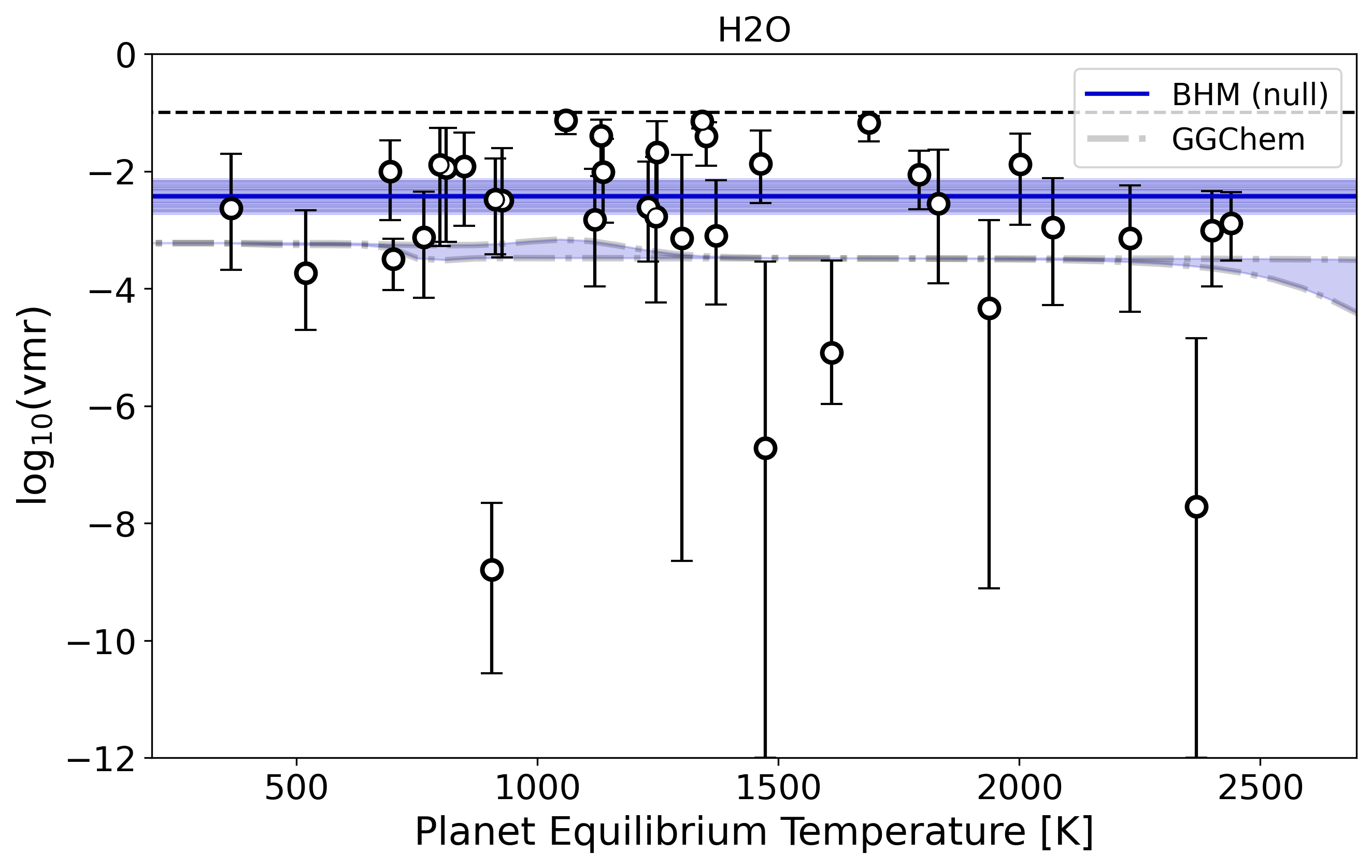}
    \includegraphics[width=0.9\columnwidth]{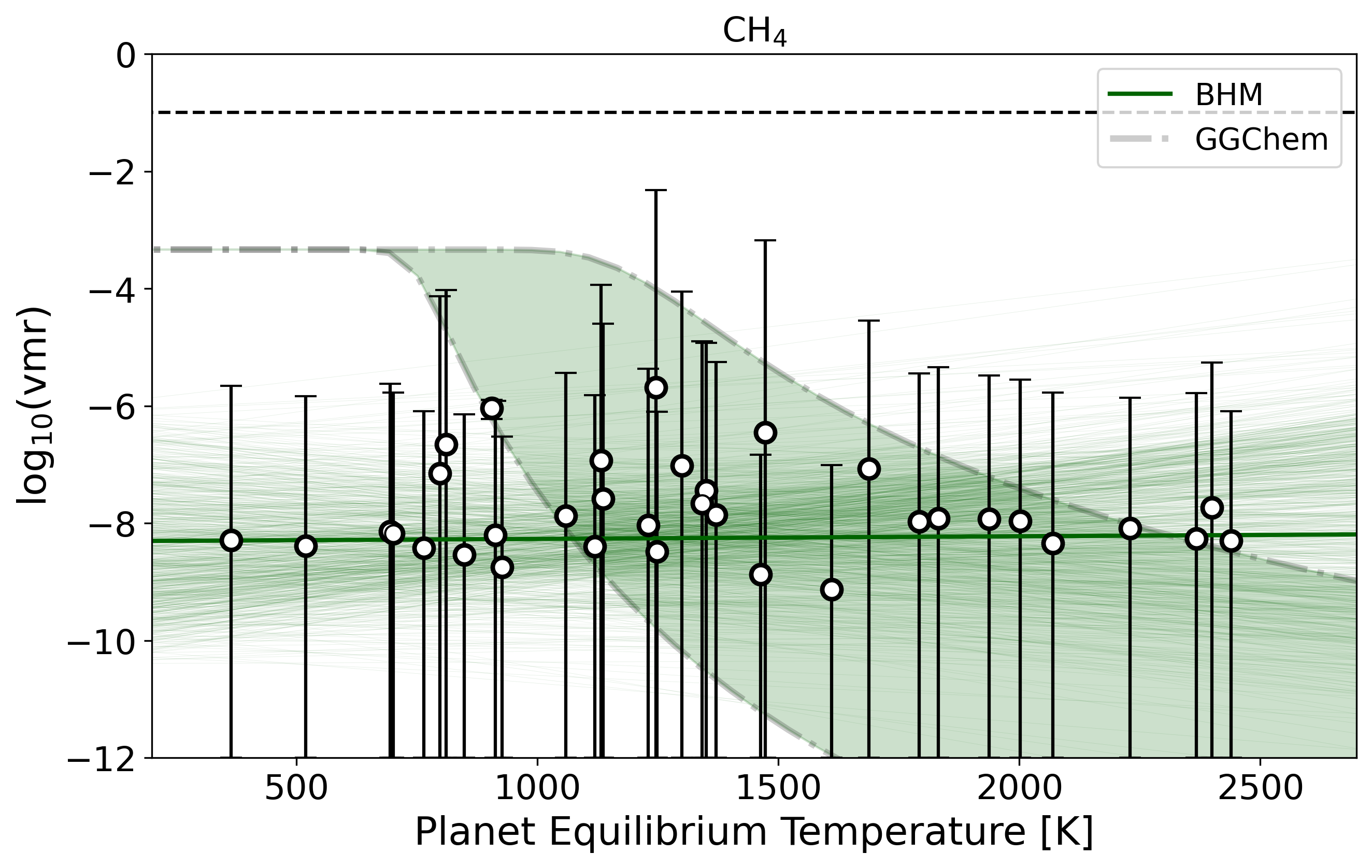}
    \includegraphics[width=0.9\columnwidth]{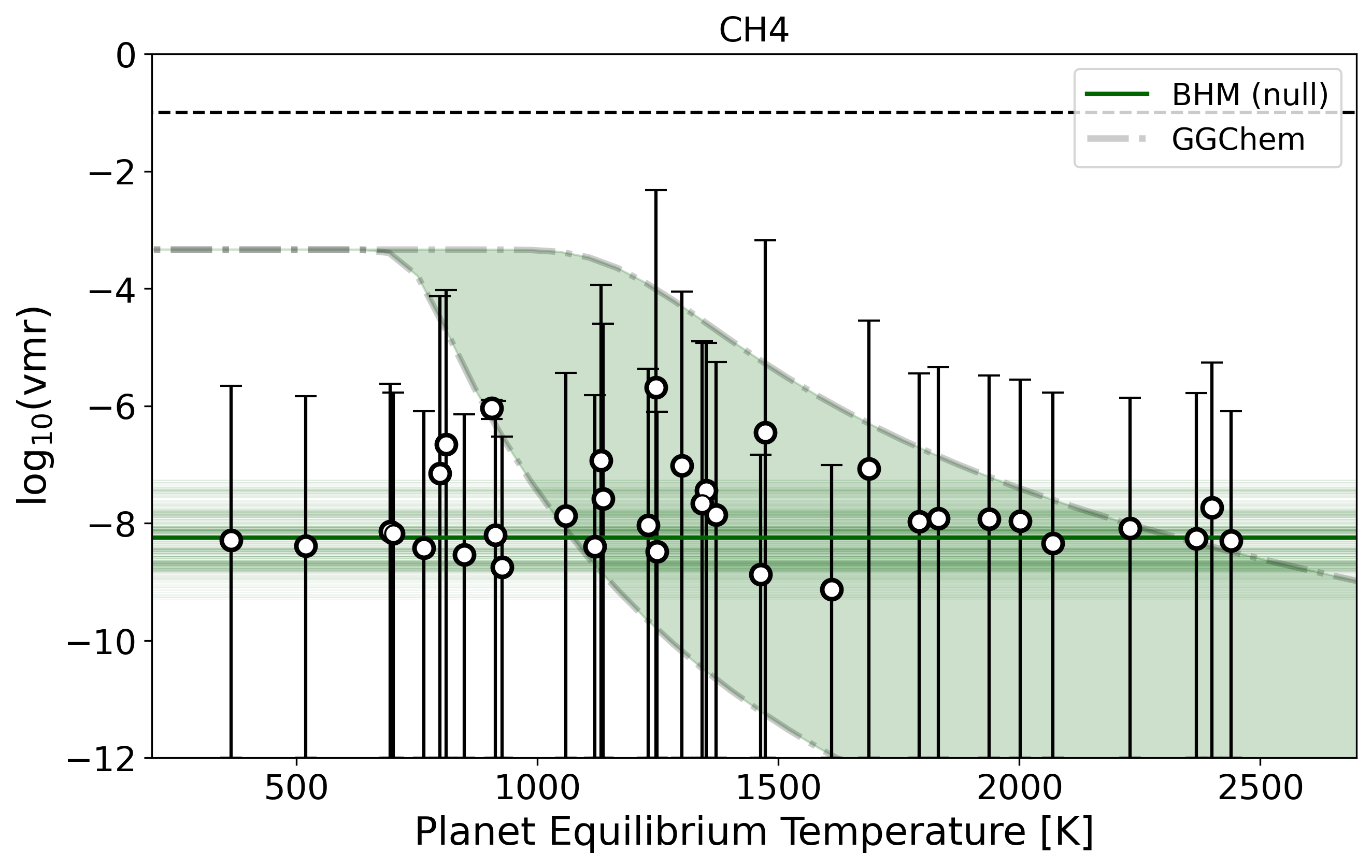}
    \includegraphics[width=0.9\columnwidth]{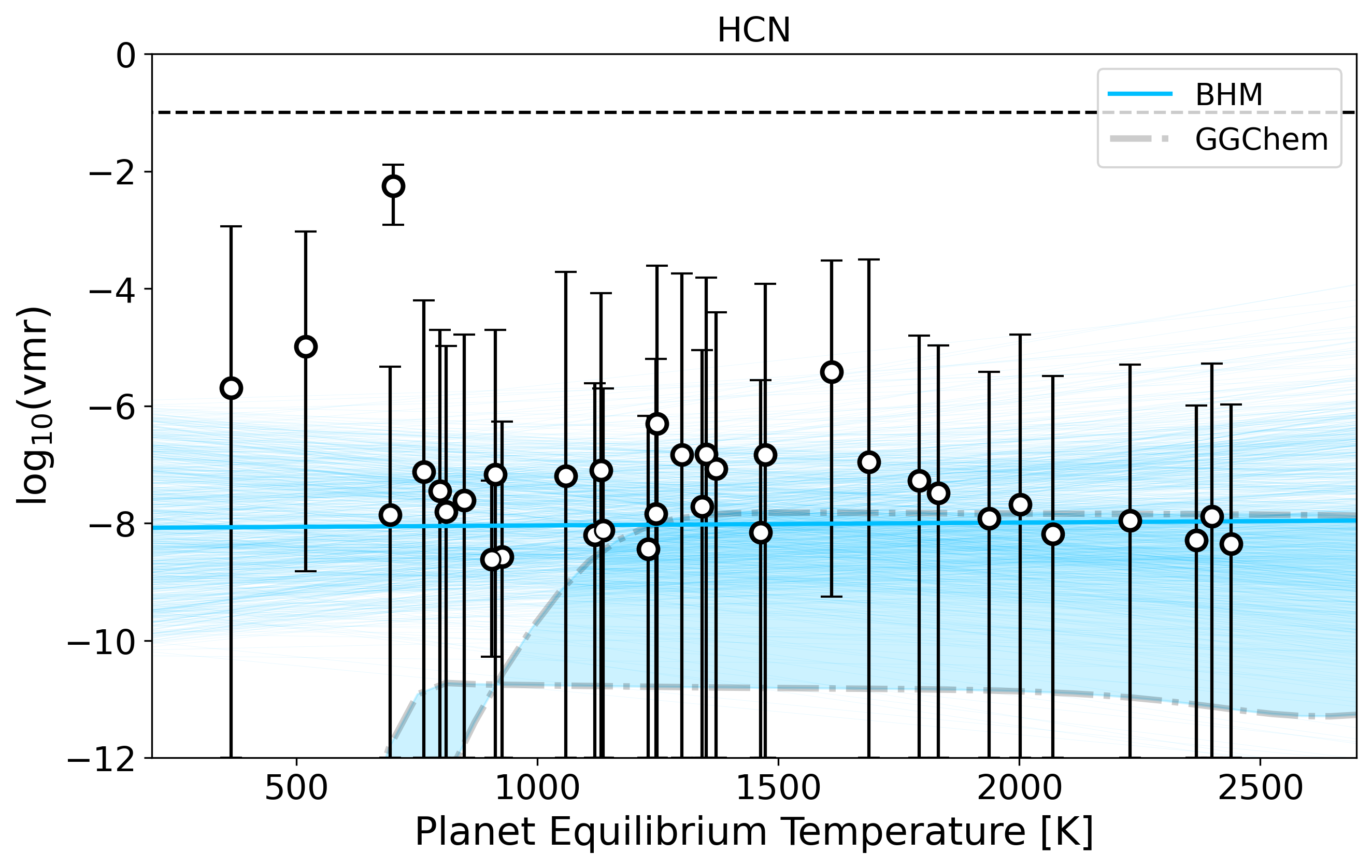}
    \includegraphics[width=0.9\columnwidth]{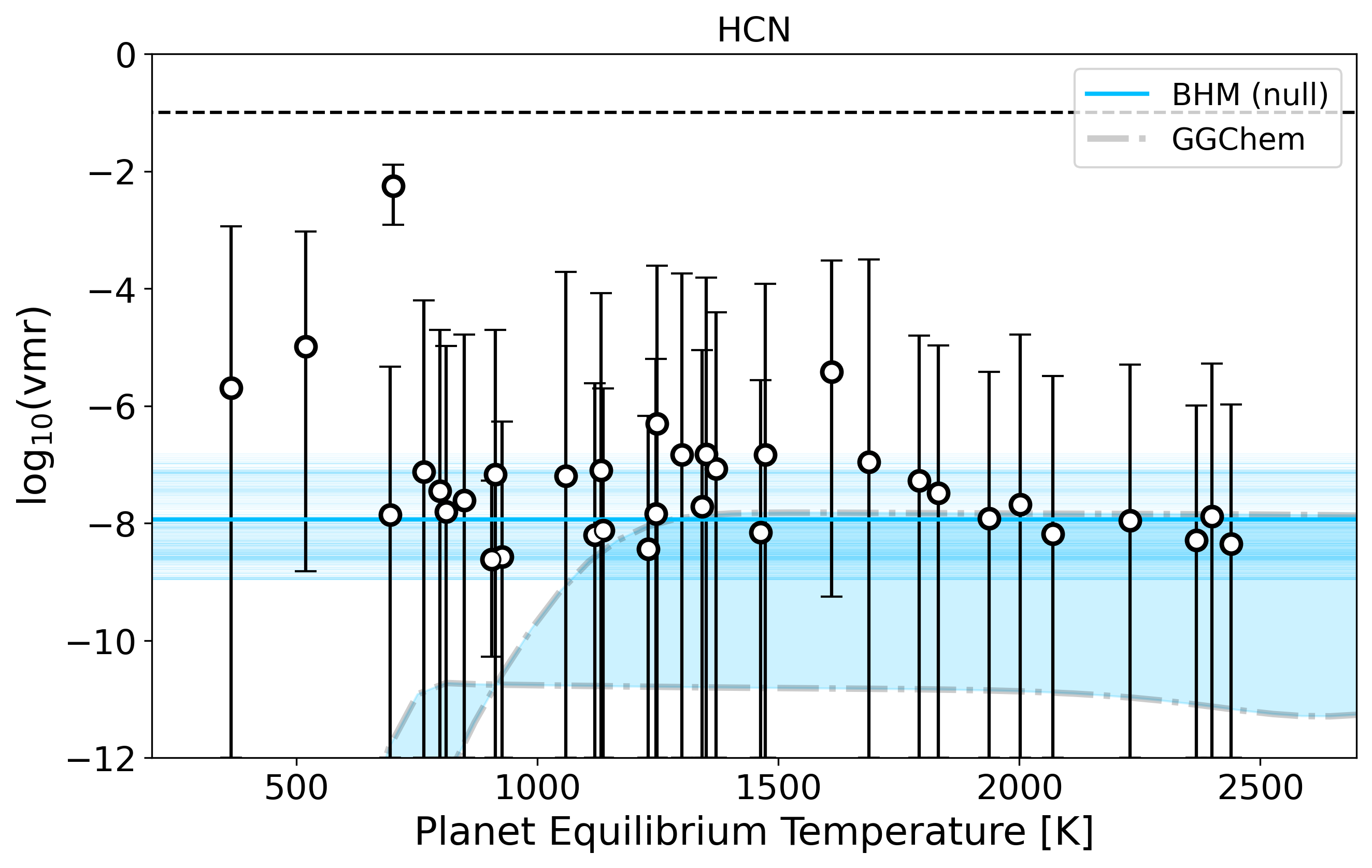}
    \includegraphics[width=0.9\columnwidth]{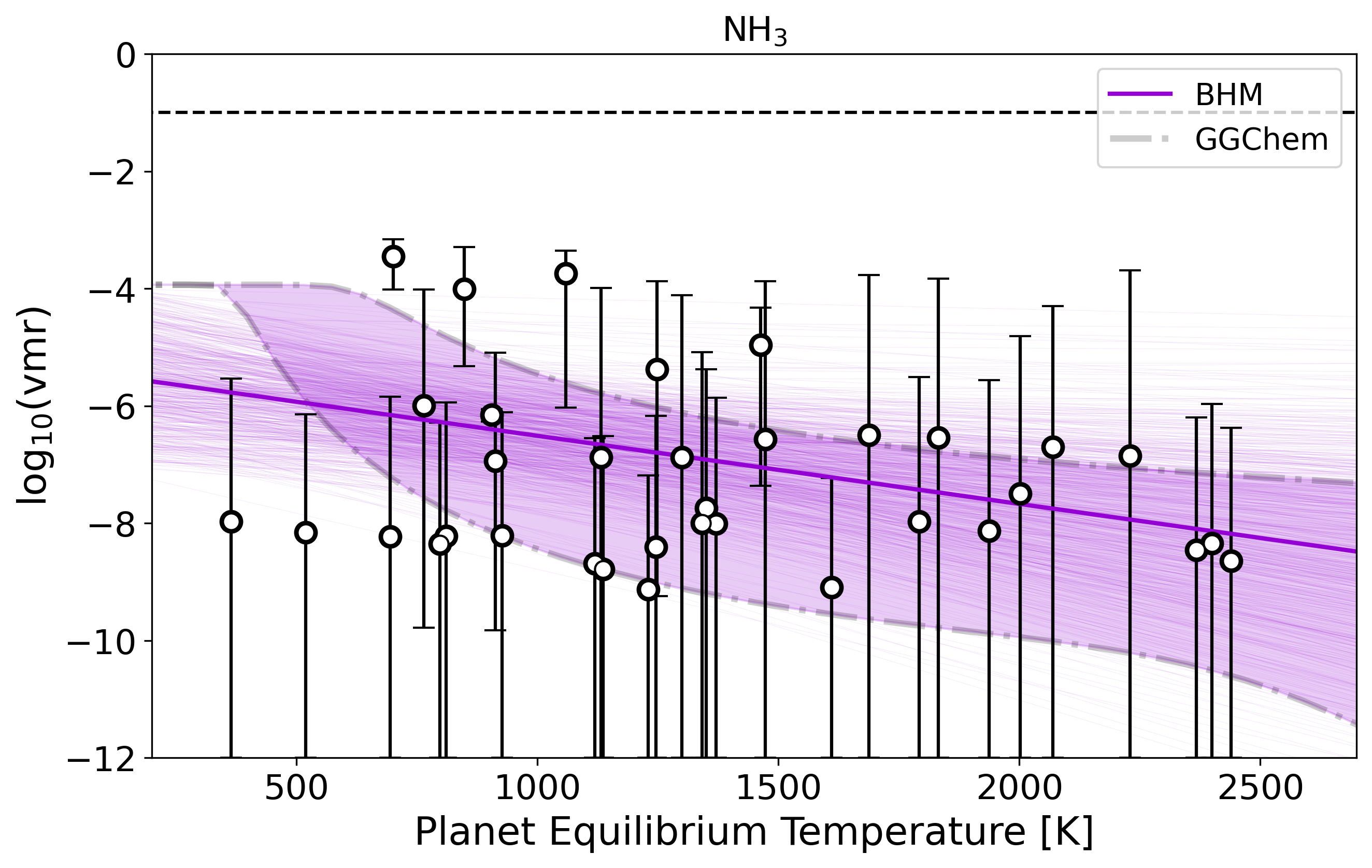}
    \includegraphics[width=0.9\columnwidth]{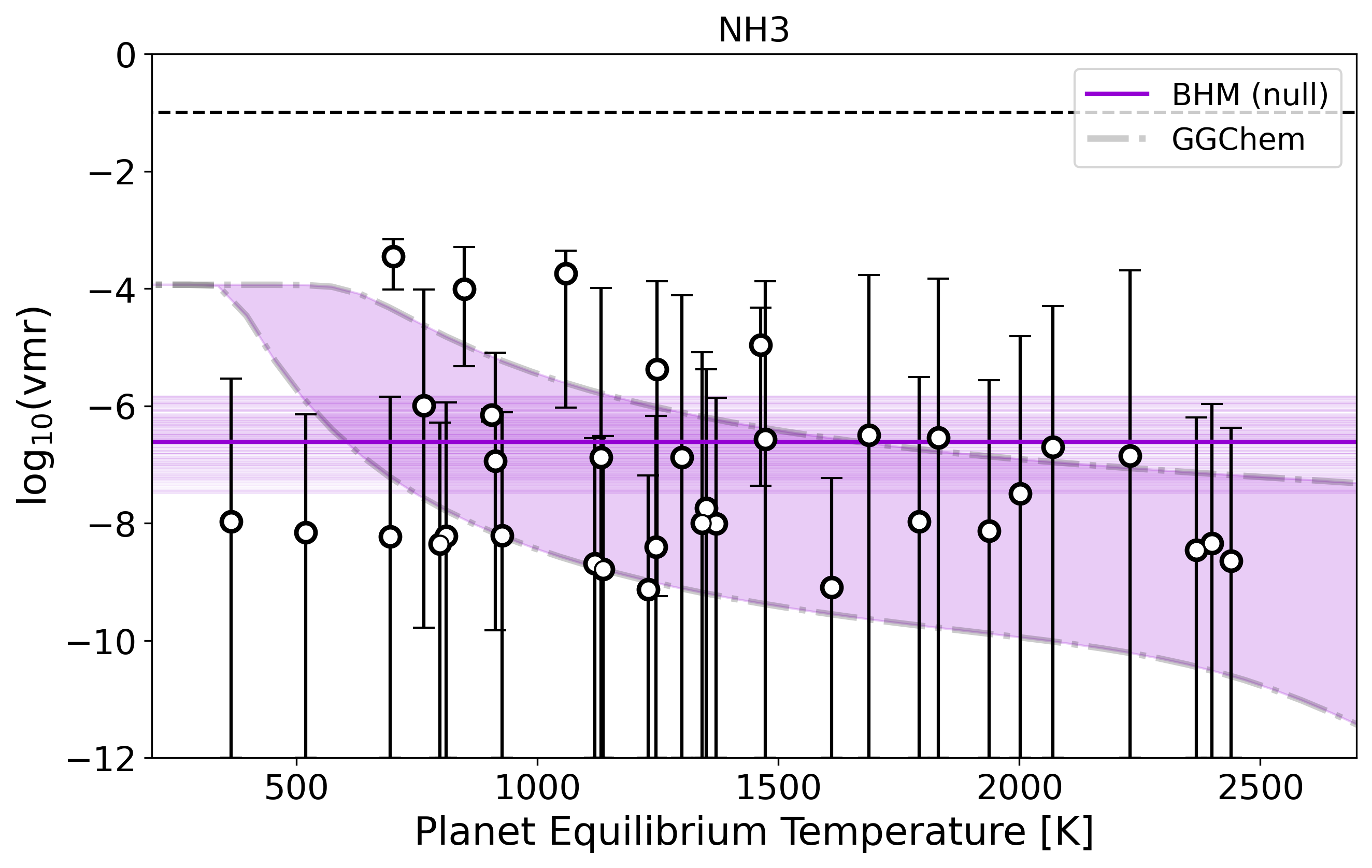}
    \caption{Retrieved abundances of H2O, CH4, HCN and NH3 against planet equilibrium temperature. In some cases, only an upper bound on the presence of the molecule could be placed and, for these, the error bar extends to log10(vmr) = -12. The filled regions bounded by dashed grey lines indicate the predicted abundances from GGchem chemical equilibrium models (assuming C/O = 0.54 and solar metallicity) across 1e2 to 1e5 Pa (1e-3 to 1 Bar). Left: The thick coloured line on each plot indicates the linear trend from the BHM while the thinner colours lines represent the traces from the fit that were within the 1$\sigma$ errors of the best-fit model. Right: the same except for the flat model (i.e. the null hypothesis). }
    \label{fig:mol_bhm}
\end{figure*}

In the main text we showed the linear models as well as providing the evidence for these and the null hypothesis. Here we show the models for these linear models again, but also show the model, as well as the traces, for the null hypothesis in each case. For H$_{\rm 2}$O, CH$_{\rm 4}$, HCN and NH$_{\rm 3}$, these are given in Figure \ref{fig:mol_bhm}. The associated log-evidence, and best-fit parameters, are shown in Table \ref{tab:ap_mol_bhm}.

\begin{table}[]
    \centering
    \begin{tabular}{ccccc}\hline \hline
   Molecule & m & c & $\sigma$ & ln(E) \\\hline 
\multirow{2}{*}{H$_{\rm 2}$O} & 0.000 $^{+ 0.001 }_{- 0.001 }$ & -2.08 $^{+ 0.84 }_{- 0.88 }$ & -0.01 $^{+ 0.12 }_{- 0.13 }$ & -213.22 \\
& - & -2.43 $^{+ 0.31 }_{- 0.31 }$ & -0.03 $^{+ 0.12 }_{- 0.11 }$ & -208.81 \\\hline 
\multirow{2}{*}{CH$_{\rm 4}$} & 0.000 $^{+ 0.001 }_{- 0.001 }$ & -8.31 $^{+ 2.27 }_{- 2.12 }$ & -1.11 $^{+ 0.8 }_{- 0.83 }$ & -185.25 \\
& - & -8.25 $^{+ 1.0 }_{- 1.05 }$ & -1.16 $^{+ 0.82 }_{- 0.78 }$ & -181.26 \\\hline 
\multirow{2}{*}{HCN}& 0.000 $^{+ 0.001 }_{- 0.001 }$ & -8.09 $^{+ 2.17 }_{- 2.19 }$ & -1.04 $^{+ 0.8 }_{- 0.84 }$ & -186.24 \\
& - & -7.94 $^{+ 1.14 }_{- 1.03 }$ & -1.1 $^{+ 0.81 }_{- 0.83 }$ & -181.94 \\\hline 
\multirow{2}{*}{NH$_{\rm 3}$} & -0.001 $^{+ 0.001 }_{- 0.001 }$ & -5.35 $^{+ 1.64 }_{- 1.61 }$ & -0.06 $^{+ 0.42 }_{- 0.44 }$ & -190.77 \\
& - & -6.61 $^{+ 0.8 }_{- 0.88 }$ & -0.1 $^{+ 0.46 }_{- 0.58 }$ & -187.37 \\\hline \hline
    \end{tabular}
    \caption{Hyperparameters for the BHM fits for each molecule. In each case, the null hypothesis (i.e. constant abundance with temperature) yielded a preferable fit to the data.}
    \label{tab:ap_mol_bhm}
\end{table}

In the main text we fitted the trends for the optical absorbers on all retrievals which, when optical absorber were included, led to an atmospheric detection of $>3\sigma$, even if the model without optical absorbers was preferred. We did this because we have no reason to expect these planets aren't drawn from the same distribution as the others and so we want to test how applicable the GGChem predictions are too all planets. However, in this Appendix we also show the fits to only those planets where the retrieval with optical absorbers gave a preferable fit. 

\begin{figure*}
    \centering
    \includegraphics[width=0.9\columnwidth]{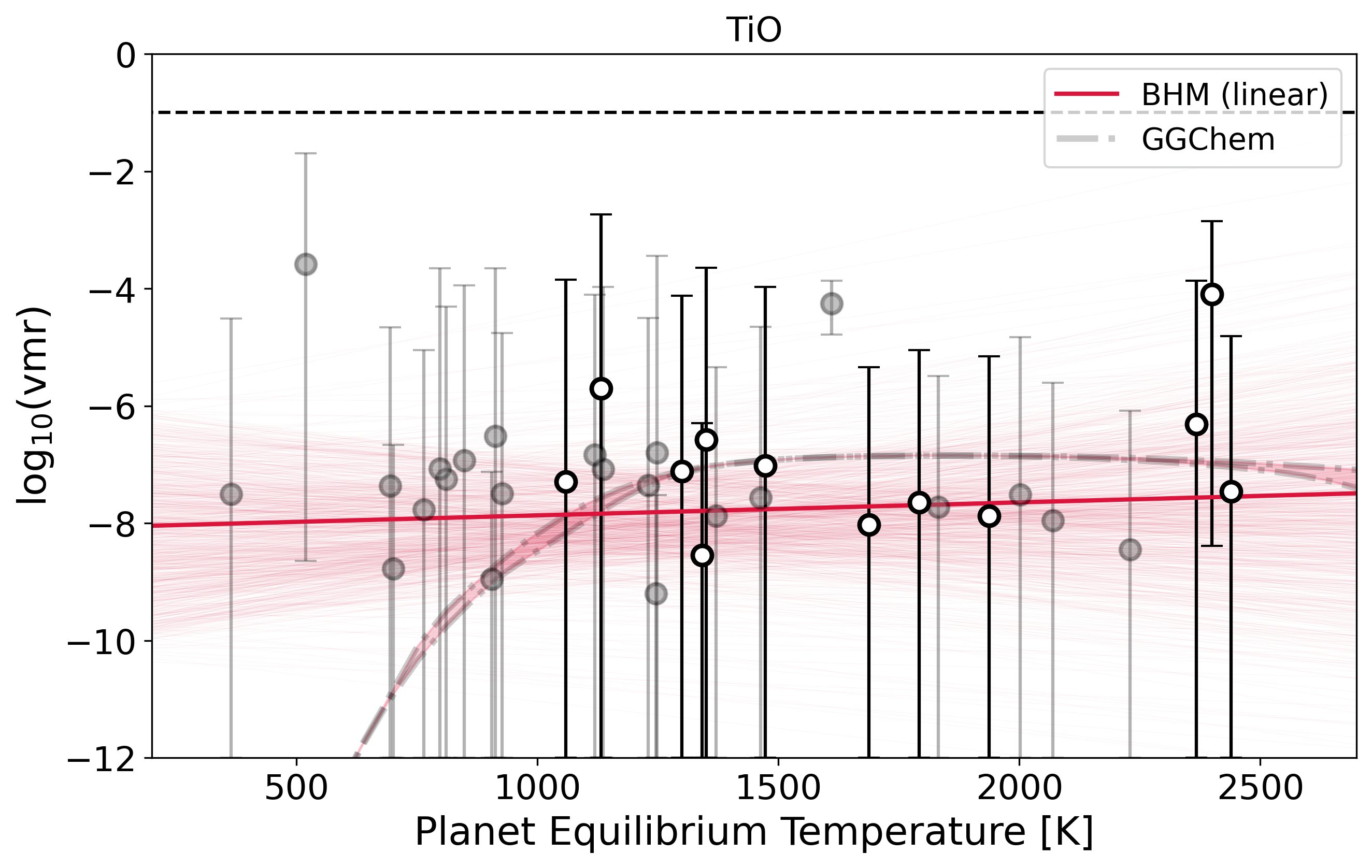}
    \includegraphics[width=0.9\columnwidth]{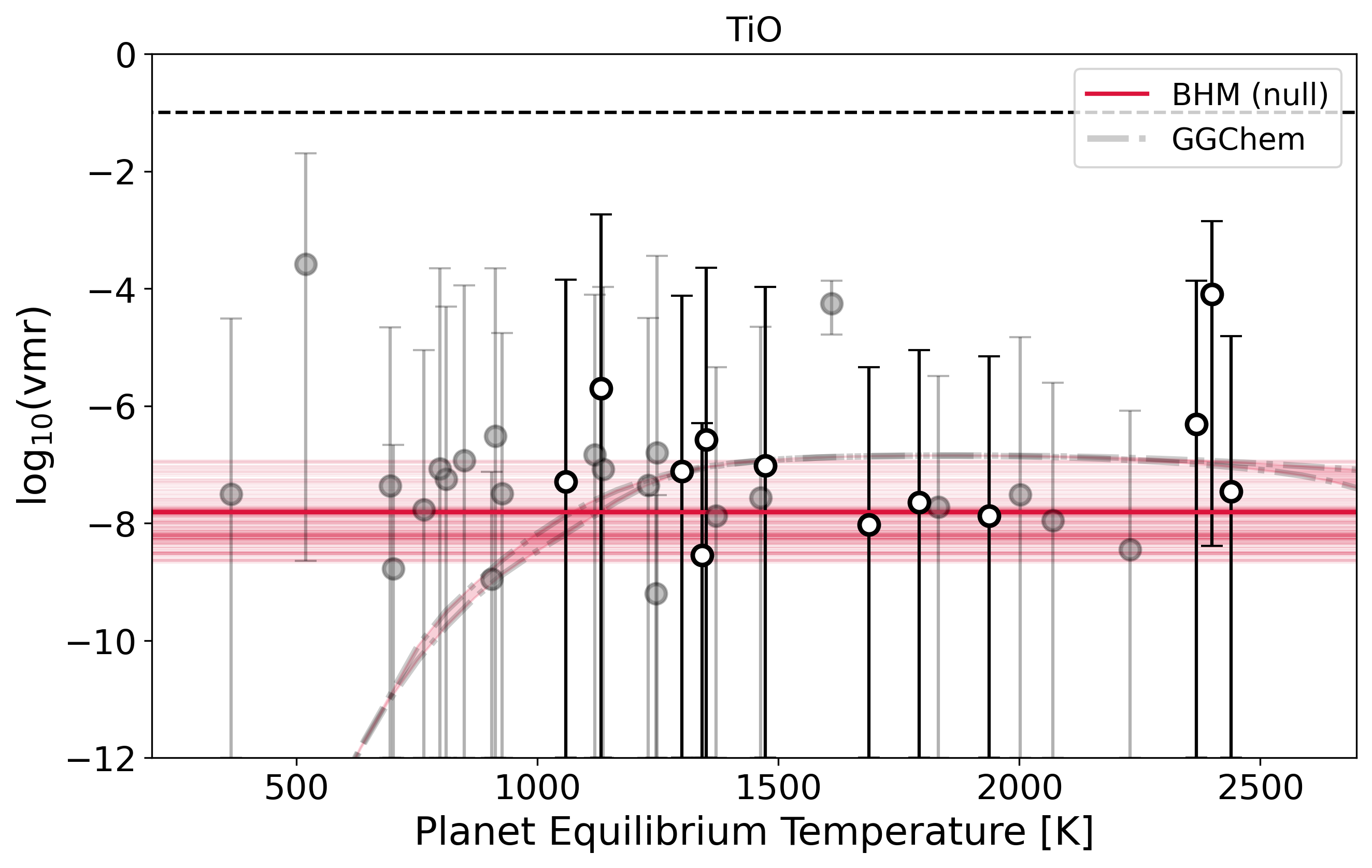}
    \includegraphics[width=0.9\columnwidth]{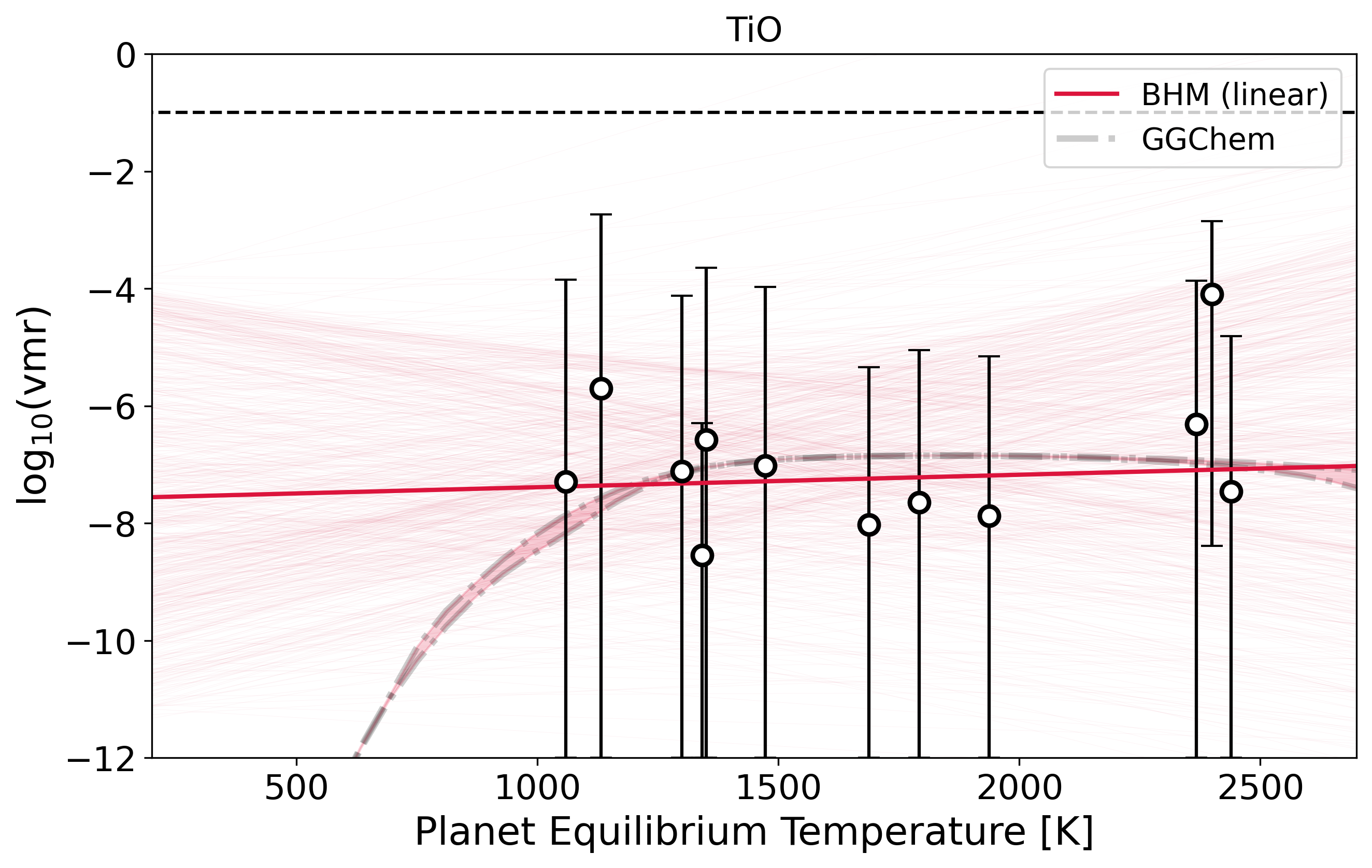}
    \includegraphics[width=0.9\columnwidth]{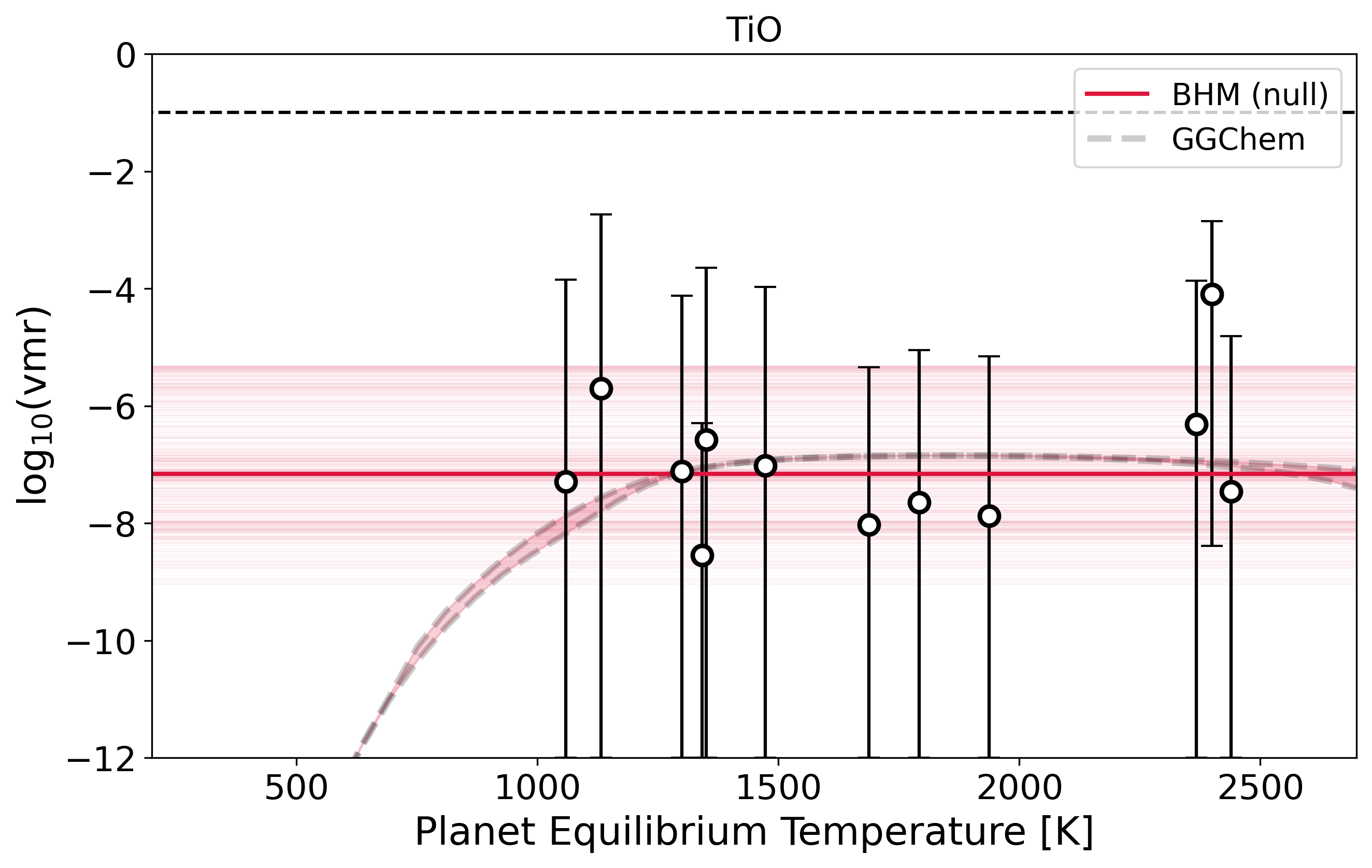}
    \caption{Retrieved abundances of TiO against planet equilibrium temperature. In some cases, only an upper bound on the presence of the molecule could be placed and, for these, the error bar extends to log10(vmr) = -12. The filled regions bounded by dashed grey lines indicate the predicted abundances from GGchem chemical equilibrium models (assuming C/O = 0.54 and solar metallicity) across 1e2 to 1e5 Pa (1e-3 to 1 Bar). Left: The thick coloured line on each plot indicates the linear trend from the BHM while the thinner colours lines represent the traces from the fit that were within the 1$\sigma$ errors of the best-fit model. Right: the same except for the flat model (i.e. the null hypothesis). Top: black data points indicate planets for which the retrieval model with optical absorbers is preferred while grey point represent those for which a preferable fit is obtained without them. In these plots, all these retrieval traces were used in the BHM. Bottom: only the retrievals for which the optical absorb model was preferred. }
    \label{fig:tio_bhm}
\end{figure*}

\begin{figure*}
    \centering
    \includegraphics[width=0.9\columnwidth]{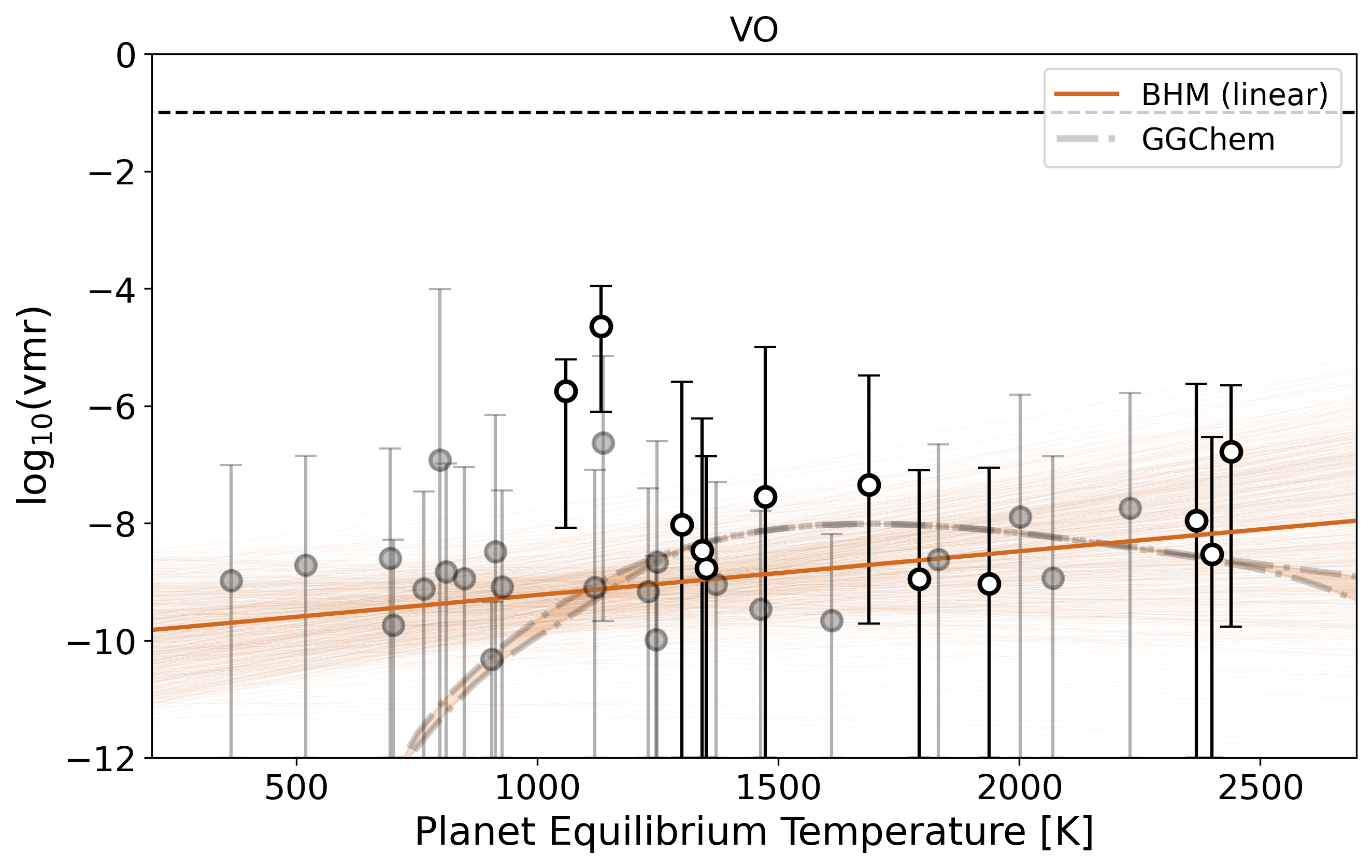}
    \includegraphics[width=0.9\columnwidth]{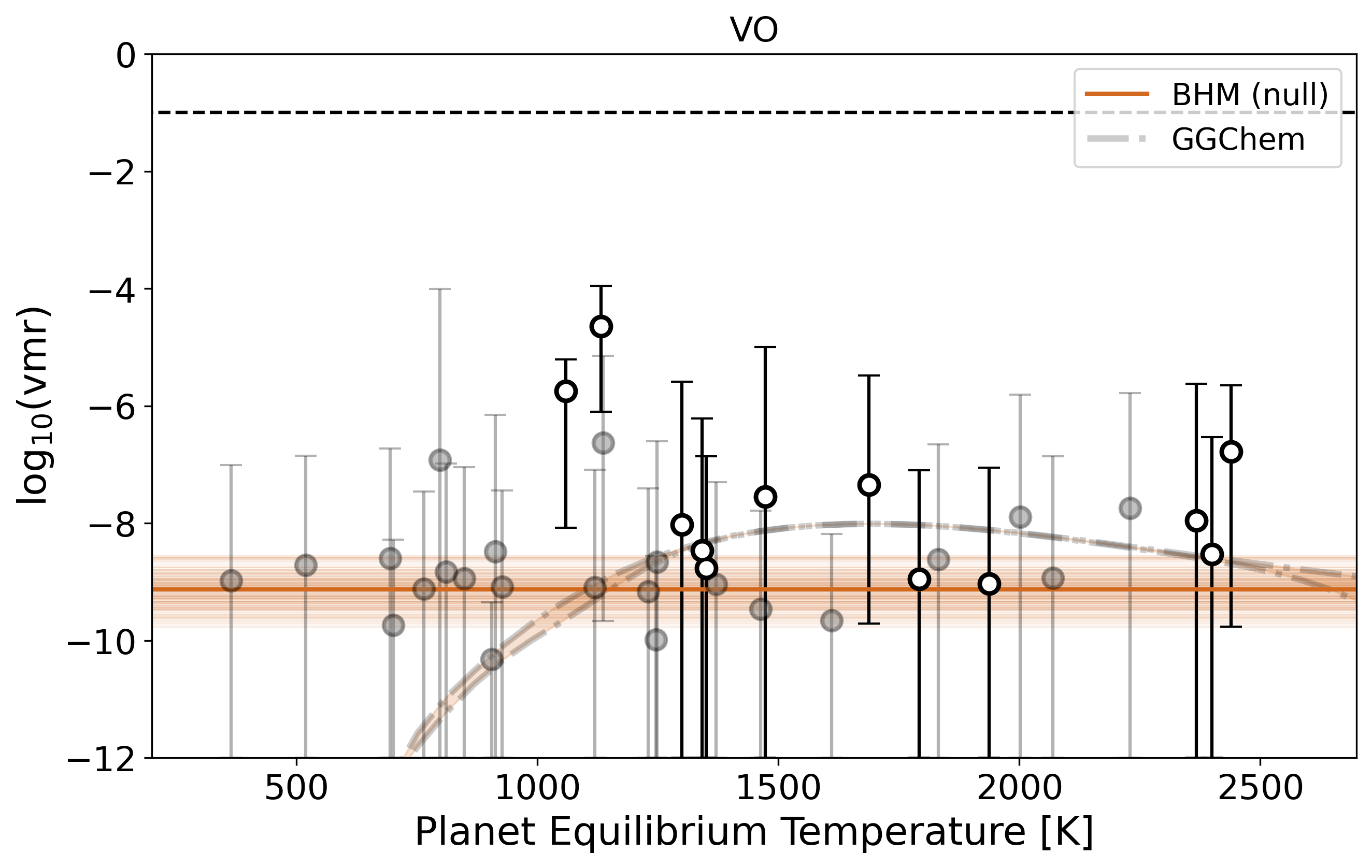}
    \includegraphics[width=0.9\columnwidth]{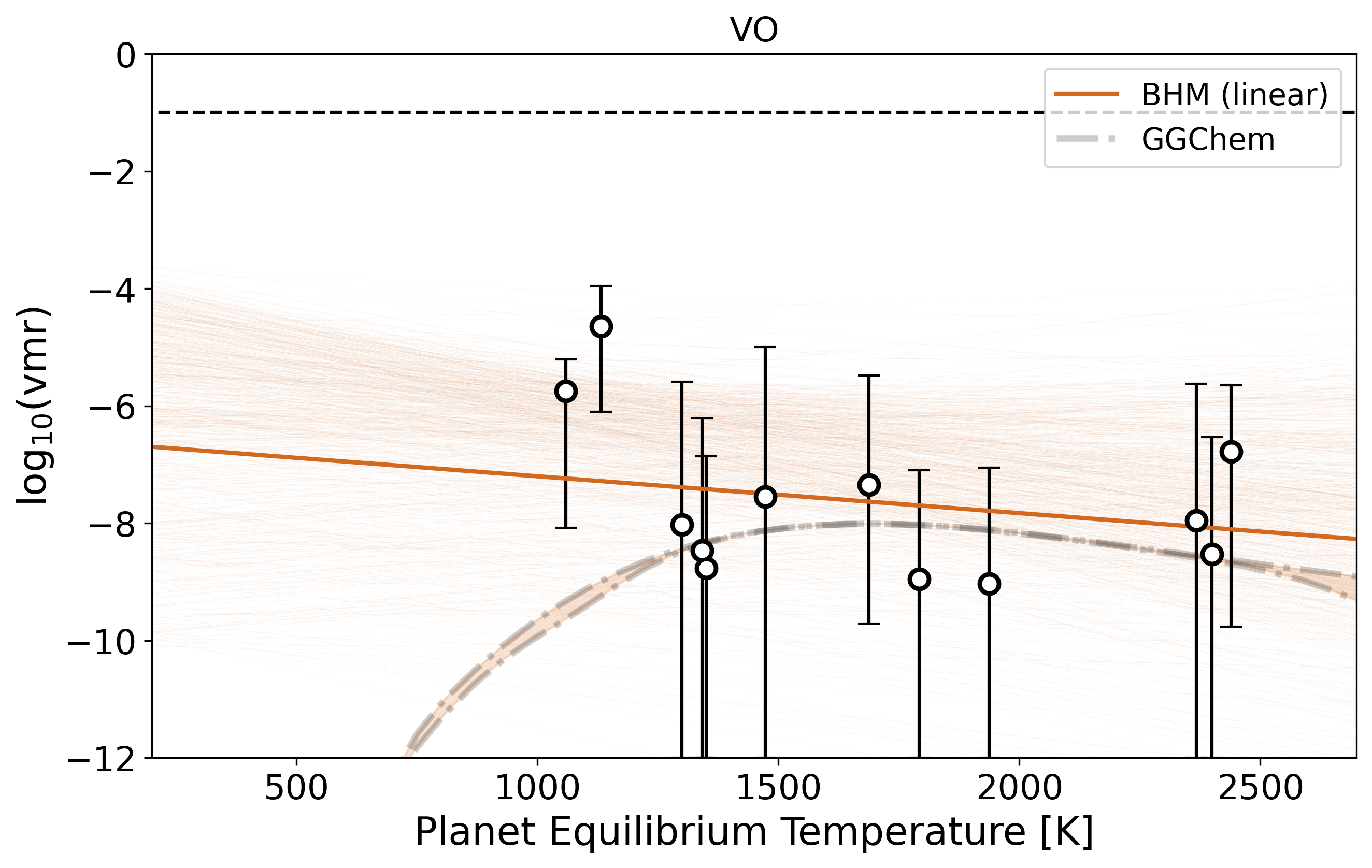}
    \includegraphics[width=0.9\columnwidth]{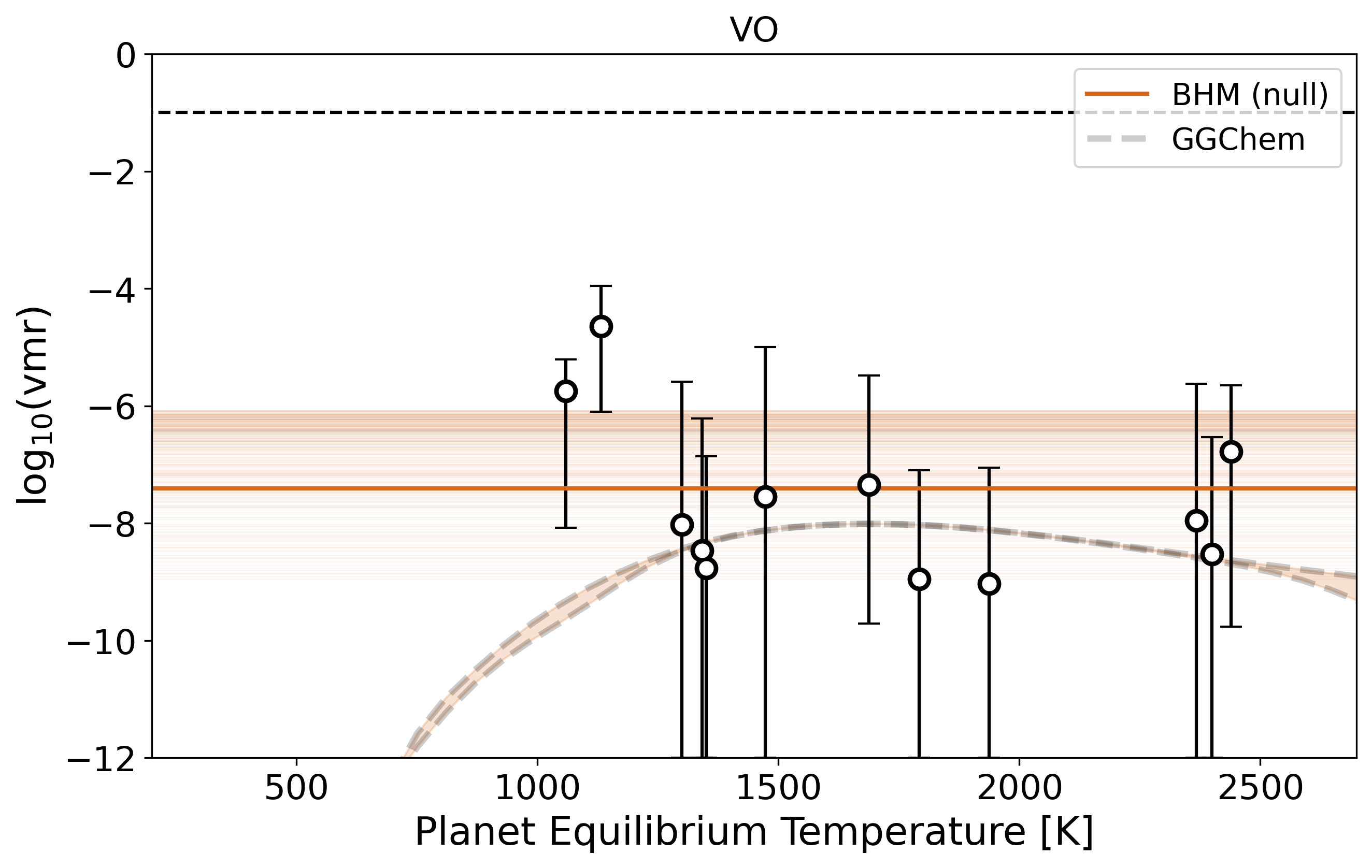}
    \caption{Same as Figure \ref{fig:tio_bhm} but for VO. }
    \label{fig:vo_bhm}
\end{figure*}

\begin{figure*}
    \centering
    \includegraphics[width=0.9\columnwidth]{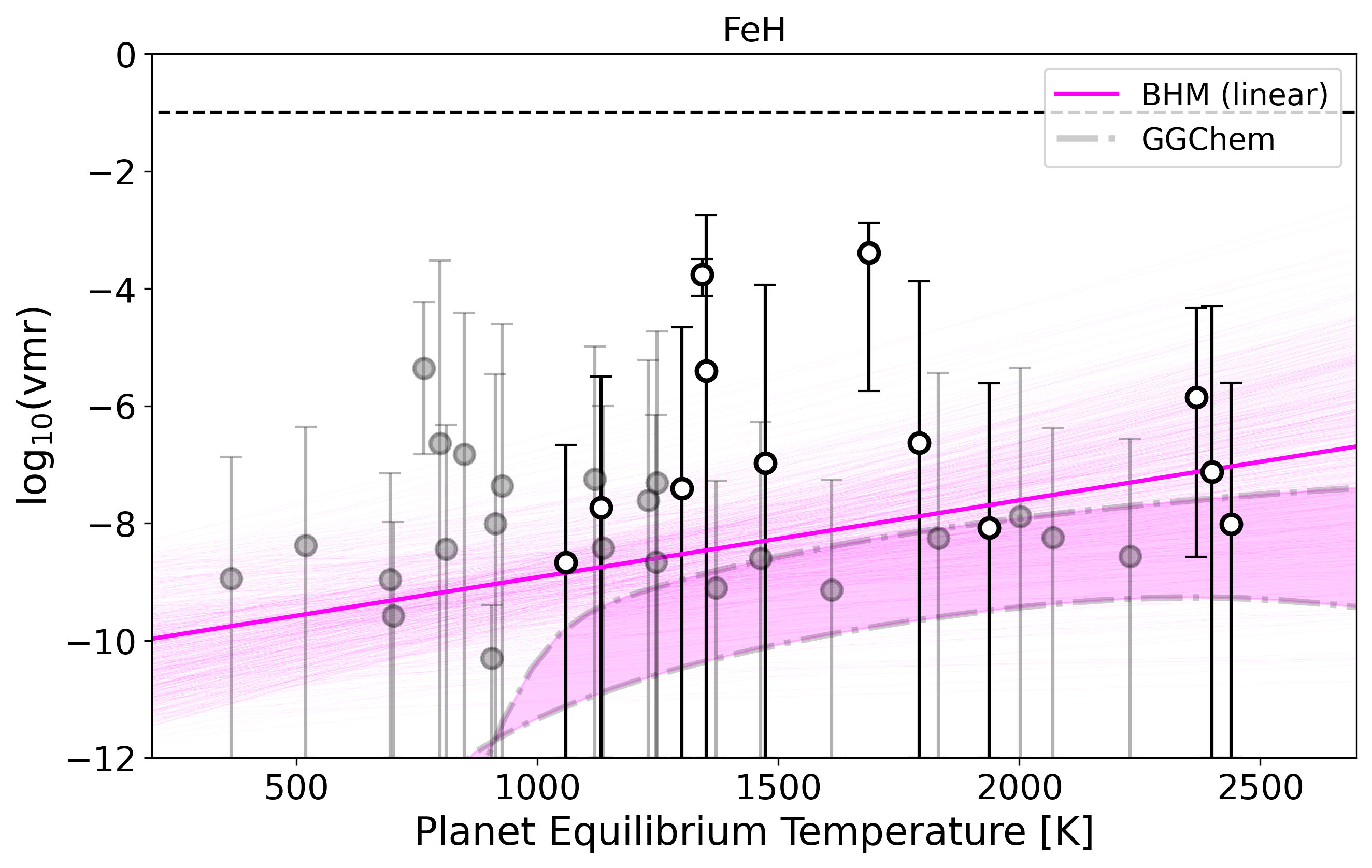}
    \includegraphics[width=0.9\columnwidth]{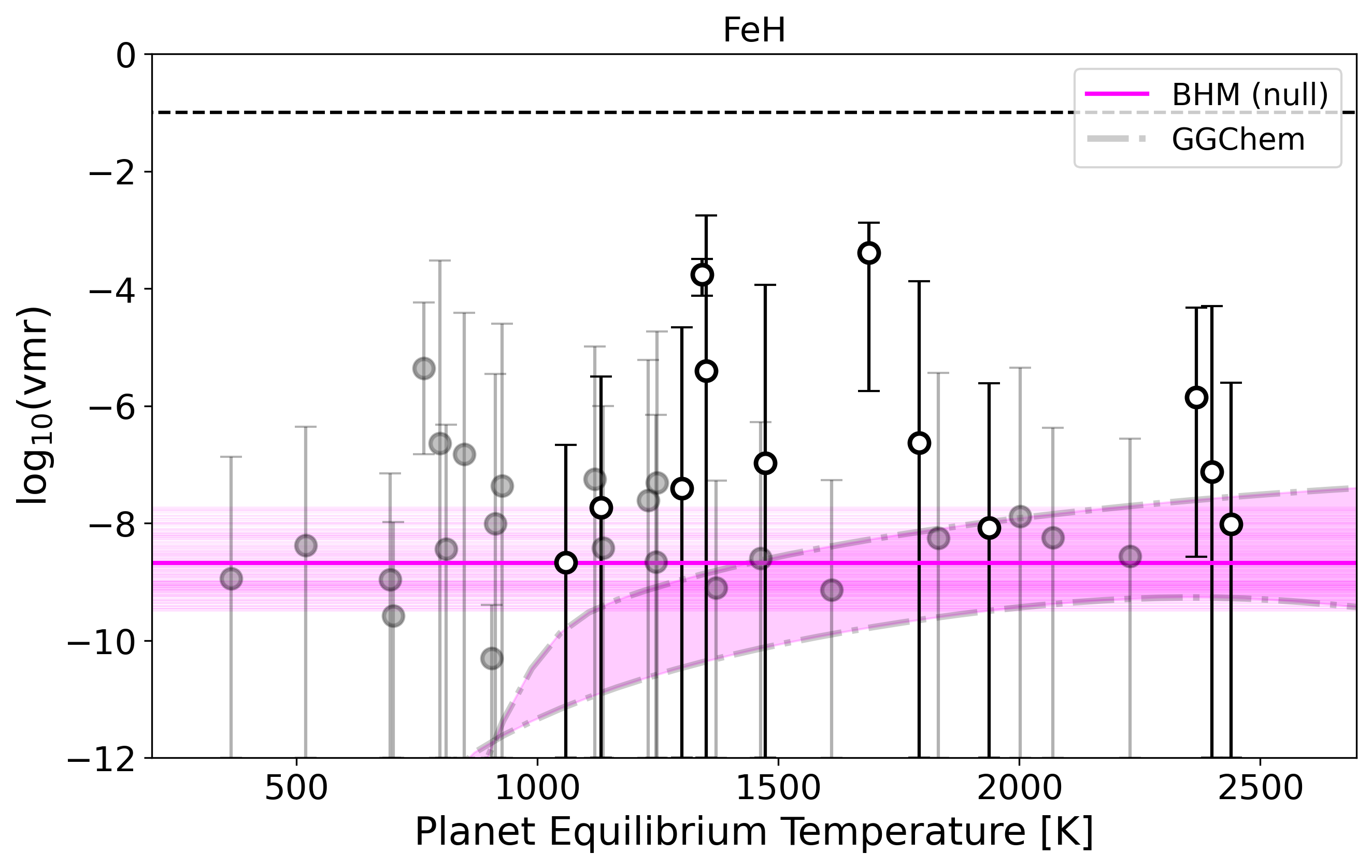}
    \includegraphics[width=0.9\columnwidth]{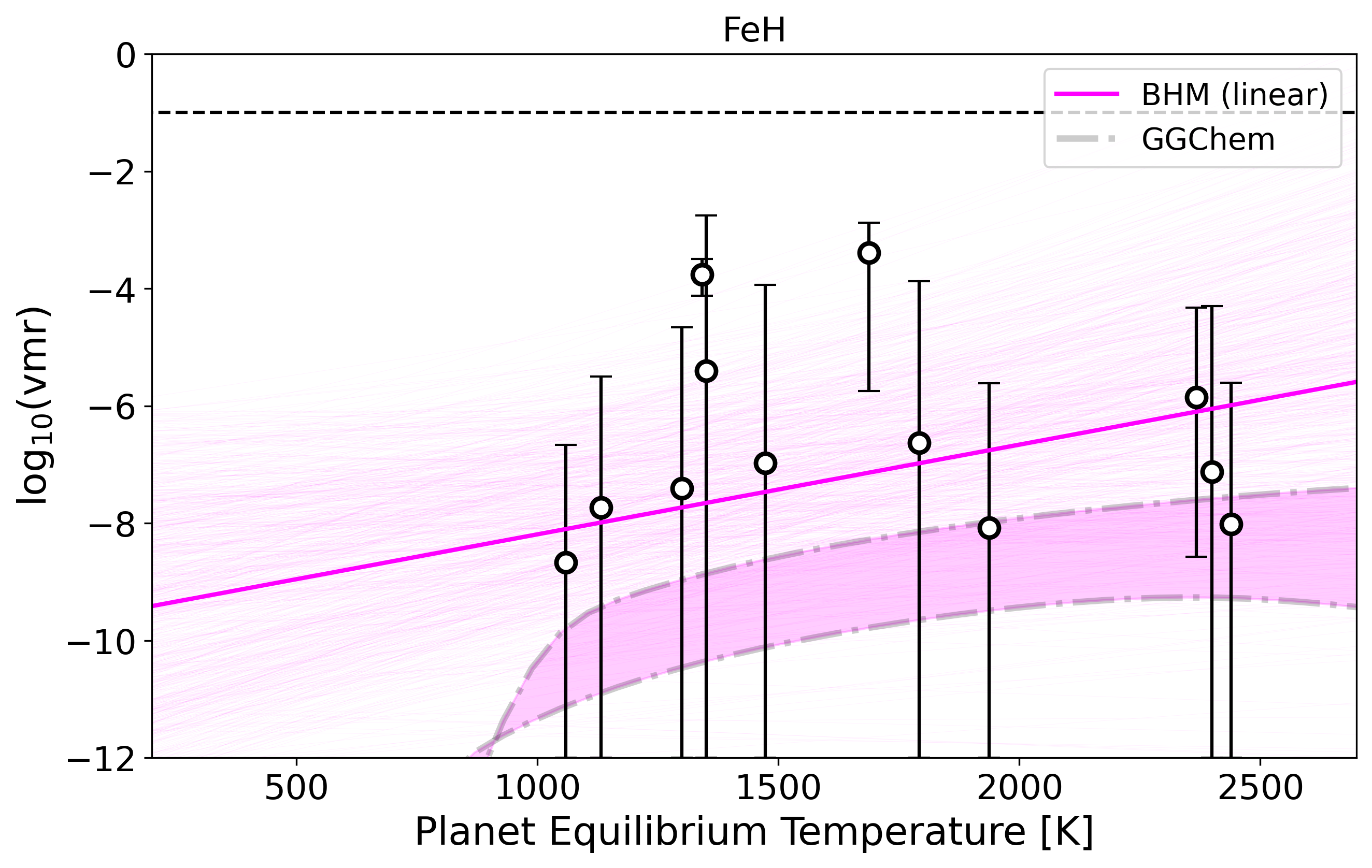}
    \includegraphics[width=0.9\columnwidth]{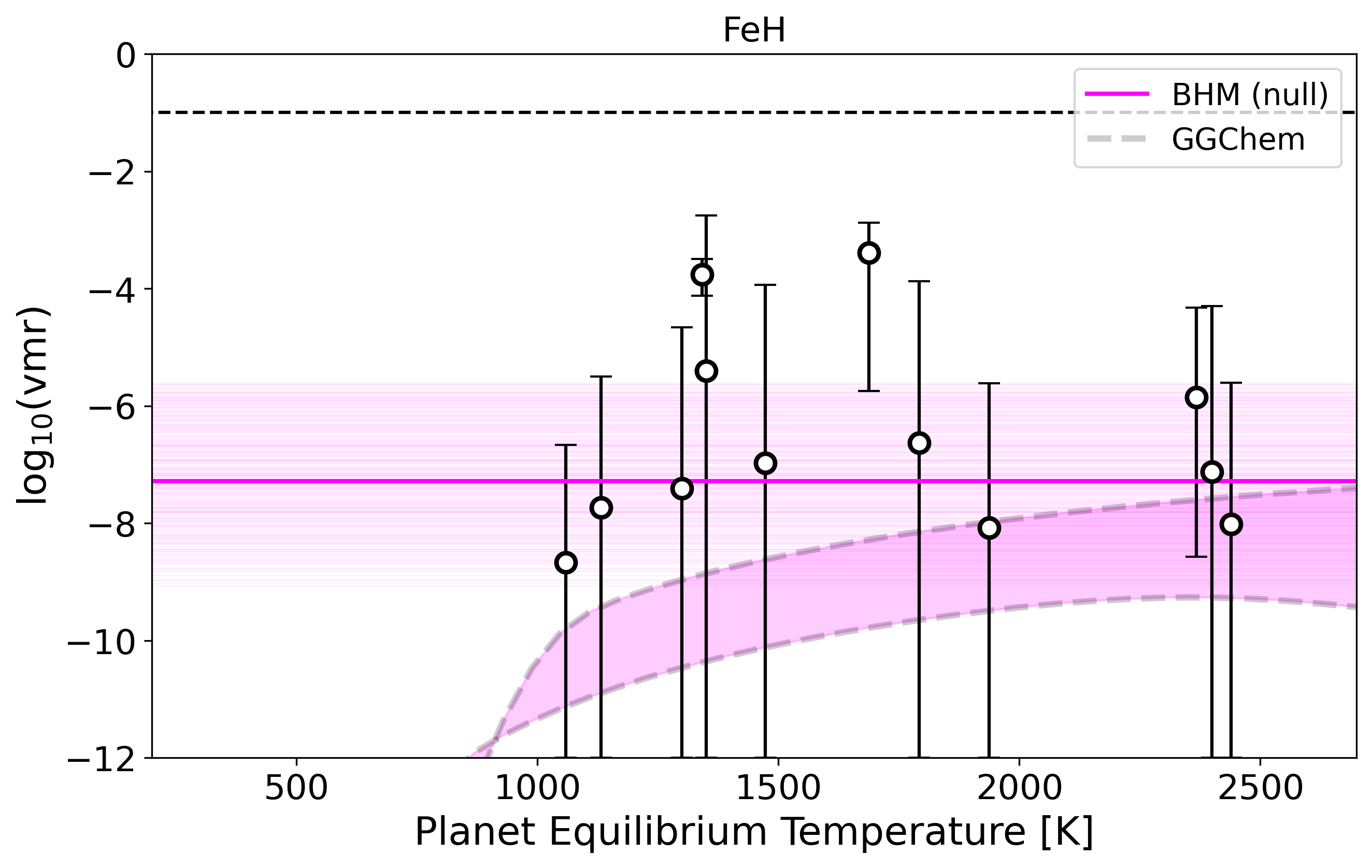}
    \caption{Same as Figure \ref{fig:tio_bhm} but for FeH.}
    \label{fig:feh_bhm}
\end{figure*}

\begin{figure*}
    \centering
    \includegraphics[width=0.9\columnwidth]{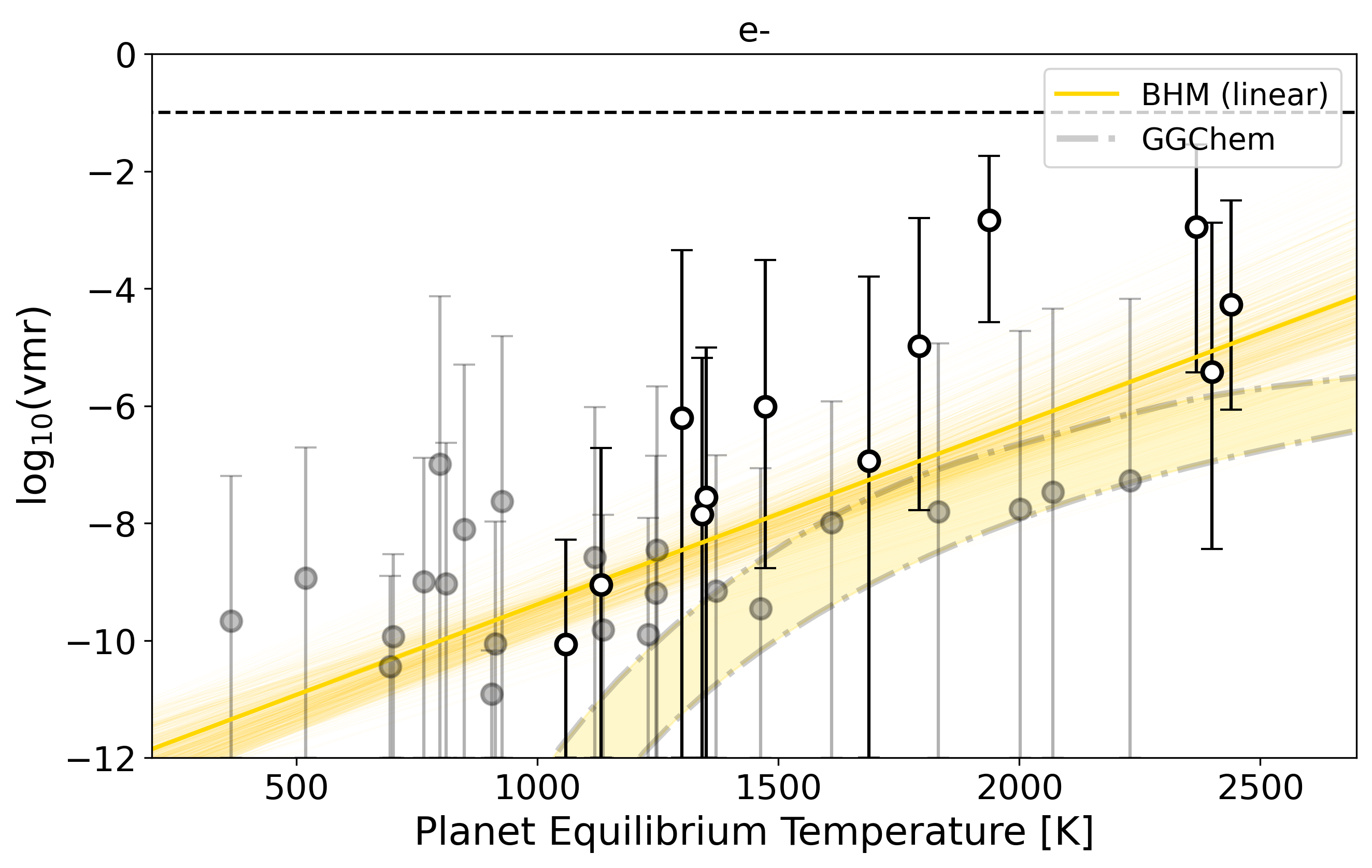}
    \includegraphics[width=0.9\columnwidth]{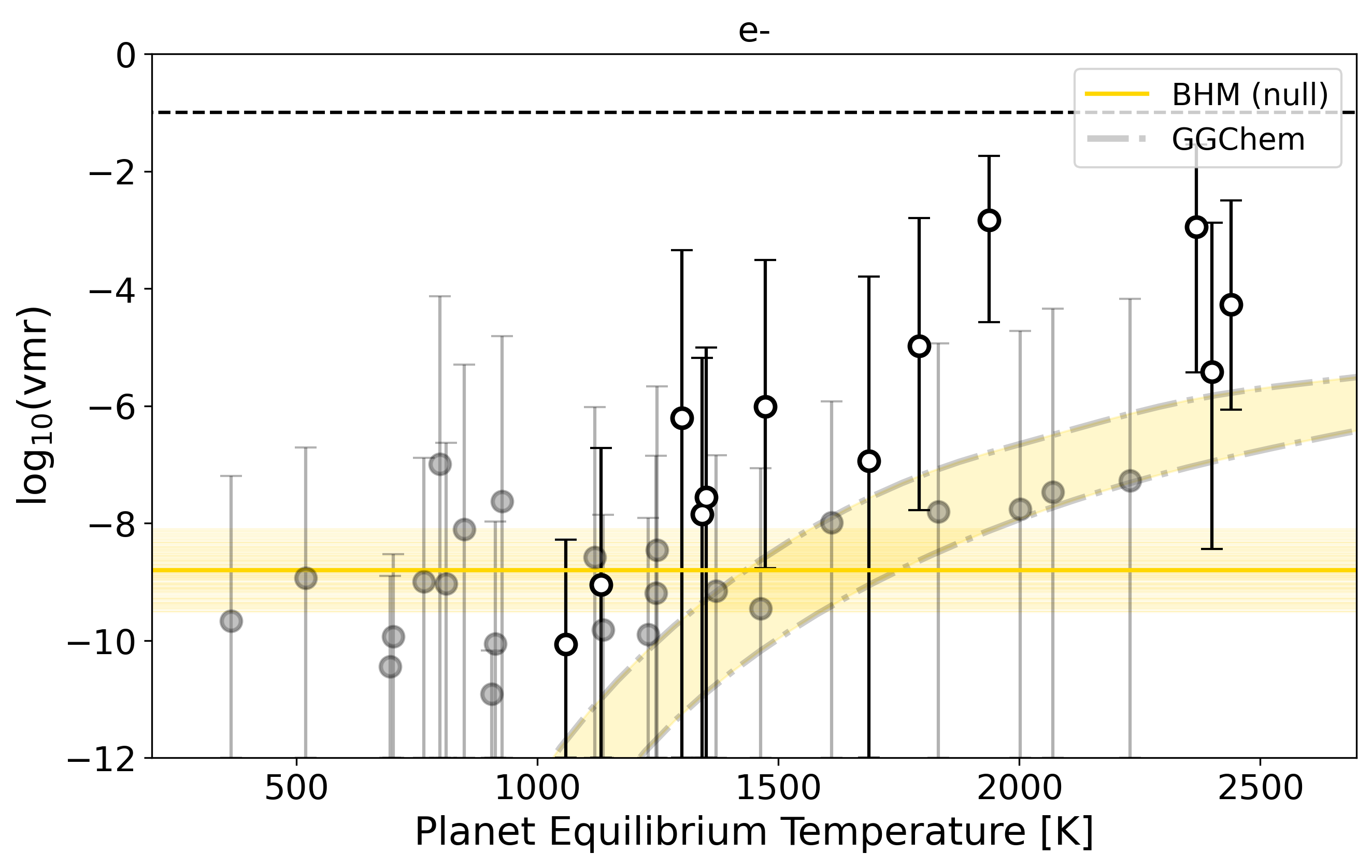}
    \includegraphics[width=0.9\columnwidth]{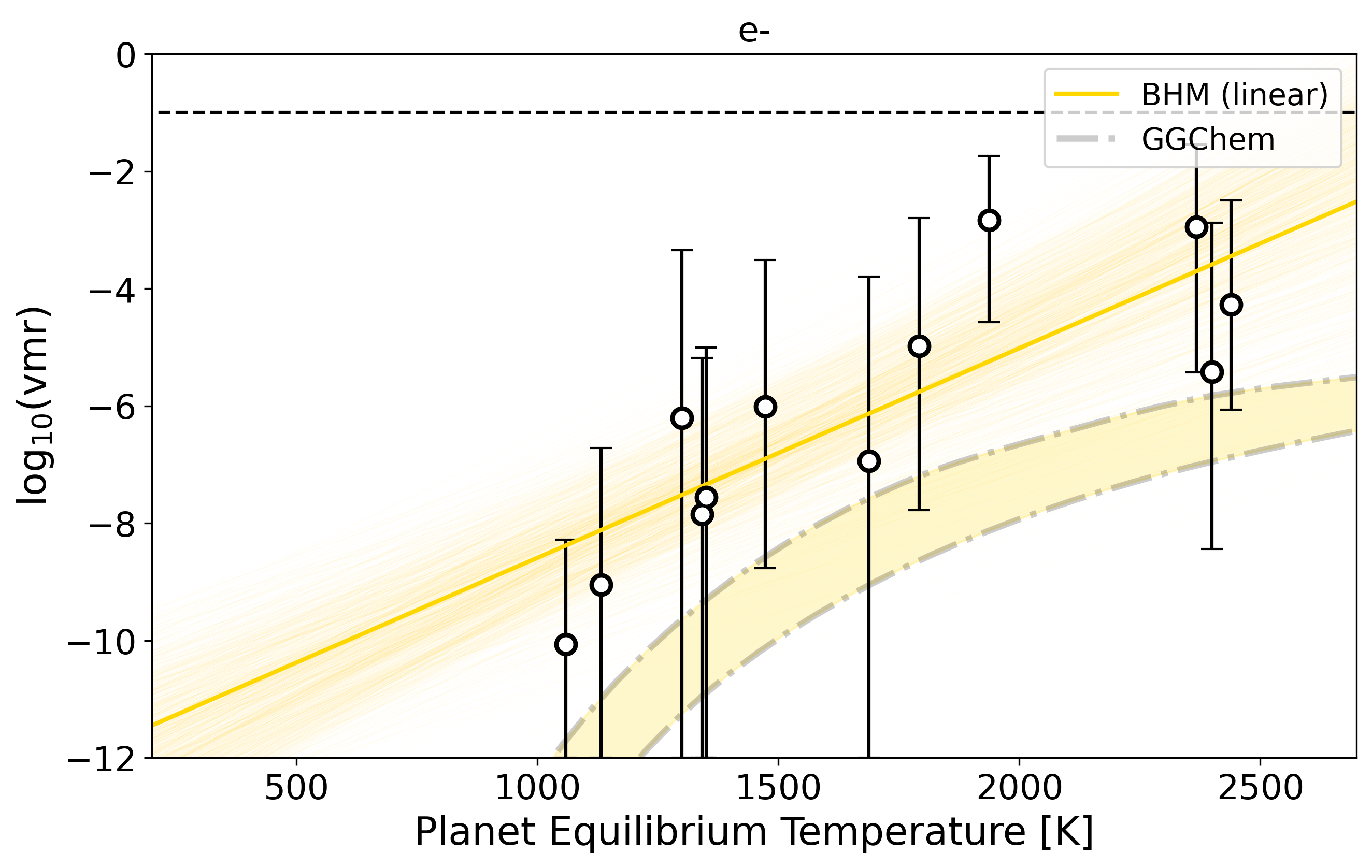}
    \includegraphics[width=0.9\columnwidth]{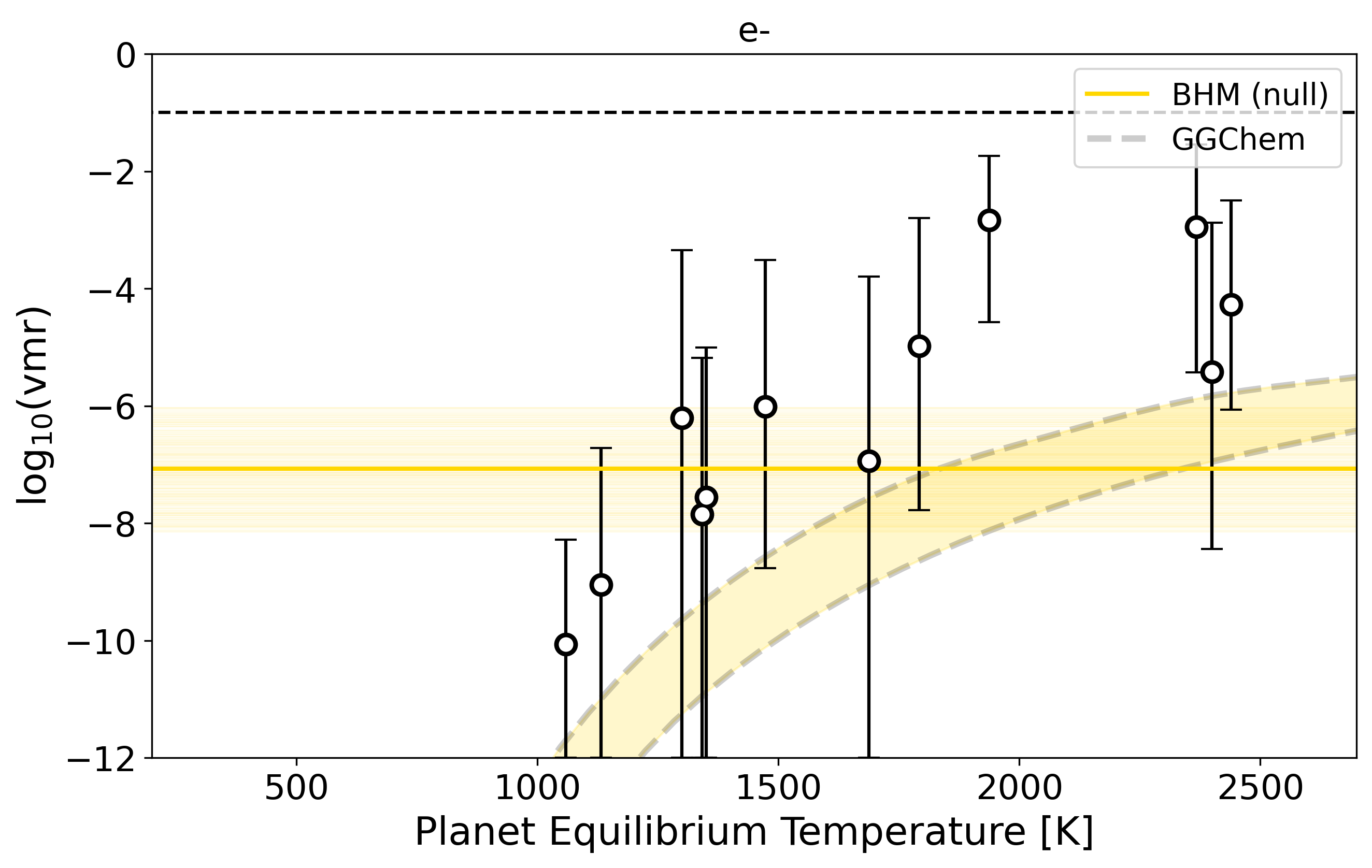}
    \caption{Same as Figure \ref{fig:tio_bhm} but for e-.}
    \label{fig:e-_bhm}
\end{figure*}

When all optical absorbers retrievals are taken, we found evidence for an increasing abundance of e- with increasing temperature. The fits to these are given in Figures \ref{fig:tio_bhm}, \ref{fig:vo_bhm}, \ref{fig:feh_bhm} and \ref{fig:e-_bhm}. Additionally, the associated evidence is given in Table \ref{tab:ap_opt_bhm}. Similarly, the fits to only retrievals which preferred the presence of optical absorbers are given in the same figures. In this case, no statistically viable trend was uncovered and the log-evidence for the models is given in Table \ref{tab:ap_opt_pref_bhm}. 

\begin{table}[]
    \centering
    \begin{tabular}{ccccc}\hline \hline
   Molecule & m & c & $\sigma$ & ln(E) \\\hline 
\multirow{2}{*}{TiO} & 0.000 $^{+ 0.002 }_{- 0.002 }$ & -8.09 $^{+ 2.25 }_{- 2.24 }$ & -1.02 $^{+ 0.83 }_{- 0.82 }$ & -189.48 \\
& - & -7.81 $^{+ 0.91 }_{- 0.87 }$ & -1.05 $^{+ 0.85 }_{- 0.86 }$ & -185.62 \\\hline
\multirow{2}{*}{VO} & 0.001 $^{+ 0.001 }_{- 0.001 }$ & -9.97 $^{+ 1.52 }_{- 1.54 }$ & -1.19 $^{+ 0.8 }_{- 0.82 }$ & -185.37 \\
& - & -9.13 $^{+ 0.59 }_{- 0.65 }$ & -1.16 $^{+ 0.81 }_{- 0.82 }$ & -181.23 \\\hline
\multirow{2}{*}{FeH} & 0.001 $^{+ 0.001 }_{- 0.001 }$ & -10.23 $^{+ 1.66 }_{- 1.69 }$ & -0.62 $^{+ 0.86 }_{- 0.97 }$ & -190.95 \\
& - & -8.68 $^{+ 0.96 }_{- 0.82 }$ & -0.55 $^{+ 0.82 }_{- 1.03 }$ & -187.19 \\\hline
\multirow{2}{*}{e-} & 0.003 $^{+ 0.001 }_{- 0.001 }$ & -12.47 $^{+ 0.88 }_{- 0.81 }$ & -1.04 $^{+ 0.78 }_{- 0.84 }$ & -186.94 \\
& - & -8.8 $^{+ 0.72 }_{- 0.72 }$ & -0.24 $^{+ 0.49 }_{- 0.63 }$ & -188.19 \\ \hline \hline
    \end{tabular}
    \caption{Hyperparameters for the BHM fits for each molecule when using all retrievals where optical abosrbers were preferred to the flat model by $>3\sigma$. Only the fit to e- provided evidence of a trend with temperature (at a 2.28$\sigma$ level). In the other cases, the null hypothesis (i.e. constant abundance with temperature) was preferred.}
    \label{tab:ap_opt_bhm}
\end{table}

\begin{table}[]
    \centering
    \begin{tabular}{ccccc}\hline \hline
   Molecule & m & c & $\sigma$ & ln(E) \\\hline 
\multirow{2}{*}{TiO} & 0.000 $^{+ 0.002 }_{- 0.002 }$ & -7.6 $^{+ 3.88 }_{- 4.01 }$ & -1.13 $^{+ 1.03 }_{- 1.04 }$ & -74.27 \\
& - & -7.16 $^{+ 1.86 }_{- 1.88 }$ & -1.02 $^{+ 0.97 }_{- 1.03 }$ & -71.01 \\\hline
\multirow{2}{*}{VO} & -0.001 $^{+ 0.002 }_{- 0.002 }$ & -6.57 $^{+ 3.17 }_{- 3.56 }$ & -1.13 $^{+ 1.04 }_{- 1.04 }$ & -74.1 \\
& - & -7.41 $^{+ 1.33 }_{- 1.56 }$ & -1.08 $^{+ 1.0 }_{- 1.01 }$ & -70.13 \\\hline
\multirow{2}{*}{FeH} & 0.002 $^{+ 0.002 }_{- 0.002 }$ & -9.72 $^{+ 3.64 }_{- 3.34 }$ & 0.01 $^{+ 0.55 }_{- 0.75 }$ & -76.95 \\
& - & -7.29 $^{+ 1.68 }_{- 1.79 }$ & 0.14 $^{+ 0.4 }_{- 0.26 }$ & -73.63 \\\hline
\multirow{2}{*}{e-} & 0.004 $^{+ 0.001 }_{- 0.001 }$ & -12.16 $^{+ 2.18 }_{- 2.05 }$ & -1.12 $^{+ 1.05 }_{- 1.07 }$ & -73.99 \\
& - & -7.07 $^{+ 1.06 }_{- 1.08 }$ & -0.17 $^{+ 0.64 }_{- 0.91 }$ & -73.48 \\ \hline \hline
    \end{tabular}
    \caption{Hyperparameters for the BHM fits for each molecule when using all retrievals where optical absorbers were preferred to the model without them. In each cases, the null hypothesis (i.e. constant abundance with temperature) was preferred.}
    \label{tab:ap_opt_pref_bhm}
\end{table}

\begin{table}[]
    \centering
    \begin{tabular}{cccc}\hline \hline
    m & c & $\sigma$ & ln(E) \\\hline
    -0.57 $^{+ 0.62 }_{- 0.6 }$ & 1.09 $^{+ 0.45 }_{- 0.47 }$ & -0.15 $^{+ 0.14 }_{- 0.13 }$ & -153.93 \\
- & 0.76 $^{+ 0.34 }$ $_{- 0.32 }$ & -0.15 $^{+ 0.14 }_{- 0.09 }$ & -152.46 \\
-0.07 $^{+ 0.17 }_{- 0.18 }$ & 0.74 $^{+ 0.36 }_{- 0.34 }$ & -0.18 $^{+ 0.14 }_{- 0.08 }$ & -204.16 \\
- & 0.66 $^{+ 0.34 }_{- 0.33 }$ & -0.2 $^{+ 0.15 }_{- 0.07 }$ & -200.41 \\
 \hline \hline
    \end{tabular}
    \caption{Hyperparameters for the BHM fits for a mass-metallicity trend when the stellar metallicity was not accounted for. In each case, the null hypothesis (i.e. constant abundance with temperature) yielded a preferable fit to the data.}
    \label{tab:mass_pl_met_ap}
\end{table}

\begin{figure*}
    \centering
    \includegraphics[width=0.9\columnwidth]{EqChemMet_Thorngren.png}
    \includegraphics[width=0.9\columnwidth]{EqChemMet_Thorngren_All.png}
    \includegraphics[width=0.9\columnwidth]{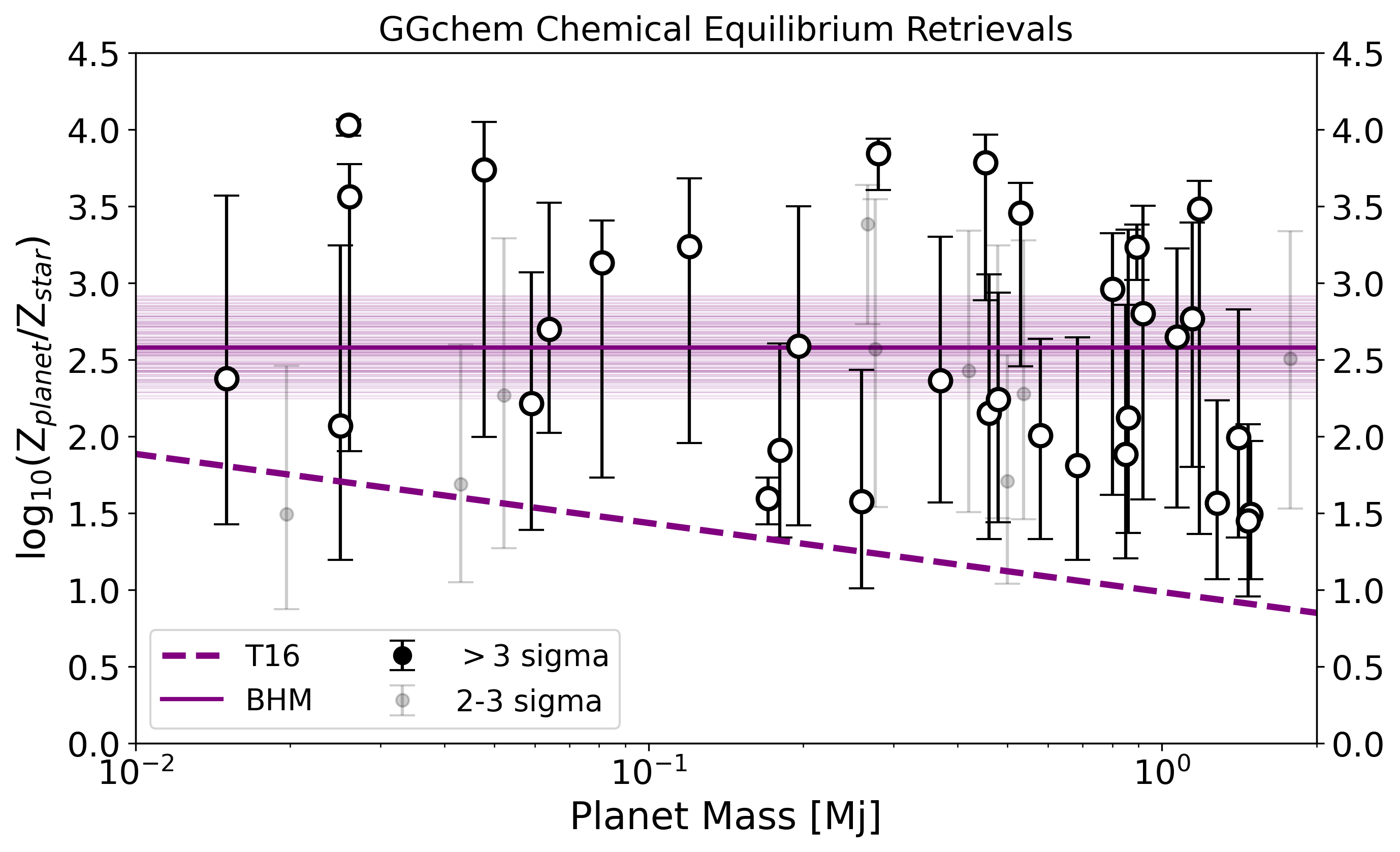}
    \includegraphics[width=0.9\columnwidth]{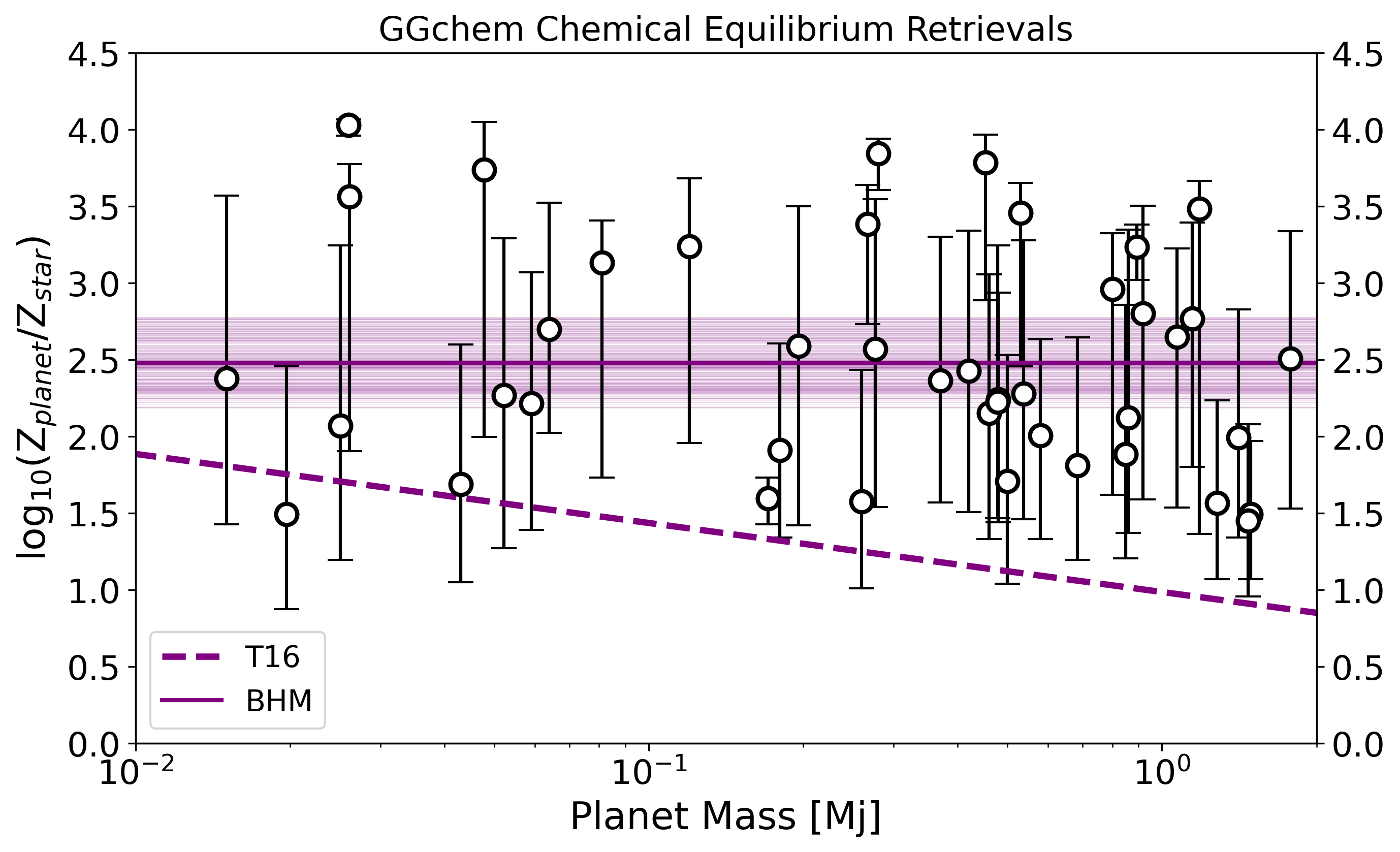}
    \caption{Retrieved metallicity from our GGChem runs. \textbf{Left:} BHM applied to retrievals which yielded a fit which was preferred to $>$3 sigma compared to the flat model. \textbf{Right:} BHM applied to retrievals which yielded a fit which was preferred to $>$2 sigma compared to the flat model. The fits for a linear trend fit (top) and flat trend (i.e. constant with mass, bottom) are shown with the thick line indicating the best-fit model and the thinner lines representing the traces from the fit that fell within 1 sigma of the best-fit model. In both cases, the flat model is preferred, implying we cannot conclude there is a trend between planet mass and the metallicity from these data.}
    \label{fig:all_met_ap}
\end{figure*}

For the mass-metallicity fits, in the main text we explored the impact of using only $>3\sigma$ atmospheric detections or using those at 2-3$\sigma$ too. We provide here the full set of models, both for the linear trend and the null hypothesis and these are given in Figure \ref{fig:all_met_ap}. 

\begin{figure*}
    \centering
    \includegraphics[width=0.9\columnwidth]{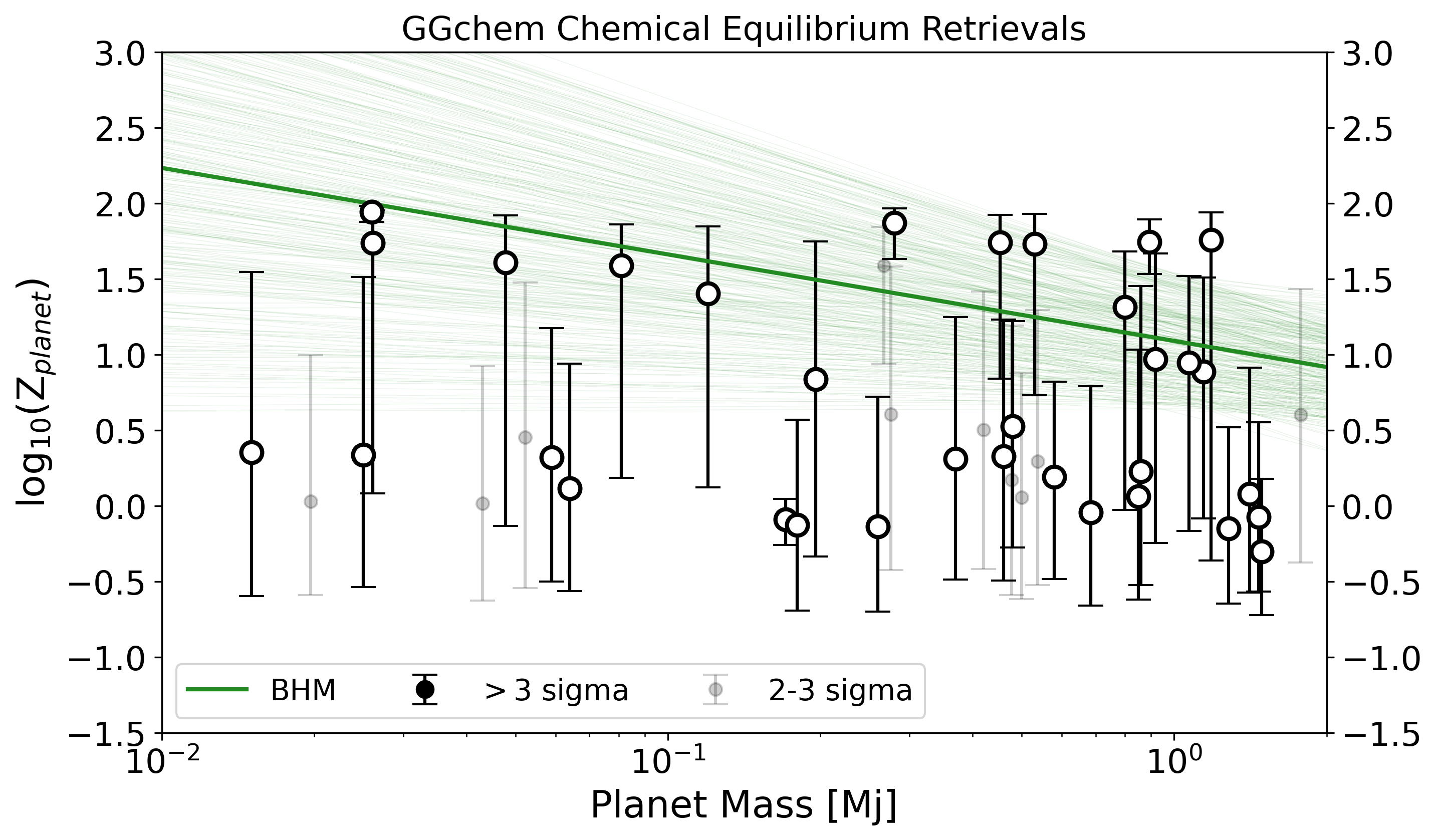}
    \includegraphics[width=0.9\columnwidth]{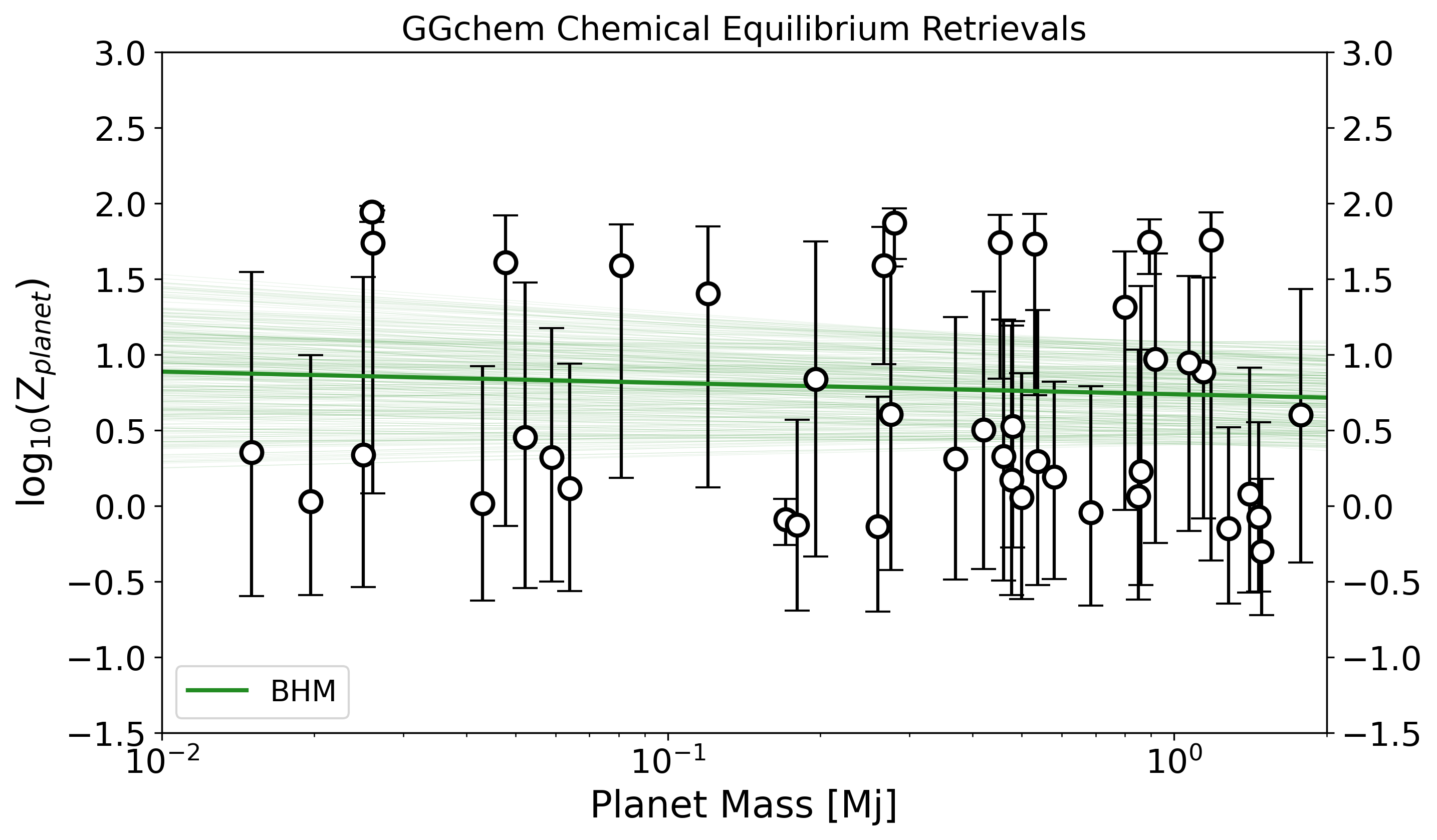}
    \includegraphics[width=0.9\columnwidth]{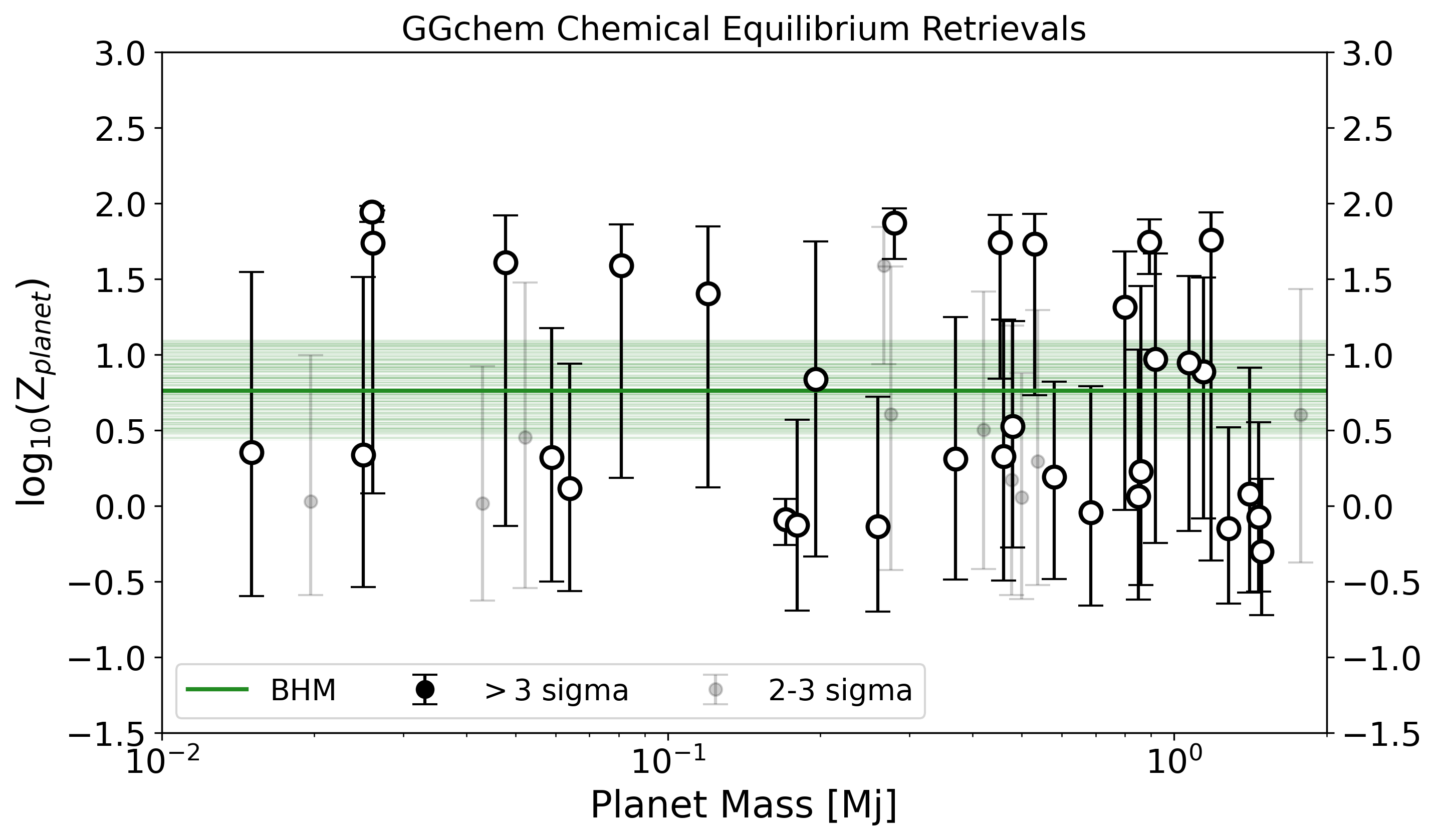}
    \includegraphics[width=0.9\columnwidth]{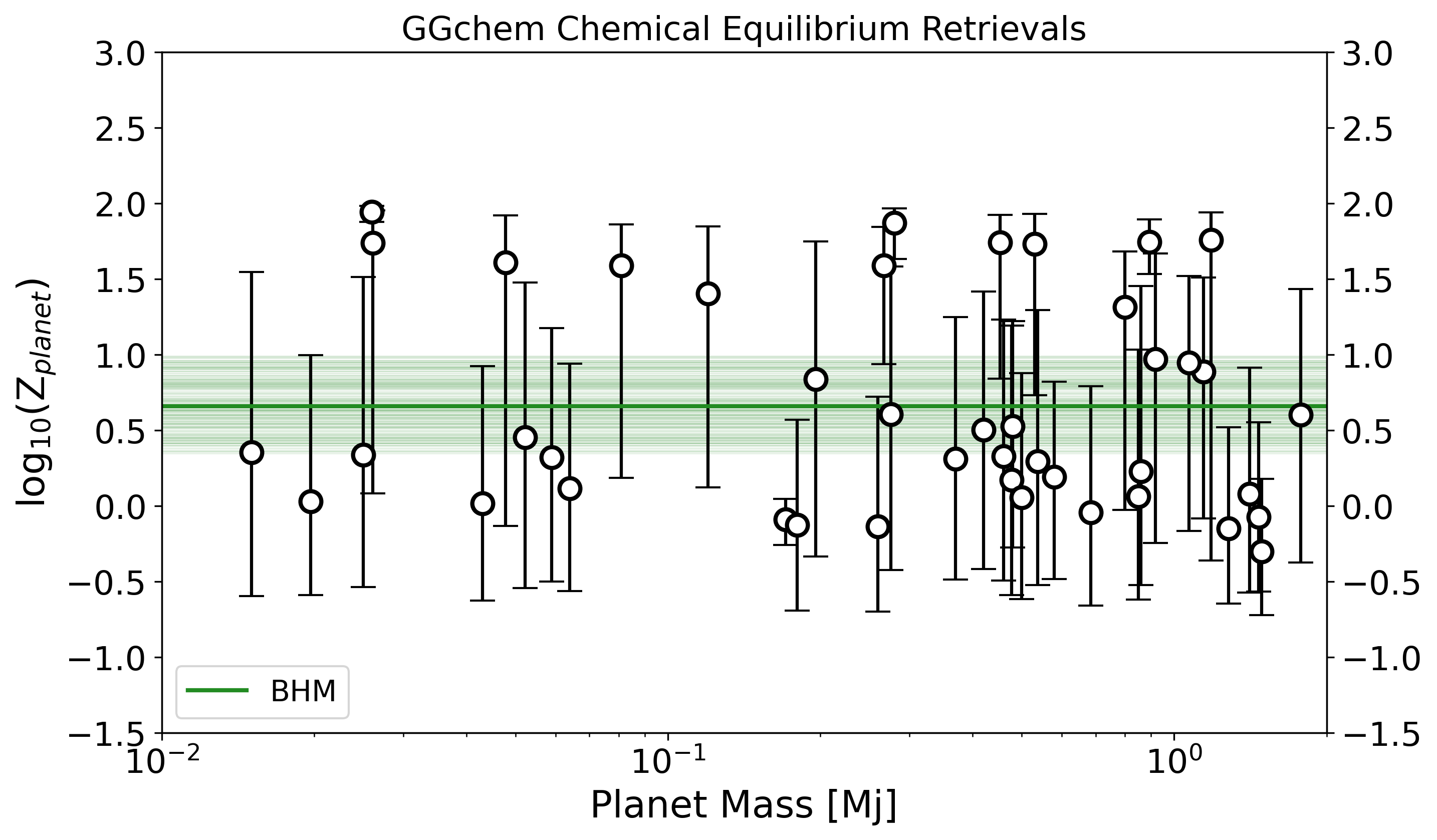}
    \caption{Same as Figure \ref{fig:all_met_ap} but without accounting for the stellar metallicity. Again there is no evidence for a mass-metallicity trend in these data as the constant metallicity model provides a preferable fit to these data.}
    \label{fig:mass_pl_met_ap}
\end{figure*}

With the metallicity for the GGChem retrievals, we used the host star's metallicity [Fe/H] to infer the relative planet-star metallicity and fit our trend to this data. Here we also explore fitting to just the recovered planet metallicity. When only retrievals which led to strong ($>3\sigma$) atmospheric detections are used, we again find the linear model has a negative slope (implying a decreasing metallicity with increasing mass, es expected). However, the null hypothesis again has a higher Bayesian evidence. Furthermore, no trend, or even a positive slope, is within the 1$\sigma$ bounds of the best fitting model. When the retrievals which provided a fit which was preferred to 2-3$\sigma$ over the flat model are utilised, the BHM shows even less evidence for a mass-metallicity trend. The results are shown in Figure \ref{fig:mass_pl_met_ap} and the hyper-parameters, as well as the log-evidence, are given in Table \ref{tab:mass_pl_met_ap}.

\begin{figure*}
    \centering
    \includegraphics[width=0.95\columnwidth]{FreeChemMet.png}
    \includegraphics[width=0.95\columnwidth]{FreeChemMet2sig.png}
    \includegraphics[width=0.95\columnwidth]{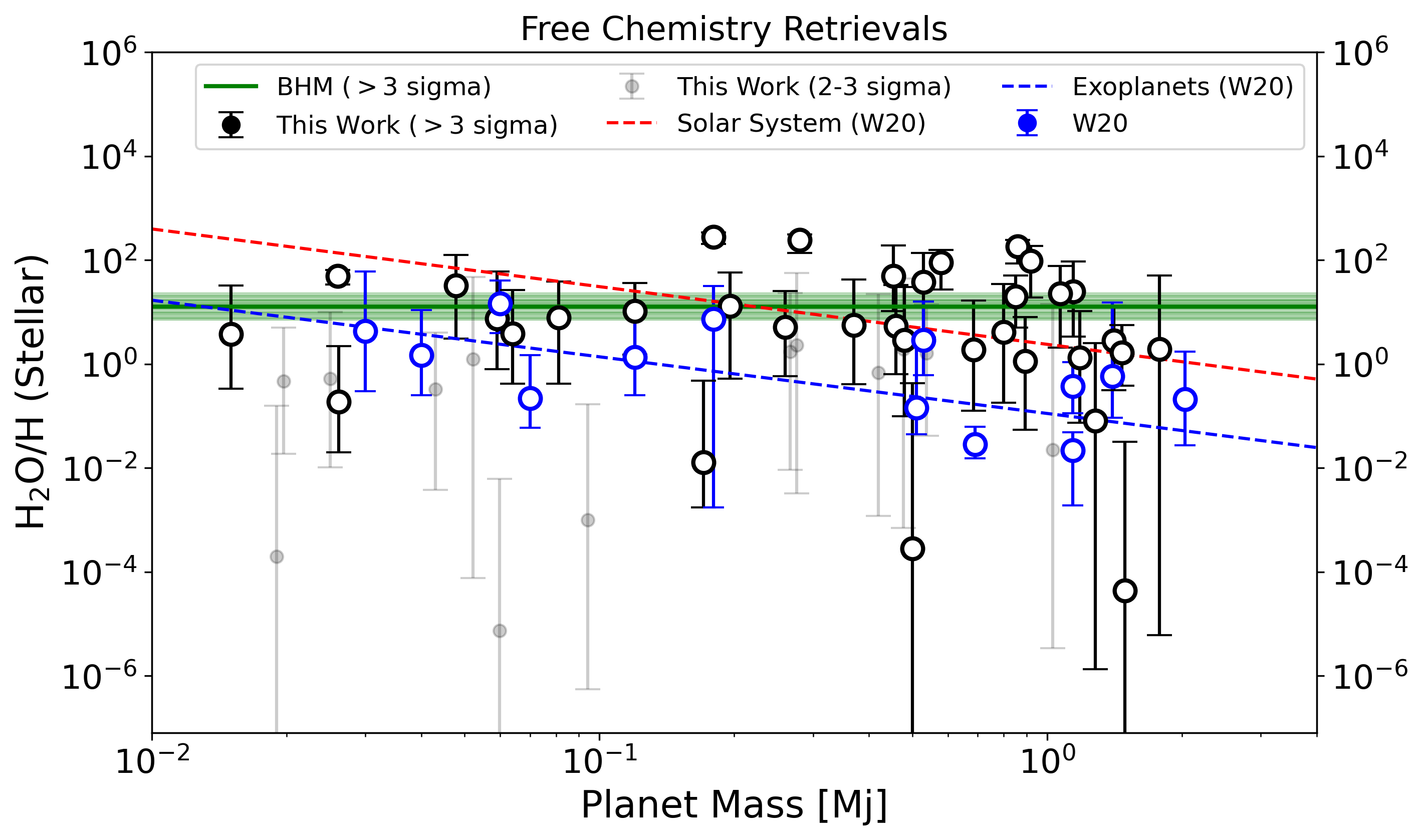}
    \includegraphics[width=0.95\columnwidth]{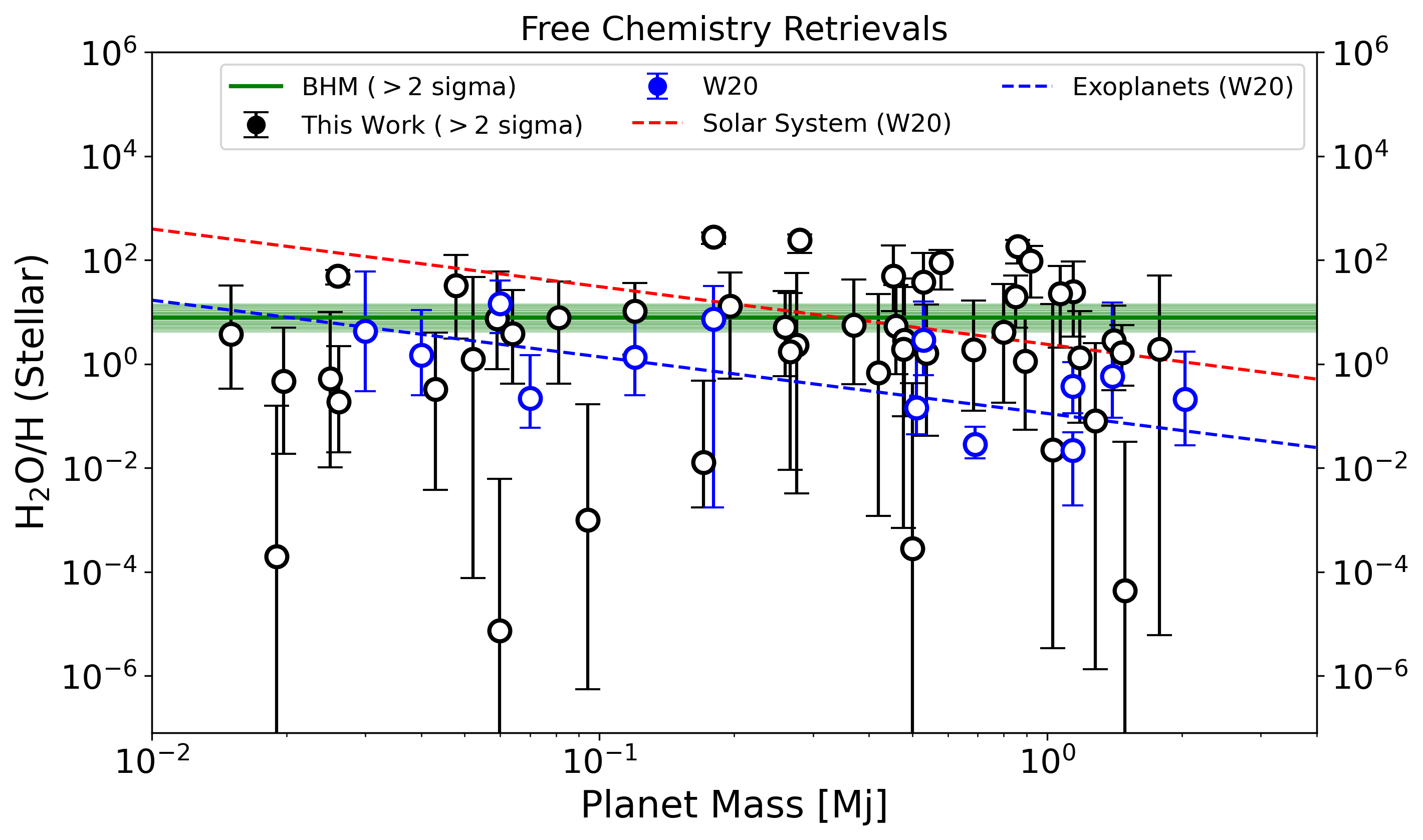}
    \caption{Fits from our BHM to the water ratio which was determined using the methods of \citet{welbanks_pop}. \textbf{Left:} BHM applied to retrievals which yielded a fit which was preferred to $>$3 sigma compared to the flat model. \textbf{Right:} BHM applied to retrievals which yielded a fit which was preferred to $>$2 sigma compared to the flat model. The fits for a linear trend fit (top) and flat trend (i.e. constant with mass, bottom) are shown with he thick line indicating the best-fit model and the thinner lines representing the traces from the fit that fell within 1 sigma of the best-fit model. In both cases, the flat model is preferred, implying we cannot conclude there is a trend between planet mass and the water-to-hydrogen ratio from these data.} 
    \label{fig:all_free_met_ap}
\end{figure*}

For completeness we also show the fits for the water-to-hydrogen case. Again, there is no evidence for a mass-metallicity trend within the data. The plots are given in Figure \ref{fig:all_free_met_ap} while Table \ref{tab:h2o_r_bhm} contains the log-evidence for each fit. We provide the retrieved metallicities and water abundances in Tables \ref{tab:chem_eq} and \ref{tab:free_chem}.

\begin{table*}[]
    \centering
    \caption{Results of our chemical equilibrium retrievals. The C/O ratio and metallicity were free parameters in these retrievals, with the planet to star metallicity ratio (log$_{10}$(Z$_{\rm P}$/Z$_{\rm S}$)) calculated from the retrieved metallicity value and the star's Fe/H as described in the text. The sigma column refers to the significance of the model with respect to the flat model. Where the preference for the atmospheric model was below 2$\sigma$, we do not report the significance and assume that, from our retrieval, there is no meaningful evidence for atmospheric modulation. For completeness, we report the C/O and metallicity in all cases.}
    \begin{tabular}{cccccccccc} \hline \hline
Planet Name & C/O & log$_{10}$(Z$_{\rm P}$) & log$_{10}$(Z$_{\rm P}$/Z$_{\rm S}$) & Sigma & Planet Name & C/O & log$_{10}$(Z$_{\rm P}$) & log$_{10}$(Z$_{\rm P}$/Z$_{\rm S}$) & Sigma\\  \hline
CoRoT-1 b & 1.09$^{+0.6}_{-0.65}$ & 0.56$^{+0.82}_{-0.96}$ &2.71$^{+0.82}_{-0.96}$ & 1.99 & LTT 9779 b & 1.39$^{+0.41}_{-0.65}$ & 0.83$^{+0.97}_{-1.03}$ &2.42$^{+0.97}_{-1.03}$ & - \\ 
 GJ 436 b & 1.04$^{+0.59}_{-0.52}$ & 0.32$^{+1.03}_{-0.81}$ &2.16$^{+1.03}_{-0.81}$ & - & TOI-270 c & 0.57$^{+0.52}_{-0.3}$ & 0.19$^{+1.21}_{-0.78}$ &2.22$^{+1.21}_{-0.78}$ & - \\
 GJ 1214 b & 0.53$^{+0.5}_{-0.28}$ & 0.03$^{+0.97}_{-0.62}$ &1.49$^{+0.97}_{-0.62}$ & 2.38 &TOI-270 d & 0.51$^{+0.39}_{-0.28}$ & 0.35$^{+1.19}_{-0.95}$ &2.38$^{+1.19}_{-0.95}$ & 3.67 \\ 
 GJ 3470 b & 0.57$^{+0.36}_{-0.3}$ & 0.02$^{+0.91}_{-0.64}$ &1.69$^{+0.91}_{-0.64}$ & 2.27 &  TOI-674 b & 0.48$^{+0.25}_{-0.52}$ & 0.12$^{+0.83}_{-0.68}$ &2.7$^{+0.83}_{-0.68}$ & 4.94 \\ 
 HAT-P-1 b & 0.64$^{+0.2}_{-0.31}$ & 1.73$^{+0.2}_{-1.0}$ &3.46$^{+0.2}_{-1.0}$ & 4.53 & TrES-2 b & 0.97$^{+0.6}_{-0.31}$ & 0.46$^{+0.91}_{-0.88}$ &2.46$^{+0.91}_{-0.88}$ & - \\ 
 HAT-P-2 b & 1.09$^{+0.56}_{-0.64}$ & 0.39$^{+1.0}_{-0.86}$ &2.1$^{+1.0}_{-0.86}$ & - & TrES-4 b & 1.18$^{+0.5}_{-0.24}$ & 0.49$^{+0.92}_{-0.94}$ &2.2$^{+0.92}_{-0.94}$ & - \\ 
 HAT-P-3 b & 1.04$^{+0.59}_{-0.54}$ & 0.51$^{+0.92}_{-0.82}$ &2.1$^{+0.92}_{-0.82}$ & - &  V1298 Tau b & 0.23$^{+0.17}_{-0.41}$ & -0.01$^{+0.36}_{-0.39}$ &1.7$^{+0.36}_{-0.39}$ & 18.48 \\ 
 HAT-P-7 b & 0.93$^{+0.64}_{-0.53}$ & 0.56$^{+0.96}_{-0.95}$ &2.15$^{+0.96}_{-0.95}$ & - &  V1298 Tau c & 0.23$^{+0.17}_{-0.45}$ & -0.01$^{+0.36}_{-0.39}$ &1.7$^{+0.36}_{-0.39}$ & 14.56 \\ 
 HAT-P-11 b & 0.47$^{+0.34}_{-0.24}$ & 1.59$^{+0.27}_{-1.4}$ &3.13$^{+0.27}_{-1.4}$ & 4.77 & WASP-6 b & 0.42$^{+0.28}_{-0.0}$ & 0.31$^{+0.94}_{-0.8}$ &2.37$^{+0.94}_{-0.8}$ & 4.99 \\
 HAT-P-12 b & 0.72$^{+0.79}_{-0.41}$ & 0.39$^{+1.07}_{-0.85}$ &2.53$^{+1.07}_{-0.85}$ & - & WASP-12 b & 0.46$^{+0.29}_{-0.24}$ & -0.07$^{+0.63}_{-0.49}$ &1.45$^{+0.63}_{-0.49}$ & 6.02 \\ 
 HAT-P-17 b & 0.82$^{+0.76}_{-0.45}$ & 0.41$^{+0.95}_{-0.84}$ &2.26$^{+0.95}_{-0.84}$ & - & WASP-17 b & 0.32$^{+0.27}_{-0.53}$ & 1.74$^{+0.18}_{-0.9}$ &3.79$^{+0.18}_{-0.9}$ & 8.77 \\ 
 HAT-P-18 b & 1.05$^{+0.65}_{-0.68}$ & 0.84$^{+0.91}_{-1.17}$ &2.59$^{+0.91}_{-1.17}$ & 3.2 & WASP-18 b & 1.12$^{+0.52}_{-0.37}$ & 0.41$^{+0.94}_{-0.9}$ &2.16$^{+0.94}_{-0.9}$ & - \\ 
 HAT-P-26 b & 0.32$^{+0.27}_{-0.16}$ & 0.32$^{+0.85}_{-0.82}$ &2.22$^{+0.85}_{-0.82}$ & 9.29 & WASP-19 b & 0.47$^{+0.26}_{-0.28}$ & 0.95$^{+0.57}_{-1.11}$ &2.65$^{+0.57}_{-1.11}$ & 4.65 \\ 
 HAT-P-32 b & 0.47$^{+0.29}_{-0.24}$ & 0.23$^{+1.23}_{-0.75}$ &2.12$^{+1.23}_{-0.75}$ & 7.31 & WASP-29 b & 1.02$^{+0.61}_{-0.23}$ & 0.27$^{+1.03}_{-0.84}$ &2.02$^{+1.03}_{-0.84}$ & - \\ 
 HAT-P-38 b & 0.51$^{+0.27}_{-0.26}$ & 1.59$^{+0.25}_{-0.65}$ &3.38$^{+0.25}_{-0.65}$ & 2.87 & WASP-31 b & 0.79$^{+0.71}_{-0.47}$ & 0.17$^{+1.02}_{-0.76}$ &2.23$^{+1.02}_{-0.76}$ & 2.4 \\ 
 HAT-P-41 b & 0.46$^{+0.25}_{-0.23}$ & 1.32$^{+0.37}_{-1.34}$ &2.96$^{+0.37}_{-1.34}$ & 4.89 & WASP-39 b & 0.32$^{+0.29}_{-0.86}$ & 1.87$^{+0.1}_{-0.24}$ &3.84$^{+0.1}_{-0.24}$ & 9.24 \\ 
 HD 3167 c & 1.36$^{+0.45}_{-0.86}$ & 1.74$^{+0.21}_{-1.66}$ &3.56$^{+0.21}_{-1.66}$ & 3.34 & WASP-43 b & 0.63$^{+0.79}_{-0.07}$ & 0.6$^{+0.83}_{-0.98}$ &2.51$^{+0.83}_{-0.98}$ & 2.6 \\ 
 HD 97658 b & 0.86$^{+0.04}_{-0.07}$ & 1.95$^{+0.04}_{-0.07}$ &4.03$^{+0.04}_{-0.07}$ & 5.3 & WASP-52 b & 0.41$^{+0.46}_{-0.48}$ & 0.33$^{+0.91}_{-0.82}$ &2.15$^{+0.91}_{-0.82}$ & 6.8 \\ 
 HD 106315 c & 0.51$^{+0.26}_{-0.26}$ & 1.61$^{+0.31}_{-1.74}$ &3.74$^{+0.31}_{-1.74}$ & 5.58 &  WASP-62 b & 0.43$^{+0.25}_{-0.43}$ & 0.19$^{+0.63}_{-0.67}$ &2.01$^{+0.63}_{-0.67}$ & 5.56 \\
 HD 149026 b & 0.69$^{+0.62}_{-0.37}$ & 0.58$^{+1.04}_{-1.01}$ &2.08$^{+1.04}_{-1.01}$ & - &  WASP-63 b & 0.7$^{+0.73}_{-0.65}$ & 0.49$^{+0.85}_{-0.88}$ &2.26$^{+0.85}_{-0.88}$ & - \\ 
 HD 189733 b & 0.5$^{+0.49}_{-0.28}$ & 0.89$^{+0.62}_{-0.97}$ &2.77$^{+0.62}_{-0.97}$ & 4.95 & WASP-67 b & 0.83$^{+0.73}_{-0.53}$ & 0.5$^{+0.91}_{-0.92}$ &2.43$^{+0.91}_{-0.92}$ & 2.33 \\
 HD 209458 b & 0.42$^{+0.29}_{-0.23}$ & -0.04$^{+0.83}_{-0.62}$ &1.81$^{+0.83}_{-0.62}$ & 6.78 & WASP-69 b & 0.42$^{+0.77}_{-0.19}$ & -0.13$^{+0.86}_{-0.57}$ &1.58$^{+0.86}_{-0.57}$ & 4.98 \\ 
 HD 219666 b & 0.89$^{+0.69}_{-0.47}$ & 0.46$^{+1.02}_{-1.0}$ &2.27$^{+1.02}_{-1.0}$ & 2.66 & WASP-74 b & 0.64$^{+0.74}_{-0.31}$ & 0.69$^{+0.84}_{-0.96}$ &2.15$^{+0.84}_{-0.96}$ & 1.93 \\ 
 HIP 41378 b & 0.87$^{+0.74}_{-0.48}$ & 0.68$^{+1.0}_{-1.06}$ &2.65$^{+1.0}_{-1.06}$ & - & WASP-76 b & 0.5$^{+0.26}_{-0.62}$ & 1.75$^{+0.15}_{-0.21}$ &3.24$^{+0.15}_{-0.21}$ & 8.11 \\
 HIP 41378 f & 0.11$^{+0.01}_{-0.0}$ & -0.49$^{+0.48}_{-0.32}$ &1.48$^{+0.48}_{-0.32}$ & 6.3 & WASP-79 b & 0.37$^{+0.24}_{-0.57}$ & 0.06$^{+0.97}_{-0.68}$ &1.89$^{+0.97}_{-0.68}$ & 5.49 \\  
 K2-18 b & 0.77$^{+0.75}_{-0.43}$ & 0.34$^{+1.18}_{-0.87}$ &2.07$^{+1.18}_{-0.87}$ & 3.05 & WASP-80 b & 0.71$^{+0.75}_{-0.63}$ & 0.29$^{+1.0}_{-0.82}$ &2.28$^{+1.0}_{-0.82}$ & 2.96 \\  
 K2-24 b & 1.14$^{+0.52}_{-0.65}$ & 0.34$^{+1.01}_{-0.86}$ &1.85$^{+1.01}_{-0.86}$ & - & WASP-96 b & 0.78$^{+0.77}_{-0.56}$ & 0.53$^{+0.7}_{-0.8}$ &2.24$^{+0.7}_{-0.8}$ & 3.84 \\ 
 KELT-1 b & 0.99$^{+0.6}_{-0.53}$ & 0.53$^{+0.91}_{-0.96}$ &2.38$^{+0.91}_{-0.96}$ & - & WASP-101 b & 1.01$^{+0.59}_{-0.68}$ & 0.06$^{+0.82}_{-0.67}$ &1.71$^{+0.82}_{-0.67}$ & 2.27 \\  
 KELT-7 b & 0.6$^{+0.53}_{-0.31}$ & -0.15$^{+0.67}_{-0.49}$ &1.57$^{+0.67}_{-0.49}$ & 3.46 & WASP-103 b & 1.03$^{+0.52}_{-0.64}$ & -0.3$^{+0.48}_{-0.42}$ &1.49$^{+0.48}_{-0.42}$ & 6.54 \\
 KELT-11 b & 0.58$^{+0.17}_{-0.19}$ & -0.09$^{+0.14}_{-0.17}$ &1.6$^{+0.14}_{-0.17}$ & 10.4 & WASP-107 b & 0.25$^{+0.18}_{-0.16}$ & 1.4$^{+0.45}_{-1.28}$ &3.24$^{+0.45}_{-1.28}$ & 9.95 \\ 
 Kepler-9 b & 0.96$^{+0.68}_{-0.55}$ & 0.64$^{+0.89}_{-1.09}$ &2.45$^{+0.89}_{-1.09}$ & 1.89 &  WASP-117 b & 0.9$^{+0.69}_{-0.54}$ & 0.61$^{+0.98}_{-1.03}$ &2.57$^{+0.98}_{-1.03}$ & 2.8 \\  
 Kepler-9 c & 1.04$^{+0.59}_{-0.56}$ & 0.25$^{+1.07}_{-0.78}$ &2.06$^{+1.07}_{-0.78}$ & - &  WASP-121 b & 0.46$^{+0.25}_{-0.26}$ & 1.76$^{+0.18}_{-2.12}$ &3.48$^{+0.18}_{-2.12}$ & 5.68 \\  
 Kepler-51 b & 1.13$^{+0.56}_{-0.62}$ & 0.58$^{+1.04}_{-1.09}$ &2.38$^{+1.04}_{-1.09}$ & - & WASP-127 b & 0.29$^{+0.21}_{-0.23}$ & -0.12$^{+0.7}_{-0.57}$ &1.91$^{+0.7}_{-0.57}$ & 18.17 \\   
 Kepler-51 d & 0.9$^{+0.63}_{-0.57}$ & 0.5$^{+1.09}_{-0.98}$ &2.3$^{+1.09}_{-0.98}$ & - & WASP-178 b & 0.57$^{+0.56}_{-0.26}$ & 0.08$^{+0.83}_{-0.65}$ &2.0$^{+0.83}_{-0.65}$ & 3.41 \\ 
 Kepler-79 d & 1.09$^{+0.58}_{-0.63}$ & 0.56$^{+1.03}_{-1.01}$ &2.49$^{+1.03}_{-1.01}$ & - & XO-1 b & 0.76$^{+0.19}_{-0.55}$ & 0.97$^{+0.7}_{-1.21}$ &2.8$^{+0.7}_{-1.21}$ & 3.28 \\ 
\hline \hline
    \end{tabular}
    \label{tab:chem_eq}
\end{table*}

\begin{table*}[]
    \centering
    \caption{Retrieved volume mixing ratios of water from our free chemistry retrievals. The H$_2$O/H ratio was subsequently derived as described in the text. The sigma column refers to the significance of the preferred model with respect to the flat model. Where the preference for the atmospheric model was below 2$\sigma$, we do not report the significance and assume that, from our retrieval, there is no meaningful evidence for atmospheric modulation. For completeness, we report the H$_2$O abundance in all cases.}
    \begin{tabular}{cccccccc}\hline \hline
    Planet Name & log$_{10}$(H$_2$O) & log$_{10}$(H$_2$O/H) & Sigma & Planet Name & log$_{10}$(H$_2$O) & log$_{10}$(H$_2$O/H) & Sigma\\ \hline
    CoRoT-1 b & -5.33$^{+2.81}_{-3.81}$ & -1.65$^{+2.81}_{-3.81}$ & 2.36 & LTT 9779 b & -8.23$^{+2.37}_{-2.33}$ & -5.12$^{+2.37}_{-2.33}$ & - \\ 
GJ 1214 b & -2.98$^{+1.03}_{-1.39}$ & -0.33$^{+1.03}_{-1.39}$ & 2.41 & TOI-270 c & -3.73$^{+1.56}_{-3.78}$ & -0.52$^{+1.56}_{-3.78}$ & - \\ 
GJ 3470 b & -3.34$^{+1.09}_{-1.93}$ & -0.48$^{+1.09}_{-1.93}$ & 2.16 & TOI-270 d & -2.63$^{+0.93}_{-1.06}$ & 0.58$^{+0.94}_{-1.06}$ & 4.0  \\ 
GJ 436 b & -6.89$^{+2.83}_{-2.86}$ & -3.87$^{+2.83}_{-2.86}$ & - & TOI-674 b & -3.12$^{+0.78}_{-1.04}$ & 0.65$^{+0.78}_{-1.04}$ & 4.96 \\ 
HAT-P-1 b & -1.67$^{+0.53}_{-1.07}$ & 1.58$^{+0.55}_{-1.07}$ & 4.4 & TrES-2 b & -6.47$^{+3.24}_{-3.07}$ & -2.94$^{+3.24}_{-3.07}$ & - \\ 
HAT-P-2 b & -7.23$^{+2.8}_{-2.83}$ & -4.0$^{+2.8}_{-2.83}$ & 2.87 & TrES-4 b & -6.85$^{+3.14}_{-2.96}$ & -3.62$^{+3.14}_{-2.96}$ & - \\ 
HAT-P-3 b & -6.72$^{+3.08}_{-3.13}$ & -3.61$^{+3.08}_{-3.13}$ & - & V1298 Tau b & -2.51$^{+0.41}_{-0.54}$ & 0.39$^{+0.41}_{-0.54}$ & 19.4 \\ 
HAT-P-7 b & -6.48$^{+3.4}_{-3.2}$ & -3.36$^{+3.41}_{-3.2}$ & - & V1298 Tau c & -1.56$^{+0.32}_{-0.52}$ & 1.64$^{+0.34}_{-0.52}$ & 14.53 \\ 
HAT-P-11 b & -1.94$^{+0.68}_{-1.26}$ & 0.89$^{+0.7}_{-1.27}$ & 4.62 & WASP-6 b & -2.83$^{+0.87}_{-1.14}$ & 0.75$^{+0.88}_{-1.14}$ & 4.88 \\ 
HAT-P-12 b & -3.72$^{+1.61}_{-3.33}$ & -0.15$^{+1.62}_{-3.33}$ & - & WASP-12 b & -2.89$^{+0.53}_{-0.63}$ & 0.22$^{+0.53}_{-0.63}$ & 8.3 \\ 
HAT-P-17 b & -6.0$^{+3.41}_{-3.44}$ & -2.96$^{+3.41}_{-3.44}$ & - & WASP-17 b & -1.87$^{+0.57}_{-0.67}$ & 1.7$^{+0.59}_{-0.67}$ & 9.15 \\ 
HAT-P-18 b & -1.89$^{+0.63}_{-1.39}$ & 1.11$^{+0.65}_{-1.4}$ & 3.57 & WASP-18 b & -6.75$^{+3.46}_{-3.27}$ & -3.46$^{+3.46}_{-3.27}$ & - \\ 
HAT-P-26 b & -2.5$^{+0.89}_{-0.97}$ & 0.88$^{+0.9}_{-0.97}$ & 9.35 & WASP-19 b & -1.88$^{+0.52}_{-1.03}$ & 1.36$^{+0.53}_{-1.04}$ & 4.56 \\
HAT-P-32 b & -1.18$^{+0.12}_{-0.32}$ & 2.27$^{+0.12}_{-0.33}$ & 7.35 & WASP-29 b & -6.51$^{+2.95}_{-3.03}$ & -3.29$^{+2.95}_{-3.03}$ & - \\ 
HAT-P-38 b & -3.07$^{+1.12}_{-2.27}$ & 0.24$^{+1.13}_{-2.27}$ & 2.25 & WASP-31 b & -3.29$^{+1.36}_{-3.44}$ & 0.28$^{+1.36}_{-3.44}$ & 2.01 \\
HAT-P-41 b & -2.55$^{+0.92}_{-1.36}$ & 0.62$^{+0.93}_{-1.36}$ & 4.72 & WASP-39 b & -1.13$^{+0.09}_{-0.24}$ & 2.39$^{+0.1}_{-0.25}$ & 9.47 \\
HD 3167 c & -3.73$^{+1.07}_{-0.97}$ & -0.73$^{+1.07}_{-0.97}$ & 4.61 & WASP-43 b & -3.14$^{+1.41}_{-5.5}$ & 0.29$^{+1.42}_{-5.5}$ & 3.63 \\
HD 97658 b & -3.49$^{+0.35}_{-0.53}$ & -0.23$^{+0.35}_{-0.53}$ & 13.25 & WASP-52 b & -2.61$^{+0.78}_{-0.93}$ & 0.73$^{+0.79}_{-0.93}$ & 7.1 \\ 
HD 106315 c & -1.92$^{+0.58}_{-1.01}$ & 1.51$^{+0.59}_{-1.02}$ & 5.61 & WASP-62 b & -1.4$^{+0.23}_{-0.51}$ & 1.95$^{+0.24}_{-0.52}$ & 5.81 \\ 
HD 149026 b & -4.54$^{+2.45}_{-4.12}$ & -1.53$^{+2.45}_{-4.12}$ & - & WASP-63 b & -3.49$^{+1.48}_{-4.05}$ & -0.2$^{+1.48}_{-4.05}$ & - \\ 
HD 189733 b & -2.01$^{+0.56}_{-0.86}$ & 1.4$^{+0.57}_{-0.87}$ & 4.98 & WASP-67 b & -3.6$^{+1.51}_{-2.75}$ & -0.17$^{+1.51}_{-2.75}$ & 2.21 \\  
HD 209458 b & -3.1$^{+0.95}_{-1.17}$ & 0.28$^{+0.95}_{-1.17}$ & 6.76 & WASP-69 b & -2.48$^{+0.69}_{-0.94}$ & 0.71$^{+0.7}_{-0.94}$ & 5.32 \\ 
HD 219666 b & -3.23$^{+1.57}_{-4.21}$ & 0.1$^{+1.58}_{-4.21}$ & 2.33 & WASP-74 b & -3.88$^{+2.13}_{-4.79}$ & -0.9$^{+2.14}_{-4.79}$ & 1.98 \\  
HIP 41378 b & -4.03$^{+2.34}_{-4.79}$ & -0.65$^{+2.35}_{-4.79}$ & - & WASP-76 b & -2.96$^{+0.84}_{-1.33}$ & 0.06$^{+0.84}_{-1.33}$ & 7.49 \\ 
HIP 41378 f & -8.8$^{+1.14}_{-1.77}$ & -5.65$^{+1.14}_{-1.77}$ & 16.77 & WASP-79 b & -2.05$^{+0.4}_{-0.6}$ & 1.3$^{+0.41}_{-0.6}$ & 6.44 \\ 
K2-18 b & -3.19$^{+1.28}_{-1.71}$ & -0.28$^{+1.28}_{-1.71}$ & 2.98 & WASP-80 b & -2.97$^{+0.93}_{-1.59}$ & 0.21$^{+0.93}_{-1.59}$ & 2.69 \\  
K2-24 b & -7.83$^{+2.92}_{-2.64}$ & -5.13$^{+2.92}_{-2.64}$ & 2.46 & WASP-96 b & -2.77$^{+1.01}_{-1.47}$ & 0.47$^{+1.02}_{-1.47}$ & 3.72 \\  
KELT-1 b & -6.63$^{+3.23}_{-3.08}$ & -3.23$^{+3.23}_{-3.08}$ & - & WASP-101 b & -6.72$^{+3.18}_{-3.07}$ & -3.54$^{+3.18}_{-3.07}$ & 3.56 \\ 
KELT-7 b & -4.33$^{+1.5}_{-4.78}$ & -1.09$^{+1.5}_{-4.78}$ & 5.97 & WASP-103 b & -7.71$^{+2.86}_{-2.59}$ & -4.35$^{+2.86}_{-2.59}$ & 6.78 \\  
KELT-11 b & -5.09$^{+1.57}_{-0.87}$ & -1.89$^{+1.57}_{-0.87}$ & 10.87 & WASP-107 b & -2.0$^{+0.53}_{-0.83}$ & 1.02$^{+0.54}_{-0.84}$ & 10.33 \\ 
Kepler-9 b & -4.71$^{+2.51}_{-4.19}$ & -1.73$^{+2.51}_{-4.19}$ & - & WASP-117 b & -3.1$^{+1.38}_{-2.85}$ & 0.37$^{+1.38}_{-2.85}$ & 2.6 \\  
Kepler-9 c & -5.99$^{+2.22}_{-3.25}$ & -3.0$^{+2.22}_{-3.25}$ & 2.78 & WASP-121 b & -3.14$^{+0.9}_{-1.25}$ & 0.12$^{+0.9}_{-1.25}$ & 5.6 \\ 
Kepler-51 b & -6.1$^{+3.13}_{-3.32}$ & -3.11$^{+3.13}_{-3.32}$ & - & WASP-127 b & -1.14$^{+0.08}_{-0.12}$ & 2.44$^{+0.09}_{-0.13}$ & 18.44 \\ 
Kepler-51 d & -6.92$^{+2.9}_{-3.04}$ & -3.93$^{+2.9}_{-3.04}$ & - & WASP-178 b & -3.01$^{+0.67}_{-0.95}$ & 0.45$^{+0.67}_{-0.96}$ & 4.85 \\ 
Kepler-79 d & -6.82$^{+2.9}_{-3.04}$ & -3.71$^{+2.9}_{-3.04}$ & 2.03 & XO-1 b & -1.39$^{+0.27}_{-0.69}$ & 1.98$^{+0.29}_{-0.7}$ & 3.2 \\ 
\hline \hline
    \end{tabular}
    \label{tab:free_chem}
\end{table*}

\clearpage

\section*{Appendix 4: Fitting The WFC3 G141 Spectral Feature Size}
\label{app:feature_size}

The key spectral feature within the HST WFC3 G141 band is the 1.4 $\mu$m water feature and, instead of preforming retrievals in an attempt to recover the abundance of this molecule, several studies have instead measured the size of the feature in relation to other bands within WFC3's spectral range \citep{Fu_2017,wakeford_clouds}. While the information gained in the approach is limited, we nonetheless explored its use to compare our results to other studies which have previously used this metric. We achieved this by first modelling a cloud-free atmosphere where water in the only molecular opacity. We also include scattering due to Rayleigh and Collision Induced Absorption (CIA). The model is then scaled using:

\begin{equation}
    \label{eq:feature_size}
    y(\lambda) = ax(\lambda) + bx + c,
\end{equation}
where $x(\lambda)$ is the original spectrum model, $a$ is a scaling factor, $b$ is the wavelength coefficient for the baseline slope, and $c$ is a constant offset \citep{Fu_2017}. The scaling factor, $a$, moderates the size of the water feature, acting as proxy for clouds while $b$ can be used to account for slopes within the spectrum (e.g. due to H- opacity). We fit the model using Nestle\footnote{\url{https://github.com/kbarbary/nestle}} \citep{skilling_2004,mukherjee_2006,shaw_2007} and we follow the methodology of \citet{stevenson_wfc3}, computing the difference between the minimum in the J band (1.22-1.30) and the maximum in the H$_2$O band (1.36-1.44) from the preferred model spectrum after removing the slope.

By dividing the feature size by the expected transit depth modulation from one scale height of atmosphere, one can determine the magnitude of atmospheric absorption in an way which easily facilitates comparisons across a broad parameter space. The change in the transit depth due to one scale height of atmosphere was determined by: 
\begin{equation}
    \Delta D = \frac{2HR_p}{R^2_s}
\end{equation}
where $R_p$ is the planet's radius, $R_s$ is the radius of the host star and $H$ is the atmospheric scale height, calculated from:

\begin{equation}
    H = \frac{k T}{\mu g}
\end{equation}
where $k$ is the Boltzmann constant, $T$ is the planet's equilibrium temperature, $\mu$ is the atmospheric mean molecular weight (set to 2.3), and $g$ is the planet's surface gravity. 

\begin{figure}
    \centering
    \includegraphics[width=0.95\columnwidth]{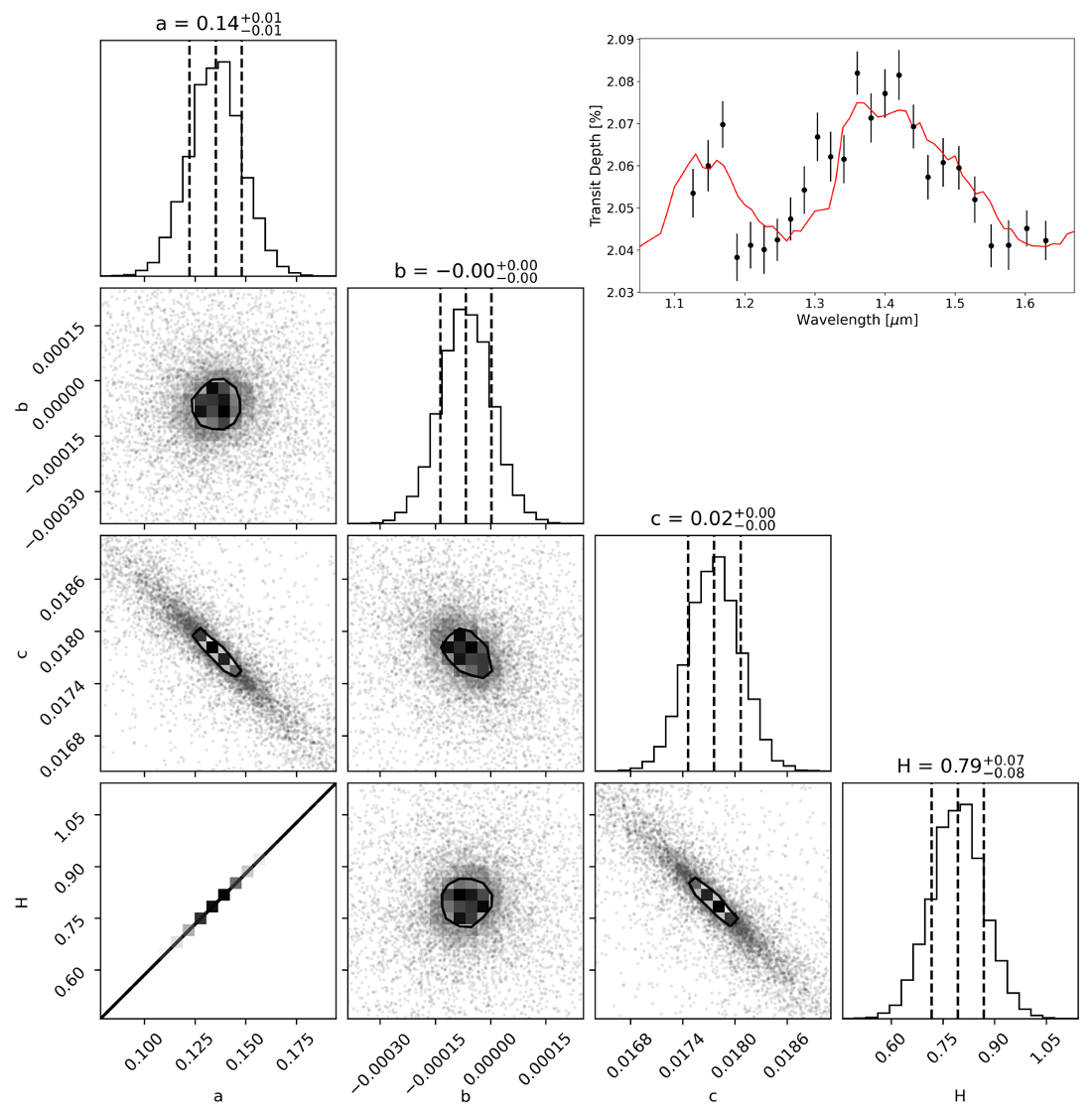}
    \caption{Posteriors and best-fit spectrum for amplitude fitting of WASP-107\,b.}
    \label{fig:amp_w107}
\end{figure}

An example fit for WASP-107\,b is given in Figure \ref{fig:amp_w107}. We provide the feature size for all planets studied here in Table \ref{tab:feature_size}. The spectral feature size has often been used to infer the presence of clouds and proposed as a way of guiding observers as to the spectral modulation that could be expected when planning future observations. Across the population, we recover an average feature size of 0.92 scale heights, a comparable value to previous studies: 1.4 H \citep{Fu_2017} and 0.89 H \citep{wakeford_clouds}. The amplitude of this feature is far below what would be expect from a clear, solar metallicity atmosphere. However, we note that the magnitude of the atmospheric absorption is only valid across the HST WFC3 G141 range and that other instruments, particularly those that probe further into the infrared, will see larger feature sizes. 

Extending the models used to derive the 1.4 $\mu$m feature size in the HST WFC3 G141 data, we estimate the amplitude of features seen in observations with future instruments by studying the minimum and maximum transit depth across their spectral coverage. For JWST NIRISS GR700XD \citep[0.6-2.8 $\mu$m, ][]{doyon_niriss} and JWST NIRSpec G395H \citep[2.8-5.1 $\mu$m, ][]{birkmann_nirspec}) we predict average feature sizes of 2.52 and 1.65 scale heights respectively. Combining both instruments, observing with the JWST NIRSpec PRISM (0.6-5.3 $\mu$m), or with Twinkle \citep[0.5-4.5 $\mu$m, ][]{edwards_exo}), gives an amplitude of 2.57 scale heights. Finally, the expected amplitude across the spectral coverage of Ariel \citep[0.5-7.8 $\mu$m, ][]{tinetti_ariel}) is 3.13 scale heights.

Therefore, while clouds and hazes obviously need to be accounted for during the planning of observations with future facilities, current data shows the expected amplitude should, on average, be greater than a single scale height. However, we note also that the methodology used to measure the amplitude of absorption features is somewhat flawed. While the parameters in Equation \ref{eq:feature_size} allow the spectrum to be modulated to fit the data, and account for the features seen, the final fit does not provide a robust analysis of the nature of the atmosphere. The presence, and effect, of clouds can be better understood by fitting physical models to the data via atmospheric retrievals.

\begin{table}[]
    \centering
    \caption{1.4$\mu$m feature size, in scale heights, of the planets studied here. In this table we report all fitted values but, in the figures, we only plot those where the uncertainty of the feature size is less than 2 scale heights.}
    \begin{tabular}{cccc} \hline \hline
     Planet Name & Feature Size & Planet Name & Feature Size\\ \hline
CoRoT-1 b & 5.3 $\pm$ 2.08 & LTT 9779 b & 0.32 $\pm$ 0.68 \\ 
 GJ 1214 b & 0.12 $\pm$ 0.04 & TOI-270 c & 0.27 $\pm$ 0.366 \\ 
 GJ 3470 b & 0.68 $\pm$ 0.24 & TOI-270 d & 2.66 $\pm$ 0.  \\ 
 GJ 436 b & 0.02 $\pm$ 0.36 & TOI-674 b & 1.38 $\pm$ 0.26 \\ 
 HAT-P-1 b & 1.89 $\pm$ 0.37 & TrES-2 b & 1.5 $\pm$ 2.47 \\ 
 HAT-P-2 b & 6.04 $\pm$ 3.74 & TrES-4 b & -5.75 $\pm$ 3.61 \\ 
 HAT-P-3 b & 0.61 $\pm$ 0.62 & V1298Tau b & 10.66 $\pm$ 0.57 \\ 
 HAT-P-7 b & 3.31 $\pm$ 4.36 & V1298Tau c & 14.24 $\pm$ 0.71 \\ 
 HAT-P-11 b & 2.0 $\pm$ 0.39 & WASP-6 b & 1.78 $\pm$ 0.36 \\ 
 HAT-P-12 b & 0.46 $\pm$ 0.21 & WASP-12 b & 2.42 $\pm$ 0.34 \\ 
 HAT-P-17 b & 0.97 $\pm$ 0.61 & WASP-17 b & 1.88 $\pm$ 0.2 \\ 
 HAT-P-18 b & 0.81 $\pm$ 0.19 & WASP-18 b & -1.31 $\pm$ 3.01 \\ 
 HAT-P-26 b & 2.52 $\pm$ 0.27 & WASP-19 b & 2.29 $\pm$ 0.46 \\ 
 HAT-P-32 b & 1.89 $\pm$ 0.25 & WASP-29 b & -0.15 $\pm$ 0.36 \\ 
 HAT-P-38 b & 1.56 $\pm$ 0.61 & WASP-31 b & 1.07 $\pm$ 0.38 \\ 
 HAT-P-41 b & 2.06 $\pm$ 0.45 & WASP-39 b & 1.58 $\pm$ 0.15 \\
 HD 3167 c & 1.22 $\pm$ 0.31 &  WASP-43 b & 1.45 $\pm$ 0.43 \\
 HD 97658 b & 1.98 $\pm$ 0.64 & WASP-52 b & 1.6 $\pm$ 0.24 \\ 
 HD 106315 c & 2.42 $\pm$ 0.4 & WASP-62 b & 1.37 $\pm$ 0.26 \\ 
 HD 149026 b & 1.08 $\pm$ 0.54 & WASP-63 b & 0.64 $\pm$ 0.28 \\ 
 HD 189733 b & 2.09 $\pm$ 0.38 & WASP-67 b & 1.16 $\pm$ 0.51 \\ 
 HD 209458 b & 0.98 $\pm$ 0.15 & WASP-69 b & 0.61 $\pm$ 0.12 \\ 
 HD 219666 b & 2.05 $\pm$ 0.62 & WASP-74 b & 0.96 $\pm$ 0.44 \\ 
 HIP 41378 b & 2.25 $\pm$ 0.97 & WASP-76 b & 1.39 $\pm$ 0.19 \\ 
 HIP 41378 f & -2.41 $\pm$ 0.68 & WASP-79 b & 2.41 $\pm$ 0.38 \\ 
 K2-18 b & 2.33 $\pm$ 0.67 & WASP-80 b & 0.44 $\pm$ 0.21 \\ 
 K2-24 b & -1.04 $\pm$ 0.68 & WASP-96 b & 2.36 $\pm$ 0.66 \\ 
 KELT-1 b & -1.19 $\pm$ 4.63 & WASP-101 b & 0.11 $\pm$ 0.29 \\ 
 KELT-7 b & 1.35 $\pm$ 0.46 & WASP-103 b & 1.97 $\pm$ 0.64 \\  
 KELT-11 b & 1.12 $\pm$ 0.14 & WASP-107 b & 0.79 $\pm$ 0.08 \\ 
 Kepler-9 b & 3.26 $\pm$ 1.57 & WASP-117 b & 0.99 $\pm$ 0.31 \\
 Kepler-9 c & -0.9 $\pm$ 1.86 & WASP-121 b & 1.05 $\pm$ 0.18 \\ 
 Kepler-51 b & 0.14 $\pm$ 0.48 & WASP-127 b & 2.47 $\pm$ 0.15 \\ 
 Kepler-51 d & 0.08 $\pm$ 0.47 & WASP-178 b & 1.6 $\pm$ 0.44 \\  
 Kepler-79 d & -2.46 $\pm$ 1.58 & XO-1 b & 2.72 $\pm$ 0.69 \\ \hline \hline
    \end{tabular}

    \label{tab:feature_size}
\end{table}

\clearpage

\bibliography{main}{}
\bibliographystyle{aasjournal}

\end{document}